\documentclass[preprint,review,12pt]{elsarticle}
\usepackage[margin=1in]{geometry}

\usepackage{amssymb}
\usepackage{amsmath,amsthm,amsfonts,amscd, stmaryrd}  
\usepackage{eucal} 	 		   
\usepackage{verbatim}      
\usepackage{makeidx}       
\usepackage{psfig}         		
\usepackage{epsfig}            
\usepackage{citesort}         %
\usepackage{url}		          
\usepackage{subfigure}
\usepackage{amssymb}
\usepackage{url}
\usepackage{float}
\usepackage{enumerate}
\usepackage{color}
\usepackage{verbatim}
\usepackage{tabulary}

\usepackage{tcolorbox}
\usepackage[ruled,vlined]{algorithm2e}
\usepackage{mathtools}
\usepackage{multicol}
\usepackage{cancel}
\usepackage{graphicx}
\usepackage{multirow}
\usepackage{xcolor}
\usepackage{tabularx}
\usepackage{stackengine}
\usepackage{caption}

\newcommand\xrowht[2][0]{\addstackgap[.5\dimexpr#2\relax]{\vphantom{#1}}}

\newcolumntype{L}[1]{>{\raggedright\arraybackslash}p{#1}}
\newcolumntype{C}[1]{>{\centering\arraybackslash}p{#1}}
\newcolumntype{R}[1]{>{\raggedleft\arraybackslash}p{#1}}

\newcommand{\one}[1]{\textcolor{black}{#1}}
\newcommand{\two}[1]{\textcolor{black}{#1}}
\newcommand{\three}[1]{\textcolor{black}{#1}}
\newcommand{\more}[1]{\textcolor{black}{#1}}

\newcommand{\oneR}[1]{\textcolor{black}{#1}}
\newcommand{\twoR}[1]{\textcolor{black}{#1}}

\newcommand{\moreR}[1]{\textcolor{black}{#1}}

\usepackage{graphicx}
\usepackage{subfigure}
\usepackage[colorlinks]{hyperref}
\AtBeginDocument{%
  \hypersetup{%
    linkcolor=black,%
    citecolor=black,%
  }%
}%

\journal{}

\begin{document}

\begin{frontmatter}


\title{Cross-mode Stabilized Stochastic Shallow Water Systems Using Stochastic Finite Element Methods}

\author[inst1]{Chen Chen\corref{cor1}}
\ead{chen636489@utexas.edu}

 

\author[inst1]{Clint Dawson}

\author[inst1,inst2,inst3]{Eirik Valseth}
\cortext[cor1]{Corresponding author}

\affiliation[inst1]{organization={The Oden Institute for Computational Engineering and Sciences, The University of Texas at Austin},
            addressline={201 E. 24th St. Stop C0200}, 
            city={Austin},
            postcode={78712}, 
            state={Texas},
            country={United States of America}}

\affiliation[inst2]{organization={The Department of Mathematics, The University of Oslo},
            addressline={Moltke Moes Vei 35}, 
            city={Oslo},
            postcode={0851}, 
            country={Norway}}
\affiliation[inst3]{organization={Simula Research Laboratory},
            addressline={Kristian Augusts gate 23}, 
            city={Oslo},
            postcode={0164}, 
            country={Norway}}



\begin{abstract}
The development of surrogate models to study uncertainties in hydrologic systems requires significant effort in the development of sampling strategies and forward model simulations. Furthermore, in applications where prediction time is critical, such as prediction of hurricane storm surge, the  predictions of system response and uncertainties can be required within short time frames. Here, we  develop an efficient stochastic shallow water model to address these issues. To discretize the physical and probability spaces we use a Stochastic Galerkin  method and an Incremental Pressure Correction  Scheme to advance the solution in time. To overcome discrete stability issues, we propose cross-mode stabilization methods which employs existing stabilization methods in the probability space by adding stabilization terms to every stochastic mode in a modes-coupled way. We extensively verify the developed method for both idealized shallow water test cases and hindcasting of past hurricanes. We subsequently use the developed and verified method to perform a comprehensive statistical analysis of the established shallow water surrogate models. Finally, we propose a predictor for hurricane storm surge under uncertain wind drag coefficients and demonstrate its effectivity for Hurricanes Ike and Harvey. 
\end{abstract}


\end{frontmatter}


\clearpage

\section{Introduction} \label{sec:intro}

Over the past few decades, uncertainties of computational models have been recognized and studied by researchers from a wide range of fields, e.g., environmental engineering~\cite{najm2009uncertainty,calder1986stochastic,rodriguez1987some}, geosciences~\cite{huh2001stochastic,ghiocel2002stochastic,guo2018reliability}, and in coastal engineering~\cite{rajabalinejad2010stochastic,sochala2020polynomial}. A series of sampling-based non-intrusive methods~\cite{metropolis1949monte,minasny2006conditioned,chamoin2008non} have been developed to quantify the uncertainty of certain computational models. The advantage of such methods is that the deterministic model can remain as-is and can be considered as a black box. However,  to formulate a surrogate over any quantity of interest of a model, these methods usually require  sample outputs which are obtained from a significant number of deterministic model computations. Thus, collecting sample outputs requires significant computational resources and \moreR{time}.
In  time sensitive forecast models, such as storm surge models, it is  important to develop fast uncertainty quantification. In this work, we propose a novel model called the Stochastic Shallow Water Model (SSWM) to  forecast and quantify the associated uncertainties of storm surge. 

\moreR{To develop} this model, we apply the Spectral Stochastic Finite Element Method (SSFEM) \moreR{with the aim to} achieve real-time uncertainty forecasts for two dimensional shallow water equations (SWE). The
introduction of uncertainties into the SWE \one{may lead} to a numerical stability issue \one{due to} the \one{coupled modes introduced by the } SSFEM. 
To overcome this stability issue, we propose and implement a series of stabilization methods. We subsequently verify and validate the resulting SSWM surrogates by using well known numerical test cases for verification and two historic hurricane events in the Gulf of Mexico for validation.

Statistical analysis of uncertain model outputs via non-intrusive methods has been the subject of significant research~\cite{wan2006multi,ng2012multifidelity,eldred2009recent,hadigol2018least,hu2019parametric} and allows users to obtain consistent mean and variance information. \moreR{However,} higher order statistical moments, reliable probability density functions (PDFs), as well as the support of the output random process are more difficult to ascertain.
\one{The difficulty of such analyses} is exaggerated further  when the information is needed in a short time frame in real-time forecasting systems. The support of the output random variable (i.e., the range of random variable in which its value falls) is of great importance to  reliability analysis in coastal engineering~\cite{xiu2010numerical}.
\one{The use of various intrusive stochastic  methods to study uncertainty in shallow water flows has also been the subject of several studies. In~\cite{despres_robust_2013}, the authors study the propagation in  one-dimensional hyperbolic partial differential equations (PDEs) and show that the system may lose its hyperbolic nature under certain initial data.   An operator splitting into linear and nonlinear portions are introduced for the one-dimensional St.~Venant equations in the preprint~\cite{chertock_well-balanced_2015}. The splitting technique is developed to overcome difficulties arising from loss of hyperbolicity and numerical instability in stochastic Galerkin (SG) method with generalized polynomial chaos (gPC) expansions. Significant literature also exist on gPC and PC based SG methods, including~\cite{gerster_hyperbolic_2019,herty_entropies_2020,dai_hyperbolicity-preserving_2021,dai_hyperbolicity-preserving_2022,schlachter_hyperbolicity-preserving_2018}.  In these works, hyperbolic systems including one- and two-dimensional SWEs are \oneR{considered}. The loss of hyperbolicity is addressed in various fashions through careful mathematical analysis of and development of conditions to ensure the hyperbolicity of the stochastic numerical systems, e.g., in~\cite{schlachter_hyperbolicity-preserving_2018} a slope limiter is developed to ensure hyperbolicity. } \two{In~\cite{shaw2020probabilistic}, Shaw \emph{et al.} introduce an intrusive Haar wavelet finite volume scheme for SWE based probabilistic hydrodynamic modeling including critical physics such as wetting and drying.    }

\more{Our current work distinguishes itself by considering a non conservative SSWM in which the instabilities arising in the numerical scheme are resolved by stabilization across stochastic modes. These stabilization techniques are based on existing Petrov-Galerkin type stabilization, see, e.g.,~\cite{brooks1982streamline}.  }
From the developed SSWM and its surrogates, higher order moments, PDFs, and random variable support are readily available for statistical analysis.  
%

In the following, we introduce and comprehensively verify and validate a new SSWM. First, in Section~\ref{sec:SSWM} we introduce the SSWM with a particular focus on the novel stabilization methodology developed. In Section~\ref{sec:verify_validate}, we  verify and validate the SSWM. Next, in Section~\ref{sec:visual_analysis}, we perform statistical analyses of the SSWM surrogate responses. In Section~\ref{sec:hurricane_prediction}, we apply the SSWM to predict hurricane storm surge under uncertain wind drag parameters. Finally, in Section~\ref{sec:conclusion}, we draw conclusions and discuss future research directions.


\section{The Stochastic Shallow Water Model} \label{sec:SSWM}

\subsection{Mathematical Formulation}

\begin{figure}[H]
	\begin{center}
		\includegraphics[width=0.75\columnwidth]{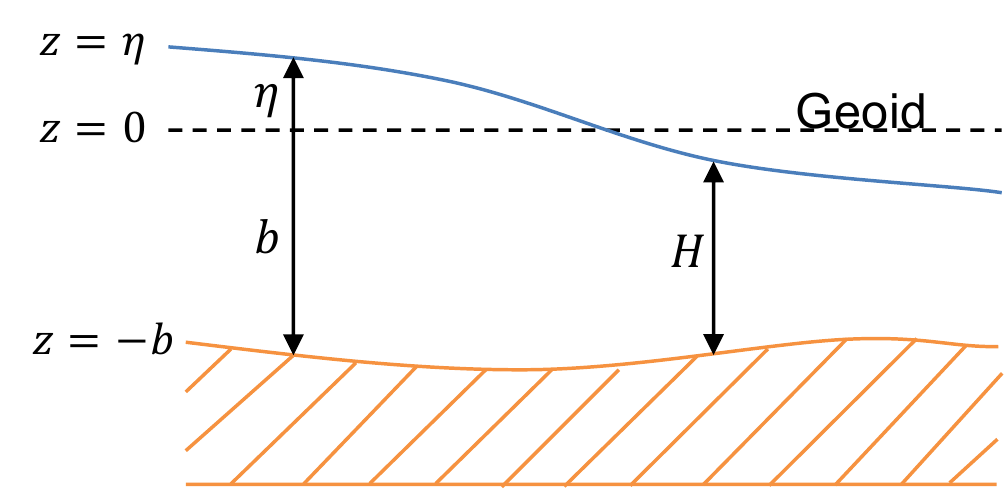}	
	\end{center}
	\caption{A illustration of the Geoid and standard shallow water quantities.}
	\label{fig:geoid}
\end{figure}

Our SSWM is based on two-dimensional deterministic SWE under standard assumptions of incompressible isotropic flow with constant density and kinematic viscosity. We also assume a hydrostatic pressure distribution and a long-wave condition so vertical fluid motion is negligible. The two-dimensional deterministic SWE~\cite{farrell2013automated,funke2014tidal} \moreR{are}:
%
%
\begin{equation} \label{eqn:swenonconserve}
\begin{aligned}
\frac{\partial \eta}{\partial t}  &+ \nabla \cdot (  H \boldsymbol{u}  )  =0 \text{ in } \Omega, \\
\frac{\partial \boldsymbol{u}}{\partial t}  + \boldsymbol{u} \cdot \nabla \boldsymbol{u} &=  - g \nabla \eta  + \nabla \cdot (\nu \nabla \boldsymbol{u}) + \boldsymbol{f} \text{ in } \Omega,
\end{aligned}
\end{equation}
where \one{$\Omega$ is \moreR{a} two-dimensional domain,} $\eta(\textbf{x}, t)$ the elevation $(\textnormal{unit}: m)$ of the free surface \three{positive upwards from the geiod}, $b(\textbf{x})$  the bathymetry $(\textnormal{unit}: m)$, \three{positive downwards from the geoid}, $H(\textbf{x}, t)$  the total depth $(\textnormal{unit}: m)$ of the water column  $H = \eta + b$ (see Figure~\ref{fig:geoid} for visual representation of these quantities),  $\boldsymbol{u} = (u, v)$  the velocity field $(\textnormal{unit}: m/s)$ averaged in the vertical direction, $g$  the constant of gravitational acceleration $(\textnormal{unit}: m/s^2)$, $\nu$  the kinematic viscosity, and $\boldsymbol{f}=(f_x, f_y)$ represents the source.
\noindent We assume the driving forces of the water column motion to be atmospheric pressure, wind, and bottom friction. Hence, the source $\boldsymbol{f}$ takes the form:
\begin{equation}
\boldsymbol{f} = - g \nabla  \left( \frac{p_{atm}}{g \rho_{w}}   \right) + \frac{	C_d }{H}\frac{\rho_{a}}{\rho_{w}} \Vert  \boldsymbol{w} \Vert \boldsymbol{w} - \frac{C_b}{H} \Vert  \boldsymbol{u} \Vert \boldsymbol{u},
\end{equation}
where $C_d$ is the wind drag coefficient, $C_b$  the bottom friction coefficient, $\rho_a$  the density of air, $\rho_w$ the density of water, $\boldsymbol{w} = (w_x, w_y)$  the wind speed (typically measured at a height of 10 meters above the surface), the atmospheric pressure $p_{atm}$, and  $\Vert  \cdot \Vert$ denotes the vector magnitude.
To ensure solvability of the SWE in the SSWM, we also need propoer initial and boundary conditions. We consider homogeneous initial conditions for both velocity and elevation in all cases unless explicitly noted. We also consider the following types of boundary conditions: free-slip, specified elevation, and no-normal flow. To identify these conditions, we separate the boundary $\Gamma$ of the domain $\Omega$ into two disjoint parts: $\Gamma = \Gamma_{cl} \cup \Gamma_{op}$ \more{ and we denote by $\boldsymbol{n}$ the outward unit normal vector to the boundaries.}. A 
free-slip boundary condition is applied to the entire boundary:
\begin{equation}
	\nabla \boldsymbol{u} \cdot \boldsymbol{n} = \boldsymbol{0} \text{ on }  \Gamma.
\end{equation}
The no-normal flow (also referred to as impenetration) boundary condition is applied to the closed, or land, portion of the boundary:
\begin{equation}
\boldsymbol{u} \cdot \boldsymbol{n} = \boldsymbol{0} \text{ on }  \Gamma_{cl}. 
\end{equation}
Finally, an elevation boundary condition is applied to the open, or ocean, part of the boundary:
\begin{equation}
	\eta = r \text{ on }  \Gamma_{op}. 
\end{equation}

\subsection{Sources of Uncertainty} \label{sec:uncertain_Sources}

In shallow water systems,  uncertainty can be induced  from several sources, e.g., initial conditions, boundary conditions, bathymetry, bottom friction coefficients, and wind drag coefficients. \moreR{These} lead to uncertainty in the surface elevation and velocity field in the SWE~\eqref{eqn:swenonconserve}. 
For the general setting of the SG method, we introduce polynomial chaos representations of the uncertainties. In particular, we use the gPC expansion, see e.g.,~\cite{xiu2010numerical}, \moreR{which is} \three{   commonly used in stochastic finite element \moreR{(FE)}  analysis  and compared to PC expansions~\cite{ghanem1991stochastic}, offers more flexibility in choice of polynomials thereby easing implementational aspects.  However, the use of gPCs comes at the additional complexity of reformulating the mathematical problem and the use of a stochastic solver. A commonly used alternative to gPC expansions and stochastic solvers are Monte-Carlo methods. These non-intrusive methods require a significant number of potentially expensive model runs to provide acceptable solutions. In the context of real-time of near real-time forecasting, the Monte-Carlo methods are \moreR{often considered}  too costly, and the gPC expansions, while intrusive, require only a single forward model solve.  }
If we consider the wind drag coefficient $C_d(\boldsymbol{x},t;\boldsymbol{\xi})$ and bottom drag coefficient $C_b(\boldsymbol{x},t;\boldsymbol{\xi})$, these are represented by the following expansions:
\begin{equation} \label{eqn:cdcb}
\begin{aligned}
C_d(\boldsymbol{x},t;\three{\boldsymbol{\xi}})=\sum_{i=0}^{M-1} C_i^d(\boldsymbol{x},t)\Phi_i(\boldsymbol{\xi}),\\
C_b(\boldsymbol{x},t;\three{\boldsymbol{\xi}})=\sum_{i=0}^{M-1} C_i^b(\boldsymbol{x},t)\Phi_i(\boldsymbol{\xi}),\\
\end{aligned}
\end{equation}
where \one{$\boldsymbol{\xi}$ represents a vector of random variables}, the $C_i$'s the expansion coefficients of the $i$'th modes, and $\Phi_i,~ i=0,1,...,M-1$  the orthonormal gPC basis. The number of gPC basis functions $M$ depends on the \one{highest degree $N$ of the gPC polynomials and the dimension $d$ of the gPC basis functions in the probability space}:
\[
M = {N+d \choose N}.
\]
The relationship between $\boldsymbol{\xi}$ and $\Phi_i$ is established by the Wiener-Askey scheme from~\cite{xiu2010numerical}. Using gPC expansions, we also represent the remaining uncertain sources analogous to~\eqref{eqn:cdcb}.
The orthogonality property of the gPCs allows us to compute the mode coefficients by taking the inner product with $\Phi_k(\boldsymbol{\xi})$, e.g.,
\begin{equation}
C_k^d(\boldsymbol{x},t) = \int C_d(\boldsymbol{x},t;\three{\boldsymbol{\xi}}) \Phi_k(\boldsymbol{\xi}) \text{d}\boldsymbol{\xi}  ~\text{for } k=0,\dots, M-1.
\end{equation}

\subsection{Stochastic Formulation} \label{sec:stochastic_formulation}

To derive the stochastic formulation of our SSWM, we  use the same set of random variables as for the uncertainties in Section~\ref{sec:uncertain_Sources} to represent the uncertain outputs $\eta$ and $\boldsymbol{u}$:
\begin{equation}\label{eqn:gPC}
\begin{aligned}
\boldsymbol{u}(\boldsymbol{x},t;\boldsymbol{\xi})=\sum_{i=0}^{M-1}\boldsymbol{u_i}(\boldsymbol{x},t)\Phi_i(\boldsymbol{\xi}),\\
\eta(\boldsymbol{x},t;\boldsymbol{\xi})=\sum_{j=0}^{M-1}\eta_j(\boldsymbol{x},t)\Phi_j(\boldsymbol{\xi}).
\end{aligned}
\end{equation}
This choice allows us to again exploit the orthogonality property of gPC expansions. 
Hence, we substitute the expansions~\eqref{eqn:gPC} into the SWE~\eqref{eqn:swenonconserve}, integrate over the probability space $L_2(\Theta, \Sigma, P)$ ($\Theta$ the event space, $\Sigma$ the $\sigma$-field on $\Theta$, and $P$ the probability measure), and apply the orthogonality property to obtain the discrete stochastic formulation for each stochastic mode $k$:
\begin{subequations} \label{eqn:sswerandom}
\begin{equation} \label{eqn:sswerandom_a}
\begin{aligned}
\frac{\partial \eta_k}{\partial t}  + \nabla \cdot (  H_i \boldsymbol{u}_j  ) &\langle  ijk  \rangle =0 ~ \text{  on  } \Omega,\\
\end{aligned}
\end{equation}
\begin{equation} \label{eqn:sswerandom_b}
\begin{aligned}
\frac{\partial \boldsymbol{u}_k}{\partial t}  + \boldsymbol{u}_i \cdot \nabla \boldsymbol{u}_j \langle  ijk  \rangle =  - g \nabla \eta_k  &+ \nabla \cdot (\nu \nabla \boldsymbol{u}_k) + \boldsymbol{f}_k ~ \text{  on  } \Omega,\\
\end{aligned}
\end{equation}
\begin{equation} \label{eqn:sswerandom_c}
\begin{aligned}
u_k(\boldsymbol{x}, 0) = u_k^0(\boldsymbol{x}) ~ &\text{at time } t=0,\\ 
\end{aligned}
\end{equation}
\begin{equation} \label{eqn:sswerandom_d}
\begin{aligned}
\eta_k(\boldsymbol{x}, 0) = \eta_k^0(\boldsymbol{x}) ~& \text{at time } t=0.\\		
\end{aligned}
\end{equation}
\end{subequations}
In~\eqref{eqn:sswerandom}, $\langle ijk \rangle = \int_{\theta \in \Theta} \Phi_i (\three{\boldsymbol{\xi}}(\theta)) \Phi_j (\three{\boldsymbol{\xi}}(\theta))  \Phi_k (\three{\boldsymbol{\xi}}(\theta)) d\theta$, and $\boldsymbol{f}_k$ is:
\begin{equation}\label{eqn:force}
\boldsymbol{f}_k = - g \nabla  \left( \frac{p_{atm.}}{g \rho_{w}}   \right) \delta_{0k} + \frac{	C_k^d }{H_0}\frac{\rho_{a}}{\rho_{w}} \Vert  \boldsymbol{w} \Vert \boldsymbol{w} - \frac{C_i^b}{H_0} \Vert  \boldsymbol{u}_0 \Vert \boldsymbol{u}_j  \langle  ijk  \rangle,
\end{equation}
where $C_k^d$ is the $k^{th}$ mode of the wind drag coefficient $C_d(\boldsymbol{x},t;\three{\boldsymbol{\xi}})$, and $C_i^b$ is the $i^{th}$ mode of the bottom drag coefficient $C_b(\boldsymbol{x},t;\three{\boldsymbol{\xi}})$  obtained by gPC expansions. Note that in the definition of $\boldsymbol{f}_k$~\eqref{eqn:force}, we \moreR{have used} the mean of the total depth $H_0$ instead of the $n^{th}$-mode of total depth $H_n$ to avoid its high order summation terms in the denominators. Second, we have  linearized the stochastic nonlinear bottom shear stress and wind stress terms.
By integrating over probability space, the stochastic problem becomes deterministic:
\emph{Given the initial conditions in~\eqref{eqn:sswerandom_c} and~\eqref{eqn:sswerandom_d},  $C_k^d$, and $C_k^b$. \moreR{Then} find the stochastic modes of $\boldsymbol{u}_k(\boldsymbol{x}, t)$ and  $ \eta_k(\boldsymbol{x}, t) ,~k=0,1,\cdots,M-1$, such that the conservation laws~\eqref{eqn:sswerandom_a} and~\eqref{eqn:sswerandom_b} \moreR{are} satisfied.}

\subsection{Spatial Discretization}
Let $\boldsymbol{V}$ denote the vector-valued  trial and test function space and $Q$ the scalar-valued trial and test function space. The regularity requirements of the equivalent weak form of the SWE~\eqref{eqn:swenonconserve} lead us to the definitions:
\begin{equation} \label{eqn:func_spaces}
\begin{aligned}
\boldsymbol{V} &= \{ \boldsymbol{u} \in \boldsymbol{H}^1(\Omega)  \times (0, T):~ \boldsymbol{u} \cdot \boldsymbol{n} = 0 \text{ on }  \Gamma_{cl} \}, \\
Q &= \{  \eta \in H^1(\Omega) \times (0,T): ~ \eta = r \text{ on }  \Gamma_{op}    \}.
\end{aligned}
\end{equation}
Also denote by $\boldsymbol{v}_k \in \boldsymbol{V}$ the test function for velocity modes in the momentum equation~\eqref{eqn:sswerandom_b} and $q_k \in Q_0$ as the test function for the elevation modes in the continuity equation~\eqref{eqn:sswerandom_a}.  $Q_0$ is the restriction of $Q$ to functions that vanish on the \three{open boundary}: 
\begin{equation}
Q_0 = \{  q \in H^1(\Omega): ~ q = 0 \text{ on }  \Gamma_{\three{op}}    \}.
\end{equation}
Multiplication of~\eqref{eqn:sswerandom_a} and~\eqref{eqn:sswerandom_b} with the test functions $(\boldsymbol{v}_k,q_k)$ and subsequent integration by parts of the divergence and viscous terms leads to the weak formulation:
\begin{equation} \label{eqn:weakform}
\hspace{-0.1in}
\begin{aligned}
\text{Find } &(\boldsymbol{u_k},\eta_k) \in \boldsymbol{V}\times Q, \text{ such that } \forall  (\boldsymbol{v}_k,q_k) \in \boldsymbol{V}\times Q_0, \quad t \in (0,T): \\
&(  \frac{\partial \eta_k}{\partial t}, q_k)_\Omega   - (  H_i \boldsymbol{u}_j\langle  ijk  \rangle, \nabla q_k  )_\Omega  = 0,  \\
&(  \frac{\partial \boldsymbol{u}_k}{\partial t}, \boldsymbol{v}_k )_\Omega + ( \boldsymbol{u}_i \cdot \nabla \boldsymbol{u}_j\langle  ijk  \rangle , \boldsymbol{v}_k )_\Omega =- ( g \nabla \eta_k,  \boldsymbol{v}_k )_\Omega  - ( \nu \nabla \boldsymbol{u}_k, \nabla \boldsymbol{v}_k )_\Omega   + ( \boldsymbol{f}_k, \boldsymbol{v}_k )_\Omega,
\end{aligned}
\end{equation} 
where we use inner product notation, i.e.,  $(  \cdot,\cdot)_\Omega $ denotes the $L_2$ inner product over $\Omega$. The boundary terms from integration by parts vanish due to application of boundary conditions.


The stochastic weak formulation~\eqref{eqn:weakform} can be directly discretized in space for each mode $k$ by applying the Bubnov-Galerkin method. \one{Hence, we select discrete subspaces $\boldsymbol{V}^h\times Q^h \subset  \boldsymbol{V}\times Q$. \oneR{These discrete spaces consists of continuous functions that are Lagrange polynomials on each element $\Omega_e$ in the FE mesh $\mathcal{T}^h$ covering $\Omega$ }  of order two and one for $\boldsymbol{V}^h$ and $Q^h$, respectively:}
\begin{equation} \label{eqn:func_spacesH}
\begin{aligned}
\one{\boldsymbol{V}^h} &\one{=} \{ \one{ \boldsymbol{u}^h\oneR{|_{\Omega_e}} \in \boldsymbol{P}^2(\Omega_e)  \times (0, T):~ \boldsymbol{u}^h \cdot \boldsymbol{n} = 0 \text{ on }  \Gamma_{cl} \oneR{\cap \Omega_e}}, \; \oneR{\forall \; \Omega_e \in \mathcal{T}^h } \}, \\
\one{Q^h} &\one{=} \{ \one{\eta^h\oneR{|_{\Omega_e}} \in P^1\oneR{(\Omega_e)}\times (0,T): ~ \eta^h = r \text{ on }  \Gamma_{op} \oneR{\cap \Omega_e}}, \; \oneR{\forall \; \Omega_e \in \mathcal{T}^h } \}
\end{aligned}
\end{equation}
\one{where $P^1$ denotes the space of polynomials of order $1$ on $\Omega$ and $\boldsymbol{P}^2$ its second order vector-valued equivalent.}
This type of discretization is often referred to as a Taylor-Hood  \moreR{FE} method and the discretized weak form is shown in~\eqref{eqn:Disc_weak}. \three{For the deterministic SWE and Navier-Stokes equations, this choice is known to be a stable discretization choice, see, e.g.,~\cite{fortin1993finite}. However, due to the coupled modes in the stochastic system, further stabilization is required \moreR{unless the FE mesh is sufficiently refined}. }
In practice, we shall always augment the discretized weak formulation with stabilization \moreR{terms} to ensure stable discretizations of the stochastic weak formulation. 
\begin{equation} \label{eqn:Disc_weak}
\hspace{-0.12in}
\begin{aligned}
\text{Find } &(\boldsymbol{u_k}^h,\eta_k^h) \in \boldsymbol{V}^h\times Q^h, \text{ such that } \forall  (\boldsymbol{v}_k^h,q_k^h) \in \boldsymbol{V}^h\times Q_0^h, \quad t \in (0,T): \\
&(  \frac{\partial \eta_k^h}{\partial t}, q_k^h)_\Omega   - (  H_i^h \boldsymbol{u}_j^h\langle  ijk  \rangle, \nabla q_k^h  )_\Omega  = 0,  \\
&(  \frac{\partial \boldsymbol{u}_k^h}{\partial t}, \boldsymbol{v}_k^h )_\Omega + ( \boldsymbol{u}_i^h \cdot \nabla \boldsymbol{u}_j^h\langle  ijk  \rangle , \boldsymbol{v}_k^h )_\Omega =- ( g \nabla \eta_k^h,  \boldsymbol{v}_k^h )_\Omega  - ( \nu \nabla \boldsymbol{u}_k^h, \nabla \boldsymbol{v}_k^h )_\Omega   + ( \boldsymbol{f}_k, \boldsymbol{v}_k^h )_\Omega.
\end{aligned}
\end{equation} 

\two{In the  following sections, we introduce the time stepping scheme we use as well as the stabilization techniques we incorporate. In the developed numerical model, we do not consider the wetting and drying of the finite elements. While inclusion of this process is often critical to ensure accurate resolution of localized shallow water flows, it is beyond the scope of this work.}

\subsection{Time Discretization} \label{sec:time_discr}

As the computational cost of our stochastic system scales linearly with the number of modes used to represent those uncertainties, we employ the Incremental Pressure Correction Scheme (IPCS)~\cite{goda1979multistep,simo1994unconditional} \more{to reduce the added computational burden. }. This operator splitting scheme decouples the hyperbolic system and enables us to \oneR{compute} surface elevation and velocity independently thereby reducing the computational cost. 
\three{For the sake of brevity, we include \oneR{only} a brief overview of the key points of the ICPS algorithm for a hyperbolic system and refer the interested reader to~\cite{chenthesis}  for further details.} Note that this algorithm is compatible with any implicit time stepping method, in this work we exclusively use the backward Euler method whereas in~\cite{chenthesis} others are considered.

\three{The IPCS consists of a linearization and decoupling procedure of the  governing SWE. First, using a semi-implicit time discretization and linearization we get:
\begin{align}
\frac{1}{\Delta t}(\eta^{n+1}-\eta^{n}) &+ \nabla \cdot (H^{n} \boldsymbol{u}^{n+1})=0, \label{eqn:etaoriginal} \\
\frac{1}{\Delta t}(\boldsymbol{u}^{n+1}-\boldsymbol{u}^{n}) + \boldsymbol{u}^{*} \cdot \nabla \boldsymbol{u}^{n+1}& = -g \nabla \eta^{n+1} + \nabla \cdot \nu \nabla \boldsymbol{u}^{n+1} + \boldsymbol{f}^{n+1}, \label{eqn:uoriginal}
\end{align}
where $\boldsymbol{u}^*=\frac{3}{2} \boldsymbol{u}^n- \frac{1}{2} \boldsymbol{u}^{n-1}$. } \three{We subsequently decouple the system by replacing the surface elevation $\eta^{n+1}$ in the momentum equation~\eqref{eqn:uoriginal} by $\eta^{n}$. This momentum equation does not properly represent the velocity at the $(n+1)$th time step. Therefore, the tentative velocity $\boldsymbol{\tilde{u}}^{n+1}$ is introduced \moreR{and it} is governed by:
\begin{equation} \label{eqn:utentative}
\frac{1}{\Delta t}(\boldsymbol{\tilde{u}}^{n+1}-\boldsymbol{u}^{n}) + \boldsymbol{u}^* \cdot \nabla \boldsymbol{\tilde{u}}^{n+1} = -g \nabla \eta^{n} + \nabla \cdot \nu \nabla \boldsymbol{\tilde{u}}^{n+1} + \boldsymbol{f}^{n+1}. 
\end{equation}
However,  this tentative velocity does not satisfy the continuity equation \moreR{and to} account for this discrepancy, we define a velocity correction  $\boldsymbol{u}^c=\boldsymbol{u}^{n+1}-\boldsymbol{\tilde{u}}^{n+1}$. Subtracting~\eqref{eqn:utentative} from~\eqref{eqn:uoriginal}  gives:} \three{
\begin{equation}
\frac{1}{\Delta t}\boldsymbol{u}^c +  \boldsymbol{u}^* \cdot \nabla \boldsymbol{u}^c = -g  \nabla (\eta^{n+1}-\eta^{n}) + \nabla \cdot \nu \nabla \boldsymbol{u}^c,
\end{equation}
}\three{by neglecting higher order terms of $\boldsymbol{u}^c$, we obtain: } \three{
\begin{equation} \label{eqn:uc}
\boldsymbol{u}^c = - g  \Delta t  \nabla (\eta^{n+1}-\eta^{n}),
\end{equation}
}  and rewriting~\eqref{eqn:etaoriginal} in terms of $\boldsymbol{u}^c$ \moreR{subsequently} leads to:   \three{
\begin{equation} \label{eqn:etacorrectionbefore}
\frac{1}{\Delta t}(\eta^{n+1}-\eta^{n}) +  \nabla \cdot (H^n \boldsymbol{u}^c)= - \nabla \cdot (H^n \boldsymbol{\tilde{u}}^{n+1}).
\end{equation}
}  \three{ Substitution of~\eqref{eqn:uc} into~\eqref{eqn:etacorrectionbefore} yields the governing equation for the surface elevation at the $(n+1)$th time step: }  \three{
\begin{equation} \label{eqn:etacorrection}
\left(   \eta^{n+1} -\eta^{n}   \right)- g \Delta t^2 \nabla \cdot (H^{n} \nabla(\eta^{n+1}-\eta^{n})) = - \Delta t \nabla \cdot (H^n \boldsymbol{\tilde{u}}^{n+1}),
\end{equation}
} \moreR{and} the velocity is calculated using:
\begin{equation} \label{eqn:ucorrection}
\boldsymbol{u}^{n+1} = \boldsymbol{\tilde{u}}^{n+1} - g \Delta t  \nabla (\eta^{n+1} - \eta^n).
\end{equation}
The IPCS time discretization scheme can be summarized by the following algorithm:

\begin{algorithm}[H] 
	\caption{IPCS time discretization scheme} \label{alg:IPCS_alg}
	\SetAlgoLined
	Given $\boldsymbol{u}^0$ and $\eta^0$ \;
	\While{$t \leq T$}{
		1. Given the surface elevation $\eta^{n}$, solve for the IPCS "tentative velocity" $\boldsymbol{\tilde{u}}^{n+1}$, see~\cite{chenthesis}\;
		2. Given the tentative velocity $\boldsymbol{\tilde{u}}^{n+1}$, compute the surface elevation $\eta^{n+1}$ \;
		3. Given the surface elevation $\eta^{n+1}$, compute the velocity $\boldsymbol{u}^{n+1}$ \;
	}
	\KwResult{A time series of $\boldsymbol{u}$ and $\eta$\;}
\end{algorithm}

\subsection{Cross-mode stabilization methods} \label{sec:CM_Stabilize}

It is well known that standard Bubnov-Galerkin FE method leads to unstable numerical schemes for convection-dominated flows which exhibits itself as oscillations in the FE solution \more{ for certain choices of FE spaces}. 
\more{Here}, the stochastic weak formulation~\eqref{eqn:weakform}  leads to a deterministic modes-coupled system (see $\boldsymbol{u}_i \cdot \nabla \boldsymbol{u}_j \langle  ijk  \rangle$ term in~\eqref{eqn:weakform}), in which the resulting linear system of equations is $M$ times larger than the corresponding deterministic SWE. The effect of this coupling is stronger instabilities in the FE approximation.  To seek a stable discrete solution of each mode $\boldsymbol{u}_k^h$ and $\eta_k^h$, we propose three cross-mode stabilization methods which can be applied independently or simultaneously to the discretized stochastic weak formulation. 
The starting point for the stabilized methods is the discretized weak formulation~\eqref{eqn:Disc_weak}. For convenience, we drop the superscipt $h$ in this section as it is understood that the stabilization methods are only applied to the discrete case.  
\three{Note that these stabilization techniques are only required for the momentum equations as the IPCS leads to an elliptic equation~\eqref{eqn:etacorrection} for the continuity equation which is guaranteed to be discretely stable.}

\subsubsection{Cross-mode Streamlined Upwind Petrov Galerkin Method}

The classical streamlined upwind Petrov-Galerkin (SUPG) method for the Navier-Stokes equations was introduced by Brooks and Hughes in~\cite{brooks1982streamline}. The SUPG stabilizes the spatial FE discretization by adding artificial diffusion over element interiors along the streamline direction. 
Based on the SUPG method, we propose to add the following terms to \one{the momentum equation in}~\eqref{eqn:Disc_weak}:
\begin{equation}
\sum\limits_{e \in N_e} \tau_{SUPG} ( \mathbf{R}_k(\boldsymbol{\tilde{u}}^{n+1}_k),~~ \boldsymbol{u}_k^* \cdot \nabla \boldsymbol{v}_k     )_{\Omega_e}, ~~ k = 0,1,\cdots,M-1,
\end{equation}
where  $\boldsymbol{u}_k^*=\frac{3}{2} \boldsymbol{u}_k^n- \frac{1}{2} \boldsymbol{u}_k^{n-1}$, and $\boldsymbol{\tilde{u}}_k^{n+1}$ is the IPCS tentative velocity of the \three{$k$}-th mode at $(n+1)$th time step (see details of both quantities in the introduction of the IPCS scheme \moreR{in Section}~\ref{sec:time_discr}).
$M$ is the total number of gPC functions, $N_e$ \one{the collection of elements in the FE mesh, $\Omega_e$ the domain of the element $e$ },  $\tau_{SUPG}$ the stabilization parameter, and $\mathbf{R}_k(\boldsymbol{\tilde{u}}_k^{n+1})$ is the residual form of the original stochastic SWE:
\begin{equation} \label{eqn:residual}
\begin{aligned}
\mathbf{R}_k(\boldsymbol{\tilde{u}}_k^{n+1}) = \frac{1}{\Delta t}(\boldsymbol{\tilde{u}}_k^{n+1}-\boldsymbol{u}_k^{n}) + 	\boldsymbol{u}_i^* \cdot \nabla \boldsymbol{\tilde{u}}_j^{n+1} \langle  ijk  \rangle + g \nabla \eta_k^{n} - \nabla \cdot \nu \nabla \boldsymbol{\tilde{u}}_k^{n+1} - \boldsymbol{f}_k^{n+1}.
\end{aligned}
\end{equation}
Based on the works of Tezduyar~\cite{tezduyar2003computation,tezduyar2002stabilization}, we select $\tau_{SUPG}$:
\begin{equation} \label{Eqn:SUPG}
\tau_{SUPG} = \left(  \frac{2}{\Delta t} + \frac{2 \Vert \boldsymbol{u}_0^n \Vert}{h_e} + \frac{4 \nu}{h_e^2}    \right)^{-1},
\end{equation}
where $h_e$ is the radius of the circumscribed circle of each element and $\boldsymbol{u}_0^n$ is the mean mode of $\boldsymbol{u}$ at $n^{th}$ time step. Note that the cross-mode SUPG preserves the consistency of the classical SUPG as $\mathbf{R}_k(\boldsymbol{\tilde{u}}_k^{n+1})$ vanishes if $\boldsymbol{\tilde{u}}_k^{n+1}$ is the solution of the stochastic PDE~\eqref{eqn:sswerandom}.


\subsubsection{Cross-mode Discontinuity Capturing Method}

\begin{figure}[H] 
	\centering
	\subfigure[Illustration of $\boldsymbol{u}_{k \Vert_{u_{1}}}^*$.]{\label{fig:u||} \includegraphics[width=0.3\columnwidth]{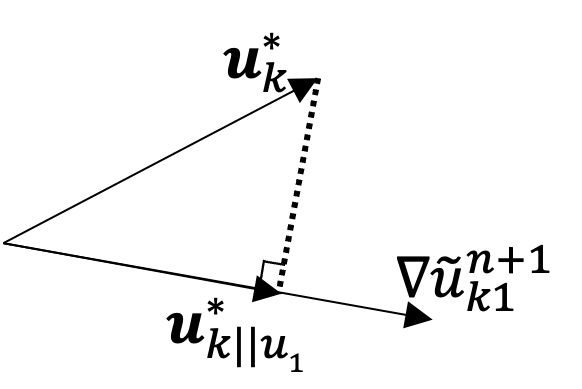}}
	\subfigure[Illustration of $\boldsymbol{u}_{k \Vert_{u_{2}}}^*$.]{\label{fig:v||} \includegraphics[width=0.3\columnwidth]{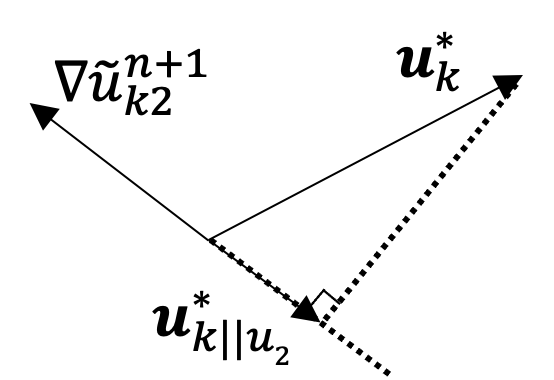}}
	\caption{Projections of velocity gradient modes.}
	\label{fig:uv||}
\end{figure}

In the numerical stabilization of the deterministic SWE using FE methods, it is often not sufficient to apply only the SUPG method to stabilize the discrete systems. Since SUPG adds diffusive effects only along the streamlines, other techniques may also be needed. This problem is further exaggerated in our SSWM due to the coupled modes. Hence, we also incorporate another residual based stabilization method, the  discontinuity capturing (DC) method of Hughes \emph{et al.}~\cite{hughes1986new}. We therefore add the following to \one{the momentum equation in}~\eqref{eqn:Disc_weak}:
\begin{equation}
\sum\limits_{e \in N_e} \tau^1_{DC} \left(   \boldsymbol{u}_{k \Vert_{u_{1}}}^* \cdot \nabla v_{k1},  ~ R_{k1}(\boldsymbol{\tilde{u}}_k^{n+1})    \right)_{\Omega_e} + \sum\limits_{e \in N_e} \tau^2_{DC} \left(   \boldsymbol{u}_{k \Vert_{u_{2}}}^* \cdot \nabla v_{k2},  ~ R_{k2}(\boldsymbol{\tilde{u}}_k^{n+1})    \right)_{\Omega_e},
\end{equation}
where $\boldsymbol{v}_k = (v_{k1}, v_{k2})$ is the test function for the $k$-th mode, $R_{k1}, R_{k2}$ are the $x, y$ components of the residual form $\mathbf{R}_k(\boldsymbol{\tilde{u}}_k^{n+1})$, see~\eqref{eqn:residual},  $\boldsymbol{u}_{k \Vert_{u_1}}^*$  the projection of $\boldsymbol{u}_k^*$ onto $\nabla \tilde{u}_{k1}^{n+1}$, $\boldsymbol{\tilde{u}}_{k \Vert_{u_2}}^*$  the projection of $\boldsymbol{u}_k^*$ onto $\nabla \tilde{u}_{k2}^{n+1}$, and $\boldsymbol{\tilde{u}}_k^{n+1} = (\tilde{u}_{k1}^{n+1}, \tilde{u}_{k2}^{n+1})$. The projection operators are illustrated in Figure~\ref{fig:uv||} and are defined by:
\begin{equation}
\boldsymbol{u}_{k \Vert_{u_{1}}}^* = \frac{ \boldsymbol{u}_k^* \cdot  \nabla \tilde{u}_{k1}^{n+1}  }{\vert \nabla \tilde{u}_{k1}^{n+1}  \vert  ^2}  \nabla \tilde{u}_{k1}^{n+1},
\end{equation}
\begin{equation}
\boldsymbol{u}_{k \Vert_{u_{2}}}^* = \frac{ \boldsymbol{u}_k^* \cdot  \nabla \tilde{u}_{k2}^{n+1}  }{\vert \nabla \tilde{u}_{k2}^{n+1}  \vert  ^2} \nabla  \tilde{u}_{k2}^{n+1}.
\end{equation}
%

\noindent $\tau^1_{DC}, \tau^2_{DC}$ are the DC stabilization parameters:
\begin{equation} \label{eqn:tau_DC}
\tau^1_{DC} = \left(  \frac{2}{\Delta t} + \frac{2 \Vert \boldsymbol{u}_{k \Vert_{u1}}^* \Vert}{h_e} + \frac{4 \nu}{h_e^2}    \right)^{-1},
\tau^2_{DC} = \left(  \frac{2}{\Delta t} + \frac{2 \Vert \boldsymbol{u}_{k \Vert_{u2}}^* \Vert}{h_e} + \frac{4 \nu}{h_e^2}    \right)^{-1}.
\end{equation}
To avoid an overly stabilized effect and therefore nonphysical solutions from both SUPG and DC, we adjust the parameters in~\eqref{eqn:tau_DC} as follows:
\begin{equation} \label{Eqn:DC}
\begin{aligned}
&\tilde{\tau}^1_{DC} = \max (0, ~~ \tau^1_{DC}  -  \tau_{SUPG} ),\\
&\tilde{\tau}^2_{DC} = \max (0, ~~ \tau^2_{DC}  -  \tau_{SUPG} ).
\end{aligned}
\end{equation}
\noindent The cross-mode DC method is also consistent, as the residual term goes to zero for the true solution of each stochastic mode.


\subsubsection{Cross-mode Continuous Interior Penalty Method}  

\three{The two preceding stabilization techniques are complementary as they provide stabilization in different directions. However, due to the coupled stochastic modes, localized discontinues in the solution also lead to stability issues. Thus, a stabilization method to penalize such discontinuities is needed to ensure stable computations.}
The last stabilization method we employ in our coupled system follows the continuous interior penalty (CIP) method~\cite{burman2007continuous}. Thus, we penalize inter-element discontinuities \three{in the jump of the gradient of the trial function} by adding the following term to \one{the momentum equation of} our discretized weak formulation~\eqref{eqn:Disc_weak}:
\begin{equation} \label{Eqn: CIP}
\sum\limits_{e \in N_e} \sigma_{CIP}  \cdot  \overline{h}_{e}^2 \cdot \text{avg}(   |   \boldsymbol{\tilde{u}}_k^{n+1} \cdot \boldsymbol{n}   |   ) 
\left(
\llbracket   \nabla \boldsymbol{v}_k \cdot \boldsymbol{n}   \rrbracket , ~
\llbracket   \nabla \boldsymbol{\tilde{u}}_k^{n+1}   \cdot \boldsymbol{n}   \rrbracket      \right)_{\partial \Omega_{e}},
\end{equation}
where $\sigma_{CIP}$ is a positive constant, $\llbracket \cdot \rrbracket$ the jump operator over adjacent elements, avg($\cdot$) represents the average operator over adjacent elements,  $\overline{h}_e$ is the maximum edge length in an element $\Omega_{e}$, \three{and  $\boldsymbol{n}$  the outward normal vector of the edge $e$ on either side of the edge}. Note that the cross-mode CIP method is also consistent  as the added jump terms vanish for sufficiently smooth stochastic solutions. 


\section{Verification of the SSWM} \label{sec:verify_validate}

To perform a verification process of the proposed SSWM framework, we will perform three idealized numerical tests that represent small scale short term shallow water flows and two realistic numerical tests that represents large scale long term applications. \more{As a validation of the deterministic version of our framework, we will compare our results against outputs from other models as well as experimentally measured data.}

\begin{figure}[H]
	\begin{center}
		\includegraphics[width=0.75\columnwidth, angle=0]{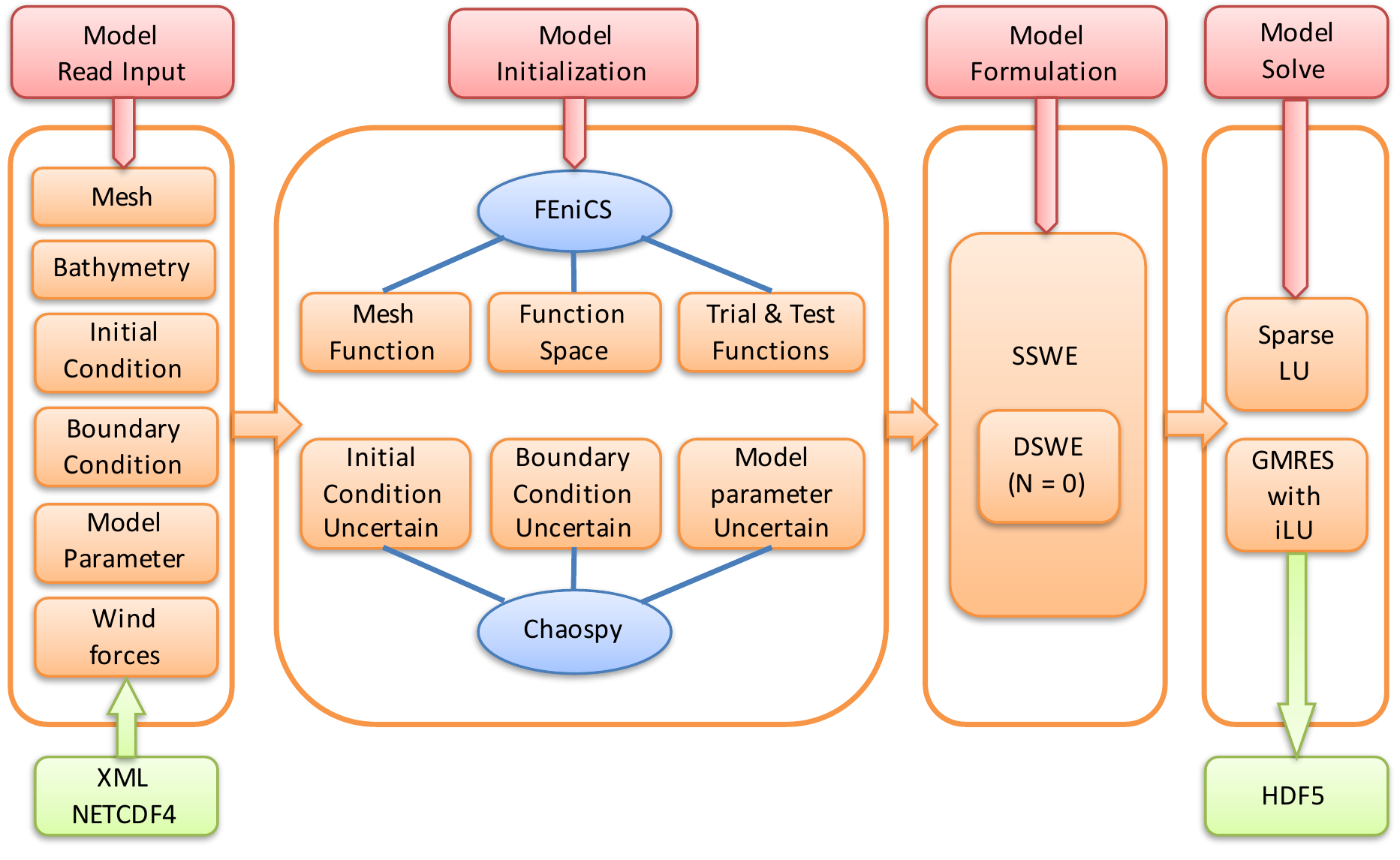}	
	\end{center}
	\vspace*{-0.5cm}
	\caption{A illustration of the SSWM code structure.}
	\label{fig:implementation}
\end{figure}

The corresponding numerical program is a python program solving the nonlinear SSWM for shallow water flows. It is built on the finite element package FEniCS~\cite{logg2012automated,logg2010dolfin,kirby2006compiler,alnaes2014unified,alnaes2012unified,kirby2004algorithm} and the statistics package Chaospy~\cite{feinberg2015chaospy}.  Chaospy is used for \one{generating the gPC  basis functions and integrating over probability space} in the SSWM discretization. \three{Within the FEniCS framework, the implicit matrix solvers required in the SSWM are all performed using a GMRES Krylov solver with an \emph{ilu} preconditioner~\cite{saad1986gmres,meijerink1977iterative}}. The program is designed in four modules and its structure is visually presented in Figure~\ref{fig:implementation}.
\vspace{-0.5cm}

To validate the numerical implementations of SSWM and demonstrate the effectiveness of the proposed cross-mode stabilization methods, we will conduct a two-step process. First, we will verify and validate the deterministic shallow water model (DSWM), which is a degraded (i.e., polynomial chaos order $N=0$) SSWM. This verification and validation step is done with respect to analytical solutions, well-established model simulation results, and experimental data. Second, we verify the SSWM surrogate by comparison to \one{its corresponding ensemble runs using DSWM}. Note that in the second step,  we  compare  the two representations of the output random variable: one represented by the SSWM surrogate; the other represented by the ensembles.
\more{Thus, this second step will be performed in multiple fashions to comprehensively verify the SSWM.} We will first compare the mean and variance of the output random variable at a specified location at a fixed time. Subsequently, the PDF will be computed and compared at the same spatial-temporal point. Lastly, because both the ensembles and the surrogate whose form is $f(\boldsymbol{x}, t; \boldsymbol{\xi})$ are functions of $\boldsymbol{\xi}$, we wish to compare them pointwise. Thus, we first fix the time $t$, and compare the SSWM surrogate over all sample grids $\boldsymbol{\xi}$, we subsequently fix the sample grids $\boldsymbol{\xi}$ and compare the SSWM surrogate over all time steps. 



\subsection{Numerical Tests} \label{sec:testcases}

In this section, we define and describe the set up of the test cases we consider.  \two{The cross-mode stabilization techniques are not used in all these test cases, in particular, the first two test cases of simple rectangular domains do not require any stabilization. However, in the more complex tests the SSWM will typically lead to the models crashing and therefore cannot be applied without all proposed stabilization techniques. }
\two{For each test case, we select a fixed stochastic order $N$ to be used in the subsequent computations. These selections were made based on extensive numerical experimentation with different orders $N$. We also include a discussion on this selection in Section~\ref{sec:verification_notes}    }


\subsubsection{Slosh Test Case With Uncertain Initial Condition} \label{sec:sloshtest}

As an initial test case, we consider a  rectangular domain with length $L = 100m$, width $W = 50m$, and constant water depth $H = 20m$. The four boundaries are closed, with a free slip boundary condition. The surface elevation is initially a west to east varying cosine shaped perturbation, with an amplitude of $0.1m$ and a wavelength of $200m$. The water velocity is initially zero everywhere, we assume inviscid flow, i.e., $\nu = 0$. No external forcing is applied, and the time step in the IPCS is set to $0.5$ seconds. The total simulation time is $50$ seconds and we use the uniform triangular mesh \oneR{with 400 elements.}
%
%
%
In this test case, the uncertainty of the initial condition is assumed to take the form $\eta = 0.1 \xi_1 \xi_2 \cos ( \pi x /100.0) $, where $\xi_1, \xi_2$ are both uniformly distributed $\xi_1 \sim U(0.8, 1.2)$, $\xi_2 \sim U(1.0, 2.0)$, i.e., two-dimensional generalized polynomial chaos, and the stochastic order $N=3$.  In this test case, no cross-mode stabilization is added to the computations as the mesh and time step are both sufficiently fine to ensure stable computations.  

\subsubsection{Hump Test Case With Uncertain Bathymetry} \label{sec:humptest}
\begin{figure}[h!]
	\centering
	\includegraphics[width=\textwidth]{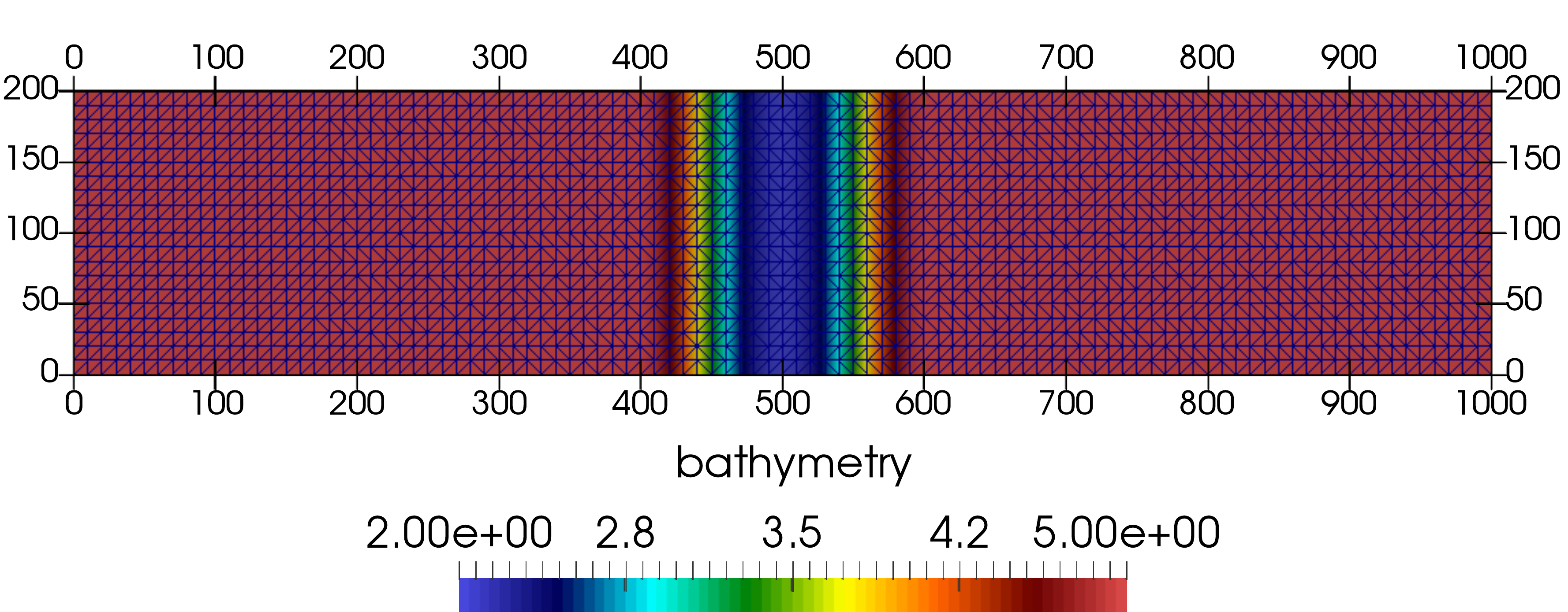}
	\vspace{-3cm}
	\caption{Hump test case: Mesh and bathymetry}
	\label{fig:humpmesh}
\end{figure}

In the second test case, we again consider a rectangular domain, in this case it is $1000m$ long and $200m$ wide. The bathymetry in this case consists of a hump, see Figure~\ref{fig:humpmesh}, and is given by:
\begin{equation}\label{eqn:bathy}
b = \left\{  
\begin{aligned}
-3.0 \left(    \frac{x-500.0}{100.0}   \right)^4 + 6.0  \left(    \frac{x-500.0}{100.0}   \right)^2 + 2.0 &, ~~ 400m < x < 600m \\
5.0 &,  ~~  else
\end{aligned}
\right.
\end{equation}
The domain is closed with free slip boundary conditions on the north, south, and west sides, whereas the east side is a tidal boundary where the sinusoidal function $0.1 \sin ( \pi t / 20)$ is prescribed. 
The initial water elevation and velocities are all zero. There is no viscosity in the domain, no external forcing  is applied, and the time step in the IPCS is set to $1.0$ seconds.  The total simulation time is $300$ seconds and we use the uniform mesh shown in Figure~\ref{fig:humpmesh}.
We further assume the uncertainty of bathymetry to be of the form:
\begin{equation} \label{eq:bathy_uncertain}
b = \left\{
\begin{aligned}
-3.0 \xi_1 \left(    \frac{x-500.0}{100.0}   \right)^4 + 
6.0  \xi_2 \left(    \frac{x-500.0}{100.0}   \right)^2 + 2.0, & ~~ 400m < x < 600m \\
-3.0 \xi_1 + 6.0 \xi_2 + 2.0,  & ~~  else
\end{aligned}
\right.
\end{equation}
where $\xi_1, \xi_2$ are assumed to be uniformly distributed,  $\xi_1 \sim U(0.8, 1.2)$ and $\xi_2 \sim U(0.9, 1.1)$. We again utilize  two-dimensional gPC and set the stochastic order $N=3$. As in the preceding case, the computations are numerically stable due to the mesh resolution and time step, and thus no cross-mode stabilization is added to the computations.

\subsubsection{Idealized Inlet Test Case With Uncertain  Boundary Condition} \label{sec:inletttest}

\begin{figure}[h!]
	\centering
\includegraphics[width=0.75\textwidth]{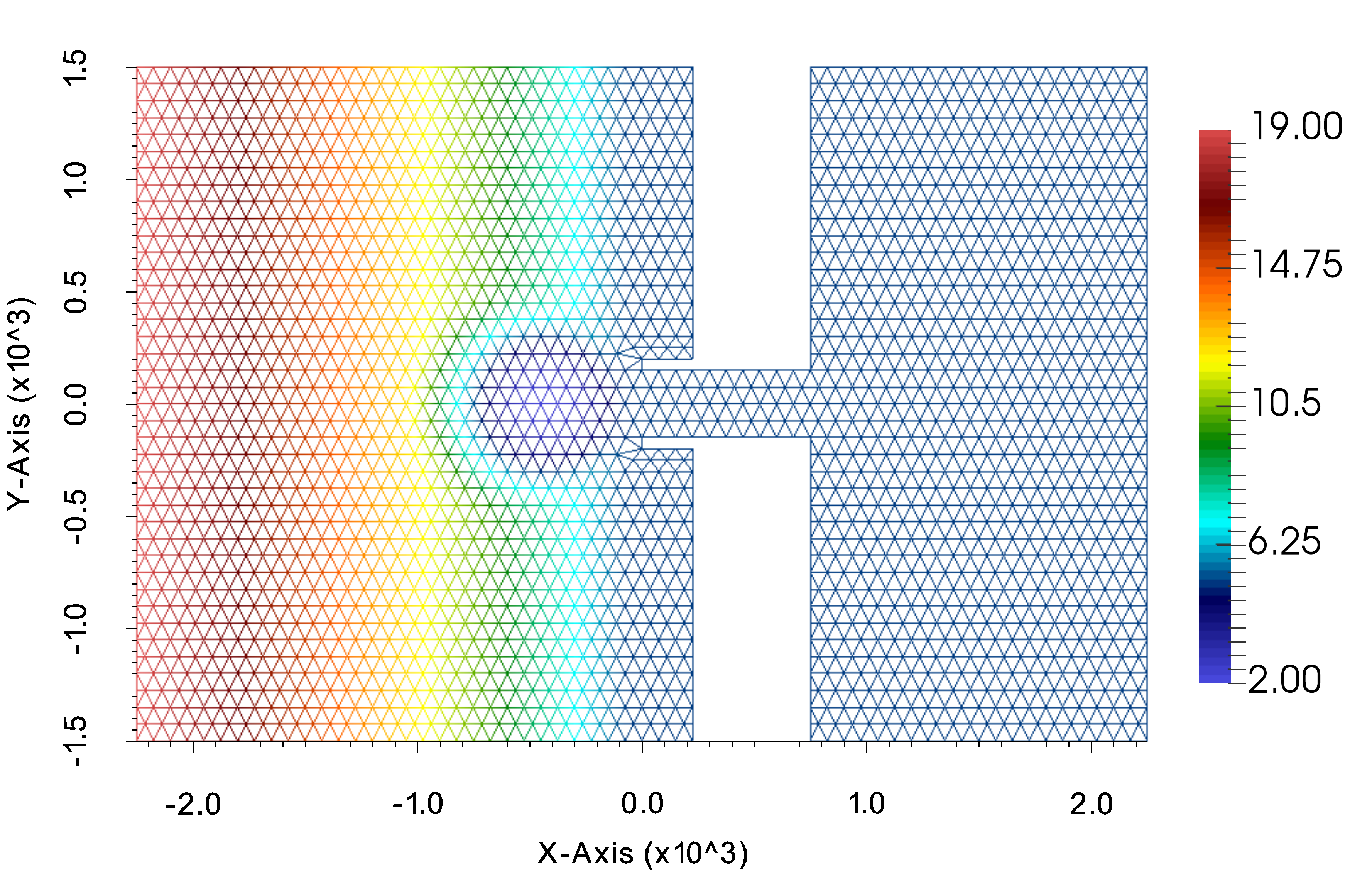}	\caption{Idealized inlet test case mesh and bathymetry.}
	\label{fig:inletmesh}
\end{figure}

To demonstrate the capability of our methodology to handle complicated  scenarios,  we consider an idealized inlet  test case. The domain consists of a rectangular harbor connected to the open ocean via a narrow  channel. The bathymetry varies linearly from $19m$ at the open ocean boundary to $5m$ at the entrance of the channel. Furthermore, there is a hump near the entrance of the channel, approximately $750m$ in diameter with a maximum height of $2m$. This hump is used to simulate the physics of an ebb shoal. These  commonly appear in coastal channels and are formed due to decelerated flows depositing transported sediments near channel exits. The mesh and bathymetry are shown in Figure~\ref{fig:inletmesh}.
All boundaries are closed except the western, which is open. To the western boundary we apply M2 tides~\cite{kubatko2006hp}, and the remaining boundaries are closed with free slip conditions. The M2 tides follow $\eta = A \sin ( \Omega t)$, where A is an amplitude of $0.75$m, $\Omega$ is an angular frequency of 1.41$\times10^{-4}$ rad/s. We apply homogeneous initial conditions, set the kinematic viscosity to $10^{-6}$, and the bottom friction coefficient is set to $C_b = 0.003$. The time step in the IPCS is set to 447 seconds and the simulation covers five M2 tidal cycles, approximately 2.5 days. To stabilize the FE discretizations in this test case, we apply all three cross-mode stabilization techniques  from Section~\ref{sec:CM_Stabilize} with $\tau_{SUPG}$ and $\tau_{DC}$, see~\eqref{Eqn:SUPG} and~\eqref{Eqn:DC} and the CIP stabilization parameter $\sigma_{CIP}$ set to $0.75$. In this test case, we assume the uncertainty of the boundary condition to be of the form $\eta = 0.2 \xi_1 \sin ( \Omega t)$, where $\xi_1$ is uniformly distributed, given as $\xi_1 \sim U(1.0, 2.0)$, i.e., one-dimensional gPC is utilized and we set the stochastic order $N=3$.


\subsubsection{Historical Hurricane Test Cases for the Gulf of Mexico and Historical Hurricanes With Uncertain Wind Drag Coefficient} \label{sec:hurricanetests}

\begin{figure}[h]
	\centering
	\subfigure{\includegraphics[width=0.45\textwidth]{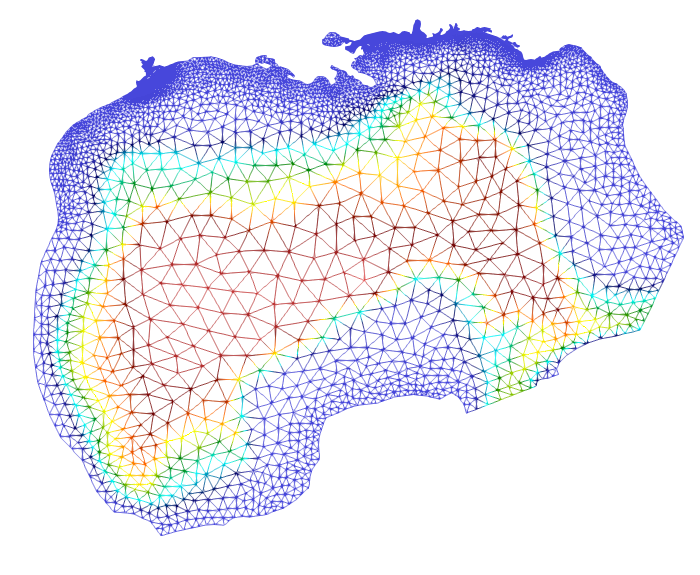}}
	\subfigure{\includegraphics[width=0.45\textwidth]{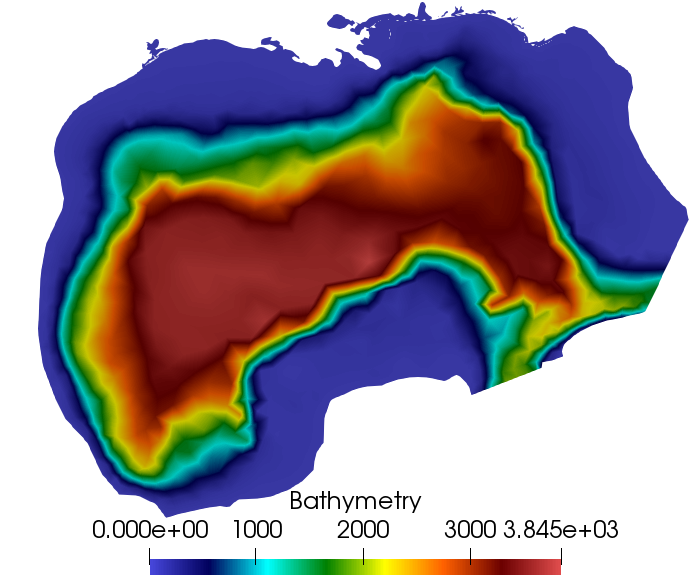}}
	\caption{Mesh and Bathymetry for  the hurricane test cases.}
	\label{fig:hurricanemesh}
\end{figure}

\begin{figure}[h]
	\centering
	\subfigure{\includegraphics[width=0.45\textwidth]{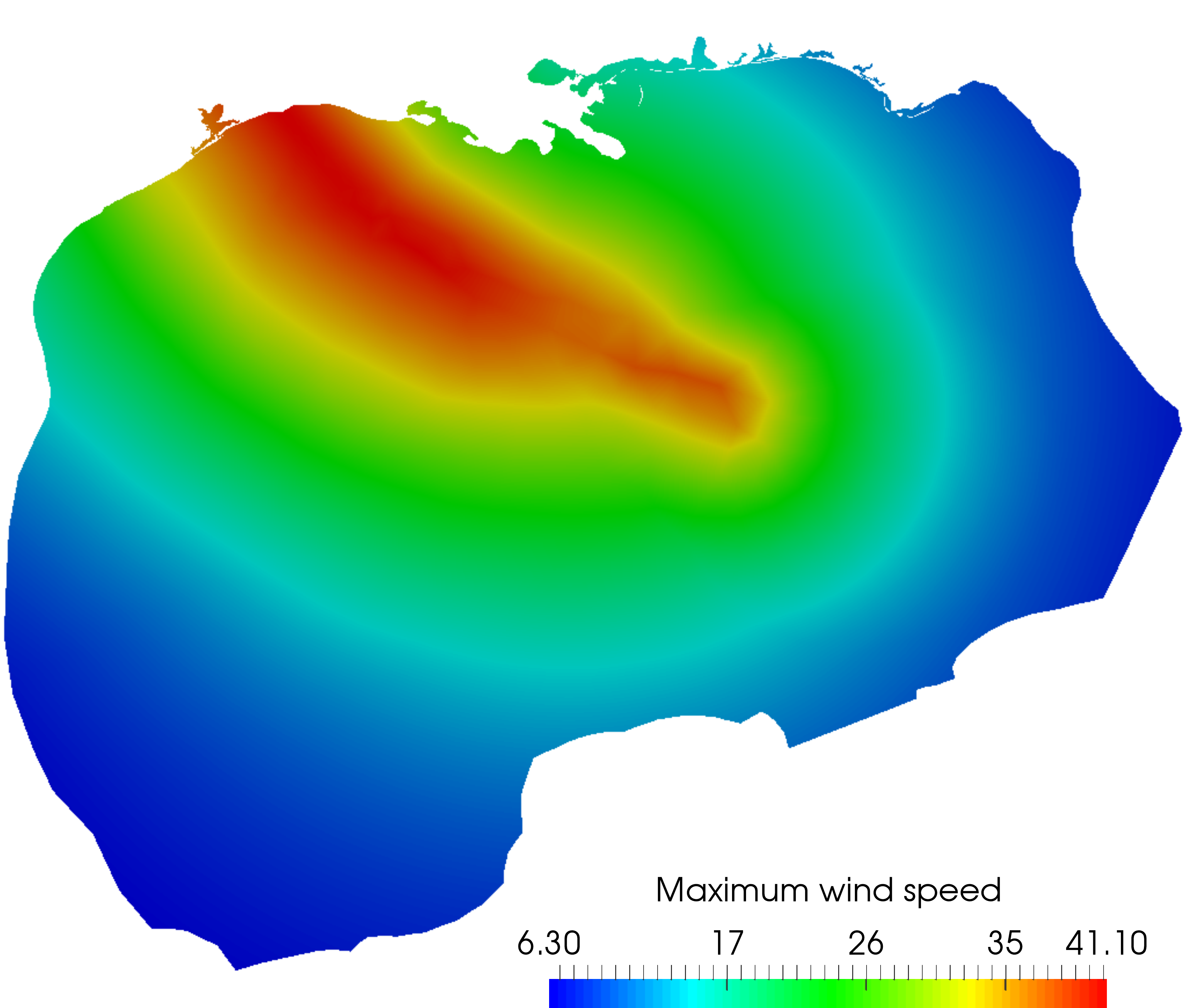}}
	\subfigure{\includegraphics[width=0.45\textwidth]{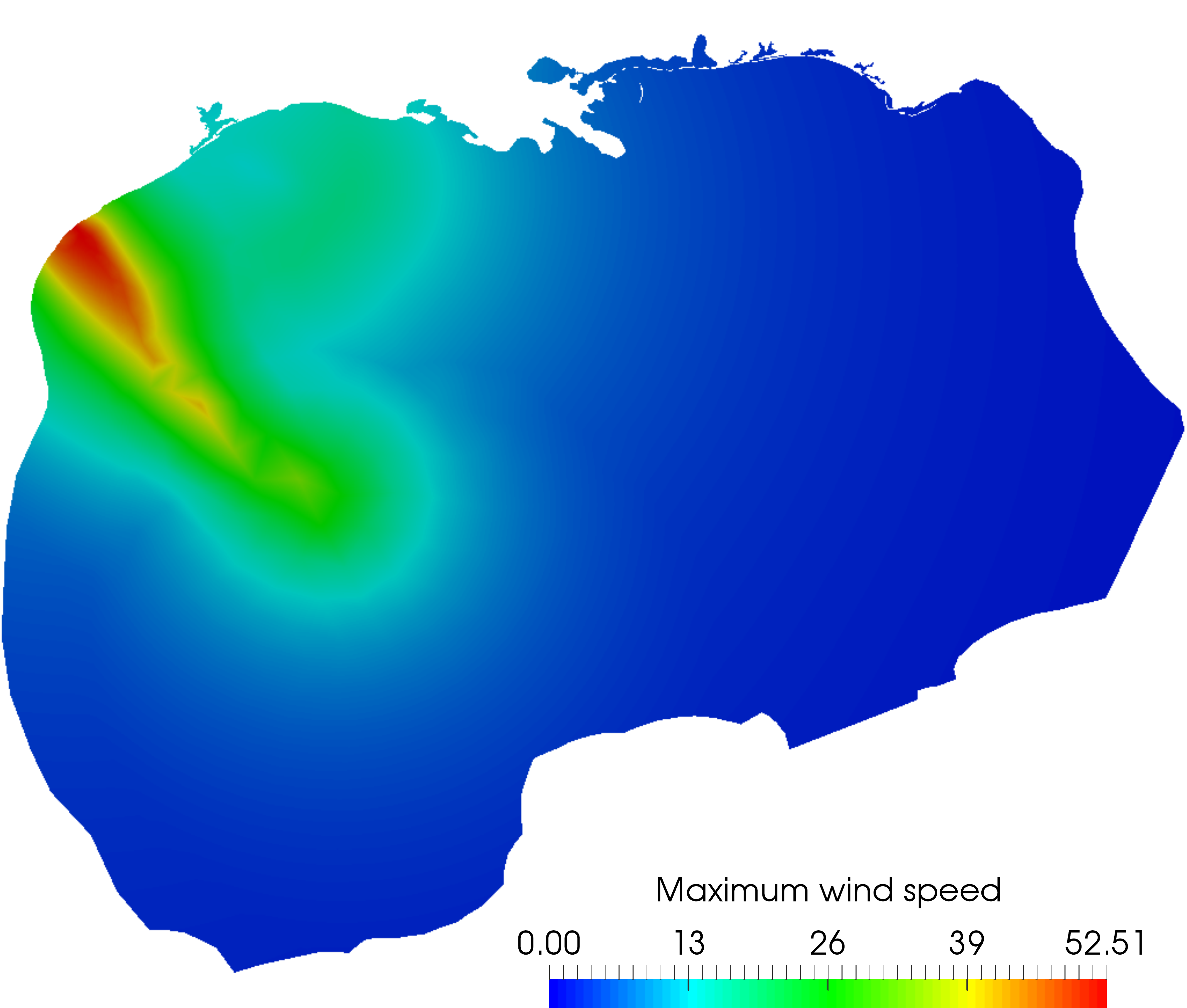}}
	\caption{Maximum wind speed during Hurricanes Ike (left) and Harvey (right).}
	\label{fig:hurricanewind}
\end{figure}

In these two tests, we choose the Gulf of Mexico as the domain of interest and the historical hurricanes Harvey (2017) and Ike (2008), see Figures~\ref{fig:hurricanemesh} and~\ref{fig:hurricanewind} for the mesh, physical domain and maximum hurricane winds. These hurricanes are selected as they are representative of two types of hurricanes in  the Gulf of Mexico, a large slow moving hurricane (Ike) and a smaller fast moving hurricane (Harvey). A closed free-slip boundary condition is applied on the entire domain and the sea water is initially at rest. The kinematic viscosity is set to $10^{-6}$ and the bottom friction coefficient is fixed: $C_b = 0.003$. In these cases, seawater motion is externally forced by the hurricane winds only. The wind fields are obtained from the National Hurricane Center's best track HURDAT2 database and we apply a Powell scheme~\cite{powell2003reduced} to determine the wind drag coefficient. The time step in the IPCS is set to 447 seconds and the simulations cover selected time spans for both hurricanes. Hence, for Hurricane Ike  $2.5$ days starting September 11 2008 12:00pm and for Hurricane Harvey $6$ days starting at August 24 2017 6:00pm (both Central Daylight Time).  To stabilize the FE discretizations in these test cases, we apply all three cross-mode stabilization techniques  from Section~\ref{sec:CM_Stabilize} with $\tau_{SUPG}$ and $\tau_{DC}$, see~\eqref{Eqn:SUPG} and~\eqref{Eqn:DC} and the CIP stabilization parameter $\sigma_{CIP}$ set to $9.0$.

In these large scale test cases, we assume that the wind drag coefficient is uncertain because this parameter is well known to significantly impact the maximum surge during  hurricane events, see e.g.,~\cite{jmse4030058}. In the Powell scheme we use to ascertain the wind drag parameter $C_d^{Powell}$,  the range of this coefficient is limited to $[0.0001, 0.0005]$ and it varies linearly with the magnitude of wind velocity in each  quadrant relative to the hurricane center. Hence, we assume the uncertain wind drag coefficient takes the same form of $C_d = \xi_1 C_d^{Powell}$, where $\xi_1$ is assumed to be uniformly distributed; i.e.,  $\xi_1 \sim U(0.8, 1.2)$, i.e., one-dimensional gPC is utilized here and we  set the stochastic order $N=1$.

\subsection{Verification of the SSWM } 

\more{For the sake of brevity, the verification experiments and results of the deterministic part of our SSWM are presented in~\ref{sec:verify_dswm}.}
To verify the SSWM, we will examine several different uncertain sources as introduced in Section~\ref{sec:uncertain_Sources}  one by one: uncertain initial condition, bathymetry,  boundary condition, and model parameter, i.e., wind drag coefficient. To accomplish such a verification, we first compare the mean and \moreR{standard} deviation of the surrogate with values computed from the deterministic realizations. \twoR{We compute the mean and standard deviation of the surrogates using the technique found in, e.g.,~\cite{xiu2010numerical}:}
\begin{equation} \label{eqn:MeanDeviation}
\mathbb{E}[f] = f_0,~~~~Var[f] = \sum\limits_{k=0}^M f_k^2.
\end{equation}
\twoR{In~\eqref{eqn:MeanDeviation}, $f$ represents either surface elevation or water velocity. Whereas the mean and standard deviation of the deterministic realizations are computed using its arithmetic counterparts.} Second, we compare the PDF by sampling 5000 grid points of the surrogate against the one given by the corresponding deterministic realizations. Also note that the output of the SSWM at each spatiotemporal point is indeed a surrogate function which connects model quantities (e.g., surface elevation) to the input random vector $\boldsymbol{\xi}$. To verify such surrogate functions, we shall perform pointwise comparisons of the surrogate at each sample grid (i.e., the grid within the support of the random vector $\boldsymbol{\xi}$) in probability space to the value given by each of the corresponding deterministic realizations. Note that this type of comparison process is similar to conducting Monte Carlo experiments. Since we in~\ref{sec:verify_dswm} comprehensively verify the deterministic part of the SSWM, we trust this DSWM as a verification tool for conducting deterministic realizations. By considering the outputs of the DSWM as benchmarks, we can compute DSWM solutions using the distributed samples of uncertain model inputs. We subsequently verify the solution function of the SSWM by comparing it pointwise with the collection of results from the DSWM. \oneR{Furthermore,  we only provide pointwise comparisons for Slosh test and hump test at a single time step. The pointwise comparisons for other tests over the probability space and time domain are relegated to~\ref{sec:verify_sswm}}.

\subsubsection{Uncertain Initial Condition - Slosh Test Case}

In the slosh test case, the uncertain initial condition is  $\eta = 0.1 \xi_1 \xi_2 \cos ( \pi x /100.0) $, where $\xi_1, \xi_2$ are both uniformly distributed given as $\xi_1 \sim U(0.8, 1.2), \xi_2 \sim U(1.0, 2.0)$. 
\two{From both $\xi_1$ and $\xi_2$, we select 20 uniformly distributed sample points   which are tensorized into a uniform grid of 400 points. These fixed $\xi_1$, $\xi_2$ grid points are subsequently used in the DSWM model to generate a set of deterministic benchmark models. } \more{To keep this presentation reasonably brief, we select only one spatial point $(25.0m, 25.0m)$ and one time step $t=20s$ to conduct the comparison over its random space and refer to \cite{chenthesis} for further details}.

\begin{figure}[h!]
	\centering
	\subfigure[$\eta_{sswm}, \eta_{benchmark}$.]{\includegraphics[width=0.5\textwidth]{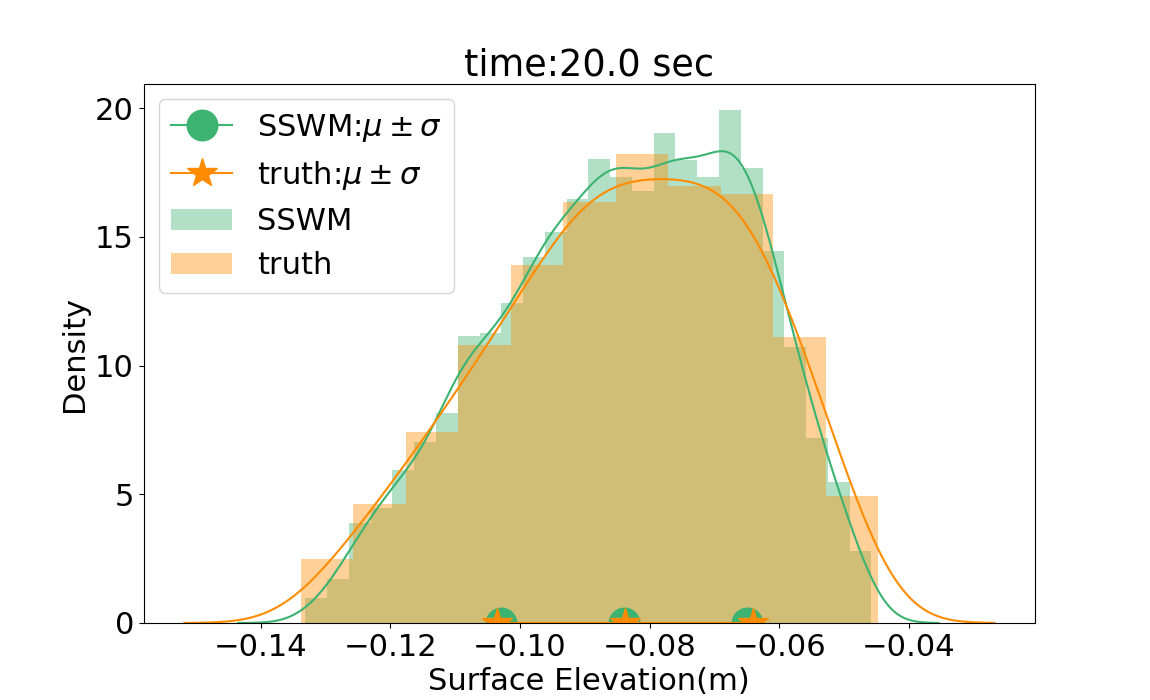}}\hfill
	\subfigure[$u_{sswm}, u_{benchmark}$.]{\includegraphics[width=0.5\textwidth]{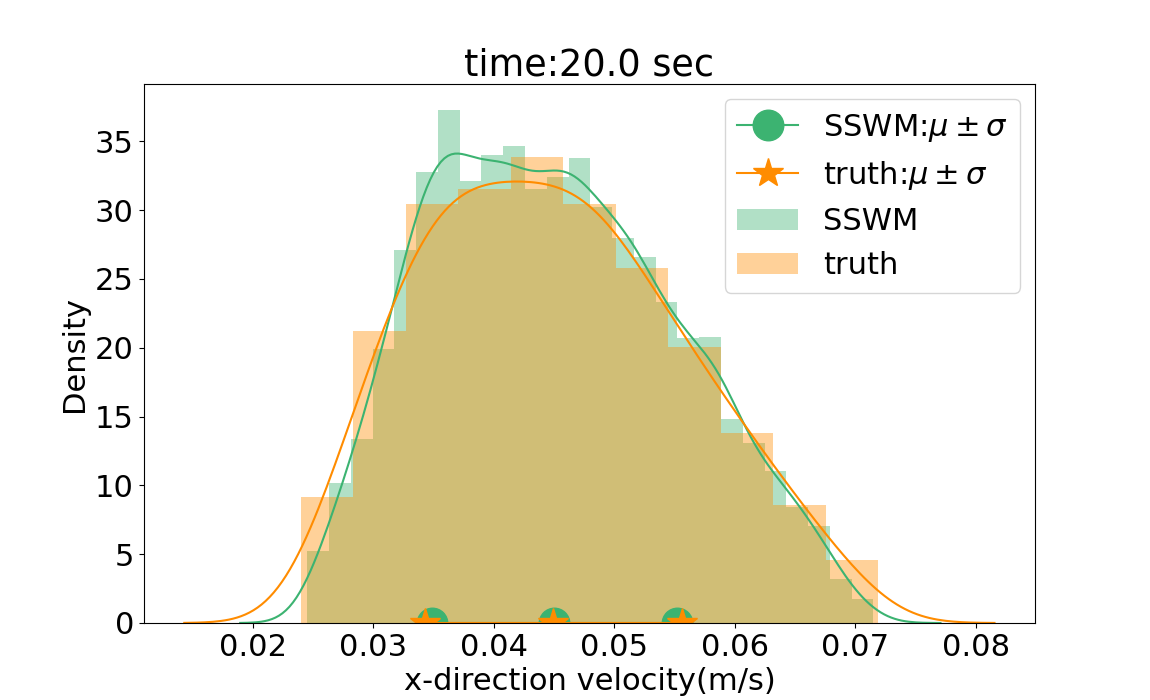}}\hfill
	\caption{Probability density function comparison at the spatial point $(25.0m, 25.0m)$ and time $t=20.0s$.}
	\label{fig:sloshpdf}
\end{figure}

\begin{figure}[h!]
	\centering
	\subfigure[$\eta_{sswm}-\eta_{benchmark}$.]{\includegraphics[width=0.5\textwidth]{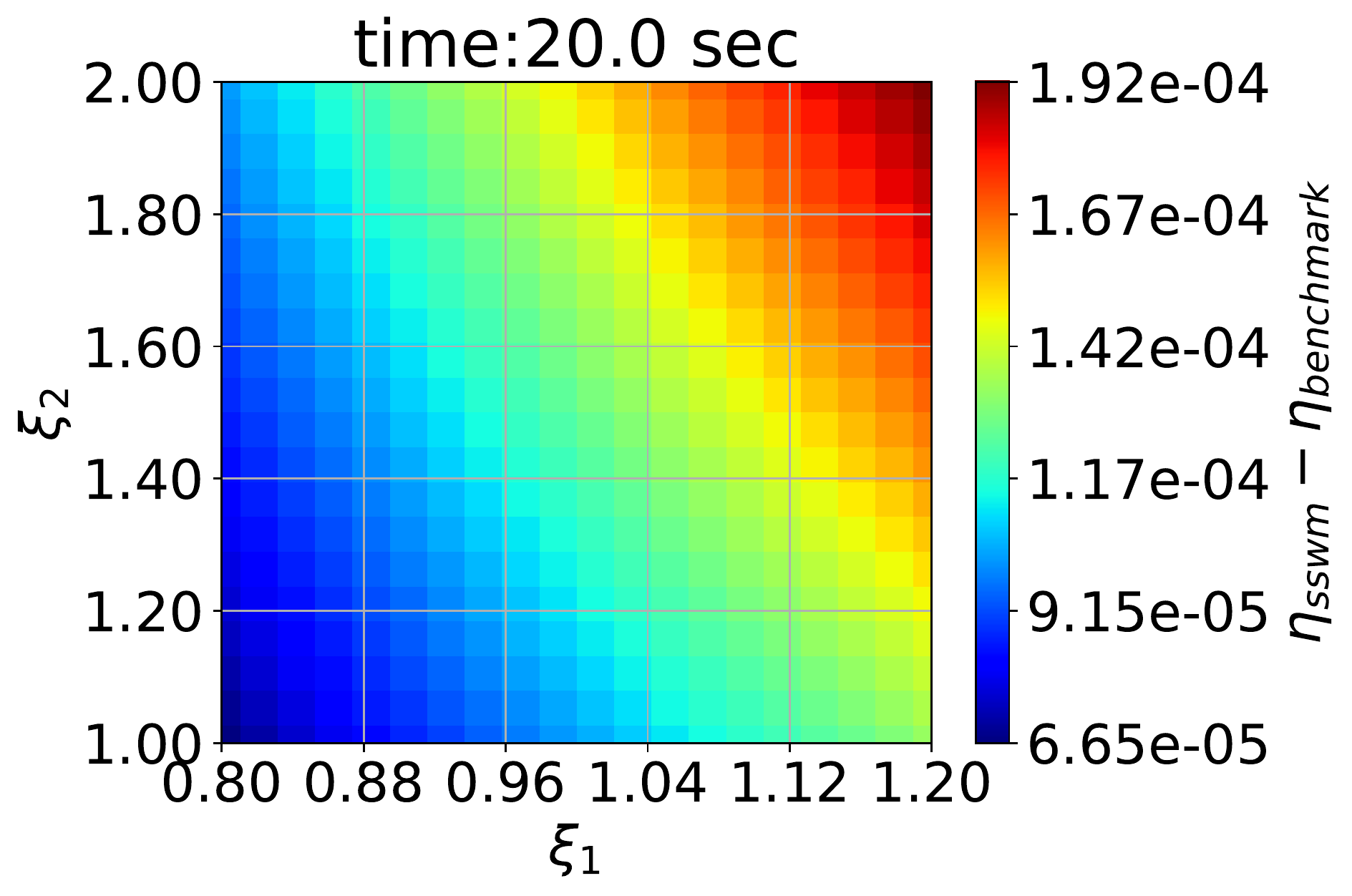}}\hfill
	\subfigure[$u_{sswm}- u_{benchmark}$.]{\includegraphics[width=0.5\textwidth]{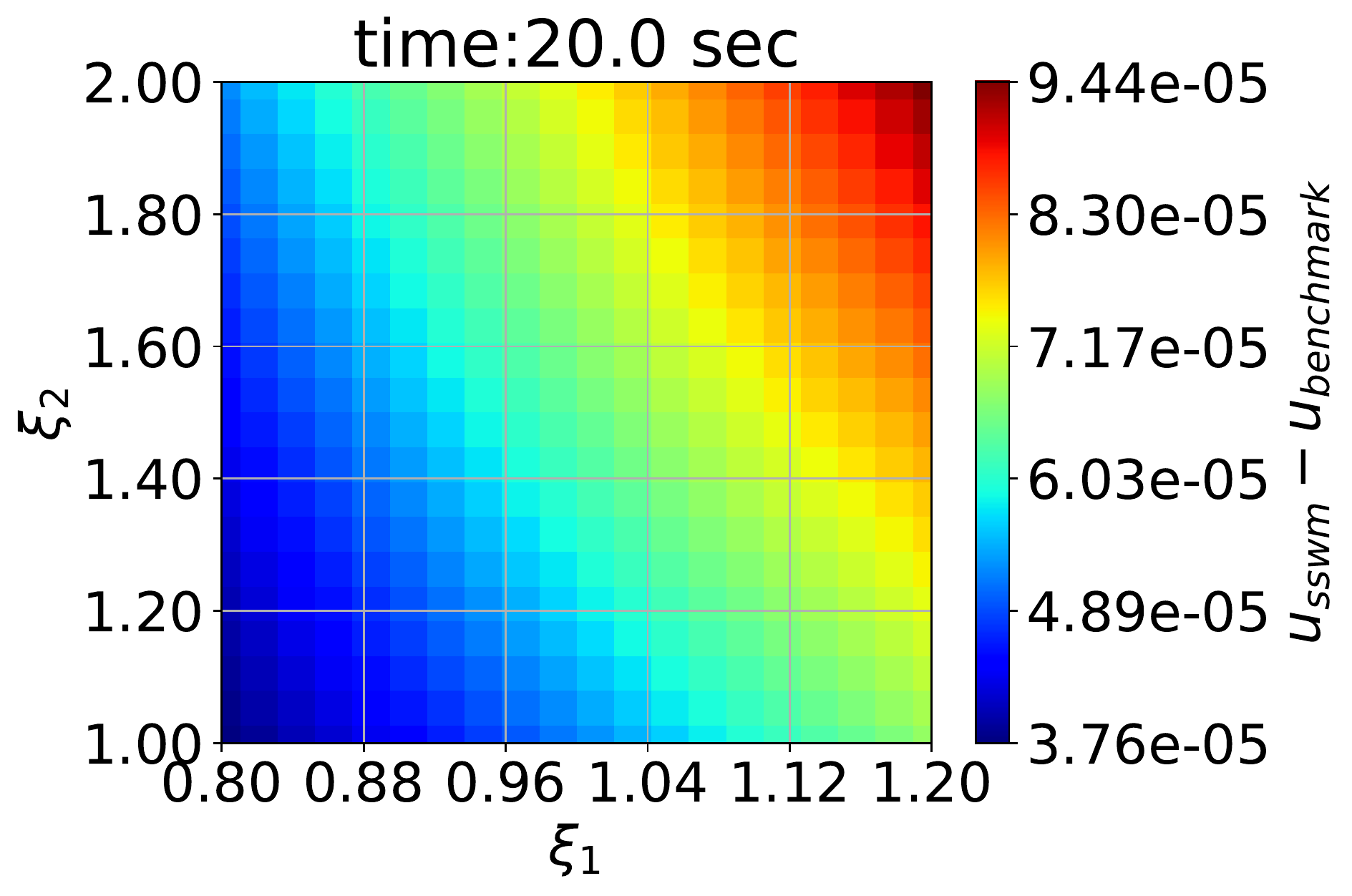}}\hfill
	\caption{Elevation surrogate comparison at the spatial point $(25.0m, 25.0m)$ and time $t=20.0s$.}
	\label{fig:slosheta}
\end{figure}

We first compare the mean and variance of our surrogate against the 400 deterministic realizations. \more{ The computed means $(\mu)$ and standard deviations $(\sigma)$ are presented in Table \ref{table:slosh-table}, where we observe that both mean and deviation for surface elevation $\eta$ and water velocity $u$ are close to each other}. Next, we can observe from Figure~\ref{fig:sloshpdf} that the PDFs also match with each other. Lastly, We compare \two{our} surrogate \oneR{pointwise} against each of the 400 benchmarks.  \moreR{In Figure~\ref{fig:slosheta},} we again observe good agreement for the surface elevation and $x$-direction velocity component, with the absolute errors in the range of $10^{-4}$ to $10^{-5}$. 
\begin{table}[h!]
\centering
\begin{tabular}{|c|c|c|c|c|}
\hline
      & $\mu_\eta$ & $\sigma_\eta$ & $\mu_u$ & $\sigma_u$ \\ \hline
SSWM  & -0.08393       & 0.01895                      & 0.04504       & 0.01019                     \\ \hline
Truth & -0.08378       & 0.01983                      & 0.04496       & 0.01066 \\ \hline                    
\end{tabular}
\caption{Mean and Standard deviation comparison at the spatial point $(25.0m, 25.0m)$ and time $t=20.0s$.}
\label{table:slosh-table}
\end{table}
\subsubsection{Uncertain Bathymetry - Hump Test Case}

\begin{figure}[h!]
	\centering
	\subfigure{\includegraphics[width=0.5\textwidth]{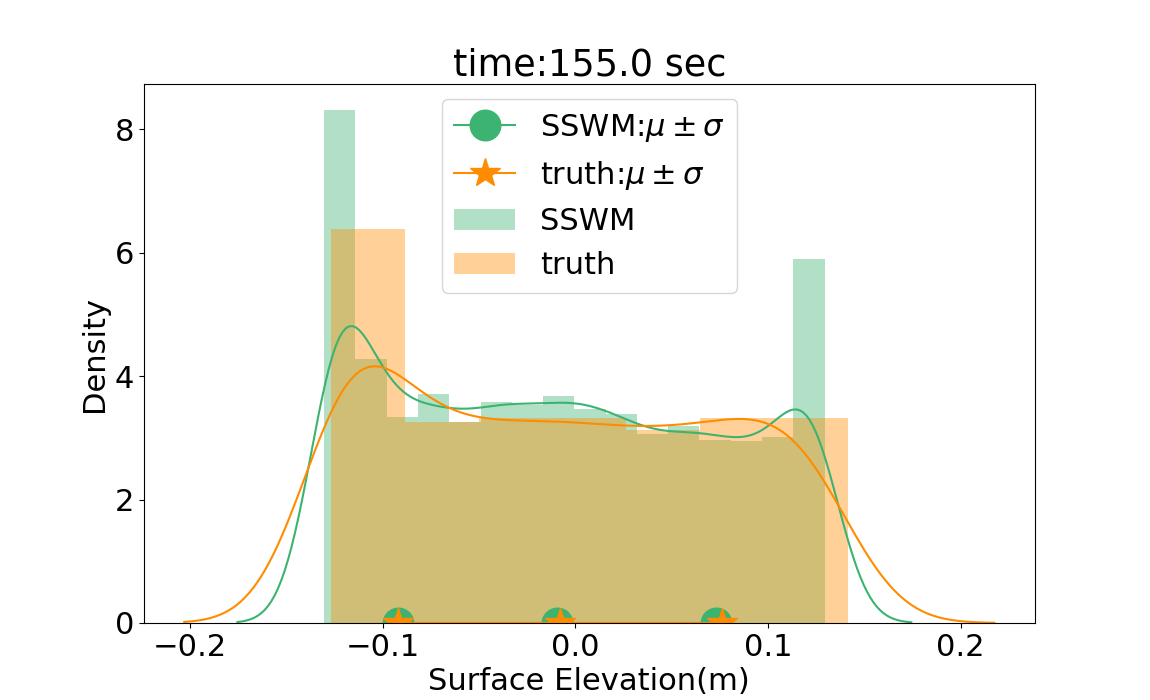}}\hfill
	\subfigure{\includegraphics[width=0.5\textwidth]{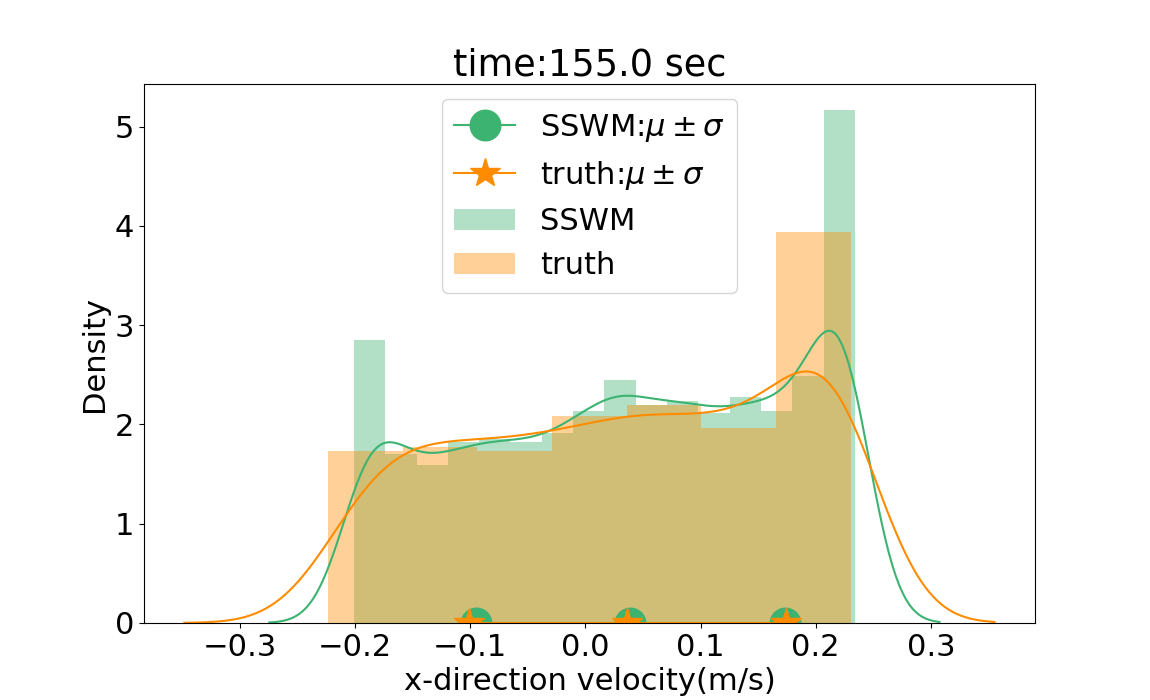}}\hfill
	\caption{Probability density function comparison for elevation and velocity at $(500.0m, 100.0m)$ and time $t=155s$.}
	\label{fig:humppdf}
\end{figure}

\begin{figure}[h!]
	\centering
	\subfigure{\includegraphics[width=0.5\textwidth]{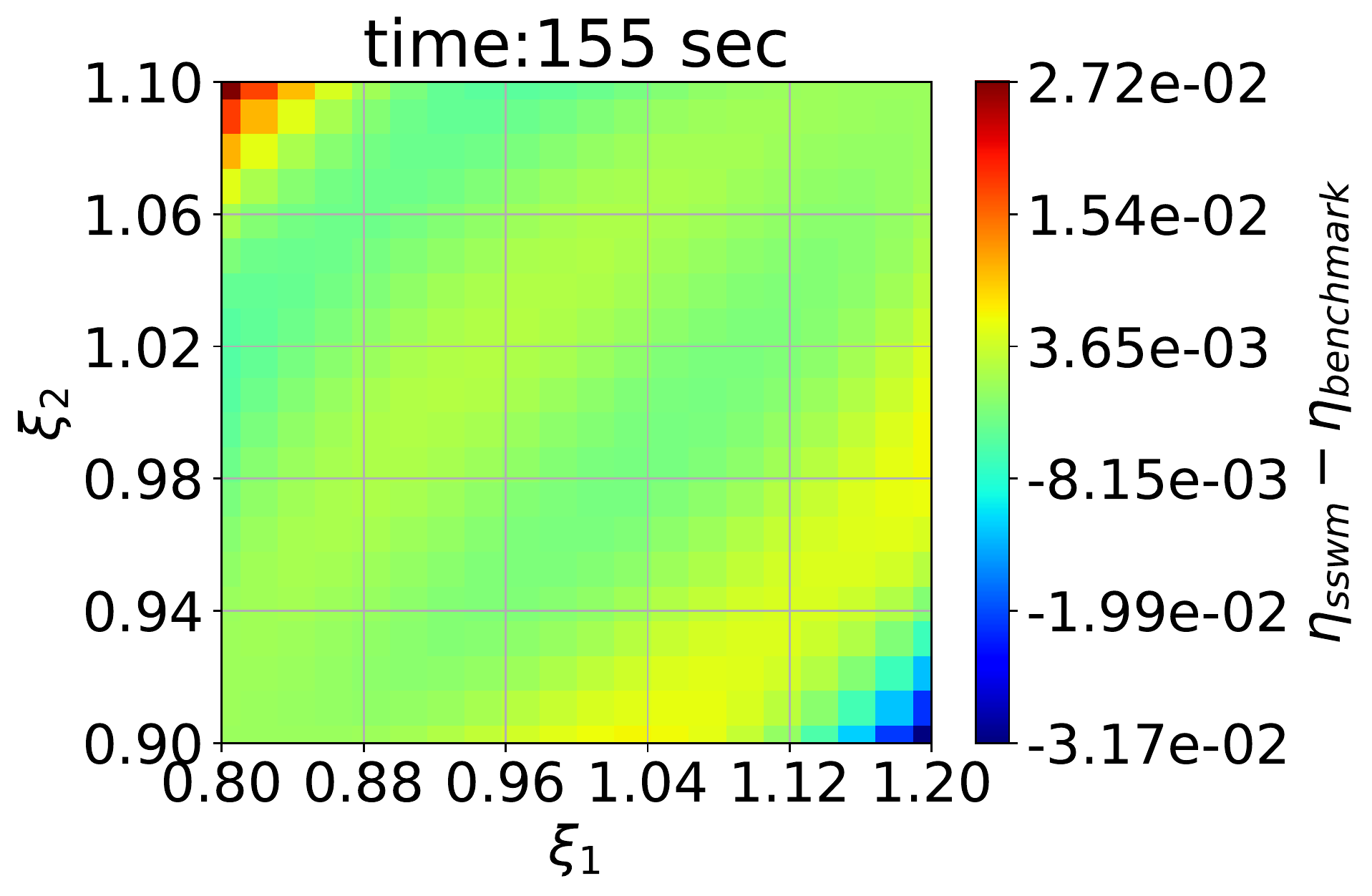}}\hfill
	\subfigure{\includegraphics[width=0.5\textwidth]{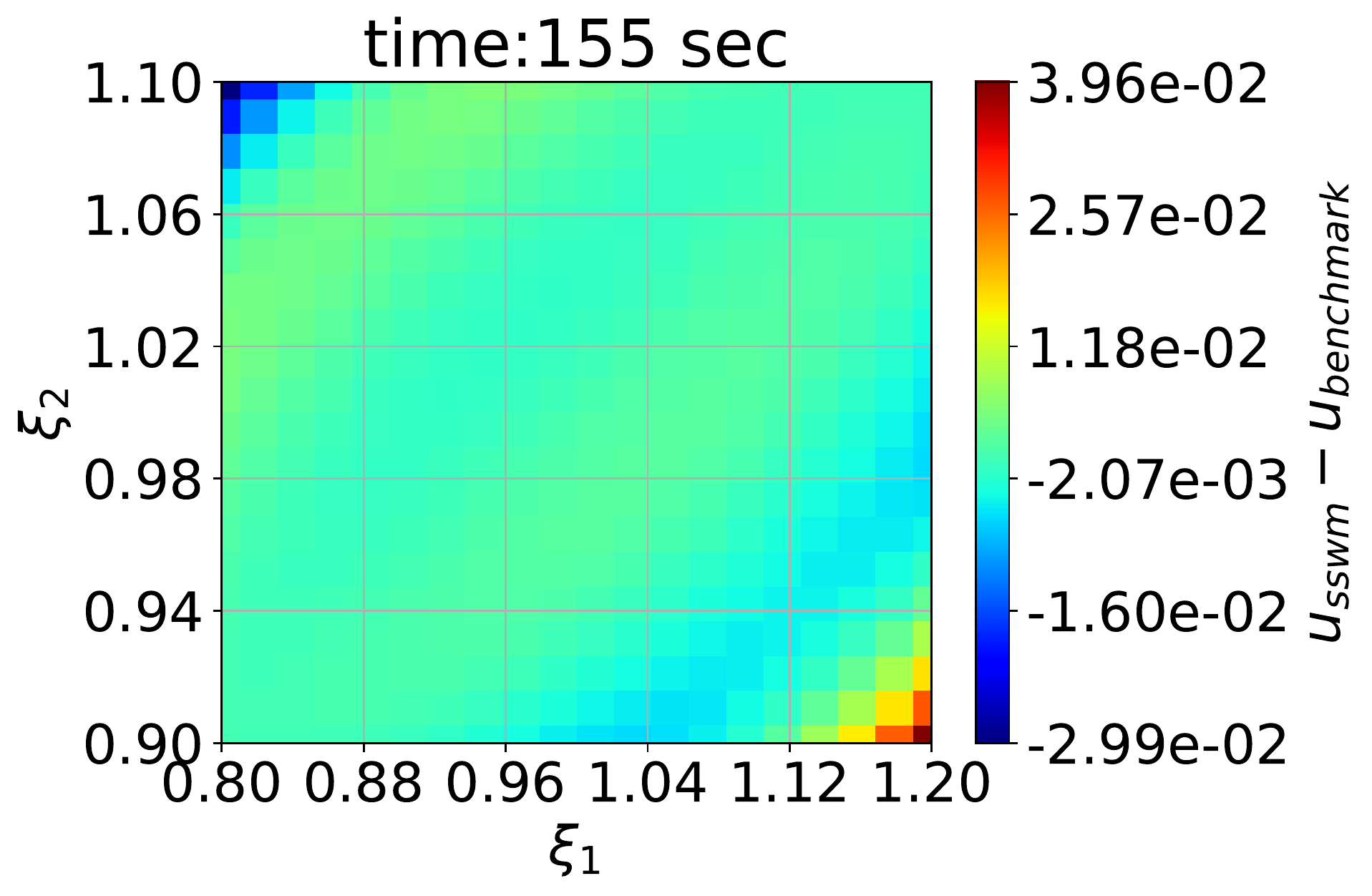}}\hfill
	\caption{Elevation and velocity surrogate at  $(500.0m, 100.0m)$ and time $t=155s$.} 
	\label{fig:humpeta}
\end{figure}

In the hump test case, for the uncertain bathymetry given in~\eqref{eq:bathy_uncertain}, we again \two{select uniformly distributed} $\xi_1$ and $\xi_2$ \two{of} 20 \two{points each} to obtain 400 sample \two{grid points and corresponding deterministic benchmarks}. The spatial point we choose here is $(500.0m, 100.0m)$ and we present PDF and pointwise comparison for both surface elevation and $x$-direction velocity component at time $t=155s$ over the random space in  Figures~\ref{fig:humppdf} and~\ref{fig:humpeta}, respectively. \more{In Table~\ref{table:hump-table}, we present the corresponding mean and standard deviations.}
\begin{table}[h!]
\centering
\begin{tabular}{|c|c|c|c|c|}
\hline
      & $\mu_\eta$ & $\sigma_\eta$ & $\mu_u$ & $\sigma_u$ \\ \hline
SSWM  & -0.0095699       & 0.08234                      & 0.03914      & 0.13452                     \\ \hline
Truth & -0.0080956       & 0.08389                      & 0.03652       & 0.13792 \\ \hline                    
\end{tabular}
\caption{Mean and Standard deviation comparison at the spatial point $(500.0m, 100.0m)$ and time $t=155s$.}
\label{table:hump-table}
\end{table}
Here, we observe good matches between the surrogate and benchmark over the random space for both surface elevation and $x$-direction water velocity, with the absolute errors in the range of $10^{-2}$ to $10^{-3}$.

\subsubsection{Uncertain Boundary Condition - Inlet Test Case}

\begin{figure}[h!]
	\centering
	\subfigure[Elevation PDF at $1.061$ days.]{\includegraphics[width=0.45\textwidth]{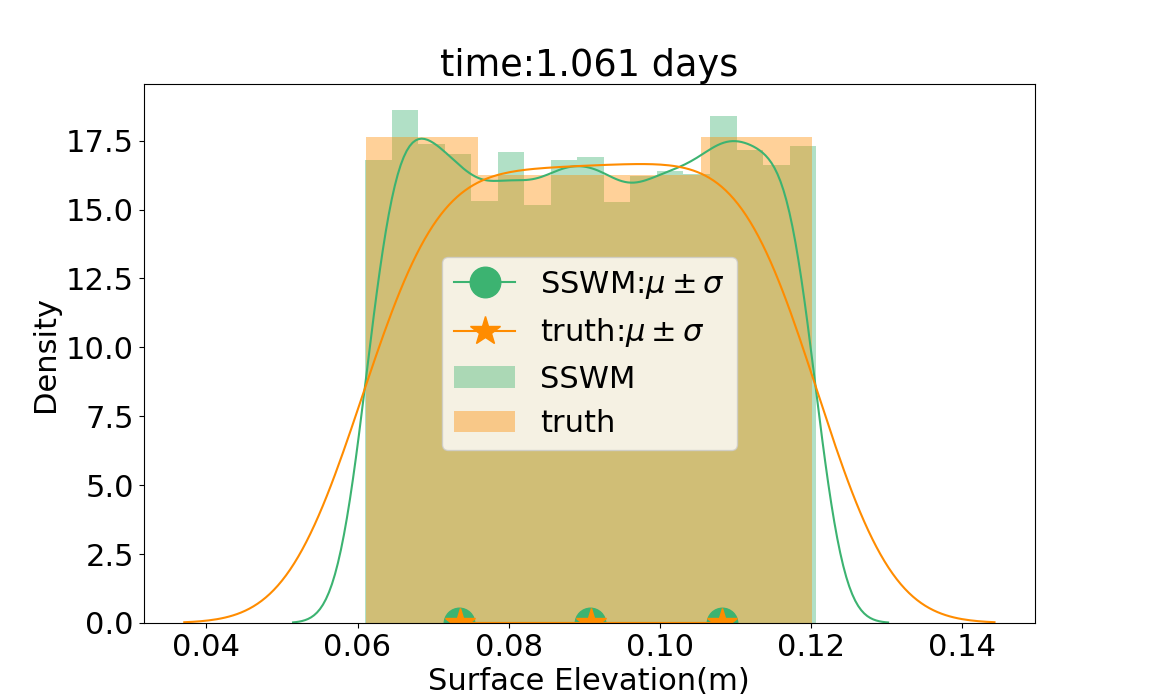}}\hfill
	\subfigure[Elevation PDF at $2.095$ days.]{\includegraphics[width=0.45\textwidth]{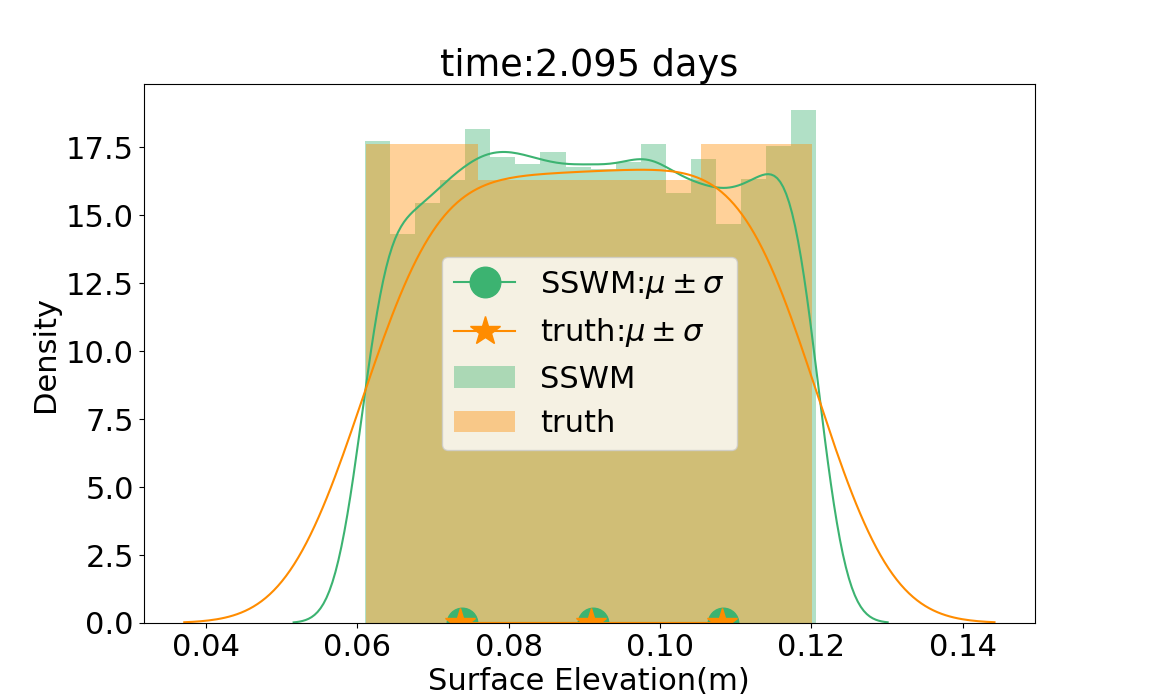}}\hfill
	\subfigure[$x$-velocity PDF at $1.061$ days.]{\includegraphics[width=0.45\textwidth]{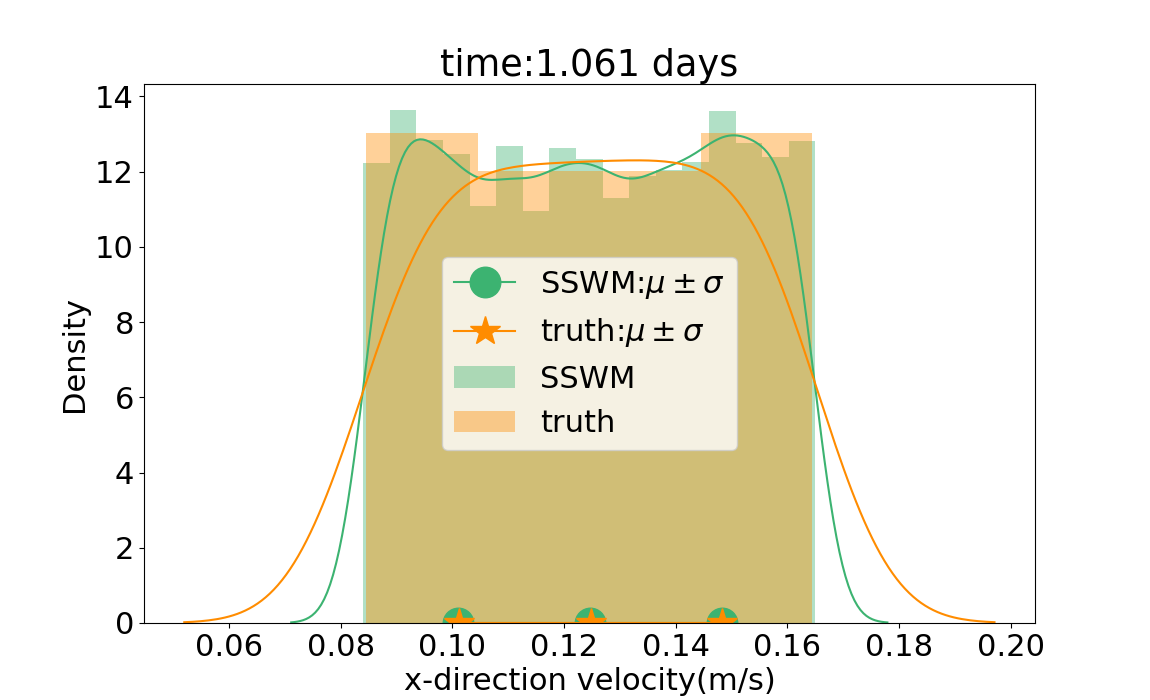}}\hfill
	\subfigure[$x$-velocity PDF at $2.095$ days.]{\includegraphics[width=0.45\textwidth]{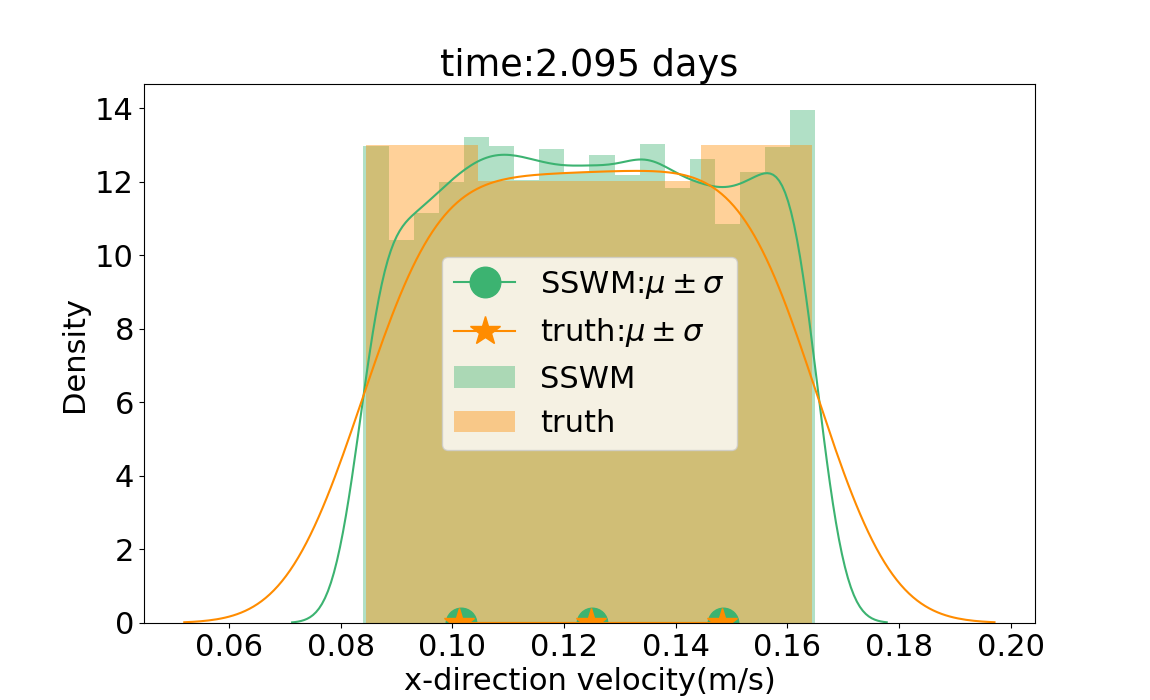}}\hfill
	\caption{Probability density function comparison for elevation and velocity at the spatial point $(0.0m, 0.0m)$ at time $t=1.061$ days and $t=2.095$ days.}
	\label{fig:inletpdf}
\end{figure}

\two{To generate deterministic benchmarks in the inlet test case, we select 50 uniform grid points from $\xi_1 \sim U(1.0, 2.0)$ for the uncertain boundary condition. } The spatial point of interest is $(0.0m, 0.0m)$, which is located at the entrance of the channel, see Figure~\ref{fig:inletmesh}, and we select two time steps, at $1.061$ and $2.095$ days, respectively. In  Figure~\ref{fig:inletpdf}, we compare the surrogate and the benchmark at the spatial point for all $\xi_1$ over probability space \more{and in Table~\ref{table:inlet-table}, the corresponding computed mean and standard deviation.} For the selected times, we observe near perfect agreement in the overall probability space for both surface elevation and $x$-direction velocity. 
\begin{table}[h!]
\centering
\begin{tabular}{|cc|c|c|c|c|}
\hline
\multicolumn{2}{|c|}{}                                      & $\mu_\eta$ & $\sigma_\eta$ & $\mu_u$ & $\sigma_u$ \\ \hline
\multicolumn{1}{|c|}{\multirow{2}{*}{t=1.061 days}} & SSWM  & 0.09082       & 0.01742                  & 0.124625      & 0.02361
            \\ \cline{2-6} 
\multicolumn{1}{|c|}{}                              & Truth & 0.09088       & 0.01738                  & 0.124773       & 0.02352                        \\ \hline
\multicolumn{1}{|c|}{\multirow{2}{*}{t=2.095 days}} & SSWM  & 0.09108       & 0.01725        
        & 0.12501      & 0.02339                    \\ \cline{2-6} 
\multicolumn{1}{|c|}{}                              & Truth & 0.09087       & 0.01737        
        & 0.12478       & 0.02353                        \\ \hline
\end{tabular}
\caption{Mean and Standard deviation comparison at spatial point $(0.0m, 0.0m)$ at time $t=1.061$ days and $t=2.095$ days.}
\label{table:inlet-table}
\end{table}

\subsubsection{Uncertain Wind Drag Parameter - Hurricane Harvey Test Case}

\begin{figure}[h!]
	\centering
	\subfigure[Elevation surrogate at $t=1.319$ days]{\includegraphics[width=0.5\textwidth]{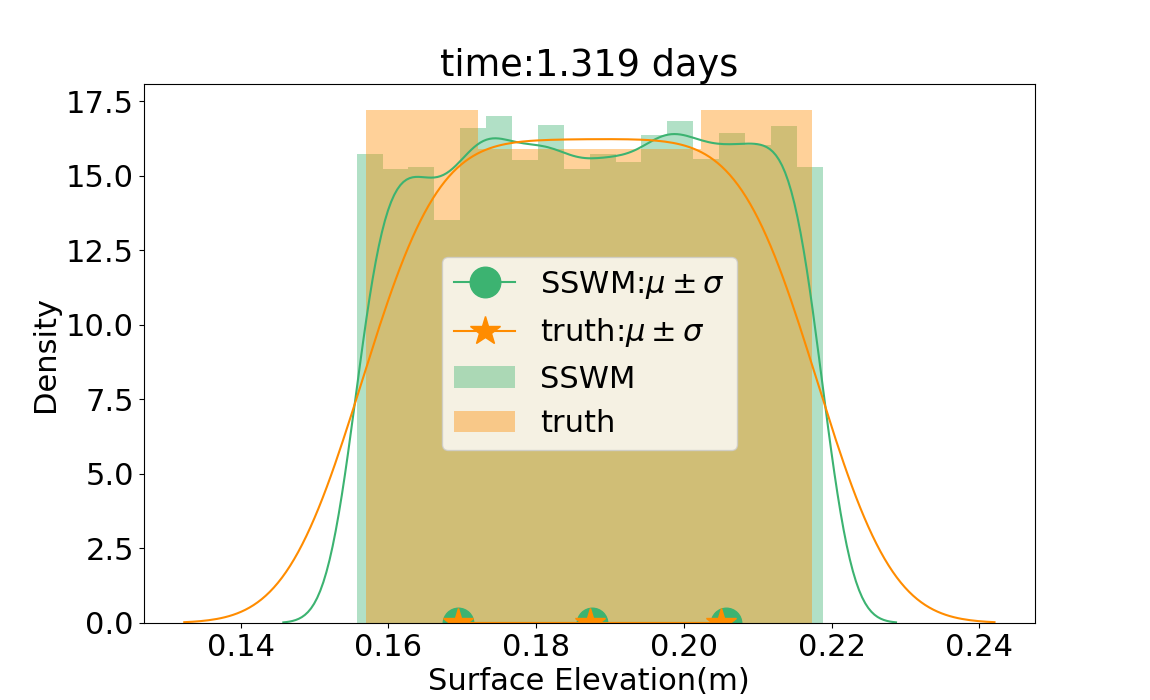}}\hfill
	\subfigure[Elevation surrogate at $t=4.139$ days.]{\includegraphics[width=0.5\textwidth]{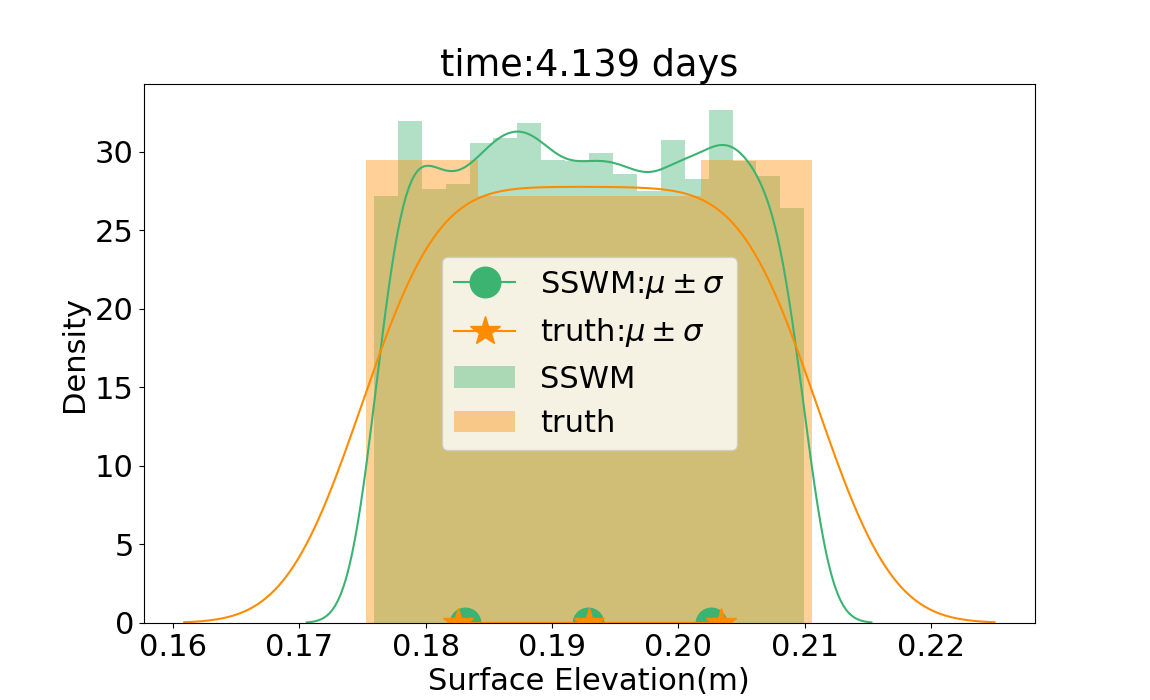}}\hfill
	\subfigure[$x$-velocity surrogate at $t=1.319$ days.]{\includegraphics[width=0.5\textwidth]{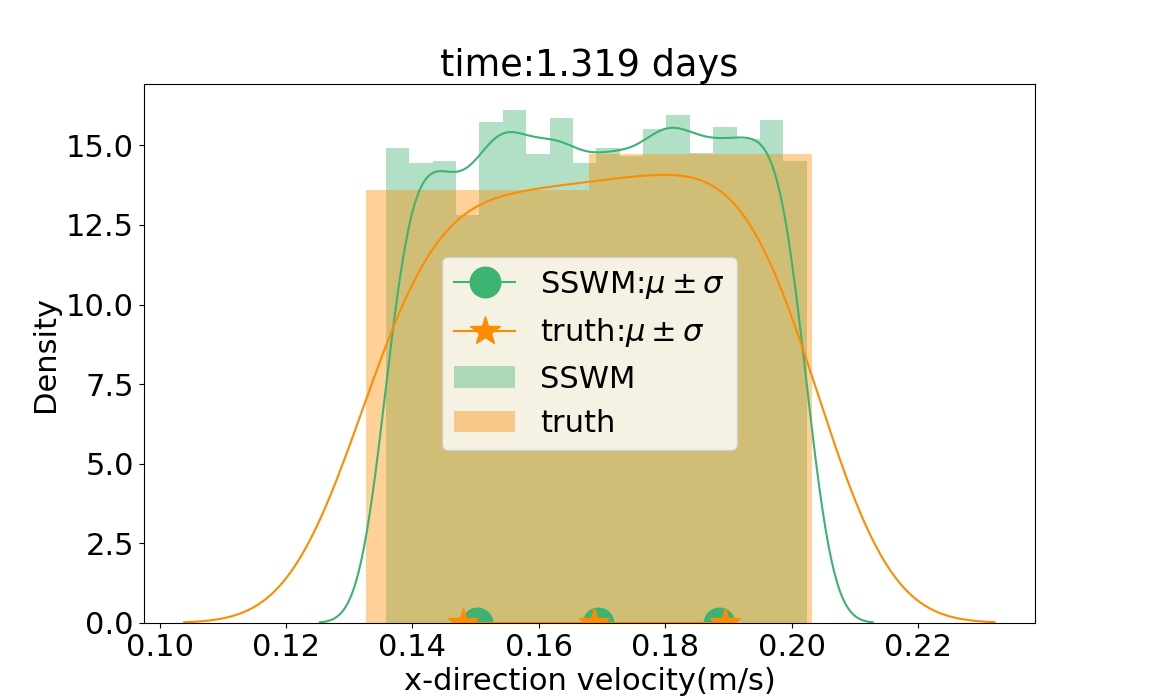}}\hfill
	\subfigure[$x$-velocity surrogate at $t=4.139$ days.]{\includegraphics[width=0.5\textwidth]{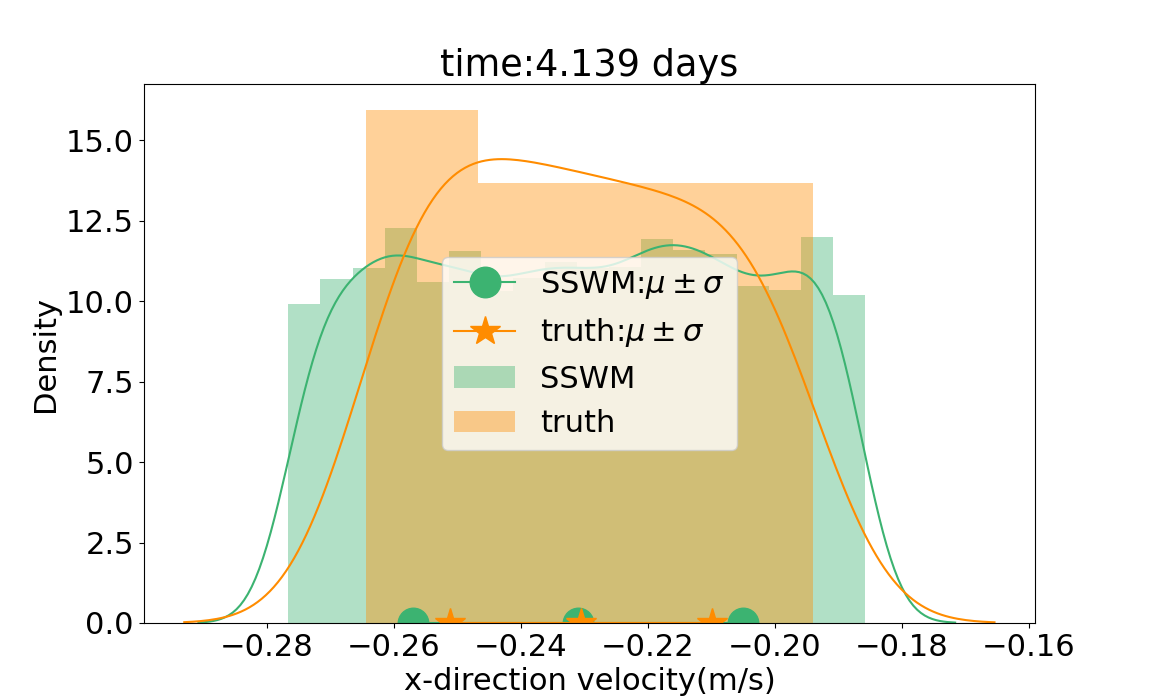}}\hfill
	\subfigure[$y$-velocity surrogate at $t=1.319$ days.]{\includegraphics[width=0.5\textwidth]{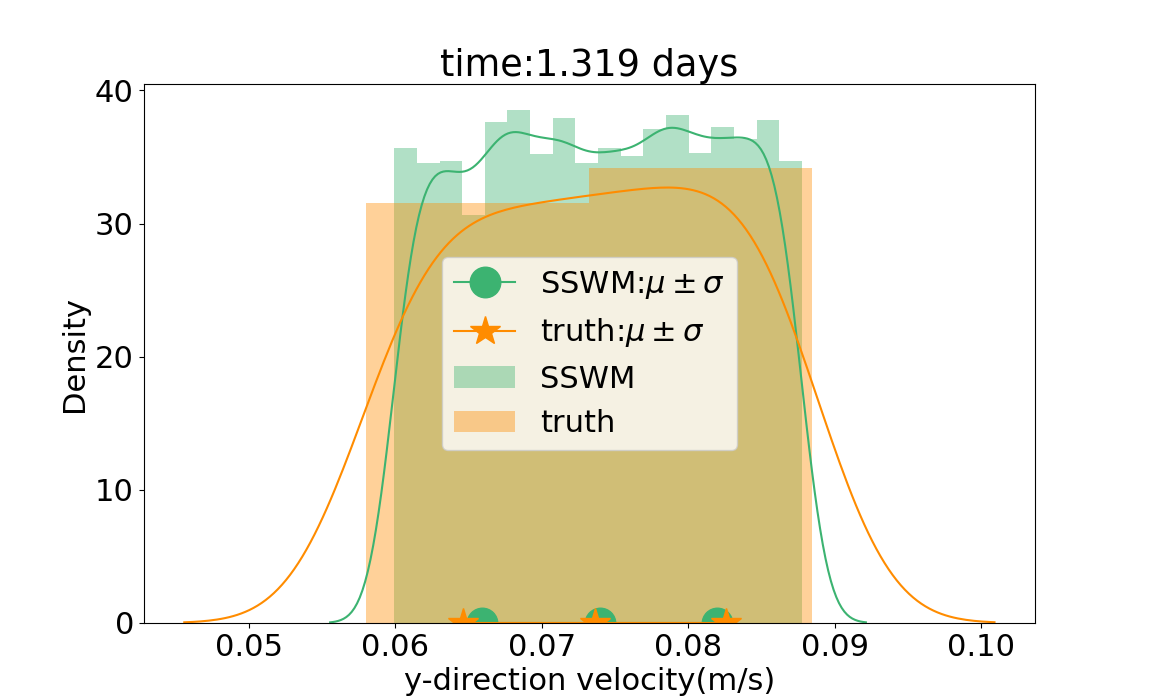}}\hfill
	\subfigure[$y$-velocity surrogate at $t=4.139$ days.]{\includegraphics[width=0.5\textwidth]{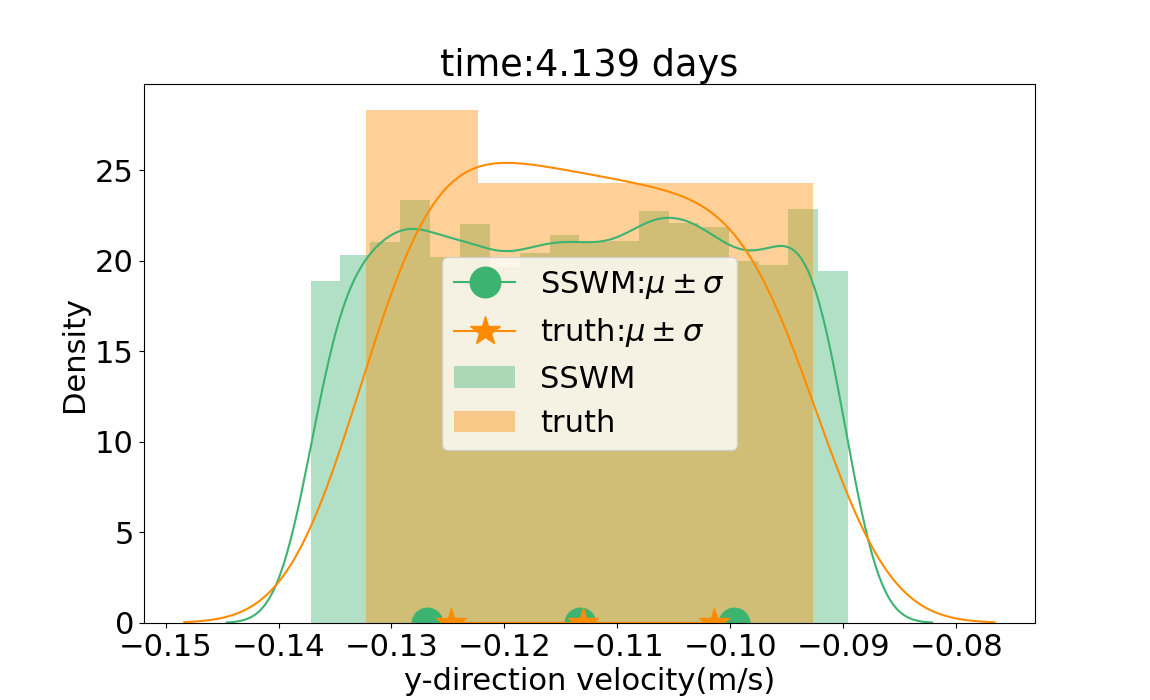}}\hfill
	\caption{Probability density function comparison for Hurricane Harvey at spatial point $(-95.24^{\circ},  28.85^{\circ})$ at time $t=1.319$ days and $t=4.139$ days.}
	\label{fig:harveypdf}
\end{figure}

As a final verification of the SSWM, we consider the Hurricane Harvey test case. We again uniformly  \two{ distribute 50 points for $\xi_1 \sim U(0.8, 1.2)$, }  the uncertain wind drag parameter and compare the resulting surrogate against the benchmarks obtained from running the deterministic model at \two{the} corresponding \two{points}. We select a spatial point located in Galveston Bay close to the ship channel with longitude and latitude $(-95.24^{\circ},  28.85^{\circ})$, and two time steps $t=1.319$ days and $t=4.139$ days. We present the comparison of elevation, $x$-direction velocity, and $y$-direction velocity over the random space with respect to $\xi_1$ in Figure~\ref{fig:harveypdf} \more{with corresponding mean and standard deviation in Table~\ref{table:harvey-table}}. The surface elevation agrees very well in the random space and we only observe minor discrepancies for both velocity components in the table and figures. 
%
%
\begin{table}[h!]
\begin{tabular}{|cc|c|c|c|c|c|c|}
\hline
\multicolumn{2}{|c|}{}                                      & $\mu_\eta$ & $\sigma_\eta$ & $\mu_u$ & $\sigma_u$ & $\mu_v$ & $\sigma_v$ \\ \hline
\multicolumn{1}{|c|}{\multirow{2}{*}{t=1.319 days}} & SSWM  & 0.18751        & 0.01816                      & 0.16938       & 0.01915                    & 0.07396      & 0.00801                     \\ \cline{2-8} 
\multicolumn{1}{|c|}{}                              & Truth & 0.18723        & 0.01778                      & 0.16871       & 0.02078                    & 0.07360      & 0.00897                     \\ \hline
\multicolumn{1}{|c|}{\multirow{2}{*}{t=4.139 days}} & SSWM  & 0.19287        & 0.009725                      & -0.23100      & 0.02593                    & -0.11327      & 0.01359 \\ \cline{2-8} 
\multicolumn{1}{|c|}{}                              & Truth & 0.19294        & 0.010392                      & -0.23050     & 0.02066                    & -0.11307      & 0.01162 \\ \hline
\end{tabular}
\caption{Mean and Standard deviation comparison at spatial point $(-95.24^{\circ},  28.85^{\circ})$ at time $t=1.319$ days and $t=4.139$ days.}
\label{table:harvey-table}
\end{table}

\subsection{Notes on the SSWM Verification \two{and Stochastic Convergence}} \label{sec:verification_notes}

We have introduced a SSWM and a set of cross-mode stabilization methods applicable to stochastic systems. For this newly developed model, we provide a comprehensive verification process of both its deterministic and stochastic versions in the appendix and preceding section. The verification process demonstrates the effectivity of the proposed cross-mode stabilization methods, as well as the correctness of the stochastic model. For both ideal and larger physically relevant test cases, we observe good agreement for the proposed SSWM and only  small discrepancies in some cases. This indicates that the approximations made in the SSWM do not deteriorate the solution, i.e., the computed surrogates. 

\begin{figure}[h!]
	\centering
	\subfigure[][\two{ PDFs with gPC dimension one.}\label{fig:Convergence_compare_dimensionA}]{\includegraphics[width=0.48\textwidth]{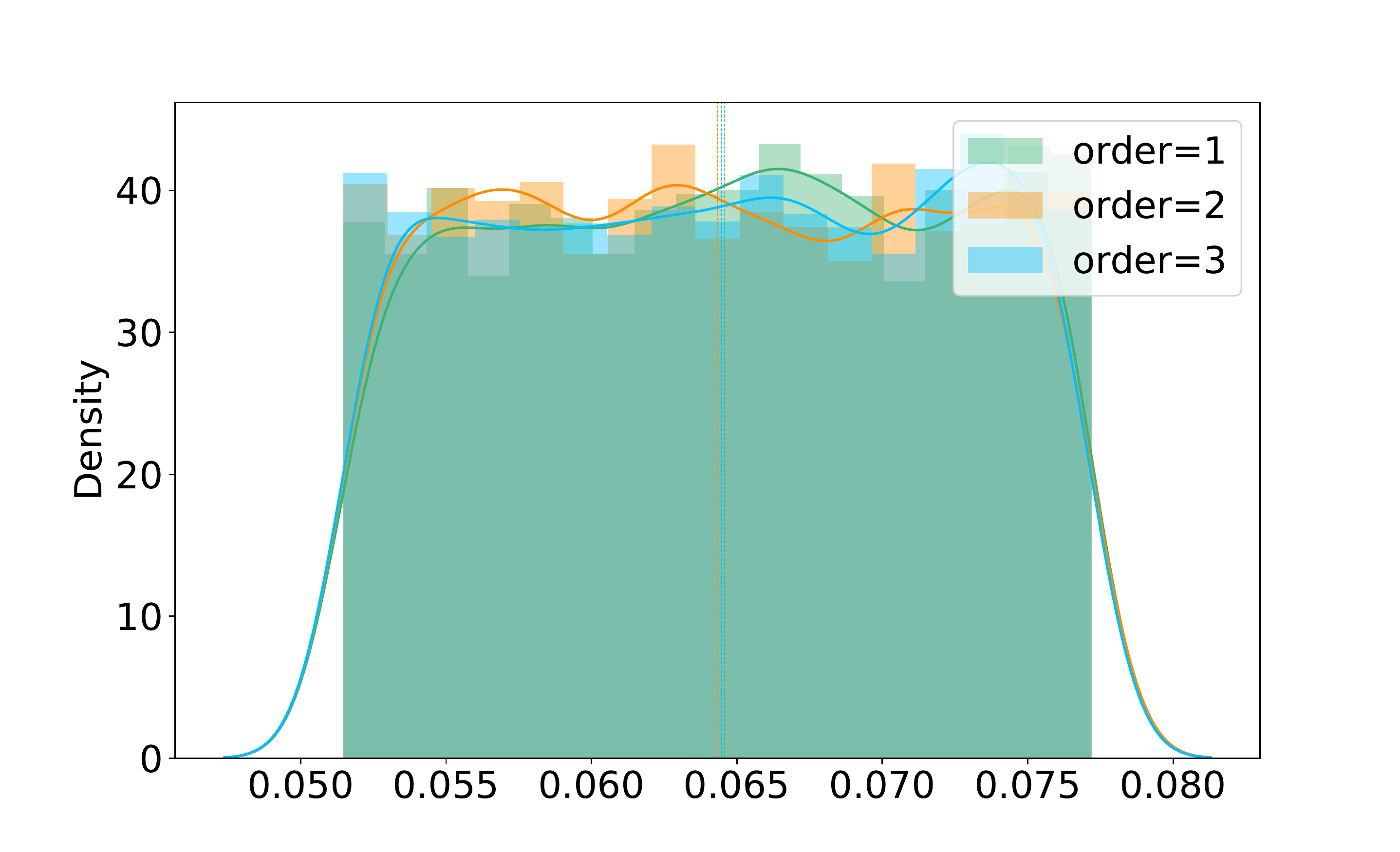}}\hfill
	\subfigure[][\two{ PDFs with gPC dimension two.\label{fig:Convergence_compare_dimensionB}}]{\includegraphics[width=0.48\textwidth]{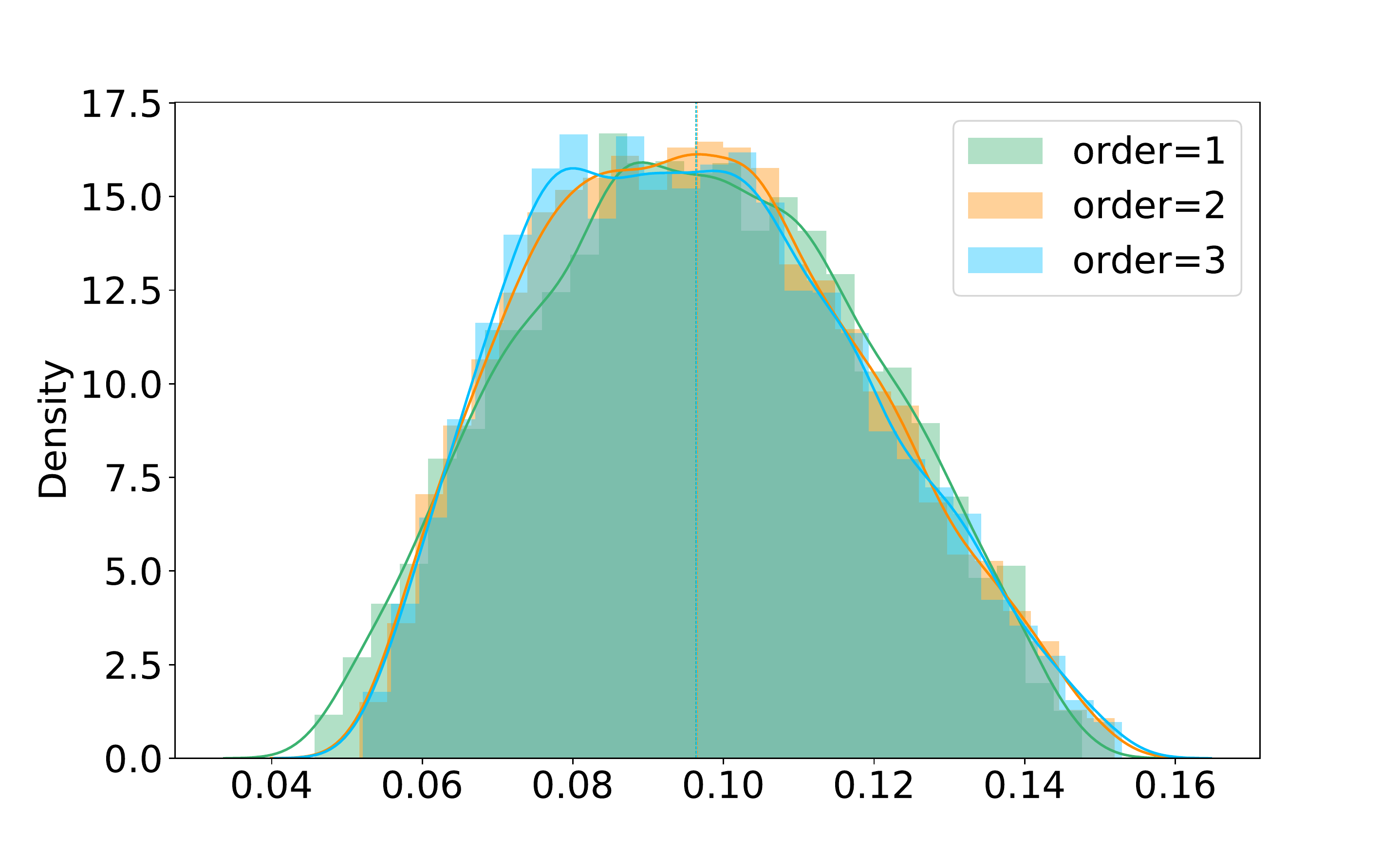}}\hfill
	\caption{\two{ PDF comparison of $\eta$ surface elevation at $(25.0m, 25.0m)$ and $t=1.0s$.}}
	\label{fig:Convergence_compare_dimension}
\end{figure}

\two{ The last point we address in this verification process is convergence  under different stochastic orders and dimensions. We highlight this by again considering the slosh test case of Section~\ref{sec:sloshtest} and focus on the surface elevation at $(25.0m,25.0m)$ and $t=1.0s$. Recall that the uncertain initial condition is of the form $\eta = 0.1 \xi_1 \xi_2 \cos ( \pi x /100.0) $. We consider cases in which the dimension of the gPCs is two as originally introduced in Section~\ref{sec:sloshtest} and one by dropping the dependence on $\xi_2$. For both dimensions of gPCs, we consider linear, quadratic, and cubic stochastic order. The resulting PDFs are shown in Figures~\ref{fig:Convergence_compare_dimensionA} and~\ref{fig:Convergence_compare_dimensionB} for gPCs of dimension one and two, respectively.  }
\two{ These figures indicate that for one-dimensional gPCs, the support of the PDFs is essentially identical for all stochastic orders. Whereas in the two-dimensional case, the support of the PDFs for second and third order cases are nearly identical. In this case we can conclude convergent with linear stochastic order for one-dimensional gPCs and quadratic stochastic order for the two-dimensional case.  This indicates that the stochastic order used must be determined on a case-by-case basis. The orders we use herein are based on extensive numerical experimentation performed in a similar fashion to this brief experiment.  }

\two{Based on these experiments and convergence results}, we conclude that the stochastic model is sufficiently verified and capable to produce reliable surrogates for the further statistical analysis.  In the following two sections, we validate the SSWM through a detailed statistical analysis as well as hindcasting of the two considered hurricanes. 



\section{Visualization and Analysis of the Second-Order Stochastic Process} \label{sec:visual_analysis}

In this section, we will explore the underlying properties of the  stochastic process utilizing the numerical tests introduced in Section~\ref{sec:testcases}. In each of the following three subsections we investigate the magnitude of the variance of the model outputs, the PDF, and the maximum variance magnitude, respectively.


\subsection{The Variation of Variance} \label{sec:var_of_variance}


Based on the verification of our SSWM in Section~\ref{sec:verify_validate}, we  now use it to compute higher order moments of the output random variables. Among  the higher-order moments, the second-order central moment, i.e., the variance, is of particular importance. Hence, we will explore the relationship between the variance in the model inputs and the variance in the model outputs as well as what affects the magnitude of the variance in the model outputs.

\begin{figure}[h!]
	\centering
	\subfigure[][Deviation of $eta$ with $\xi_2 \sim U(1.0, 2.0)$.]{\includegraphics[width=0.45\columnwidth]{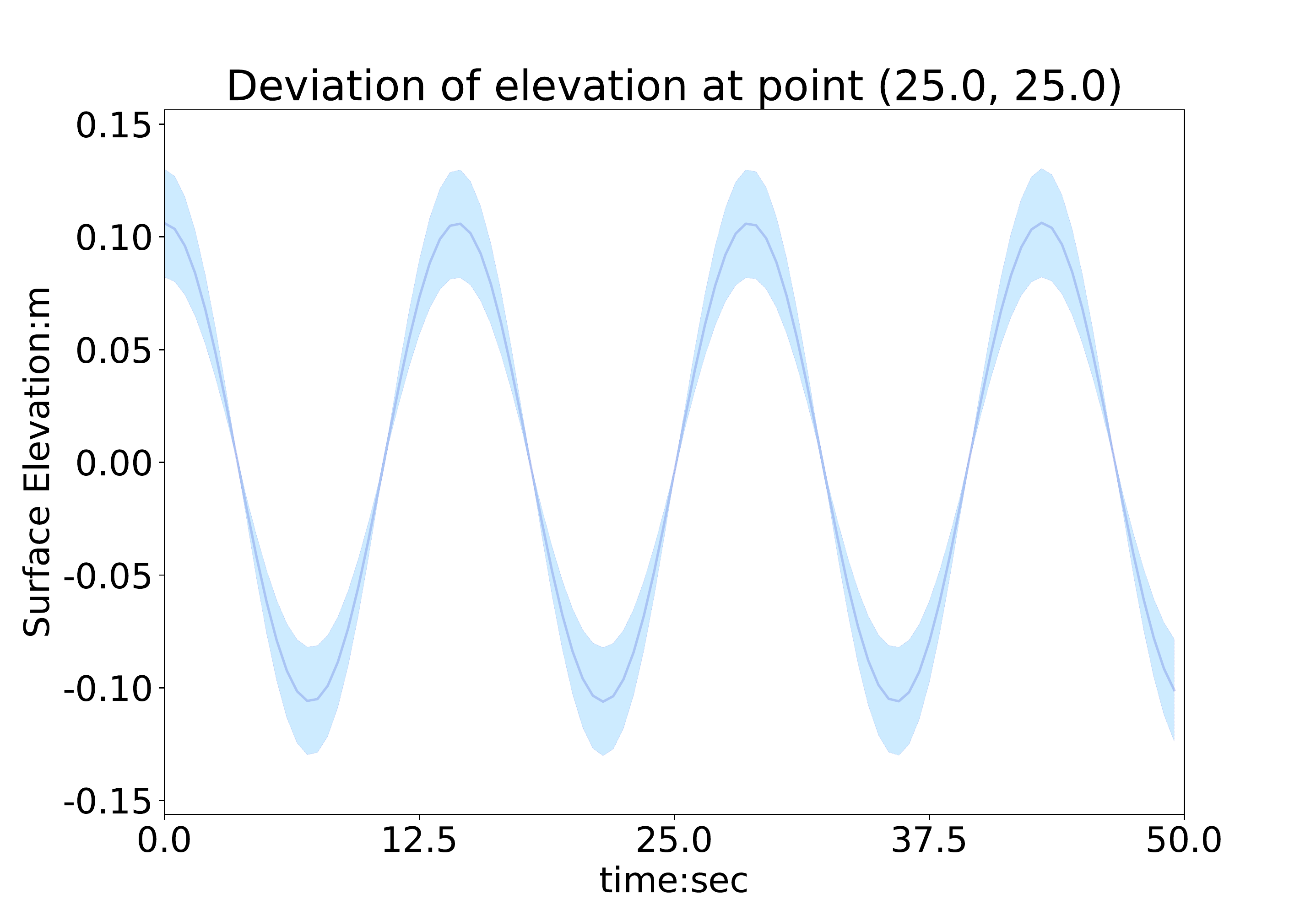}}\hfill
	\subfigure[][Deviation of $u$ with $\xi_2 \sim U(1.0, 2.0)$.]{\includegraphics[width=0.45\columnwidth]{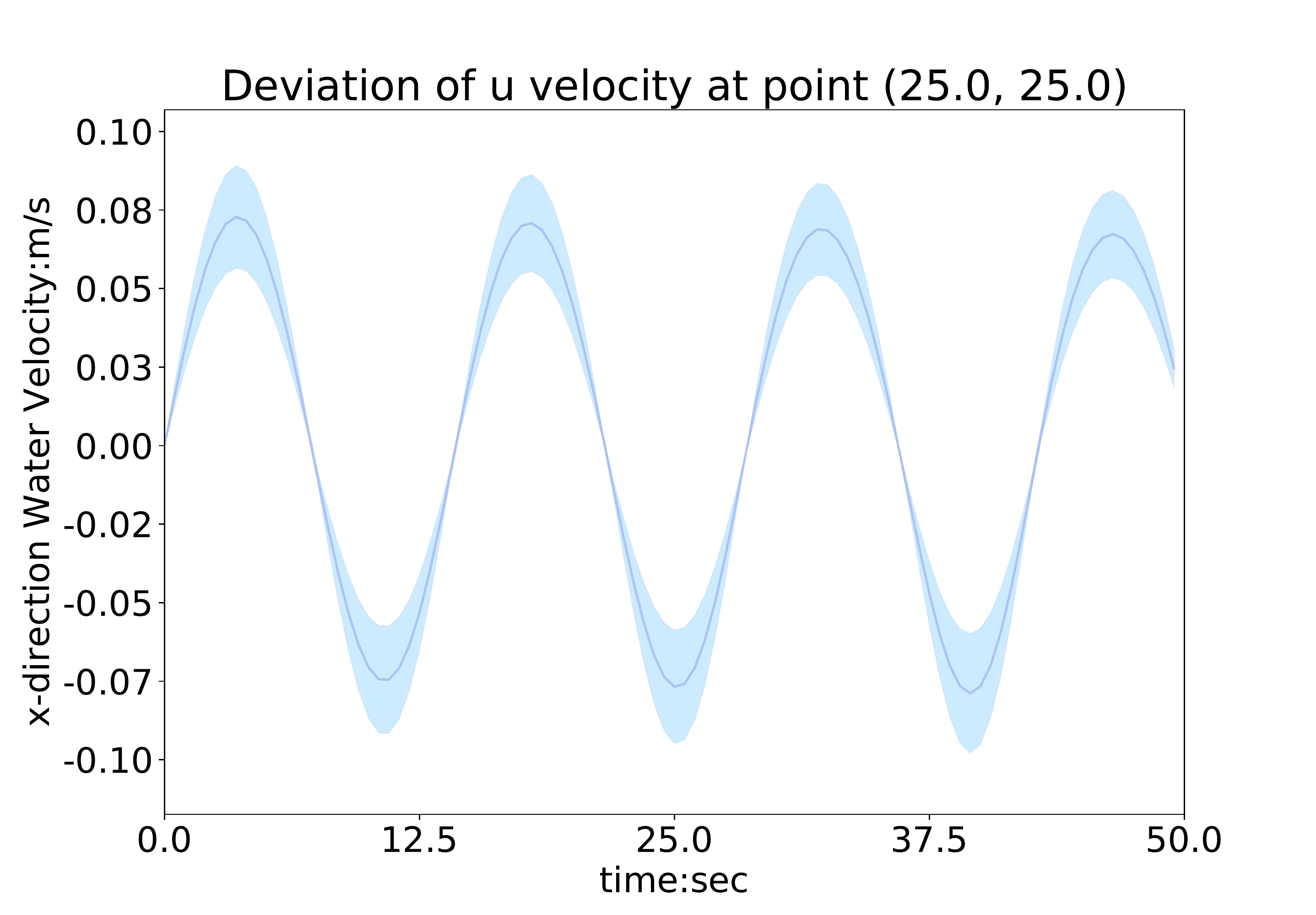}}\hfill
	\subfigure[][Deviation of $eta$ with $ \xi_2 \sim U(0.5, 2.5)$.]{\includegraphics[width=0.45\columnwidth]{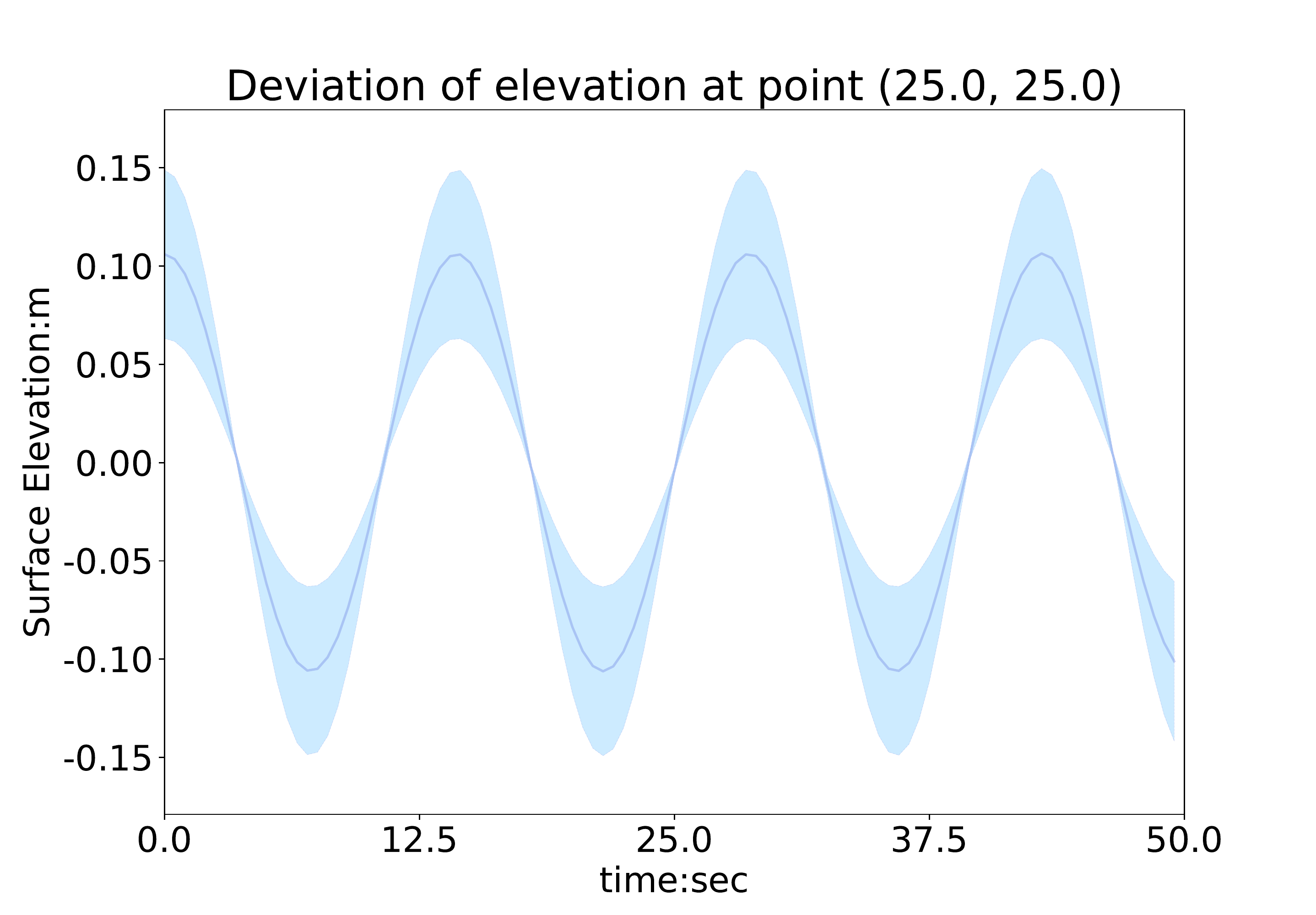}}\hfill
	\subfigure[][Deviation of $u$ with $ \xi_2 \sim U(0.5, 2.5)$.]{\includegraphics[width=0.45\columnwidth]{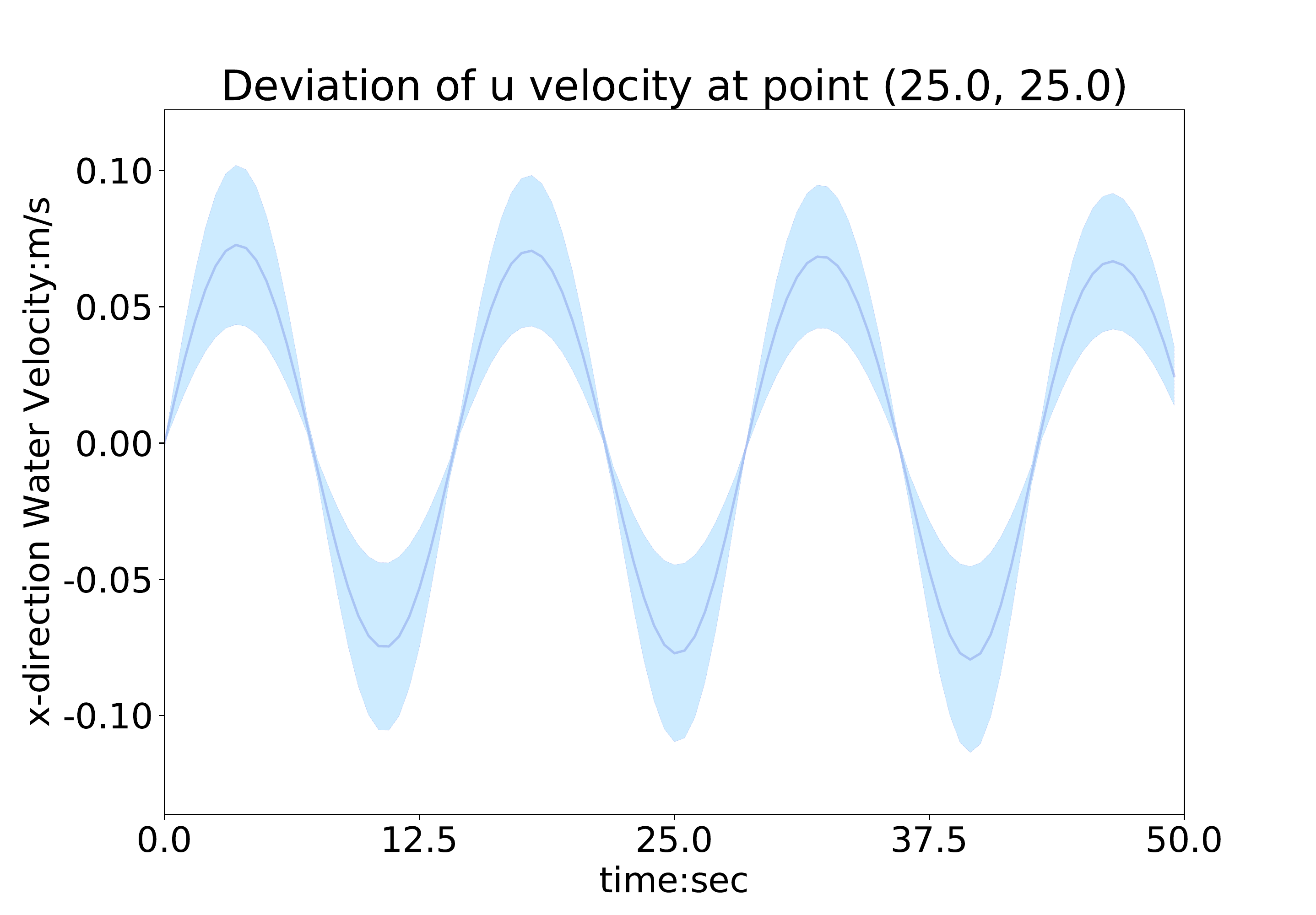}}\hfill
	\subfigure[][Deviation of $eta$ with $ \xi_2 \sim U(0.25, 2.75)$.]{\includegraphics[width=0.45\columnwidth]{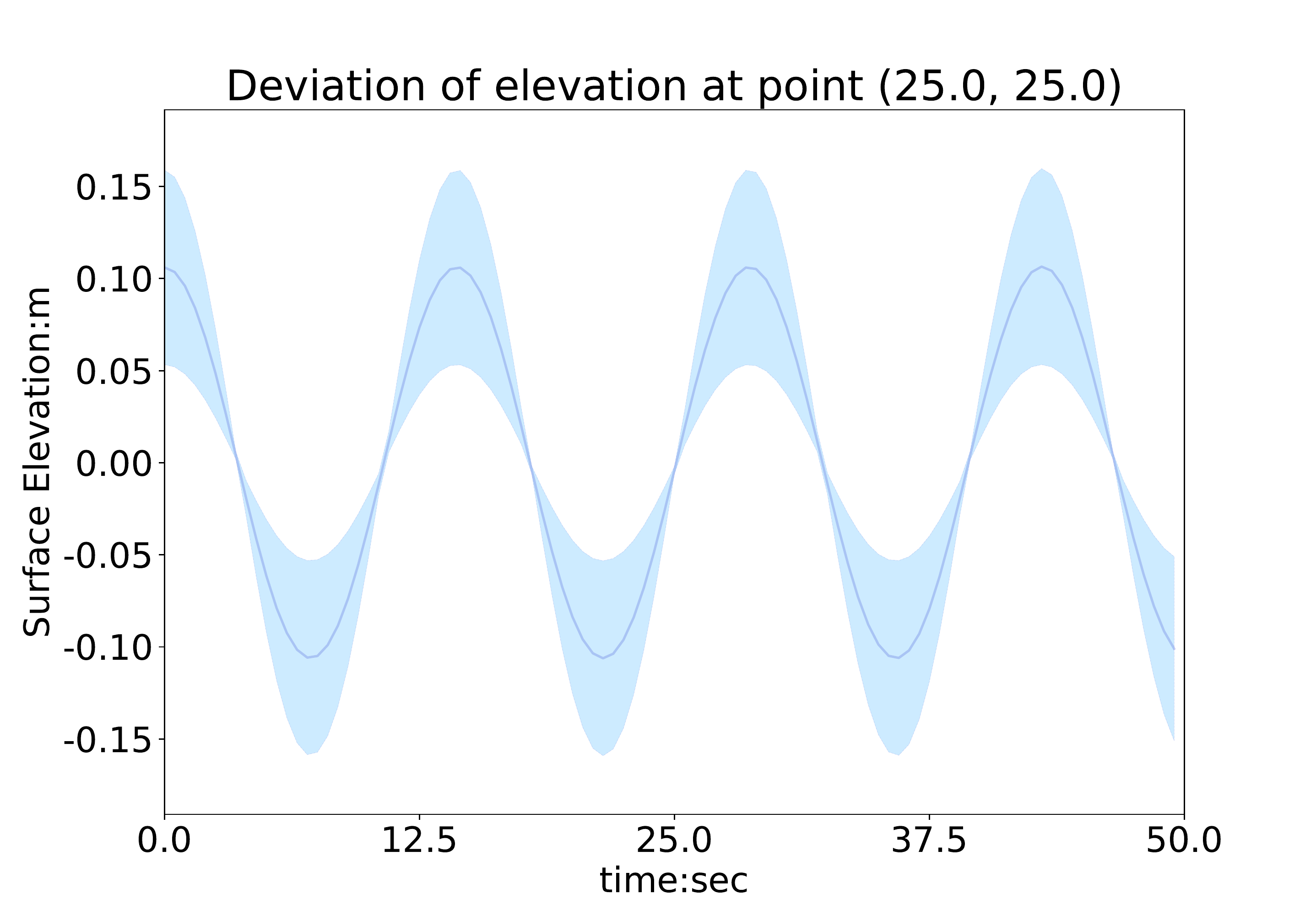}}\hfill
	\subfigure[][Deviation of $u$ with $ \xi_2 \sim U(0.25, 2.75)$.]{\includegraphics[width=0.45\columnwidth]{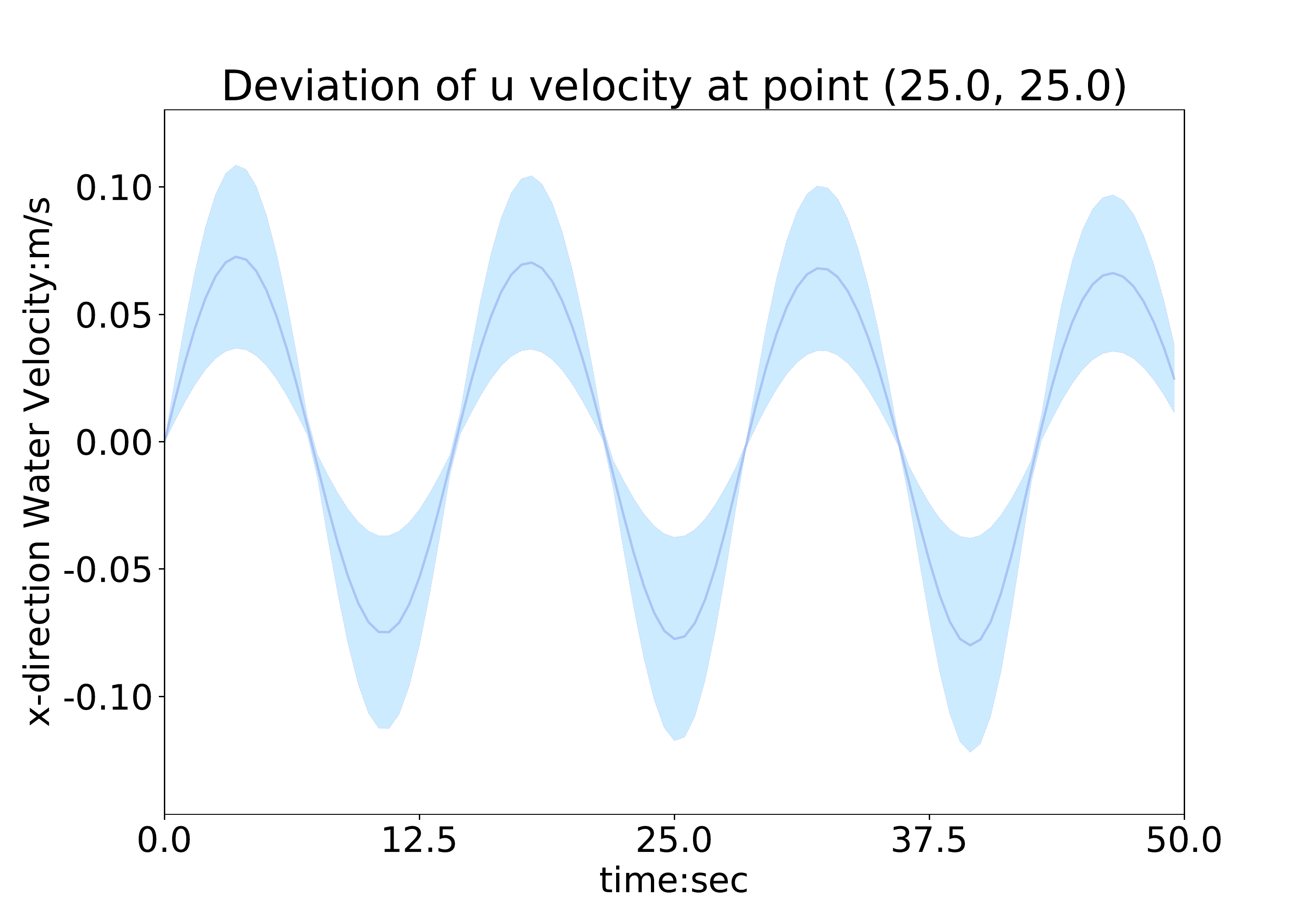}}\hfill
	\caption{Deviation of surface elevation and $x$-direction water velocity at spatial points $(25.0m, 25.0m)$. }
	\label{fig:sloshvariationcompareeta}
\end{figure}

To study the sensitivity of the variance, we consider the slosh test case described in Section~\ref{sec:sloshtest}, where the uncertain initial condition has the form $\eta = 0.1 \xi_1 \xi_2 \cos ( \pi x /100.0) $, where $\xi_1, \xi_2$ are  uniformly distributed. In this experiment, we fix $\xi_1 \sim U(0.8, 1.2)$ and vary $\xi_2$ as $U(1.0, 2.0)$, $U(0.5, 2.5)$, and $U(0.25, 2.75)$. We \oneR{select one} spatial point $(25.0m, 25.0m)$ to show the variation of variance for both surface elevation and  $x$-direction component of water velocity in Figure~\ref{fig:sloshvariationcompareeta}. The variation of variance of the other spatial point $(75.0m, 25.0m)$ can be found in~\ref{sec:Variation_pdf}. In Figure~\ref{fig:sloshvariationcompareeta}, the blue shaded area corresponds to one standard deviation at that spatial point and the central blue line represents the mean of the model solution. 
We observe in Figures~\ref{fig:sloshvariationcompareeta} that the variance in both surface elevation and water velocity increases as the uncertain range of $\xi_2$ extends. Hence, the variance of the input directly impacts the variance of the output in an intuitive fashion:  the variance of output increases as the variance of inputs increase.

\subsection{The time-varying probability density function} \label{sec:PDF}


The PDF is often used to define random variables and in this section, we investigate PDFs from the SSWM outputs and the similarities between multiple PDFs in space and time.  To visualize the predicted PDF at a fixed point in space and time, we distribute $1000$ samples on $\xi_1, \xi_2$ and collect the sample outputs based on the surrogate response. We subsequently uniformly discretize the range of sample outputs into $30$ bins and draw a histogram with its kernel-estimated PDF. With a visualized PDF, we visually inspect and investigate similarities among the different output quantities. For each of the ideal test cases introduced in Section~\ref{sec:testcases}, we ascertain PDFs at selected spatial and temporal locations for analysis. In the histograms presented in the following figures, the light blue shaded area represents the PDF distribution of a random variable and the darker blue line in these plots corresponds to a kernel density estimation of the presented distribution. \more{To keep the presentation herein compact, we only present the case of uncertain bathymetry here and the other cases are in~\ref{sec:timevar_pdf} as well as in~\cite{chenthesis}. }



\subsubsection{Uncertain Bathymetry} \label{sec:humptest_PDF}

\begin{figure}[h!]
	\centering
	\subfigure[]{\includegraphics[width=0.5\columnwidth]{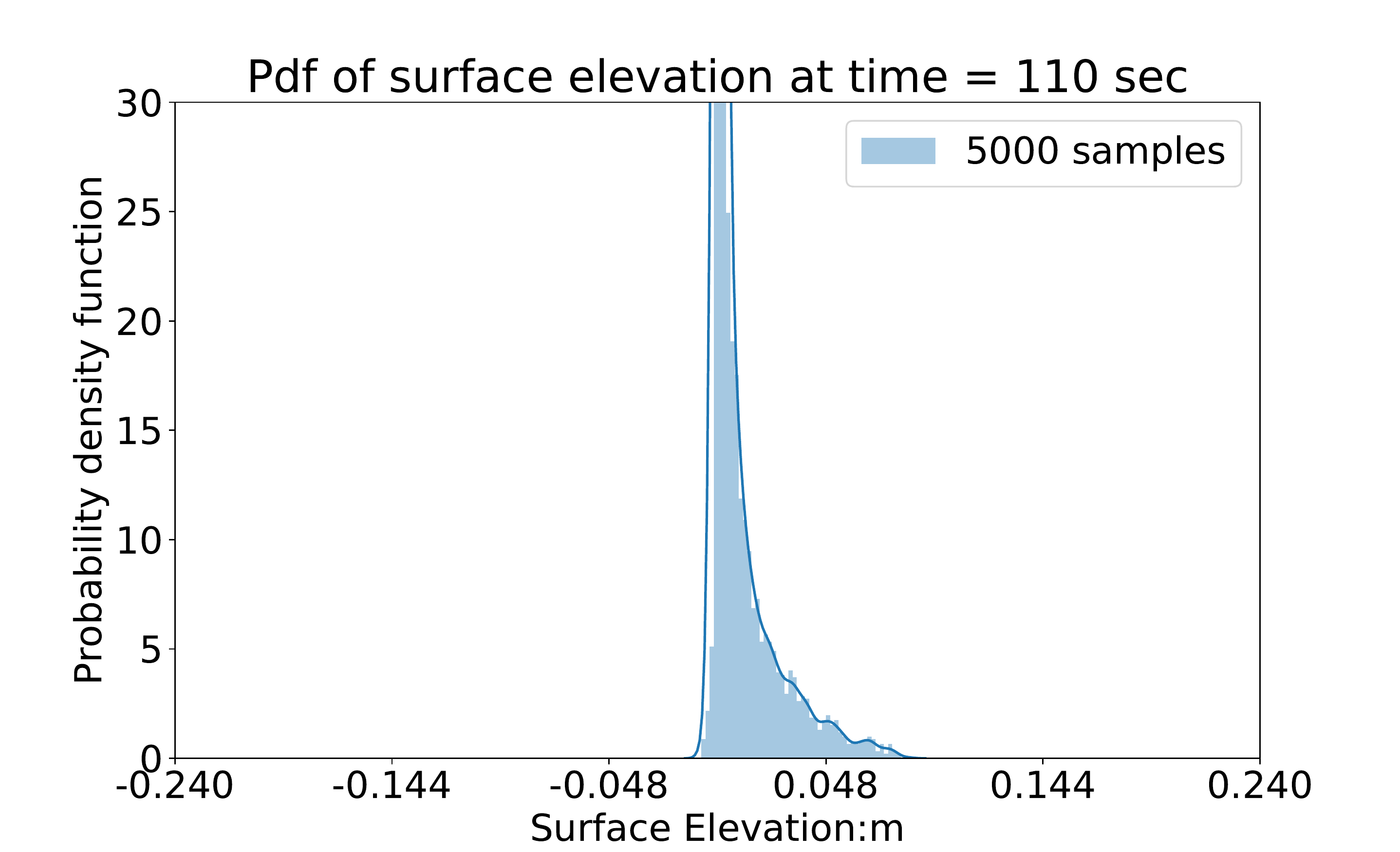}}\hfill
	\subfigure[]{\includegraphics[width=0.5\columnwidth]{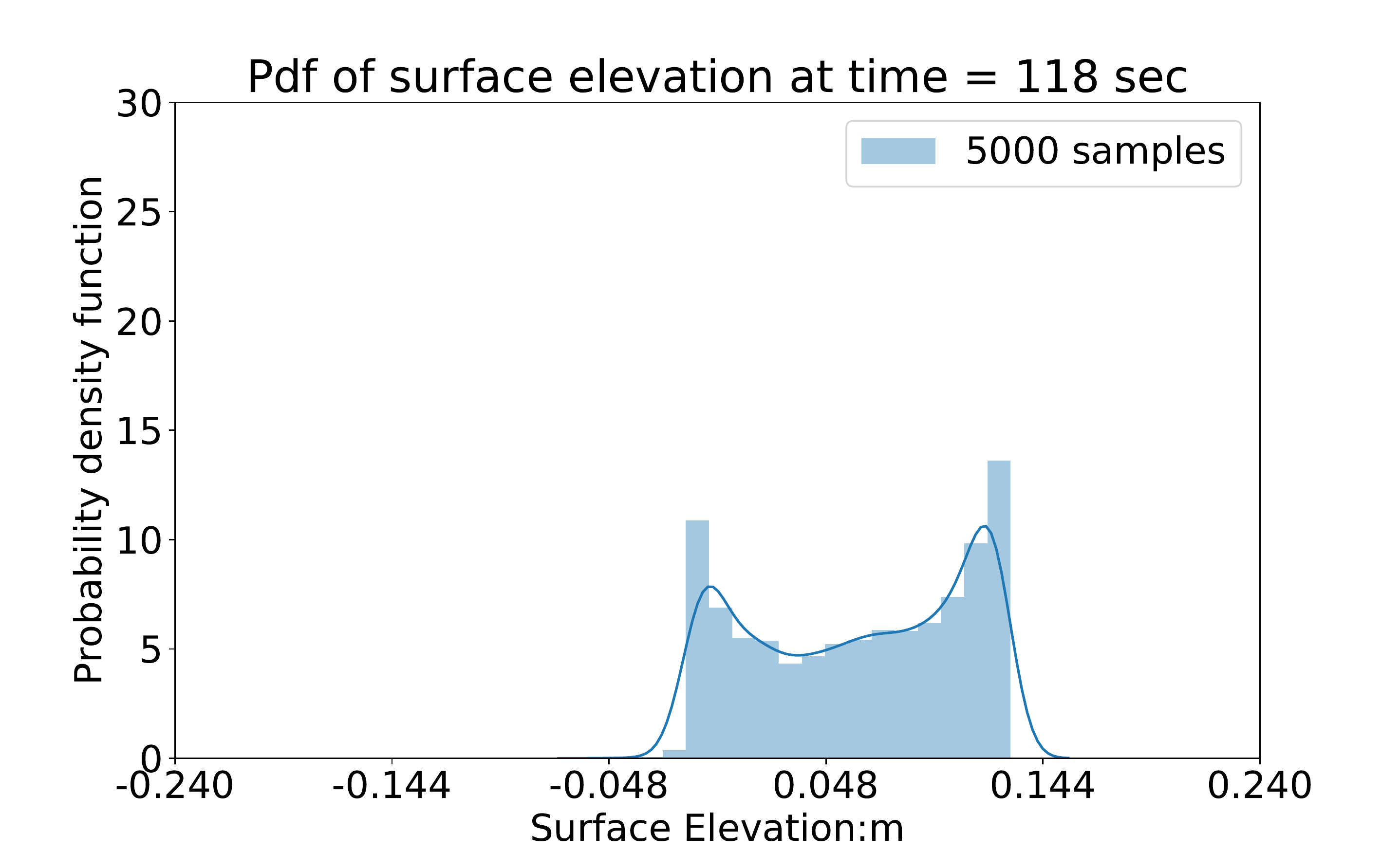}}\hfill
	\vspace{-4mm}
	\subfigure[]{\includegraphics[width=0.5\columnwidth]{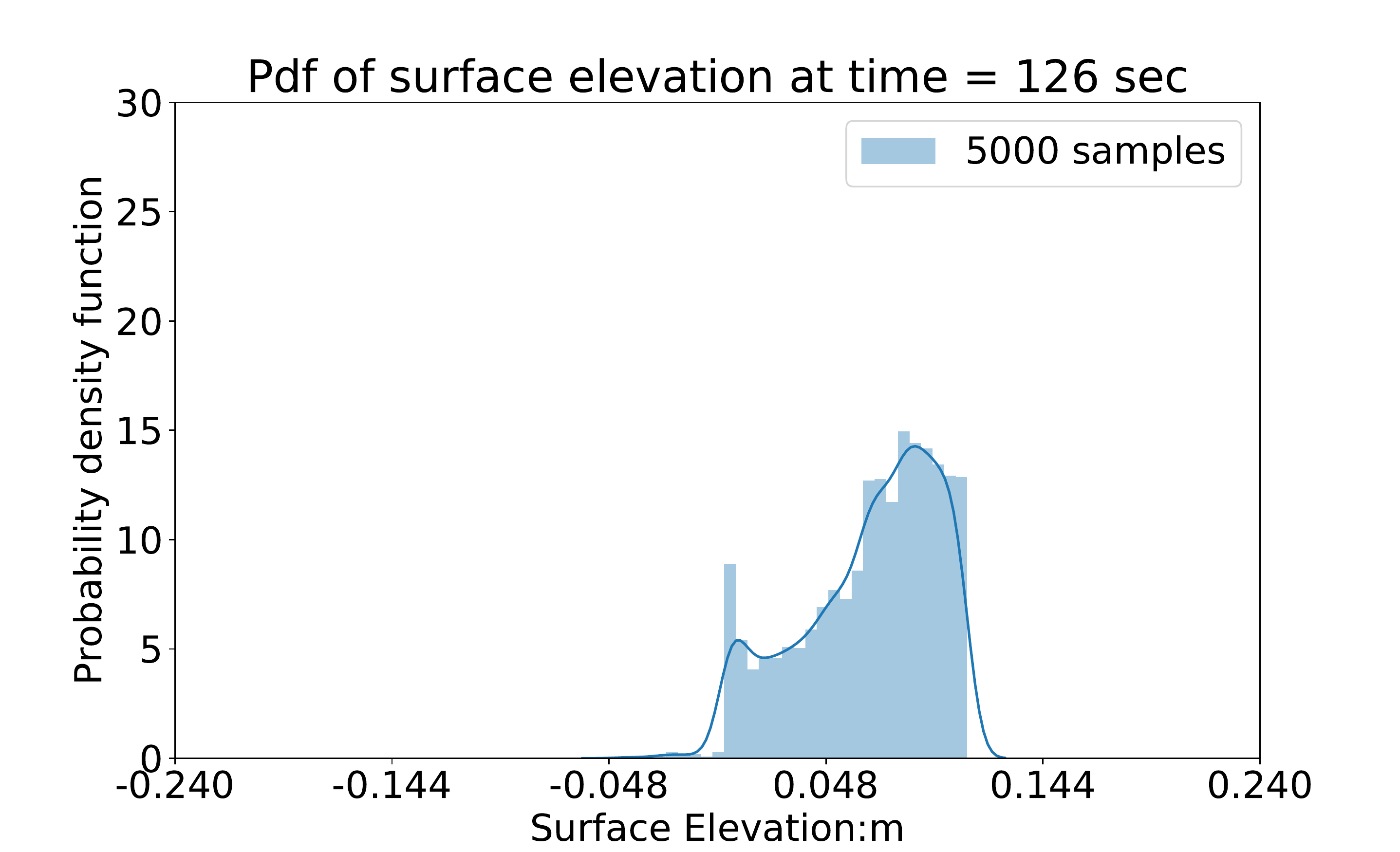}}\hfill
	\subfigure[]{\includegraphics[width=0.5\columnwidth]{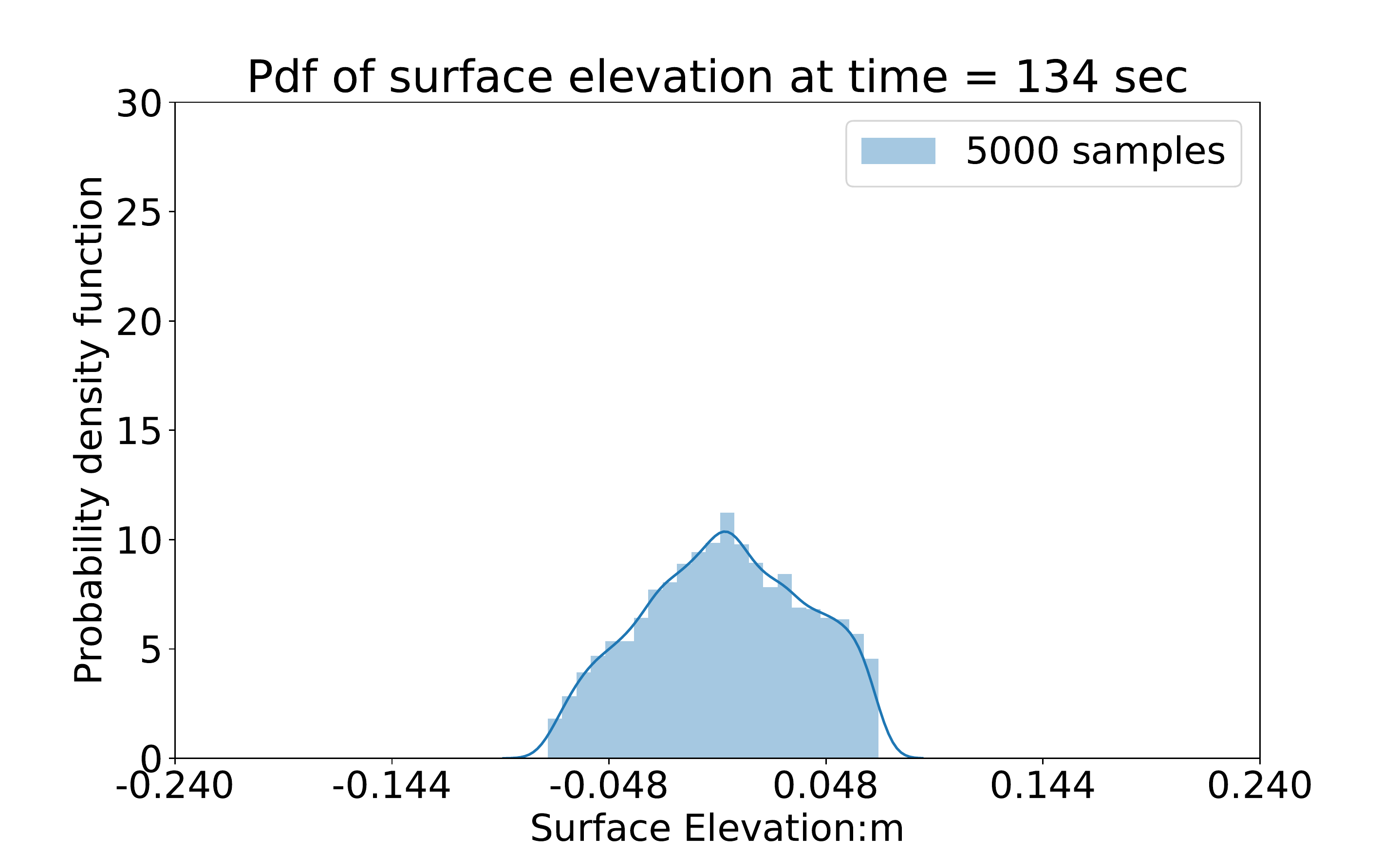}}\hfill
	\vspace{-4mm}
	\subfigure[]{\includegraphics[width=0.5\columnwidth]{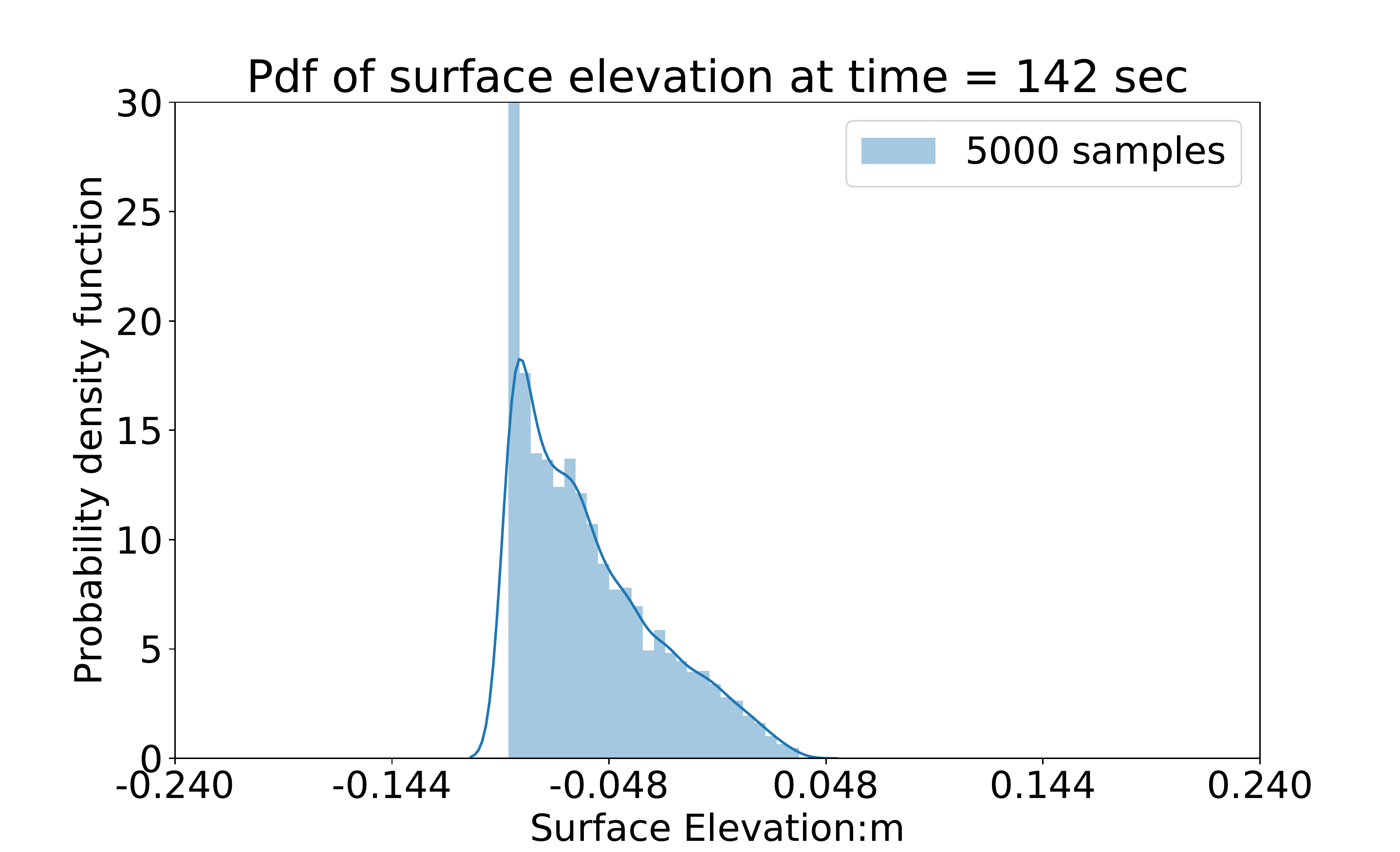}}\hfill
	\subfigure[]{\includegraphics[width=0.5\columnwidth]{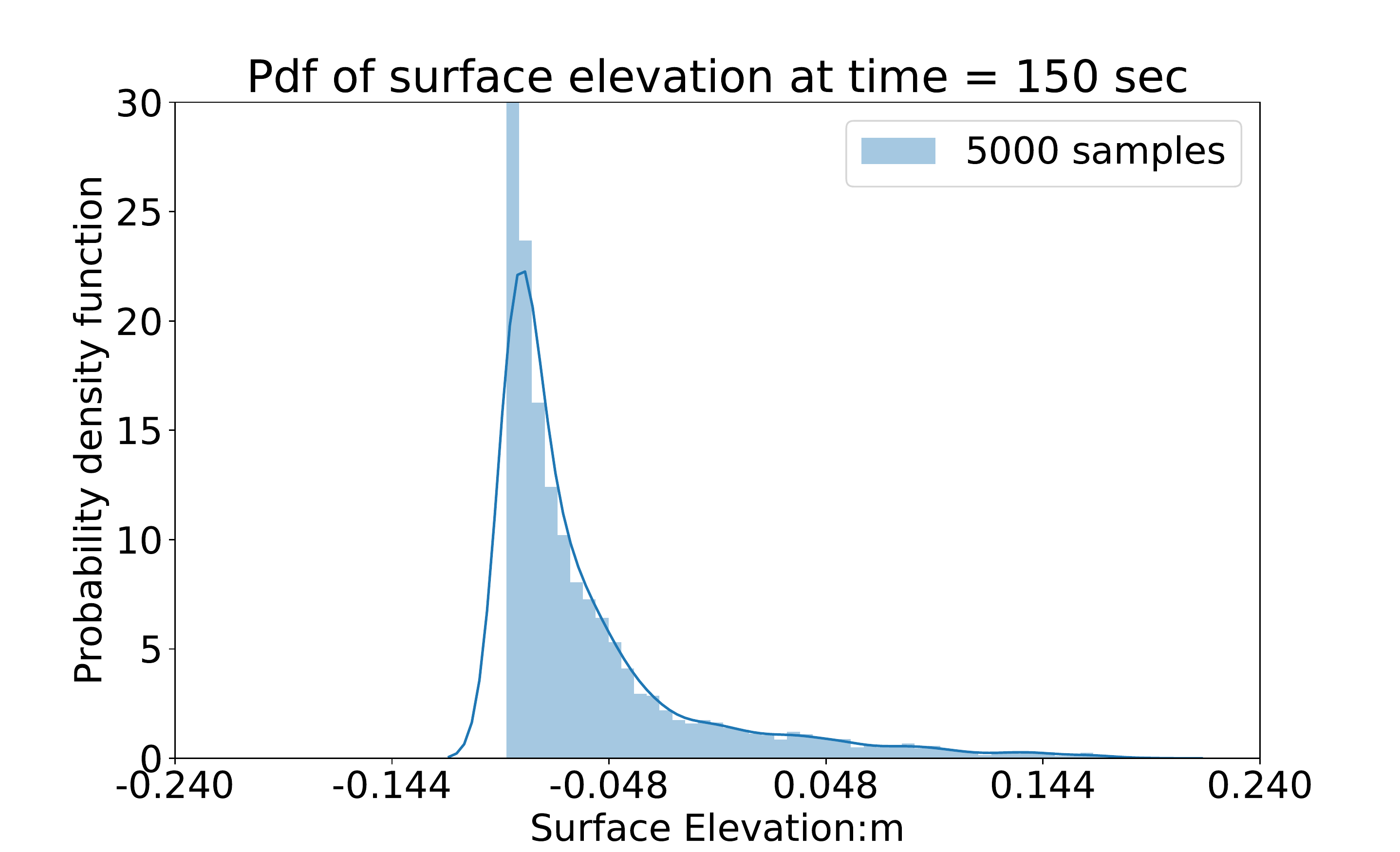}}\hfill
	\caption{Elevation PDFs at $(250.0m, 100.0m)$.}
	\label{fig:humppdfeta}
\end{figure}

\begin{figure}[h!]
	\centering
	\subfigure[]{\includegraphics[width=0.5\columnwidth]{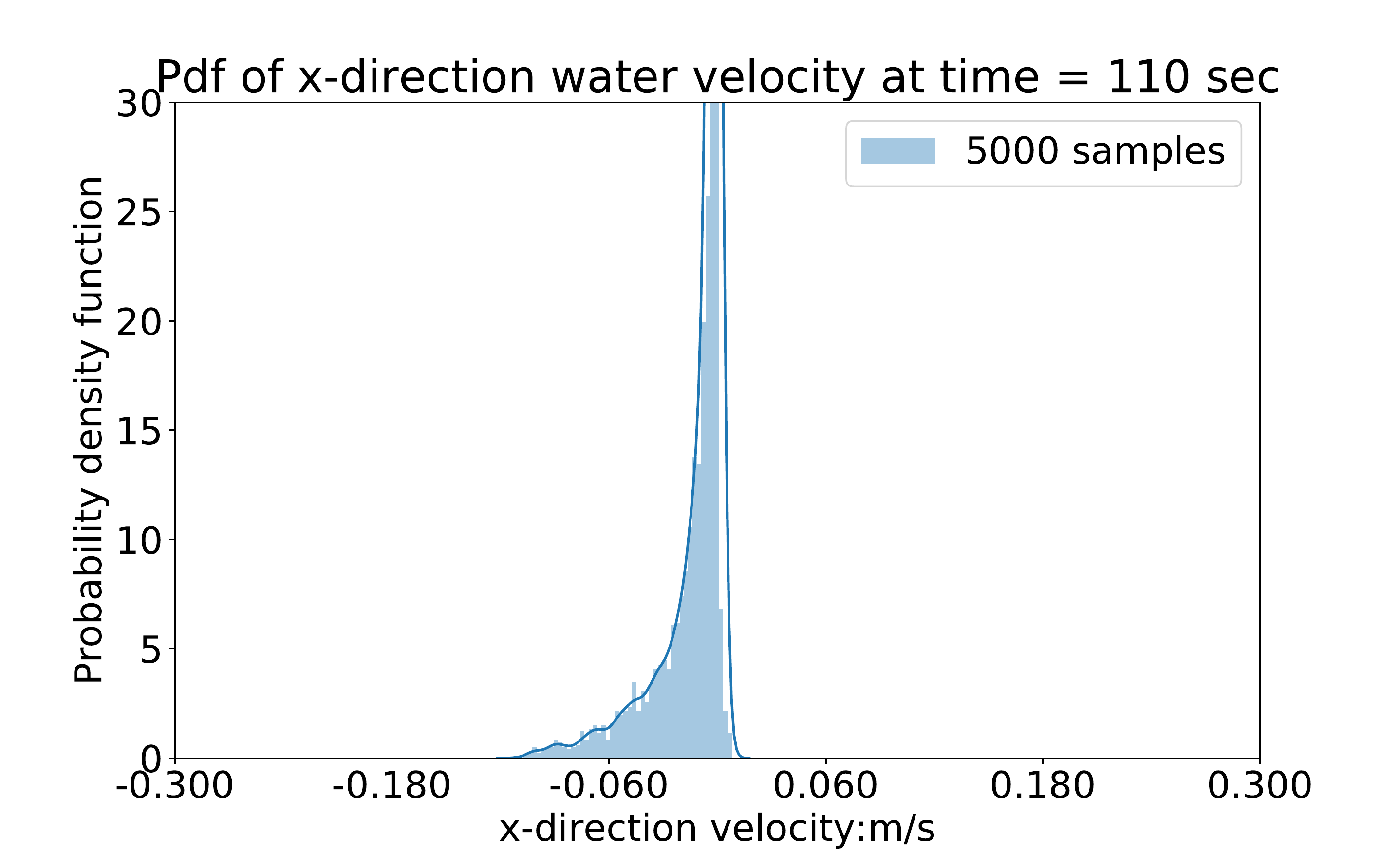}}\hfill
	\subfigure[]{\includegraphics[width=0.5\columnwidth]{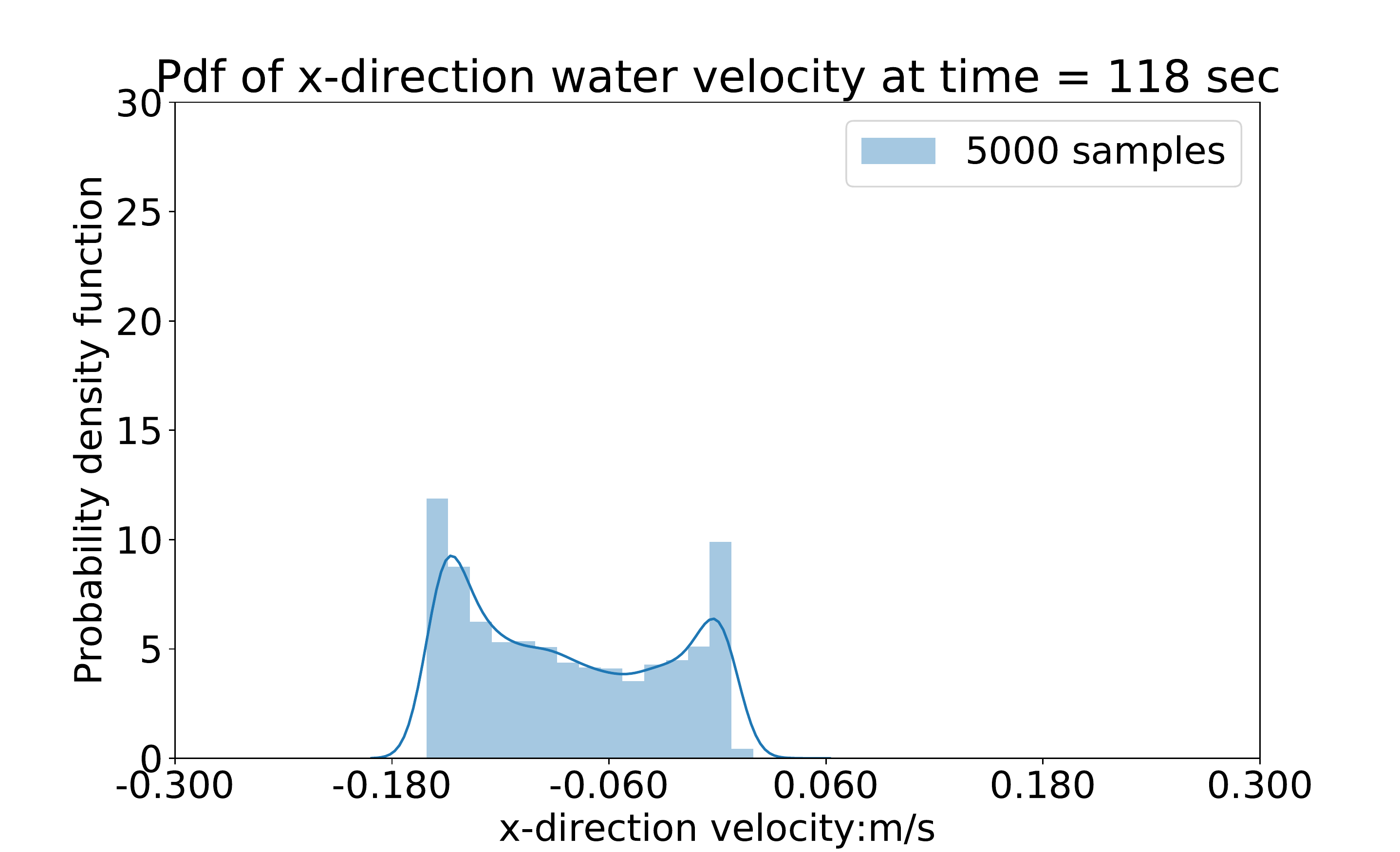}}\hfill
	\vspace{-4mm}
	\subfigure[]{\includegraphics[width=0.5\columnwidth]{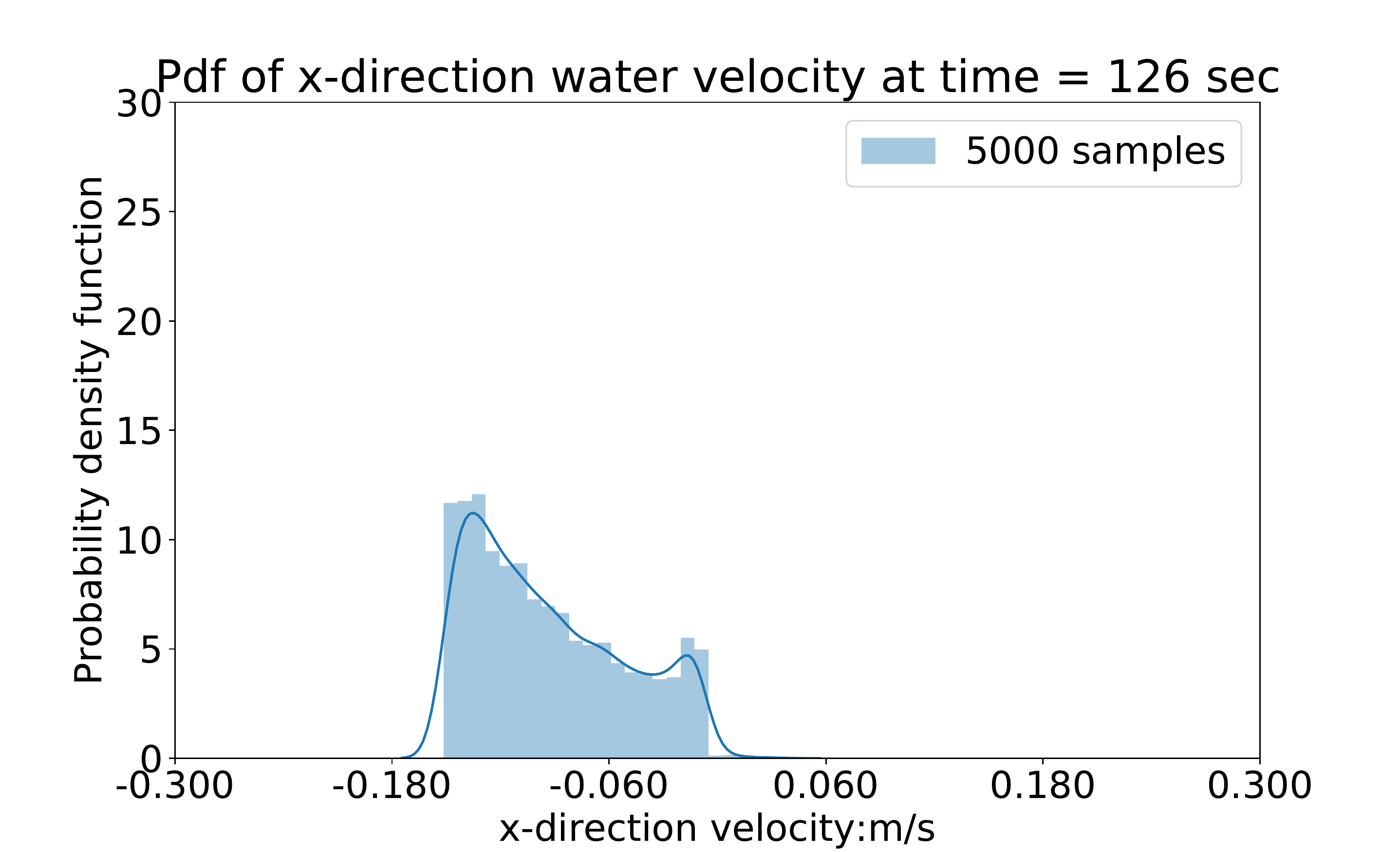}}\hfill
	\subfigure[]{\includegraphics[width=0.5\columnwidth]{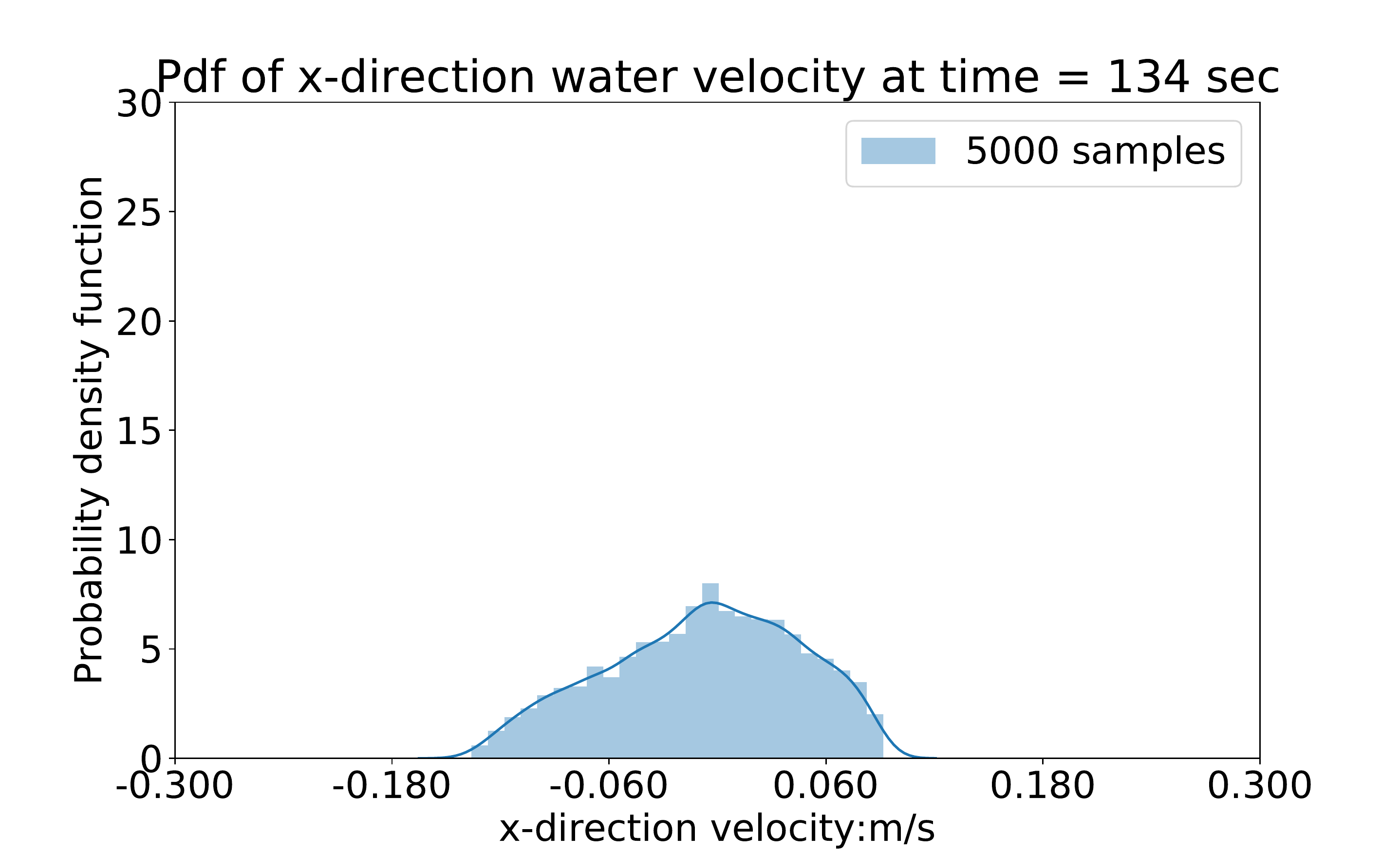}}\hfill
	\vspace{-4mm}
	\subfigure[]{\includegraphics[width=0.5\columnwidth]{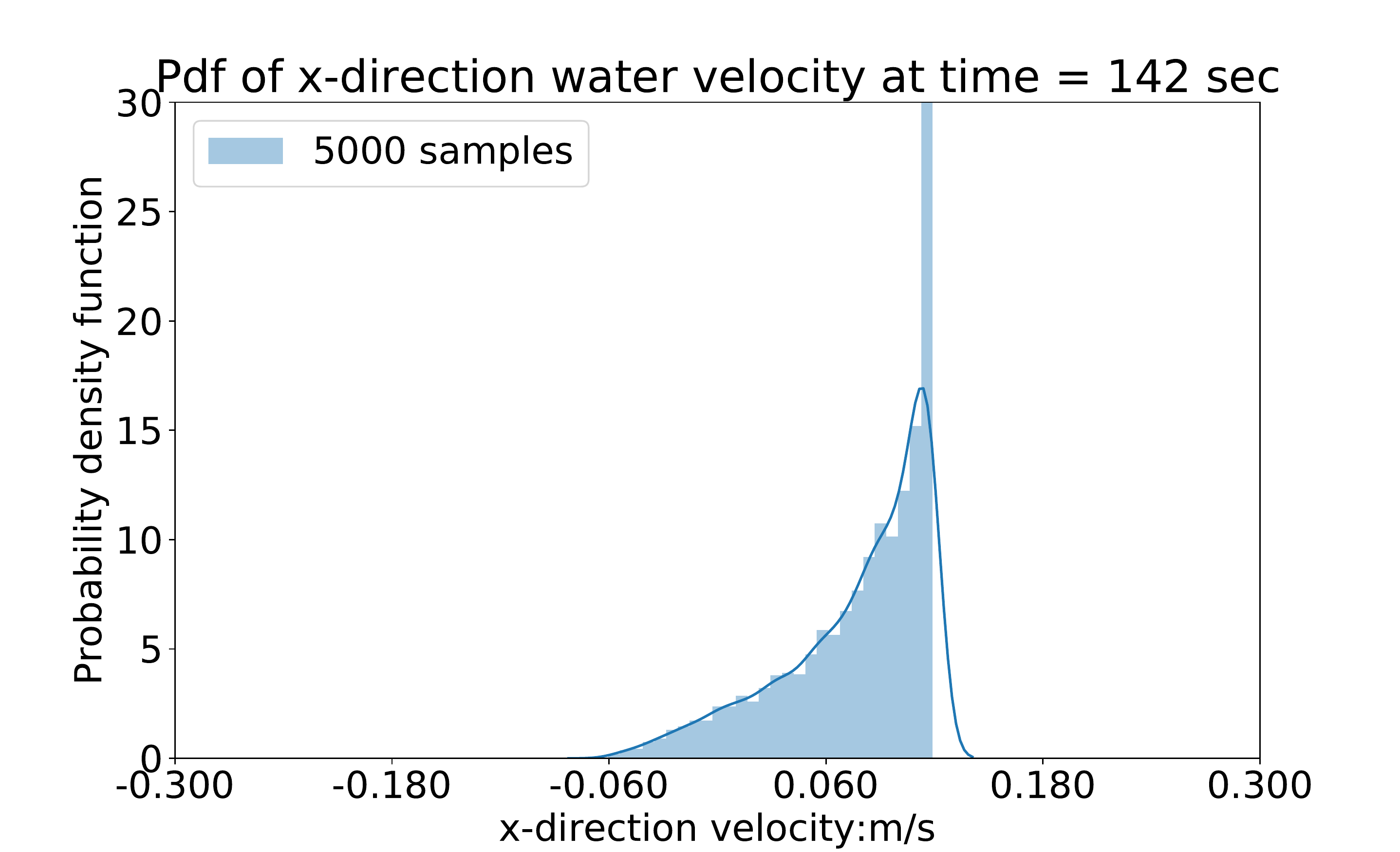}}\hfill
	\subfigure[]{\includegraphics[width=0.5\columnwidth]{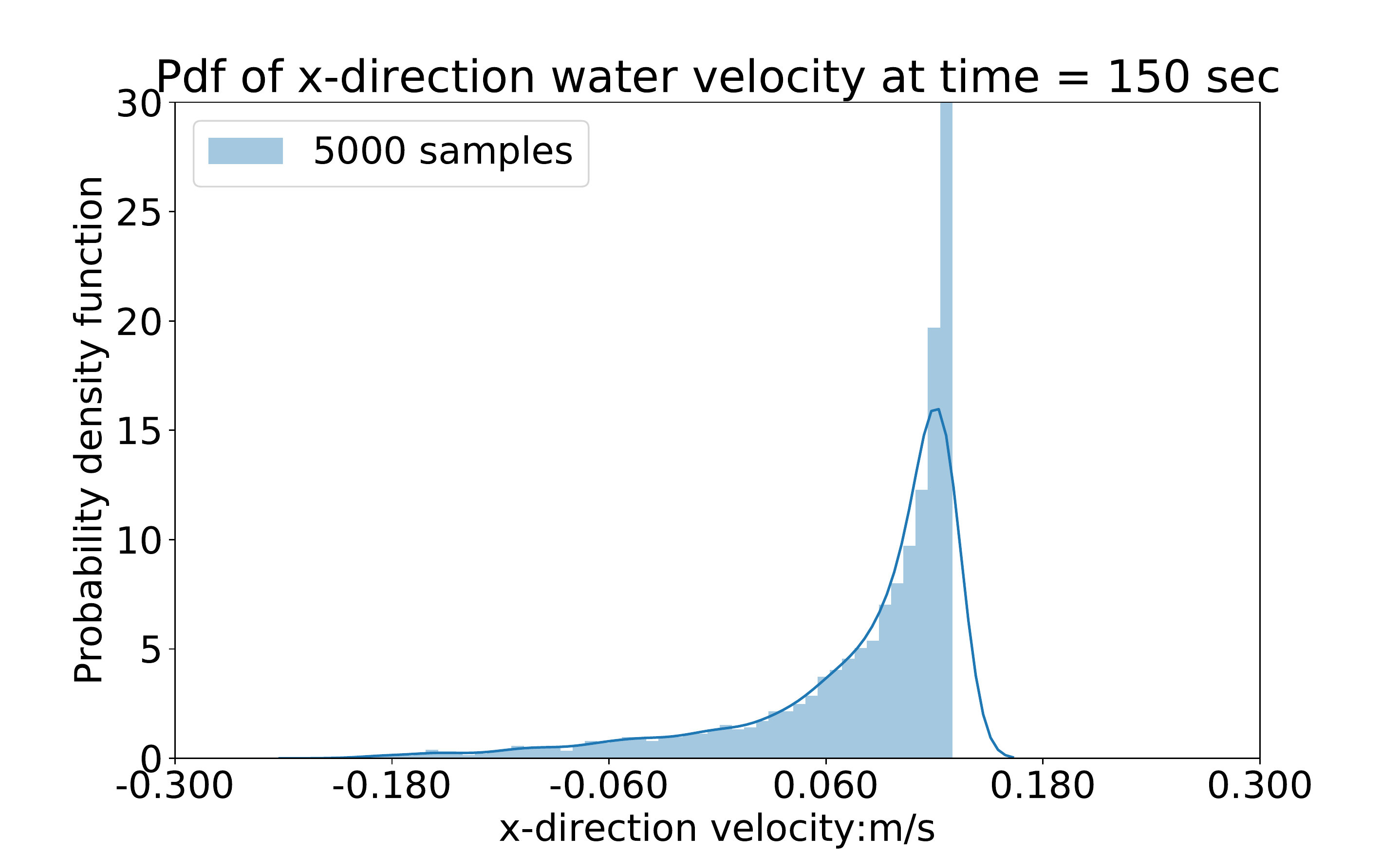}}\hfill
	\caption{ $x$-direction velocity PDFs at $(250.0m, 100.0m)$.}
	\label{fig:humppdfu}
\end{figure}

\begin{figure}[h!]
	\centering
	\subfigure[]{\includegraphics[width=0.5\columnwidth]{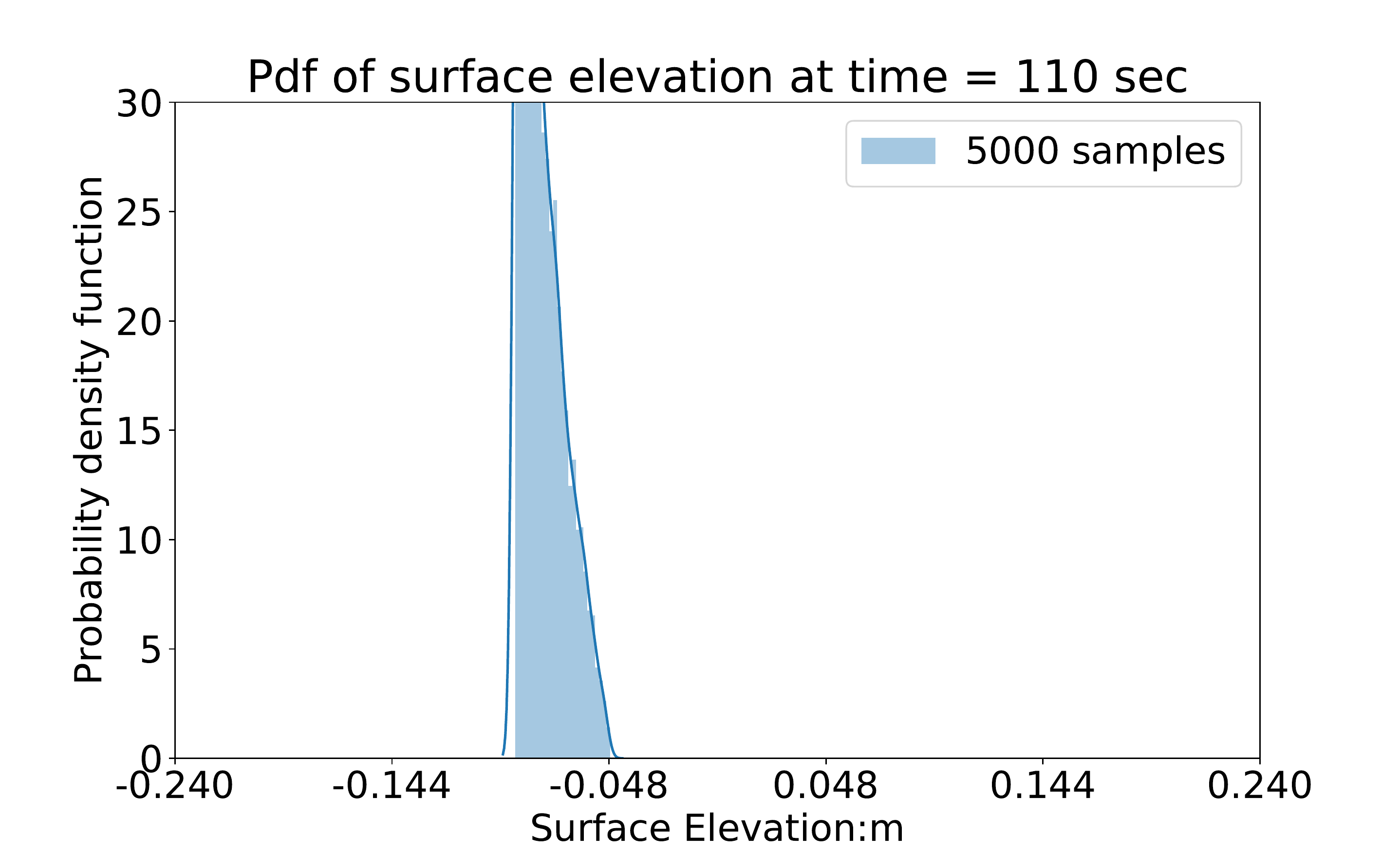}}\hfill
	\subfigure[]{\includegraphics[width=0.5\columnwidth]{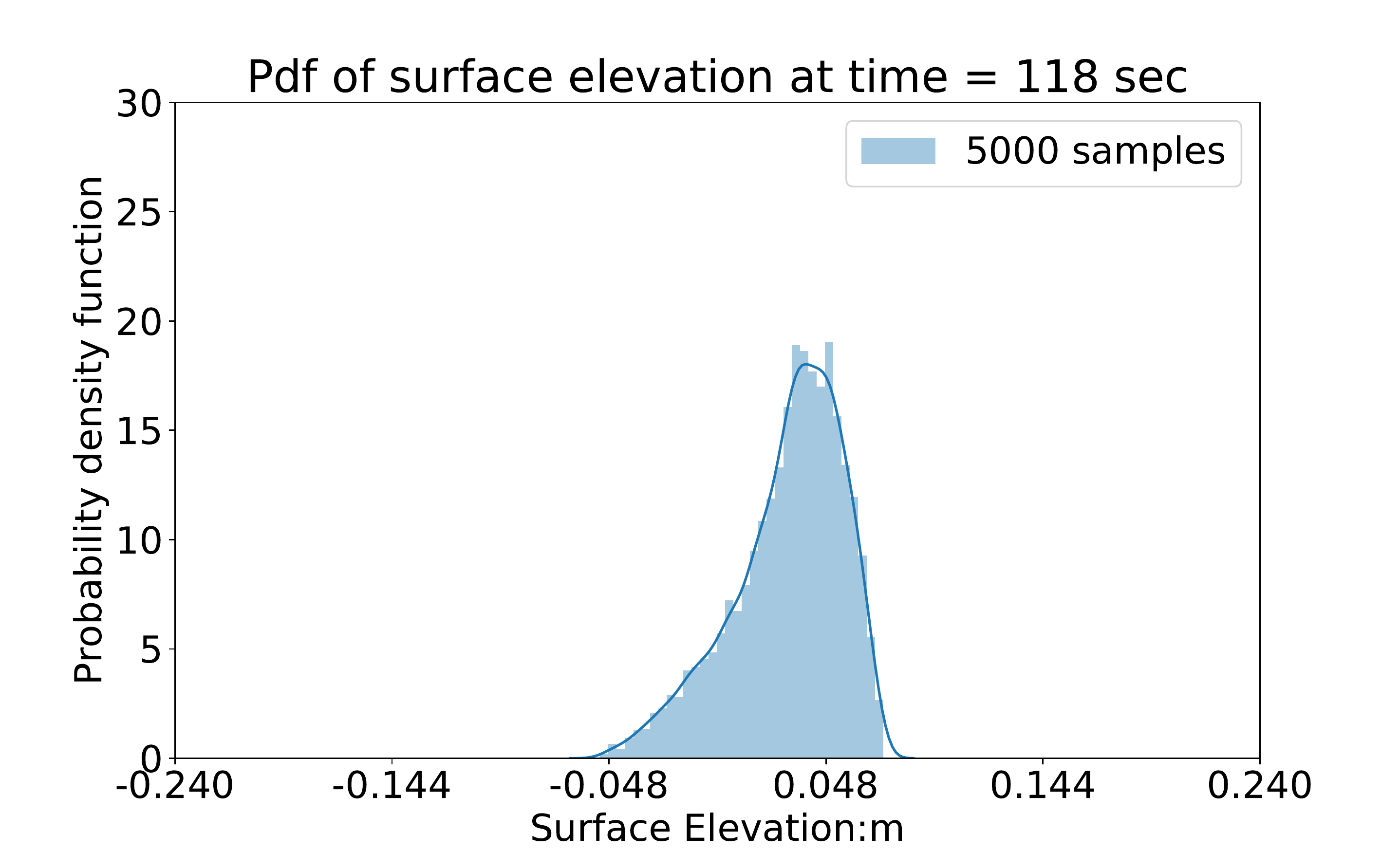}}\hfill
	\vspace{-4mm}
	\subfigure[]{\includegraphics[width=0.5\columnwidth]{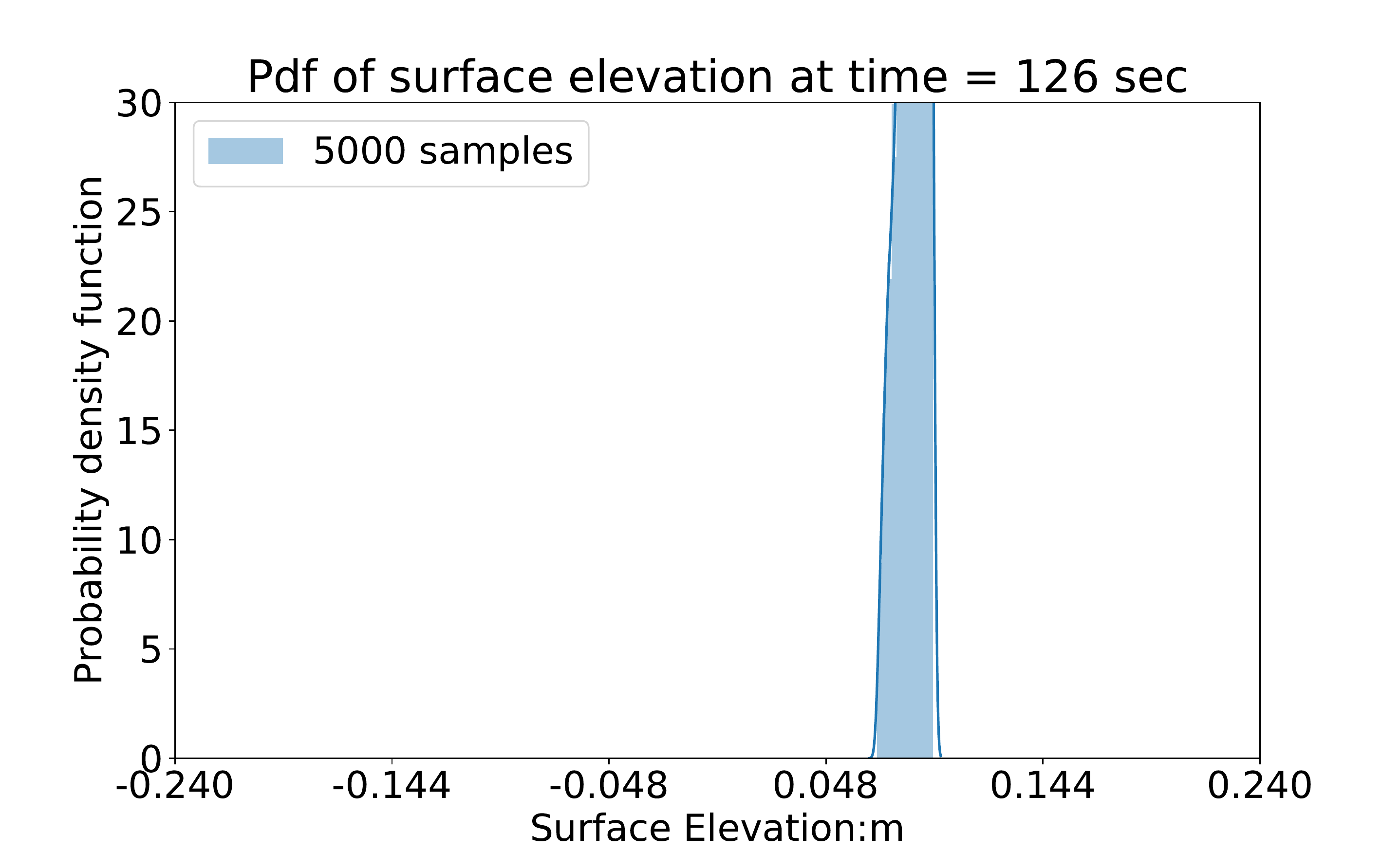}}\hfill
	\subfigure[]{\includegraphics[width=0.5\columnwidth]{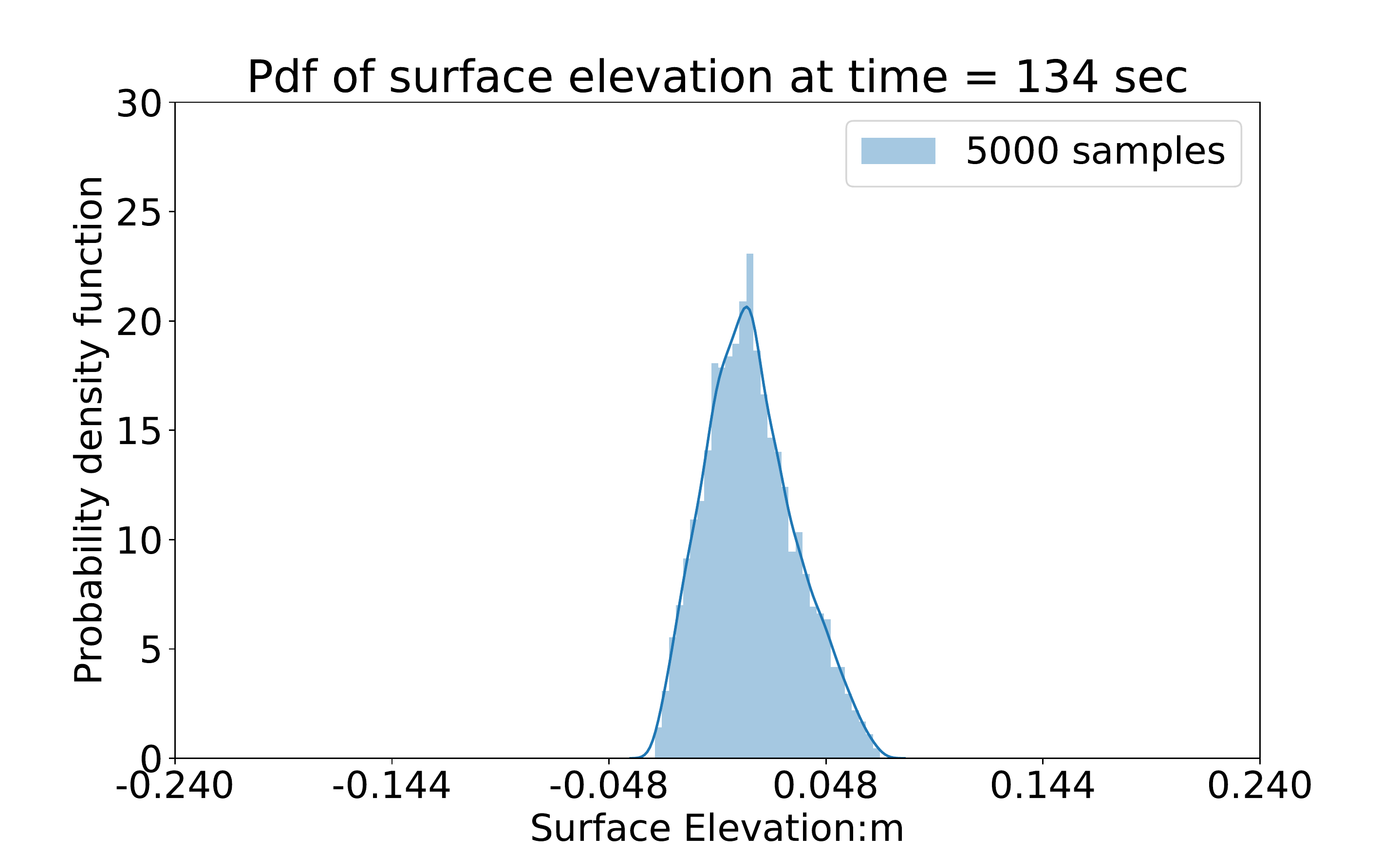}}\hfill
	\vspace{-4mm}
	\subfigure[]{\includegraphics[width=0.5\columnwidth]{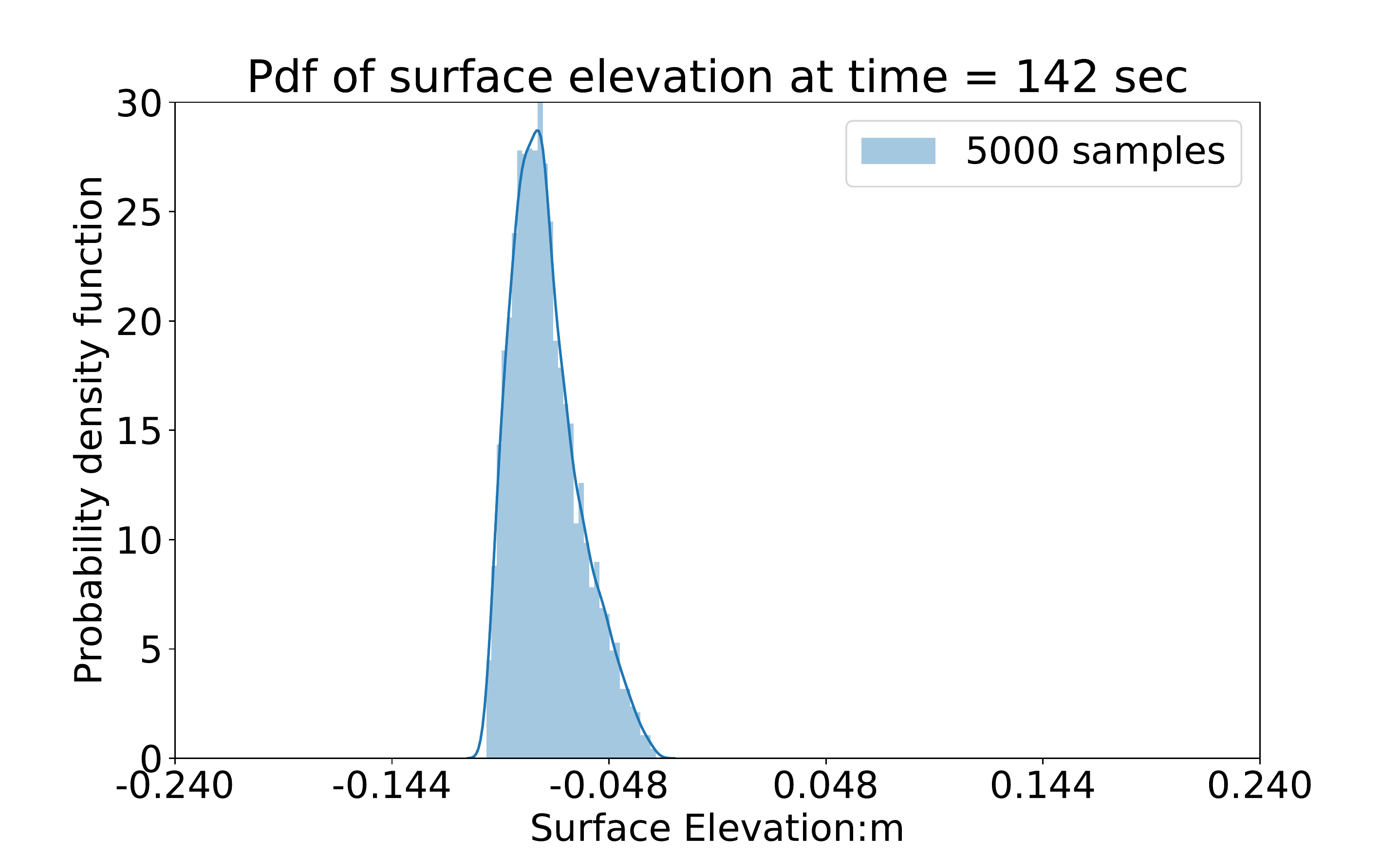}}\hfill
	\subfigure[]{\includegraphics[width=0.5\columnwidth]{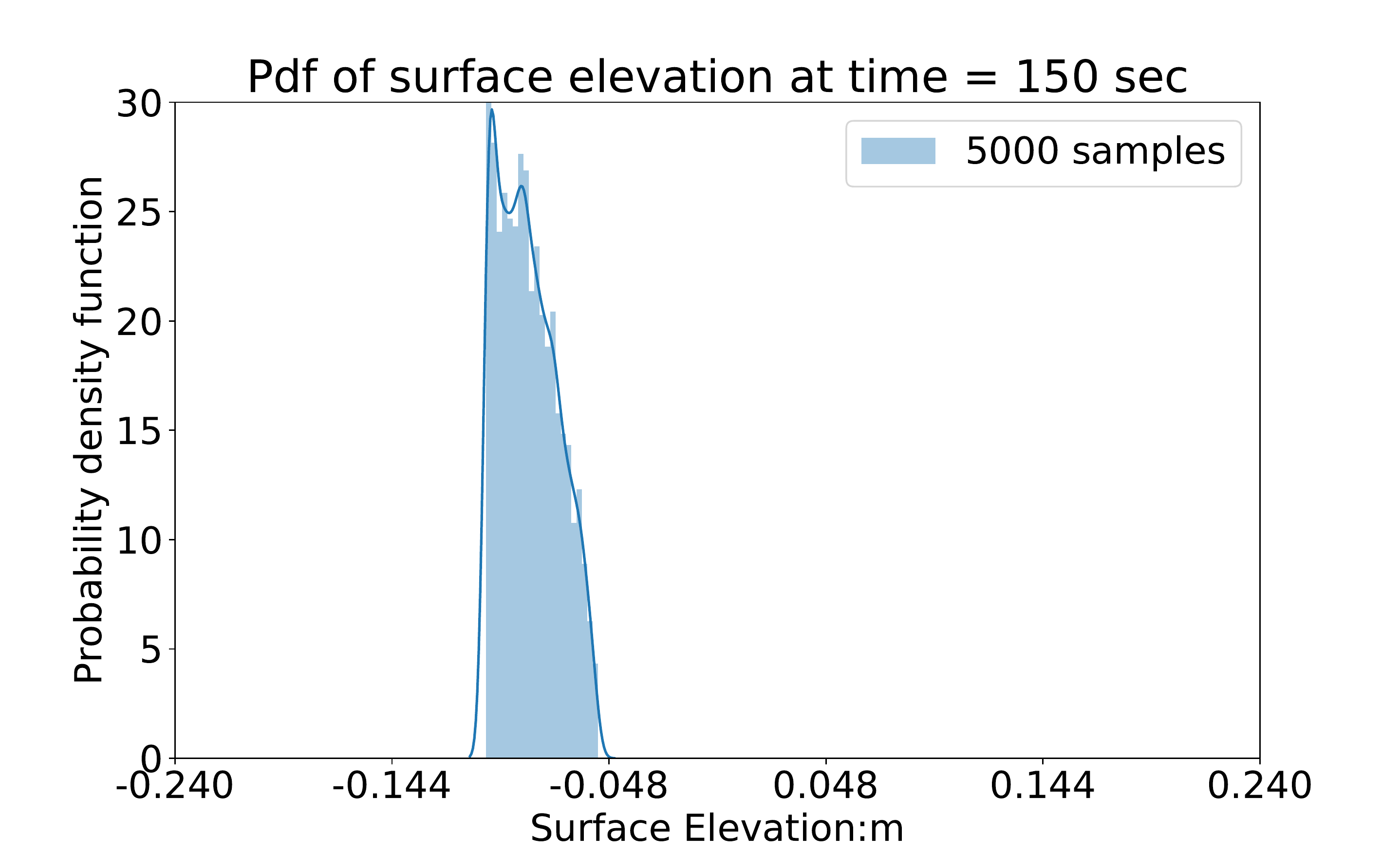}}\hfill
	\caption{Elevation PDFs at $(750.0m, 100.0m)$.}
	\label{fig:humppdfeta3}
\end{figure}

\begin{figure}[h!]
	\centering
	\subfigure[]{\includegraphics[width=0.5\columnwidth]{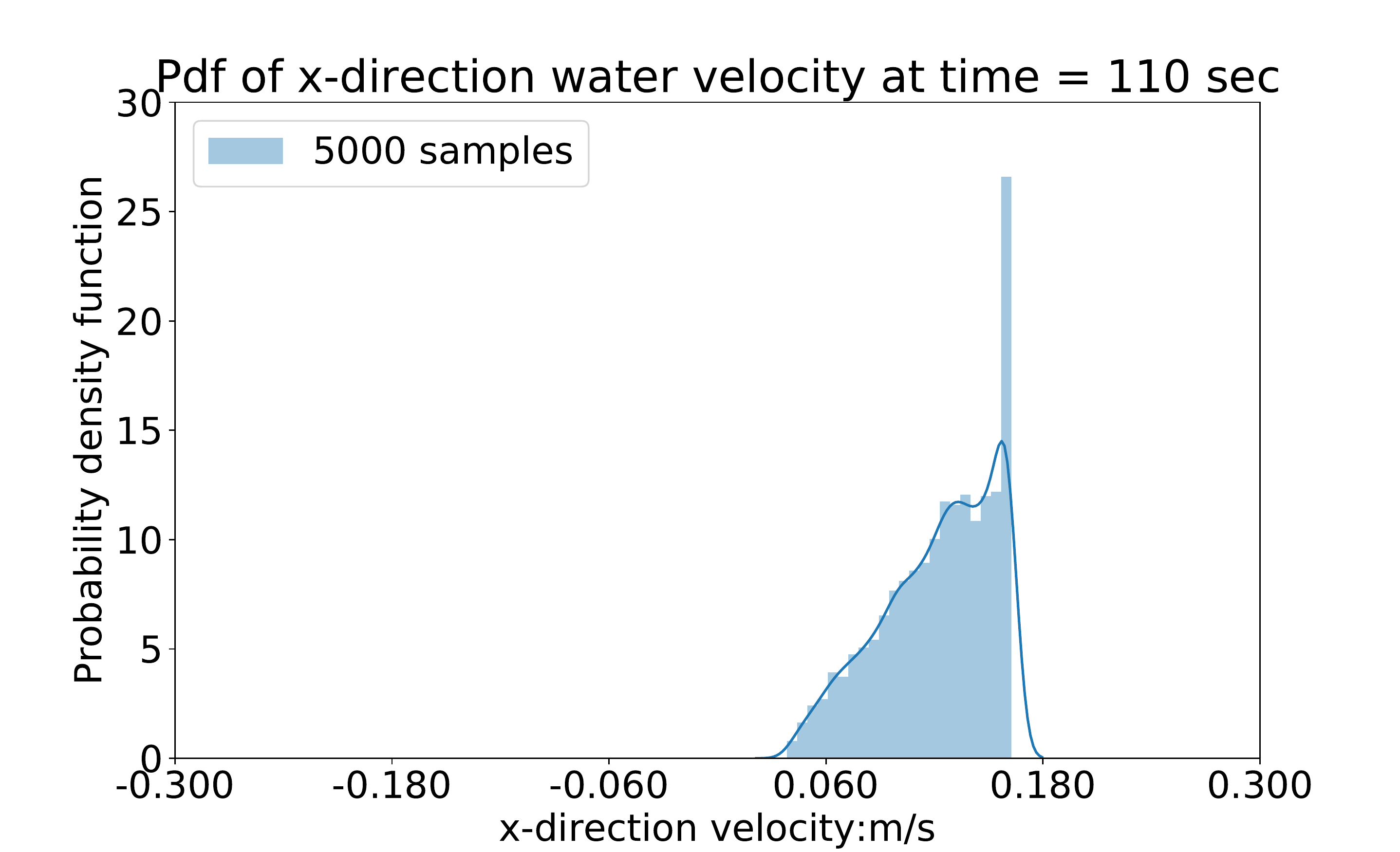}}\hfill
	\subfigure[]{\includegraphics[width=0.5\columnwidth]{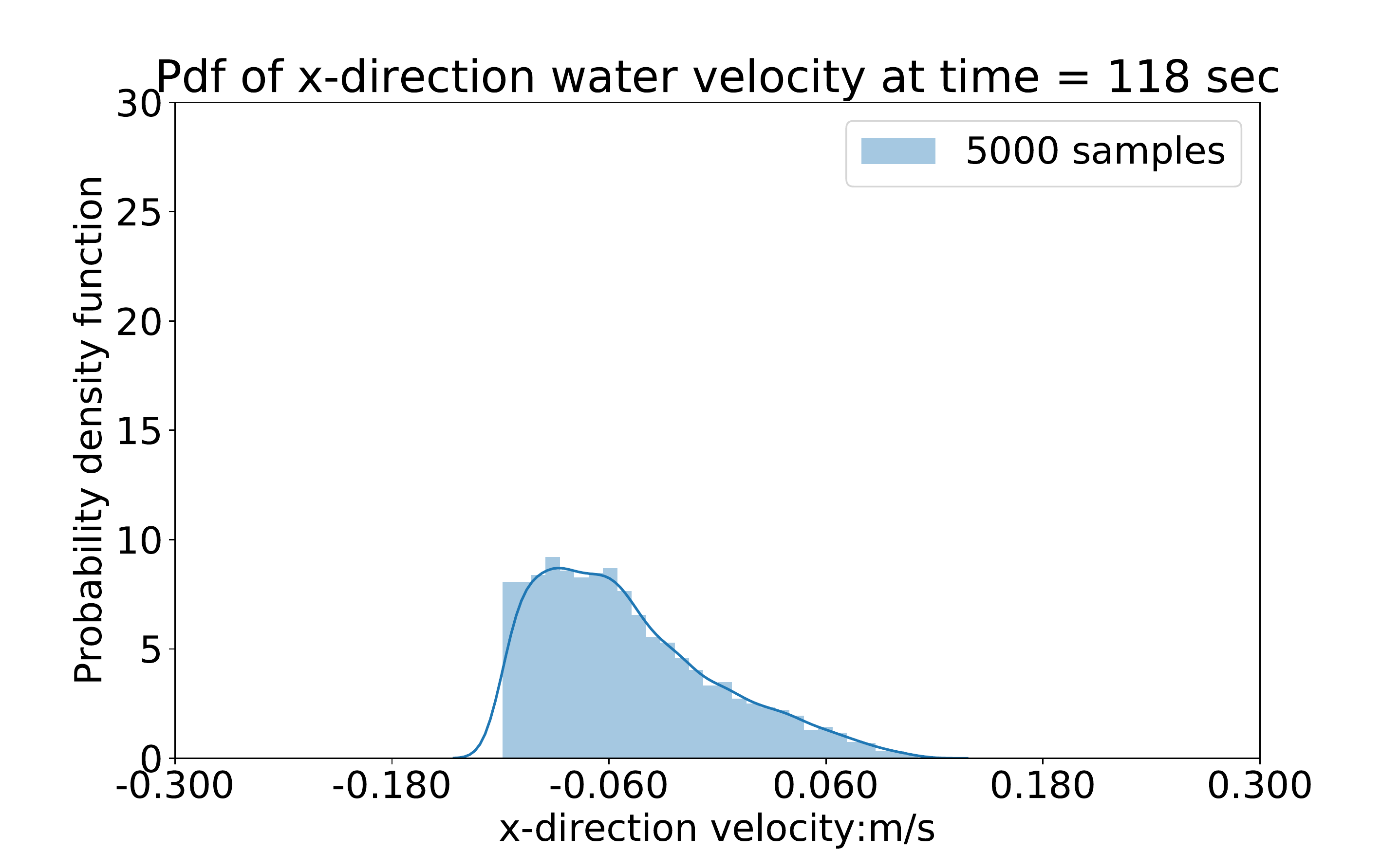}}\hfill
	\vspace{-4mm}
	\subfigure[]{\includegraphics[width=0.5\columnwidth]{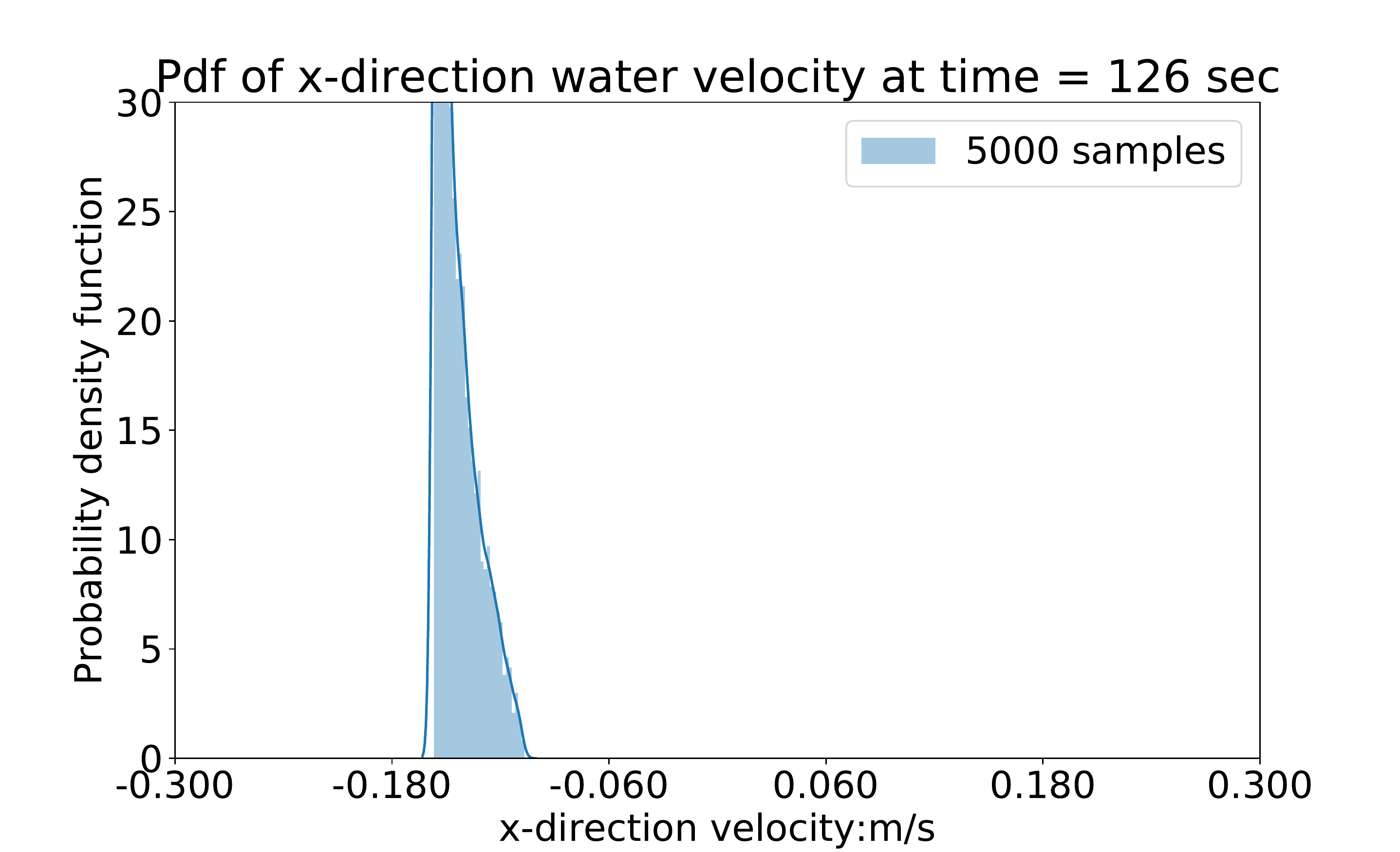}}\hfill
	\subfigure[]{\includegraphics[width=0.5\columnwidth]{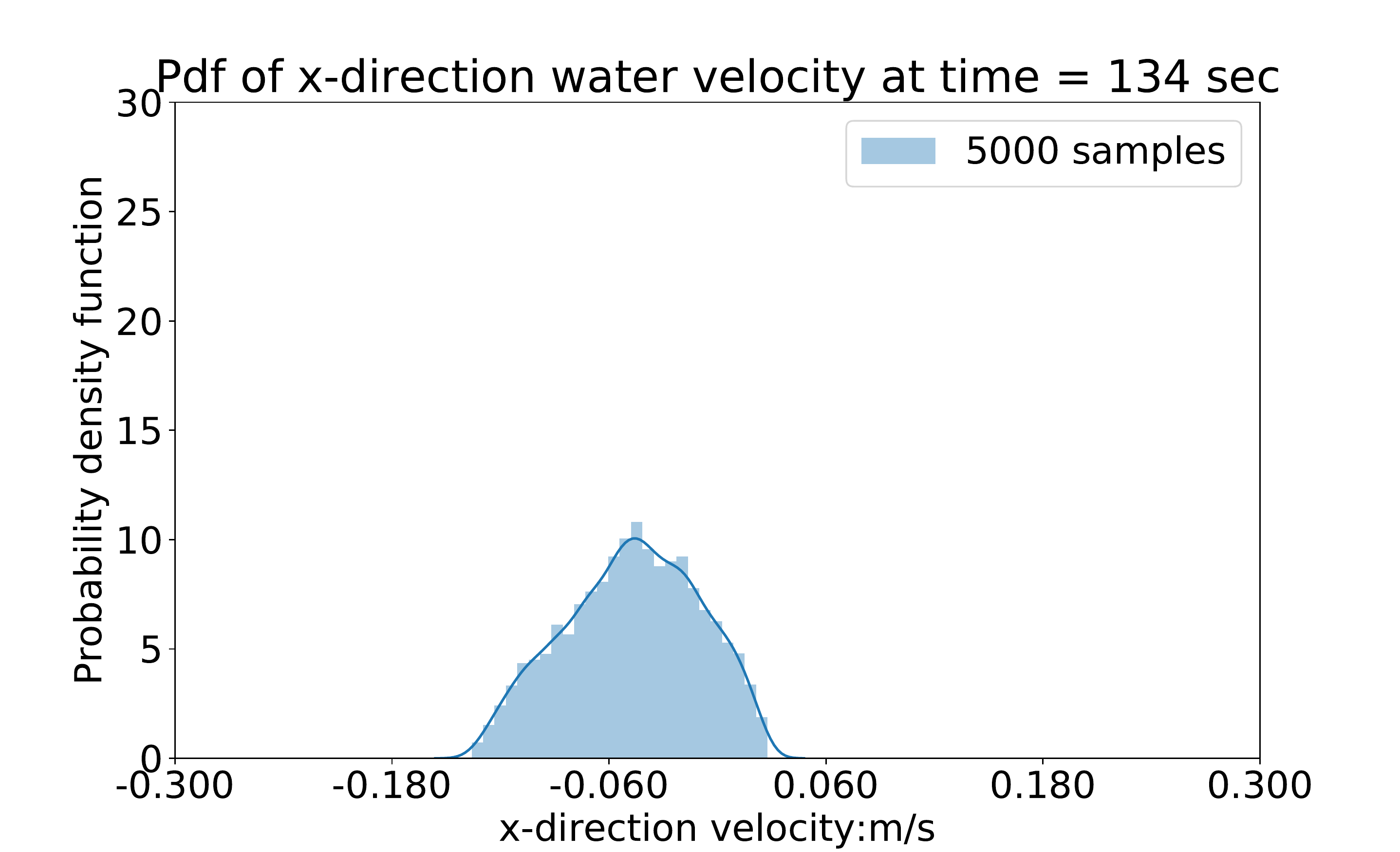}}\hfill
	\vspace{-4mm}
	\subfigure[]{\includegraphics[width=0.5\columnwidth]{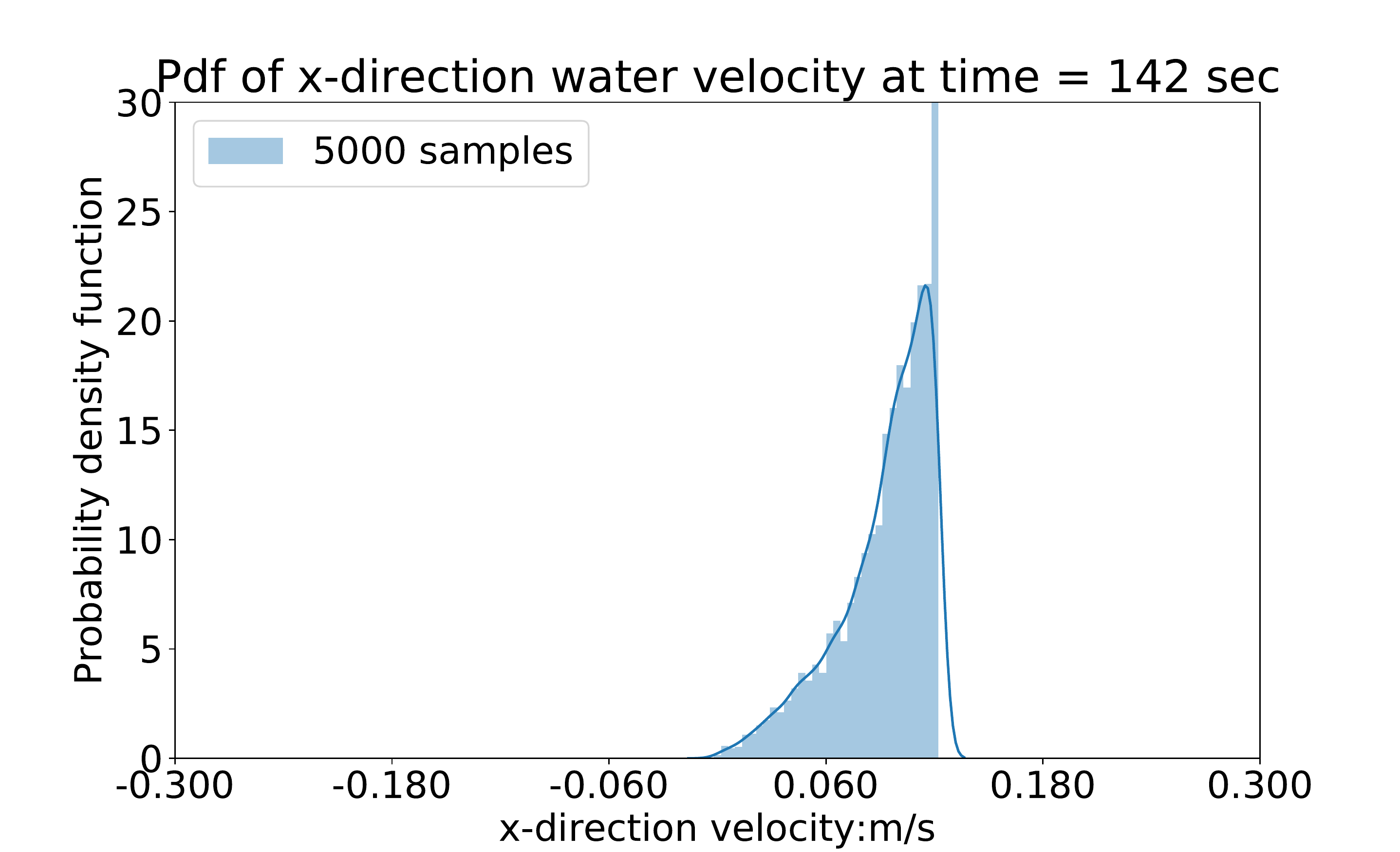}}\hfill
	\subfigure[]{\includegraphics[width=0.5\columnwidth]{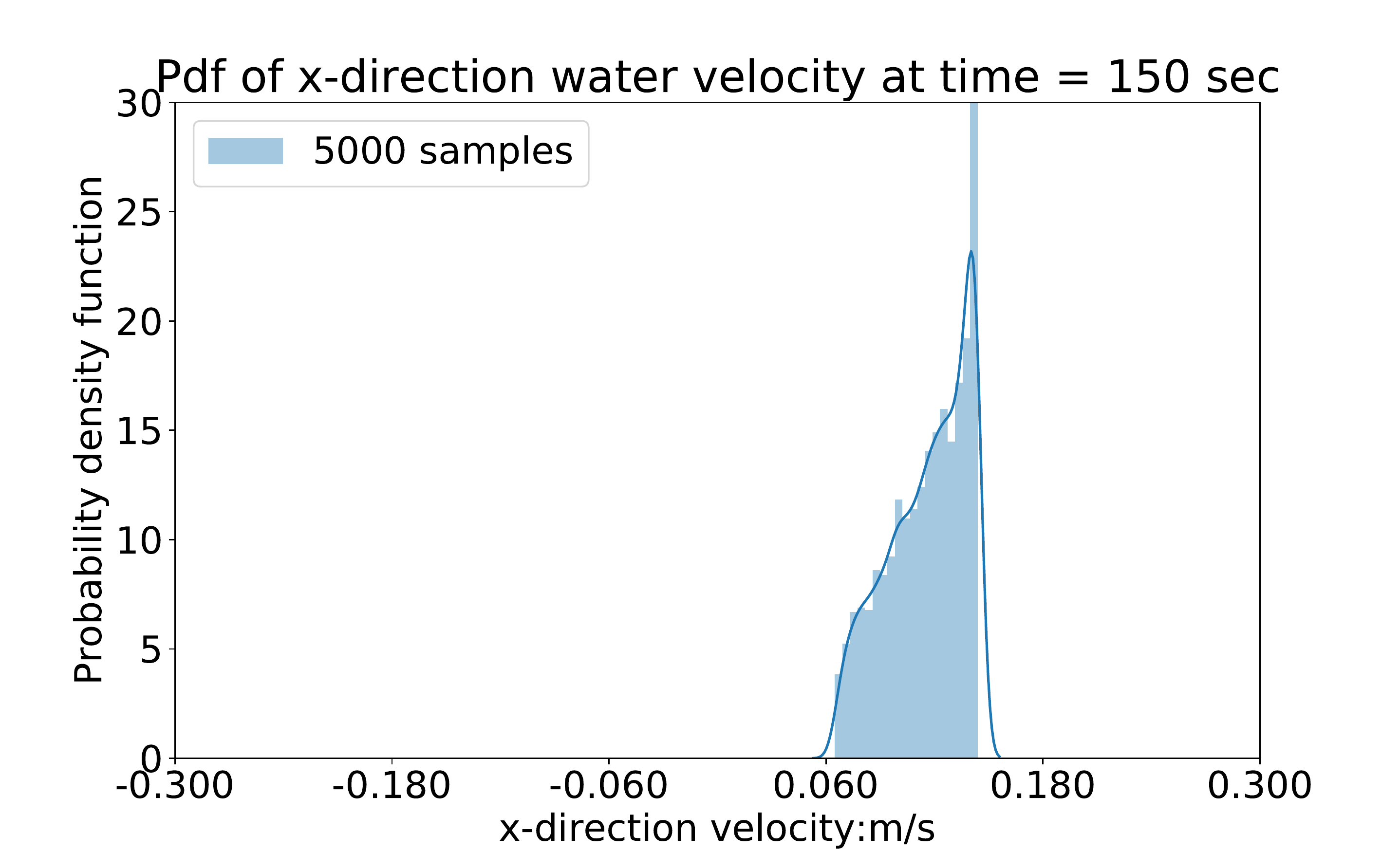}}\hfill
	\caption{$x$-direction velocity PDFs at $(750.0m, 100.0m)$.}
	\label{fig:humppdfu3}
\end{figure}

\moreR{For the case of uncertain bathymetry}, we  provide  PDFs at the points $(250.0m, 100.0m)$  and $(750.0m, 100.0m)$, which are located at one-quarter and three-quarters of the domain, respectively.  The PDFs of  surface elevation and water velocity at these spatial points at the selected time steps are shown in Figures~\ref{fig:humppdfeta} through~\ref{fig:humppdfu3}. Let us first compare the surface elevation and the water velocity PDFs at  $(250.0m, 100.0m)$ at time $t=110s$, i.e., Figure\moreR{s}~\ref{fig:humppdfeta}(a) and~\ref{fig:humppdfu}(a). Here we observe that the PDFs of both surface elevation and water velocity \one{appears to be of similar shape.} 
In fact, this phenomenon can be observed at all the six time steps, by comparison of the other \moreR{plots} in Figure\moreR{s}~\ref{fig:humppdfeta} and Figure~\ref{fig:humppdfu}. Furthermore, this trend is also observed at  $(750.0m, 100.0m)$ and all six time steps in Figures~\ref{fig:humppdfeta3} and~\ref{fig:humppdfu3}. Hence, the predicted PDFs between each output quantity ($\eta$ and $\boldsymbol{u}$) at a fixed spatial point at a fixed time are similar.
%
%
Second, let us compare the surface elevation PDFs at $(250.0m, 100.0m)$ at  each of the selected six time steps, i.e., compare all sub figures in Figure~\ref{fig:humppdfeta}. We again observe that the PDFs at these times are similar in the sense that they \one{resemble} beta distribution with changing parameters. 
This trend is also identical for the $x$-direction velocity component PDFs at $(250.0m, 100.0m)$ in Figure~\ref{fig:humppdfu} as well as for the other point in Figures~\ref{fig:humppdfeta3}, and~\ref{fig:humppdfu3}. Thus, at a fixed point, the predicted PDF for both output quantities \one{appear} similar, i.e., the stochastic process at a fixed spatial point for both output quantities \one{appear similarly distributed}.  And these observed trends for the selected points are identical for other points in the problem domain (omitted here for brevity, see~\cite{chenthesis}).

The observed trends in the PDFs are expected due to our use of the same polynomial chaos expansions for both $\boldsymbol{u}$ and $\eta$. As the task of the SSWM is to compute the stochastic modes at each  point and time, i.e., $\boldsymbol{u}_i(\boldsymbol{x},t)$ and $\eta_j(\boldsymbol{x},t)$, it is expected to observe the same type of distribution over the domain throughout the simulation time.

\subsection{The Time Varying variance field} \label{sec:time_varying_variance}


The computed surrogates from the SSWM can also be used to establish time series information on the largest uncertainty at fixed geographic locations. Alternatively, the surrogates can do the converse at a fixed time. To investigate the maximum values of the variance as well as the relationship between it and the extreme values of the mean, we consider the idealized inlet test case. 

\begin{figure}[h!]
	\centering
	\subfigure[]{\includegraphics[width=0.5\columnwidth]{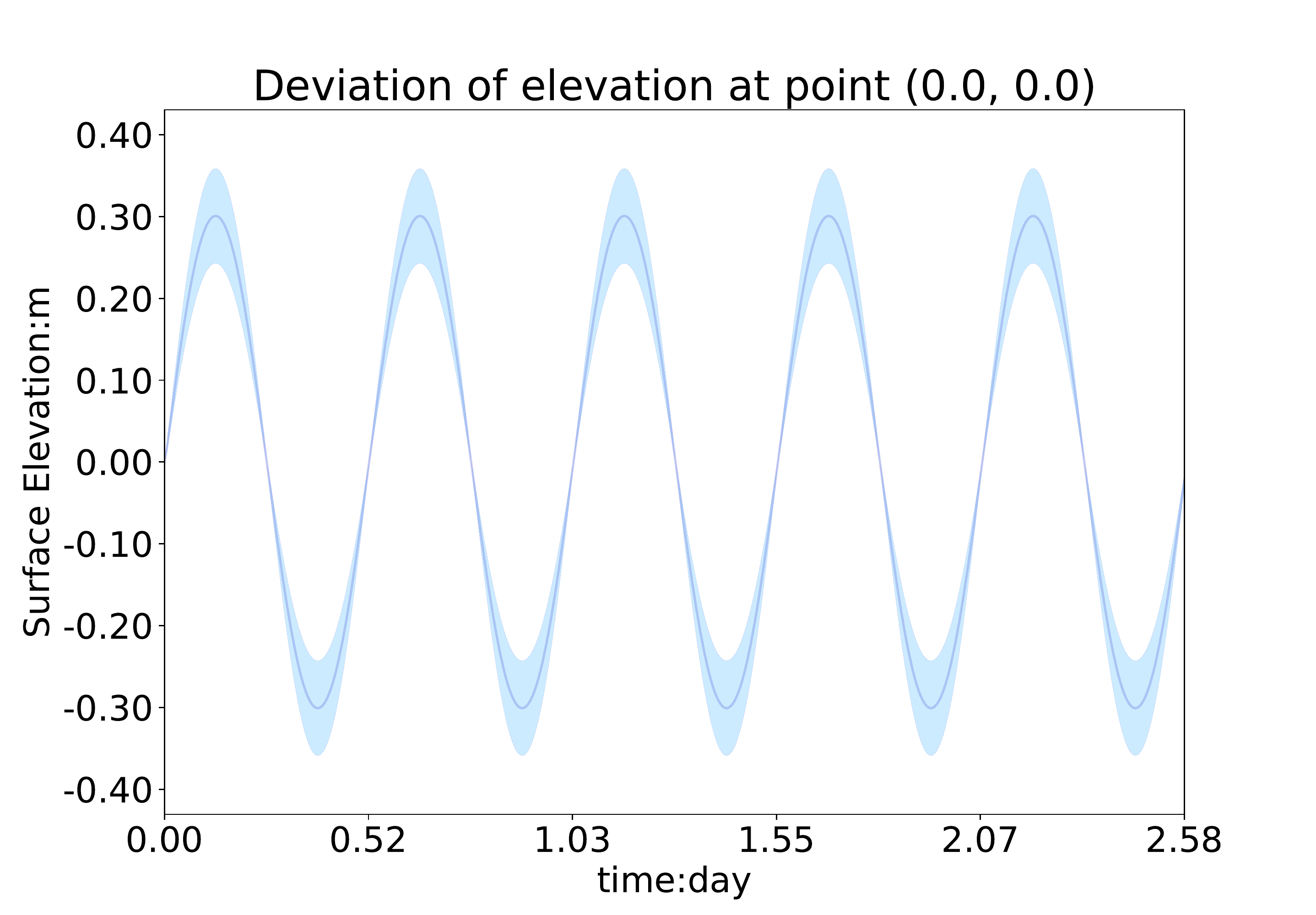}}\hfill
	\subfigure[]{\includegraphics[width=0.5\columnwidth]{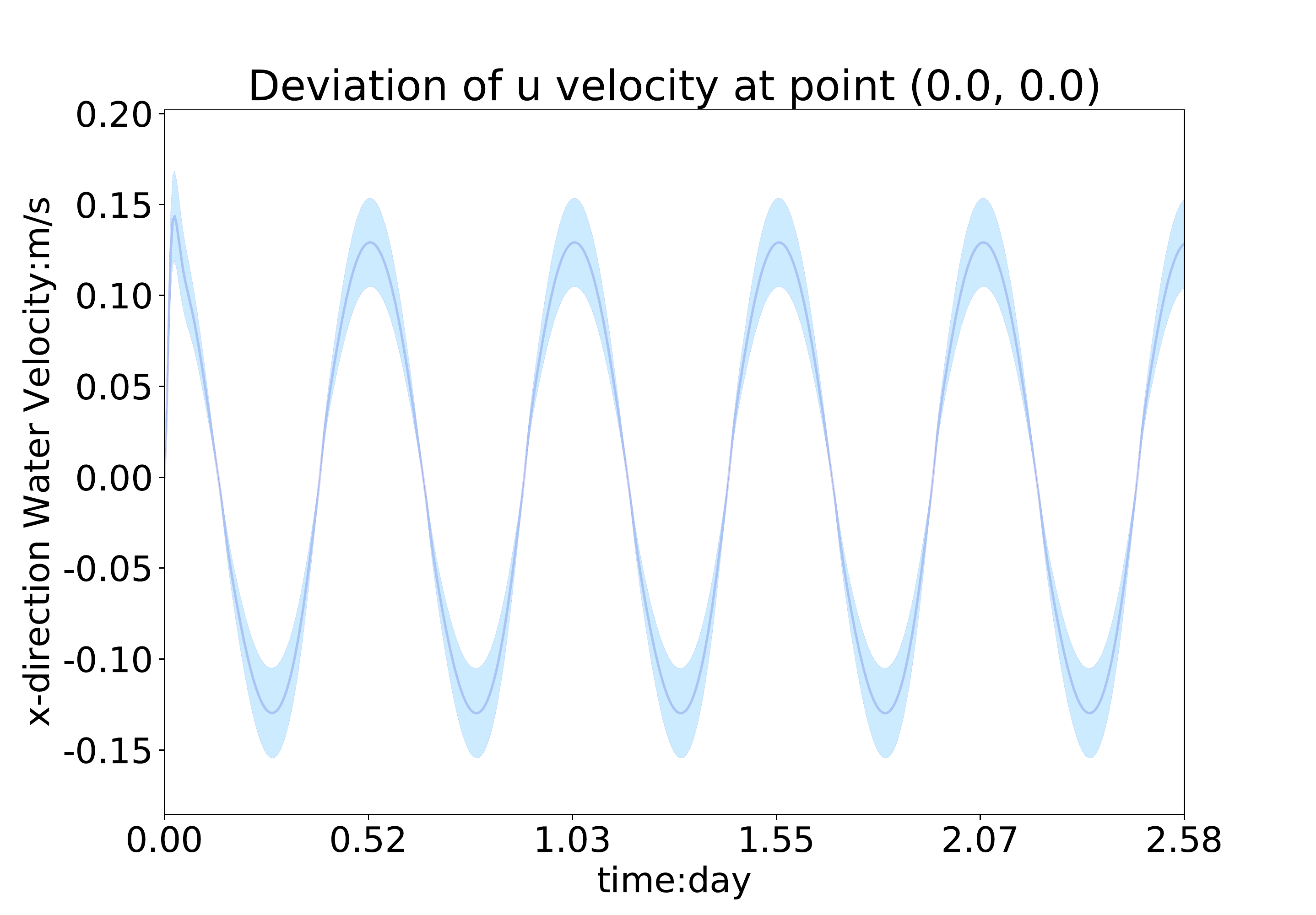}}\hfill
	\caption{ Mean and variance of surface elevation and $x$-direction of the velocity for the inlet test case at \one{a} selected point. }
	\label{fig:inletlineeta}
\end{figure}

First, we present the mean and variance time series plots \one{at a}  fixed spatial location:  $(0.0m, 0.0m)$. In Figure~\ref{fig:inletlineeta}, these are plotted for both surface elevation and $x$-direction velocity.
Here we observe that maximum variance occurs at the extreme value of the mean for both surface elevation and velocity. This indicates that the largest variance occurs at the extreme mean over time for both surface elevation and water velocity. \more{Similar trends are observed for other points in the domain and are omitted here for brevity.}

\begin{figure}[h!]
	\centering
	\subfigure{\includegraphics[width=0.85\columnwidth]{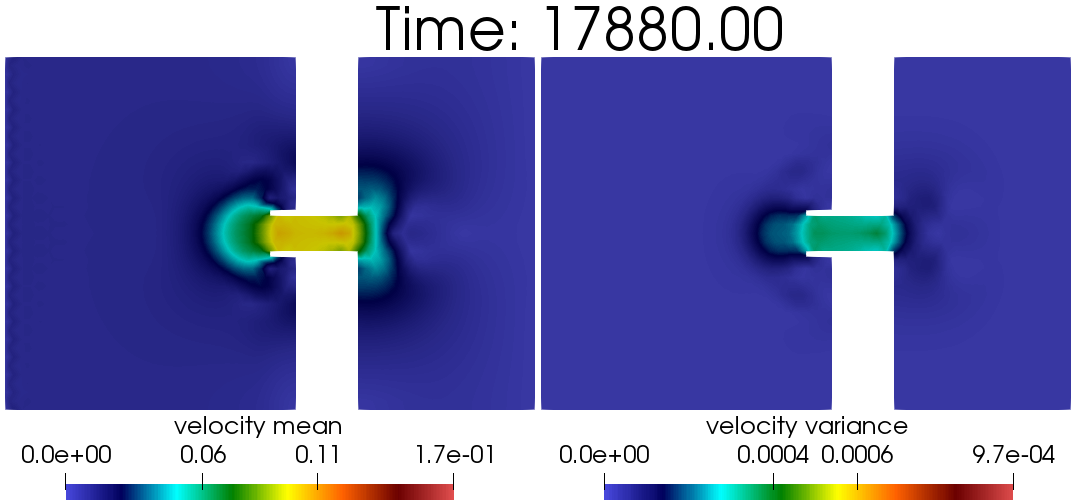}}\hfill
	\subfigure{\includegraphics[width=0.85\columnwidth]{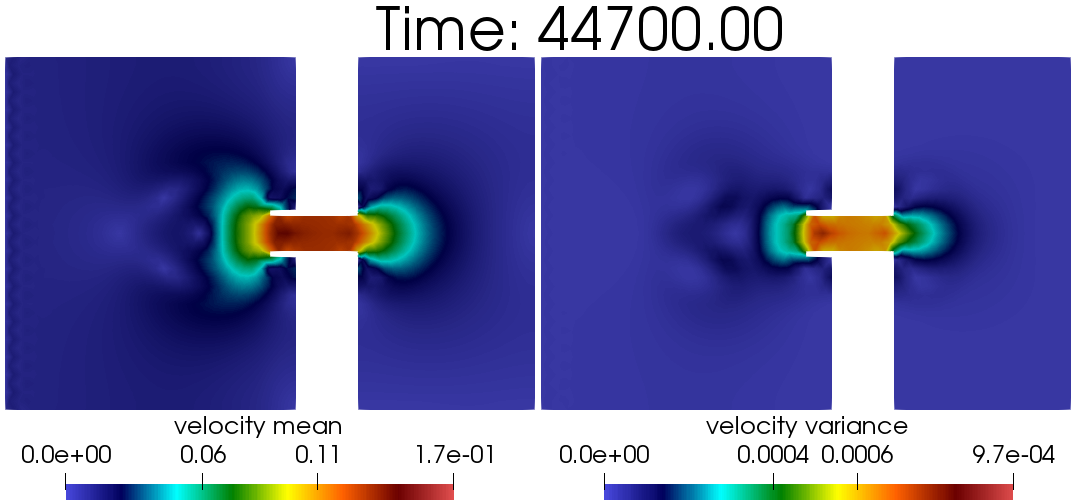}}\hfill
	\subfigure{\includegraphics[width=0.85\columnwidth]{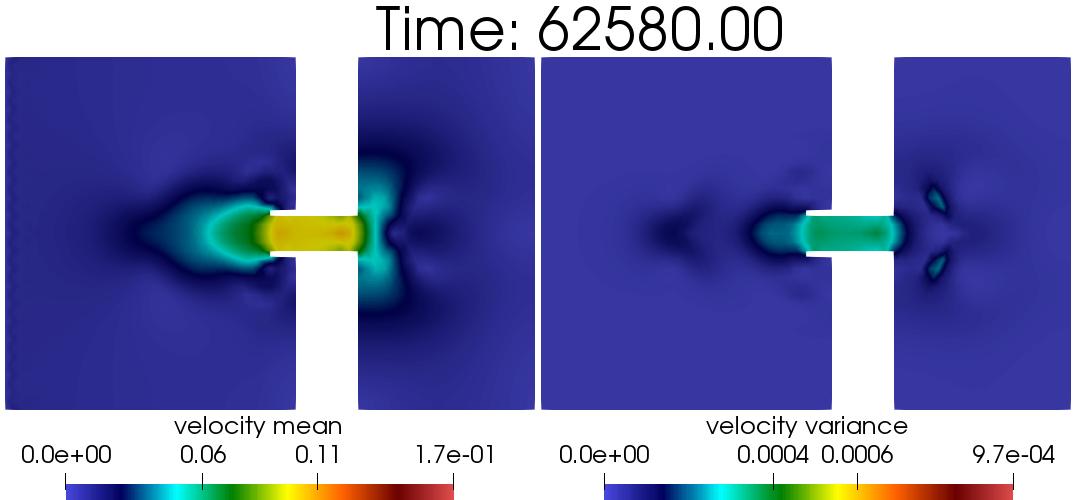}}\hfill
	\caption{Mean and variance of the velocity magnitude in the inlet test case.}
	\label{fig:inletvaru}
\end{figure}

Second, we consider the converse situation and explore the mean and variance at fixed times for the full domain. To this end, we select \more{three} times to show the mean and  variance of the velocity magnitude over the domain, shown in Figure~\ref{fig:inletvaru}. These variances are calculated by summation of the variance in both $x$- and $y$-direction velocities.
%
%
In these figures, we observe that the maximum variance of water velocity occurs at the extreme value of the mean field at all considered times. Further indicating that the largest variance occurs at the extreme mean over space for water velocity. In~\cite{chenthesis}, an identical trend is observed for the water surface elevation for the Hurricane Ike test case. Hence, we conclude that maximum variance occurs at the extreme mean for both surface elevation and water velocity. 

\clearpage

\section{Prediction of Hurricane Storm Surge under Uncertain Wind Drag Coefficient During Hurricane Ike} \label{sec:hurricane_prediction}

\begin{figure}[h!]
	\centering
	\includegraphics[width=0.9\textwidth]{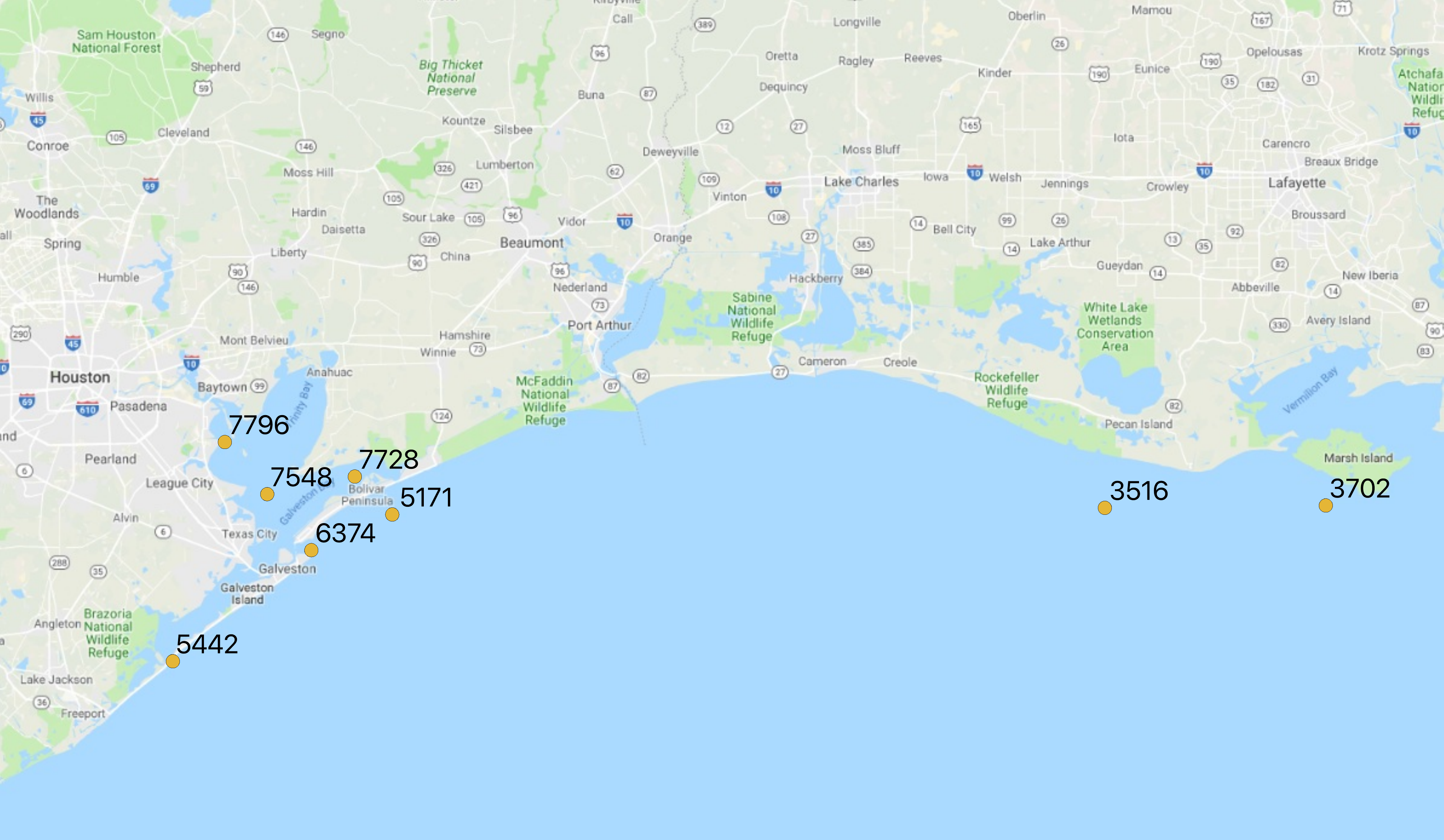}
	\caption{Spatial points for model comparison during Hurricane Ike 2008.}
	\label{fig:ikechoose}
\end{figure}

In this section, we use \moreR{the H}urricane \moreR{Ike} test case introduced in Section~\ref{sec:hurricanetests} to show the reliability of the full stochastic model SSWM. For comparison purposes, we also apply the ADCIRC model to the same cases using identical inputs and FE meshes. The only difference between these models are the wind drag coefficients. In the SSWM it is assumed to be uncertain, with the form of $C_d = \xi_1 C_d^{Powell}$, $\xi_1$  uniformly distributed $\xi_1 \sim U(0.8, 1.2)$; whereas in ADCIRC it is  deterministic, i.e., $C_d =C_d^{Powell}$. 
To present a comparison of the SSWM and the deterministic ADCIRC model, we perform three sets of numerical experiments in each of the following three sub sections. These comparisons will be performed by considering specific points on the Texas and Louisiana coasts, shown in Figure~\ref{fig:ikechoose}. Note that the numbering of these locations is based on the numbering of the nodes in our FE mesh. \more{In the following Sections~\ref{sec:com_ADC_PDF_storm} and~\ref{sec:com_ADC_PDF_stormtime}, we only consider Hurricane Ike, and selected results for Hurricane Harvey are presented  in~\ref{sec:hurricane_prediction2} from which similar conclusions may be drawn.  } 


\subsection{Time Series Surface Elevation Comparison}
\label{sec:com_ADC_PDF_stormtime}

\begin{figure}[h!]
	\centering
	\subfigure[]{\includegraphics[width=0.4\columnwidth]{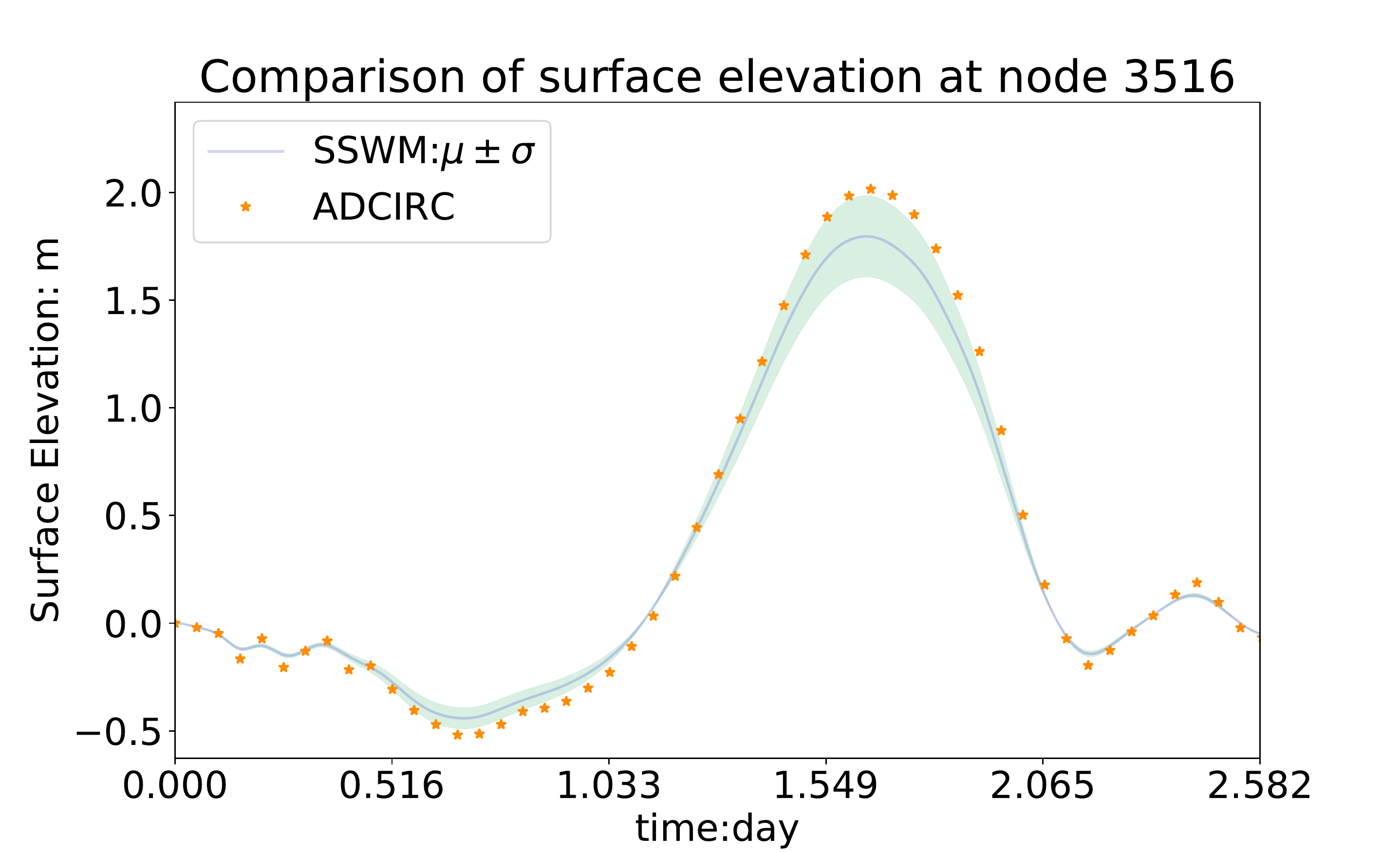}}\hfill
	\subfigure[]{\includegraphics[width=0.4\columnwidth]{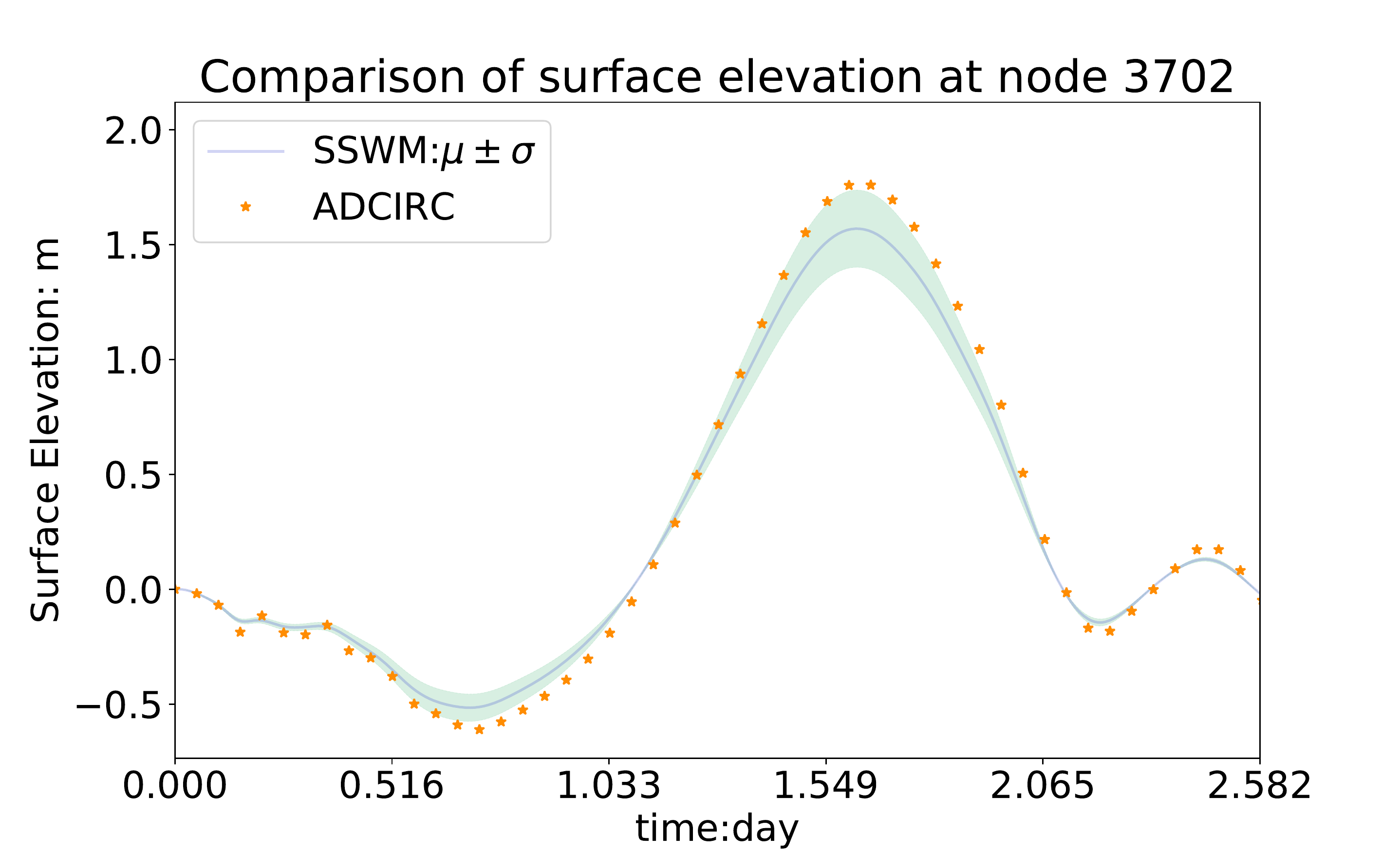}}\hfill
	\subfigure[]{\includegraphics[width=0.4\columnwidth]{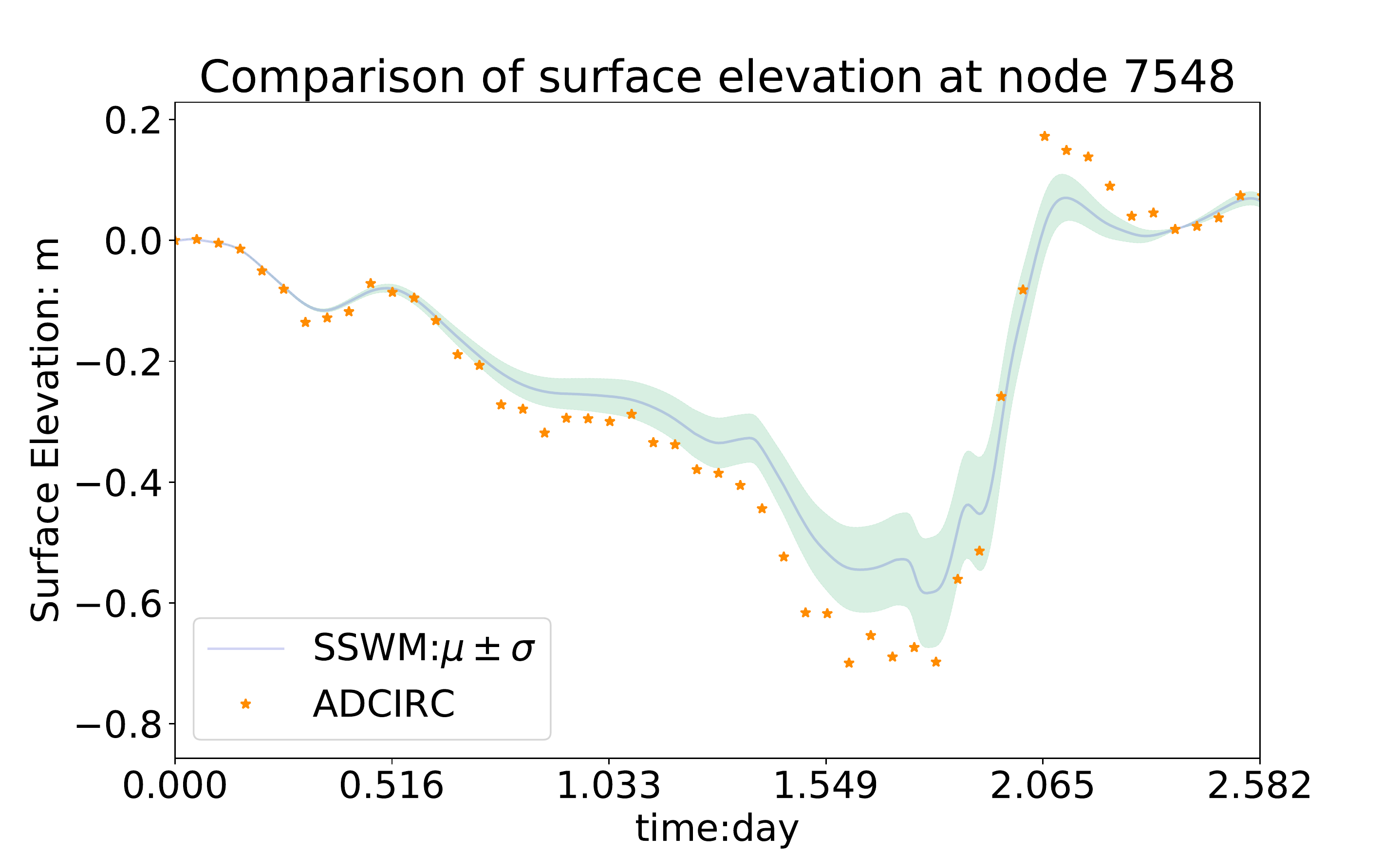}}\hfill
	\subfigure[]{\includegraphics[width=0.4\columnwidth]{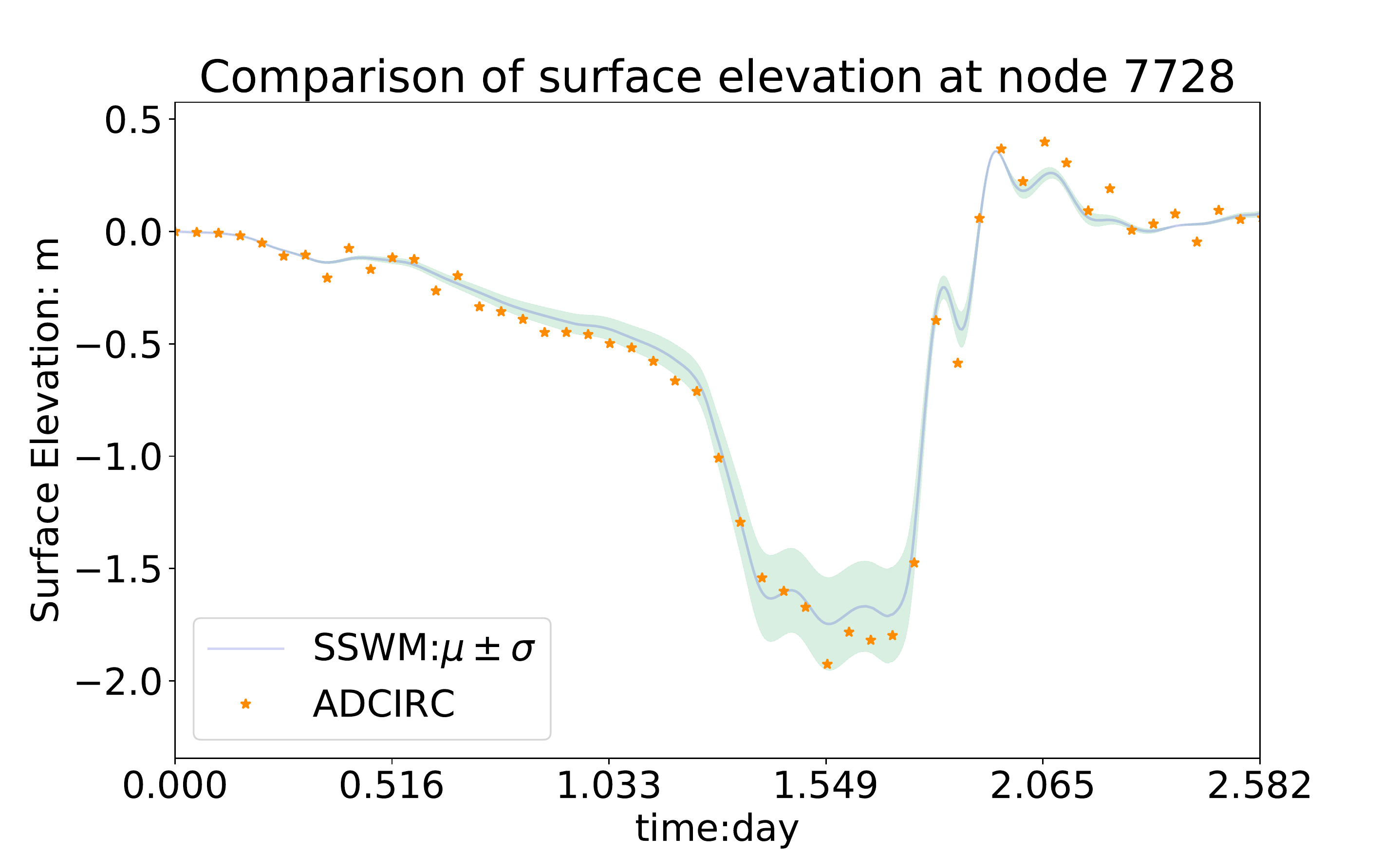}}\hfill
	\subfigure[]{\includegraphics[width=0.4\columnwidth]{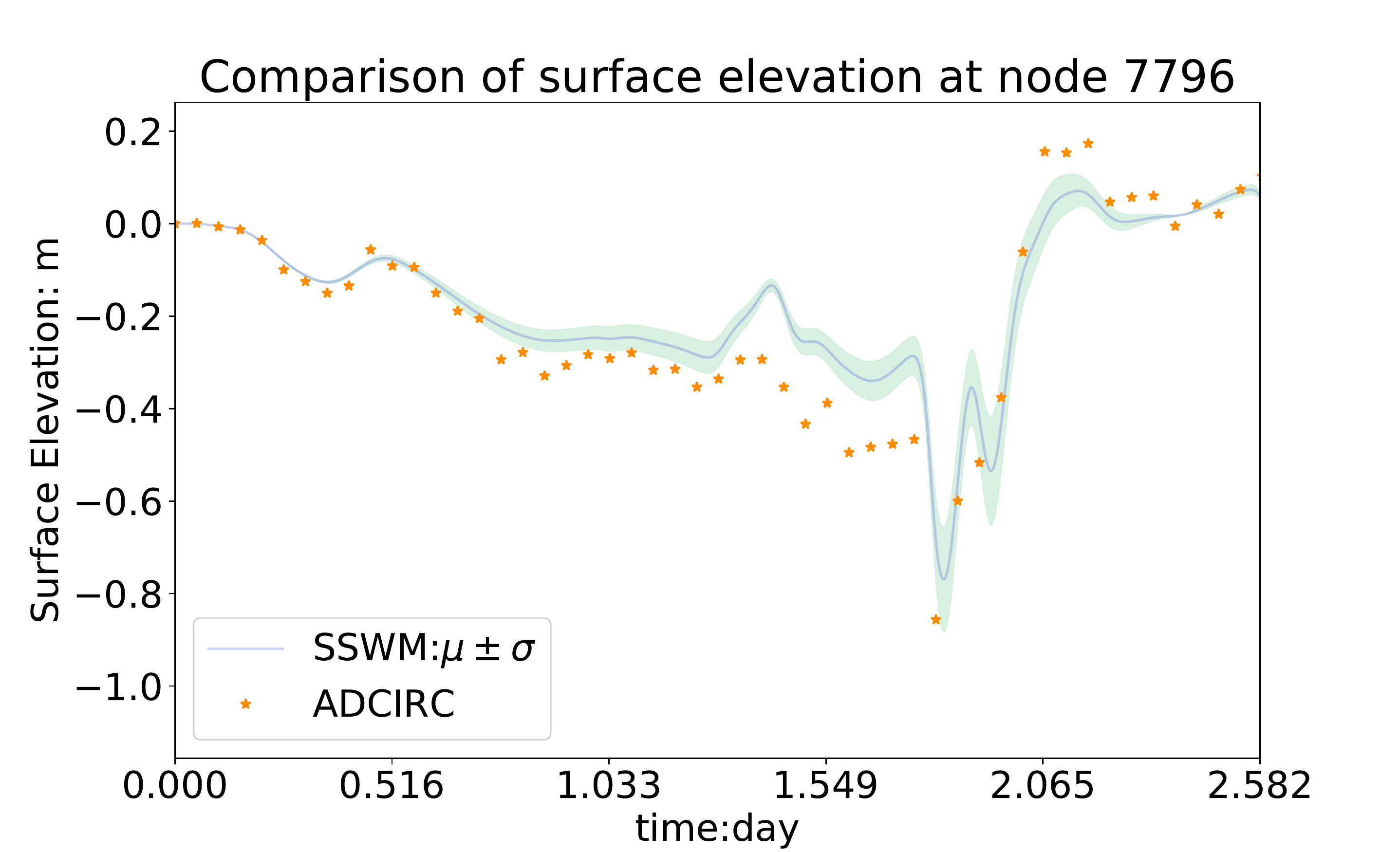}}\hfill
	\subfigure[]{\includegraphics[width=0.4\columnwidth]{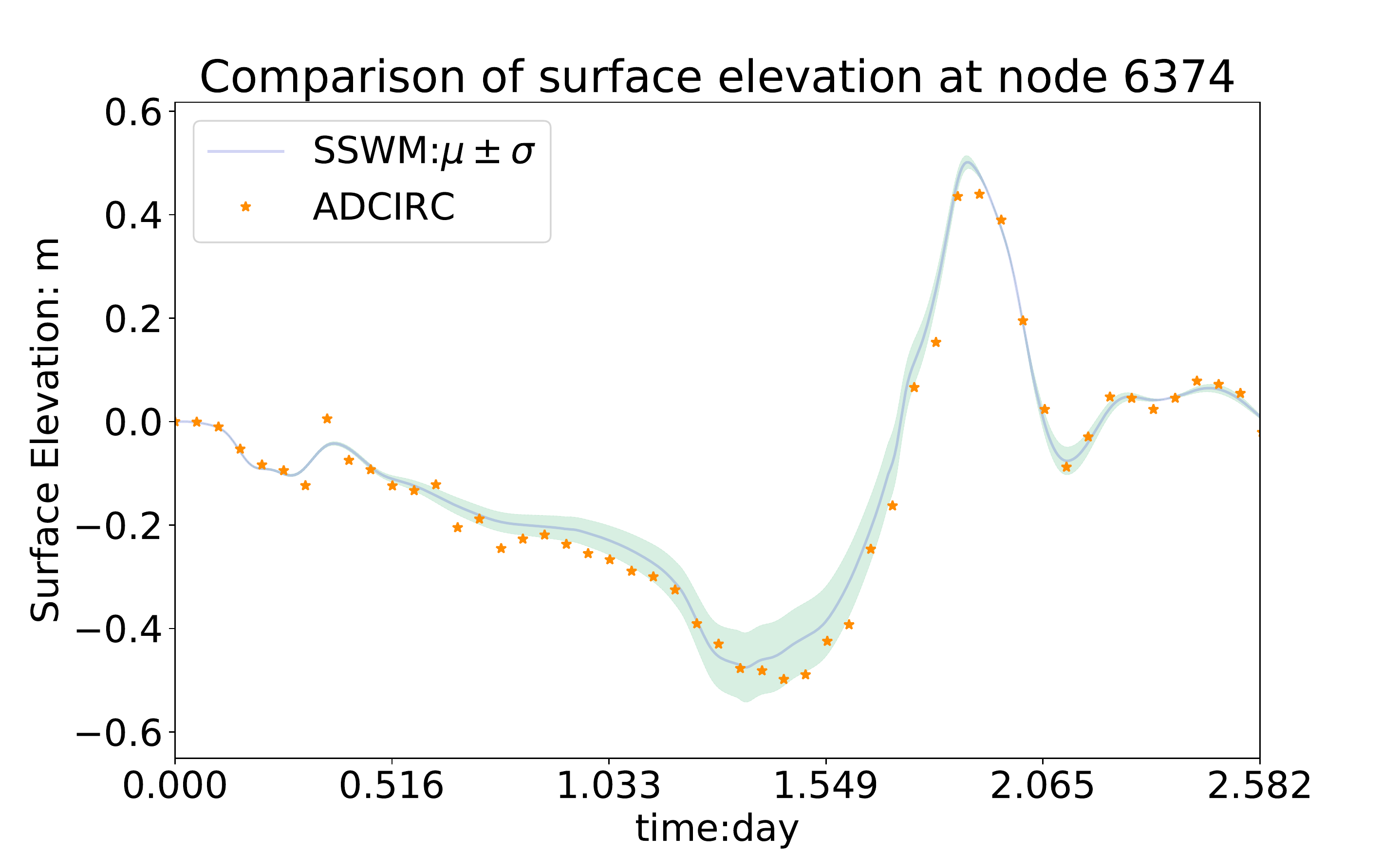}}\hfill
%
	\subfigure[]{\includegraphics[width=0.4\columnwidth]{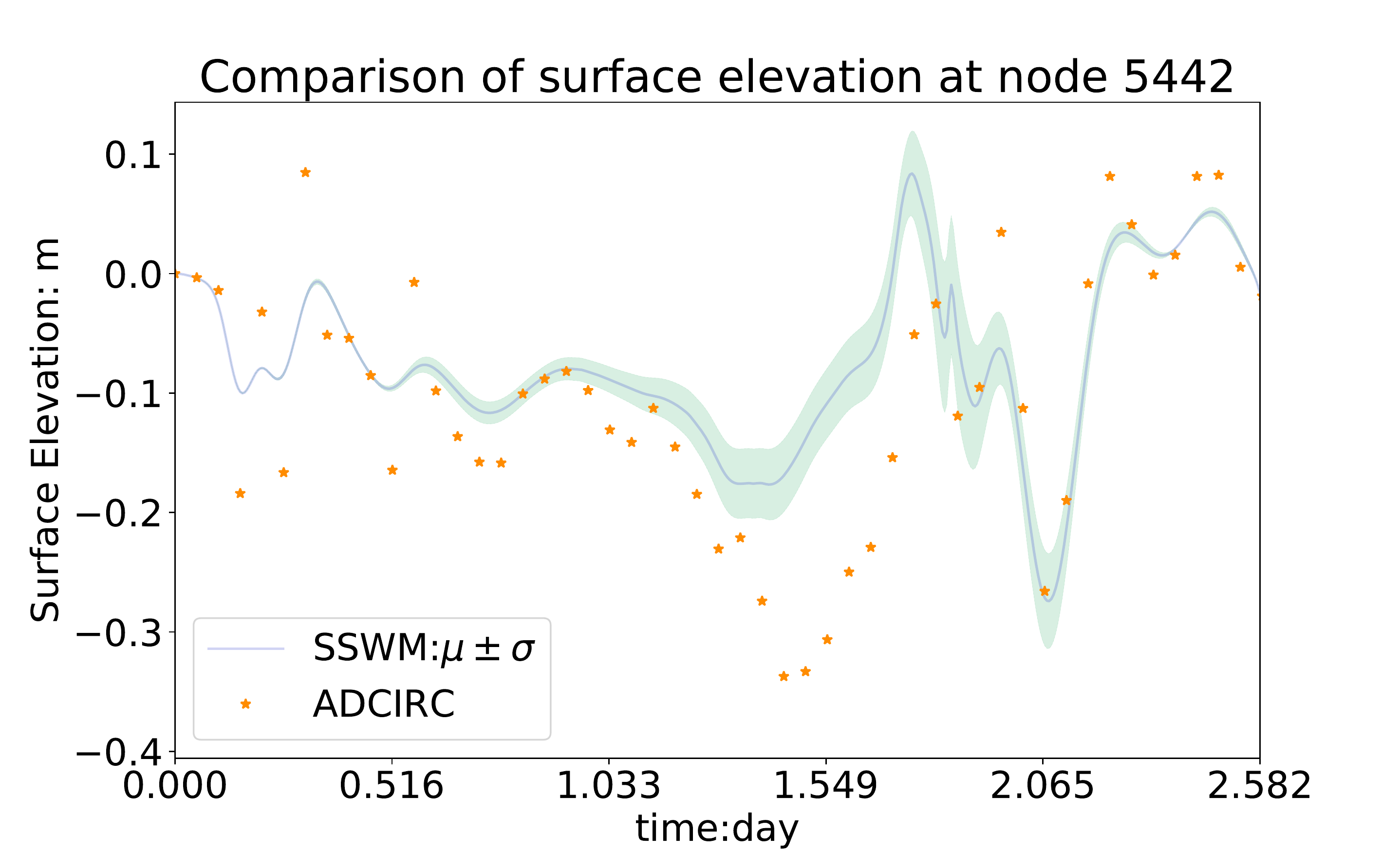}}\hfill
	\subfigure[]{\includegraphics[width=0.4\columnwidth]{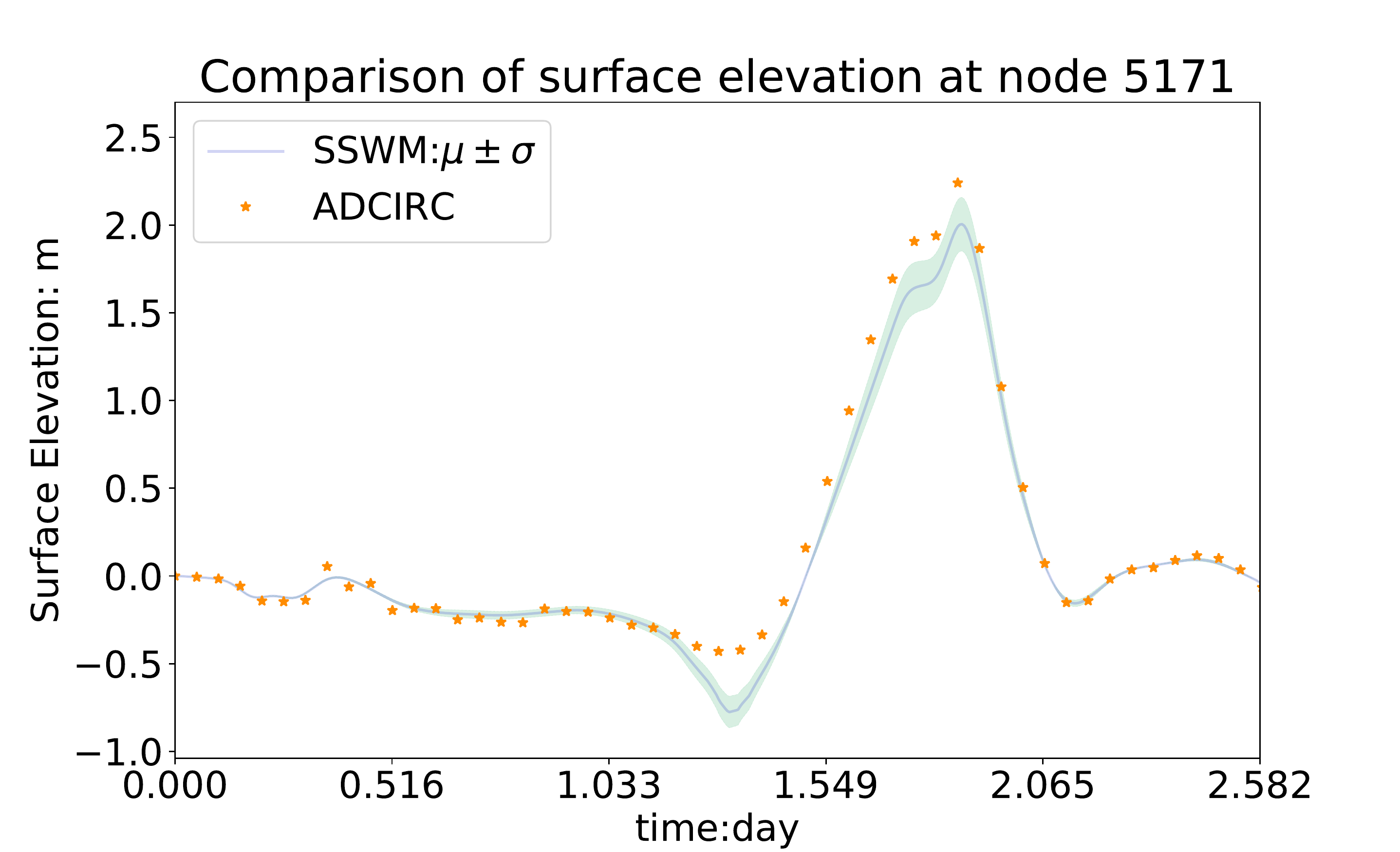}}\hfill
	\caption{SSWM Surface elevation surrogate compared to deterministic ADCIRC output for Hurricane Ike.}
	\label{fig:ikelineeta}
\end{figure}

We first compare the surface elevations between ADCIRC and the SSWM at specific locations on the Texas coast. The SSWM provides a predicted PDF at each point in space and time and we therefore visually present the mean and variance by plotting the interval $[\mu - \sigma, \mu + \sigma]$.  To this end we select eight points along the coast to  compare the time series surface elevation against the benchmark ADCIRC solution. The results are presented in Figure~\ref{fig:ikelineeta}, 
%
%
where we observe that the SSWM mean solution agrees very well with the benchmark ADCIRC solution, which demonstrates that the trend of the predicted mean solution is reliable compared to that of ADCIRC. \more{Note that the discrepancies seen in Figure~\ref{fig:ikelineeta}(g) at node 5442  are likely due to the strong localized effects this location experiences as it is close to a tidal inlet during Hurricane Ike.} Also observe the good agreement in the timing of the peak elevation which is of great importance when producing forecasts. This demonstrates that the timing of the peak surge given by the predicted mean solution is reliable compared to that of ADCIRC. 
We also observe that the benchmark ADCIRC solution around the local peak surge falls within the green shaded area at most of the points in Figure~\ref{fig:ikelineeta}. Note that the predicted range of elevation must be a superset of this interval which indicates that the predicted range also contains the benchmark ADCIRC solution. Finally, we note that the variance reaches its maximum in both troughs and crests of the time series.
%

\clearpage
\subsection{Comparison of the SSWM Predicted PDF and ADCIRC} \label{sec:com_ADC_PDF_storm}

As demonstrated in the previous subsection,  $[\mu - \sigma, \mu + \sigma]$ produces a solution range that is reliable enough  near the hurricane's landfall compared to ADCIRC. However, it cannot completely represent the statistical information (i.e., the predicted range or support of the PDF) given by the SSWM. To further demonstrate the reliability of the SSWM, we also compare the predicted PDF at specific points on the coast against the benchmark ADCIRC solution. To plot the  predicted PDF, we will use 50,000 samples for three selected spatial points in each scenario.
%

\begin{figure}[h!]
	\centering
	\subfigure{\includegraphics[width=0.4\columnwidth]{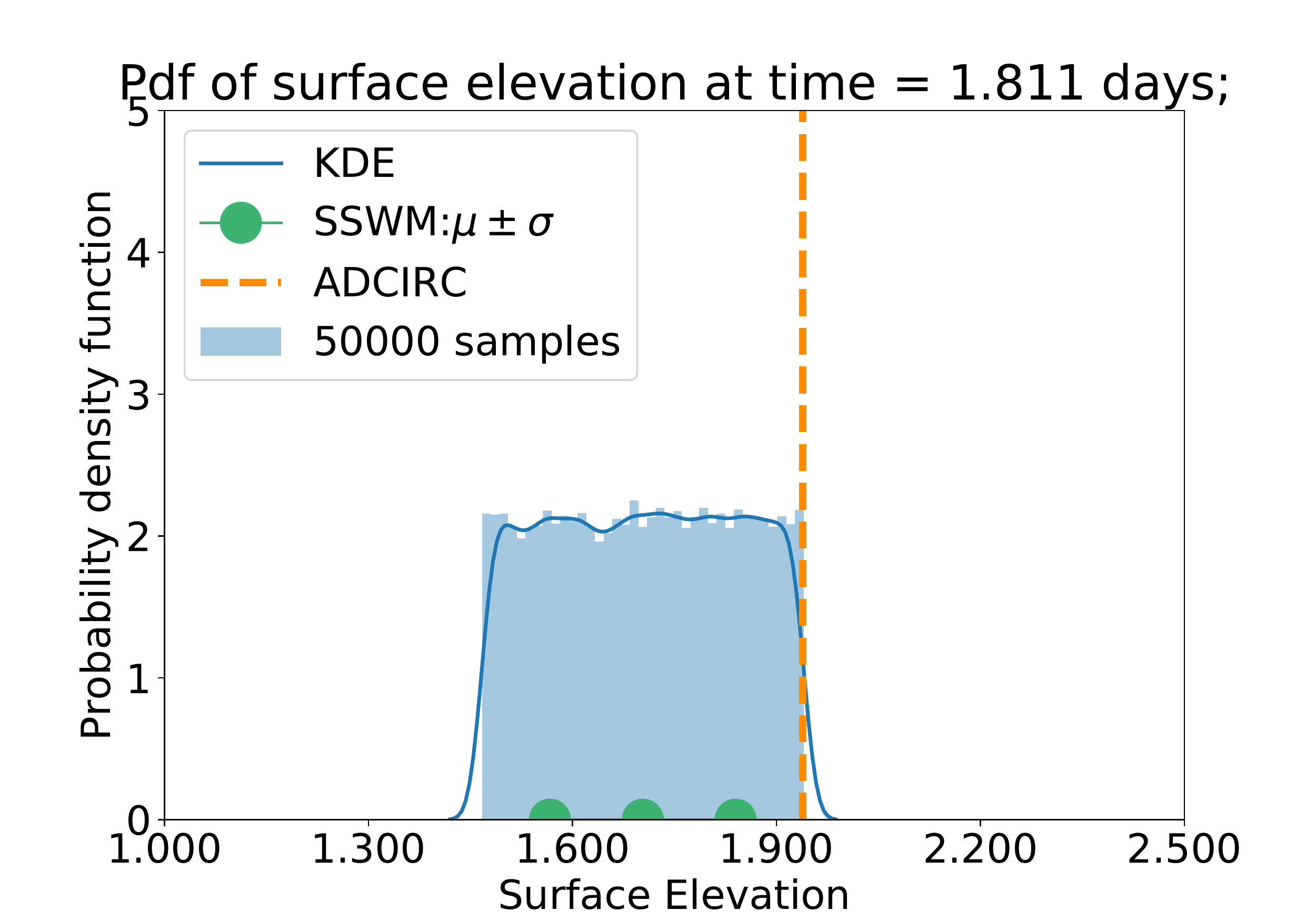}}\hfill
	\subfigure{\includegraphics[width=0.4\columnwidth]{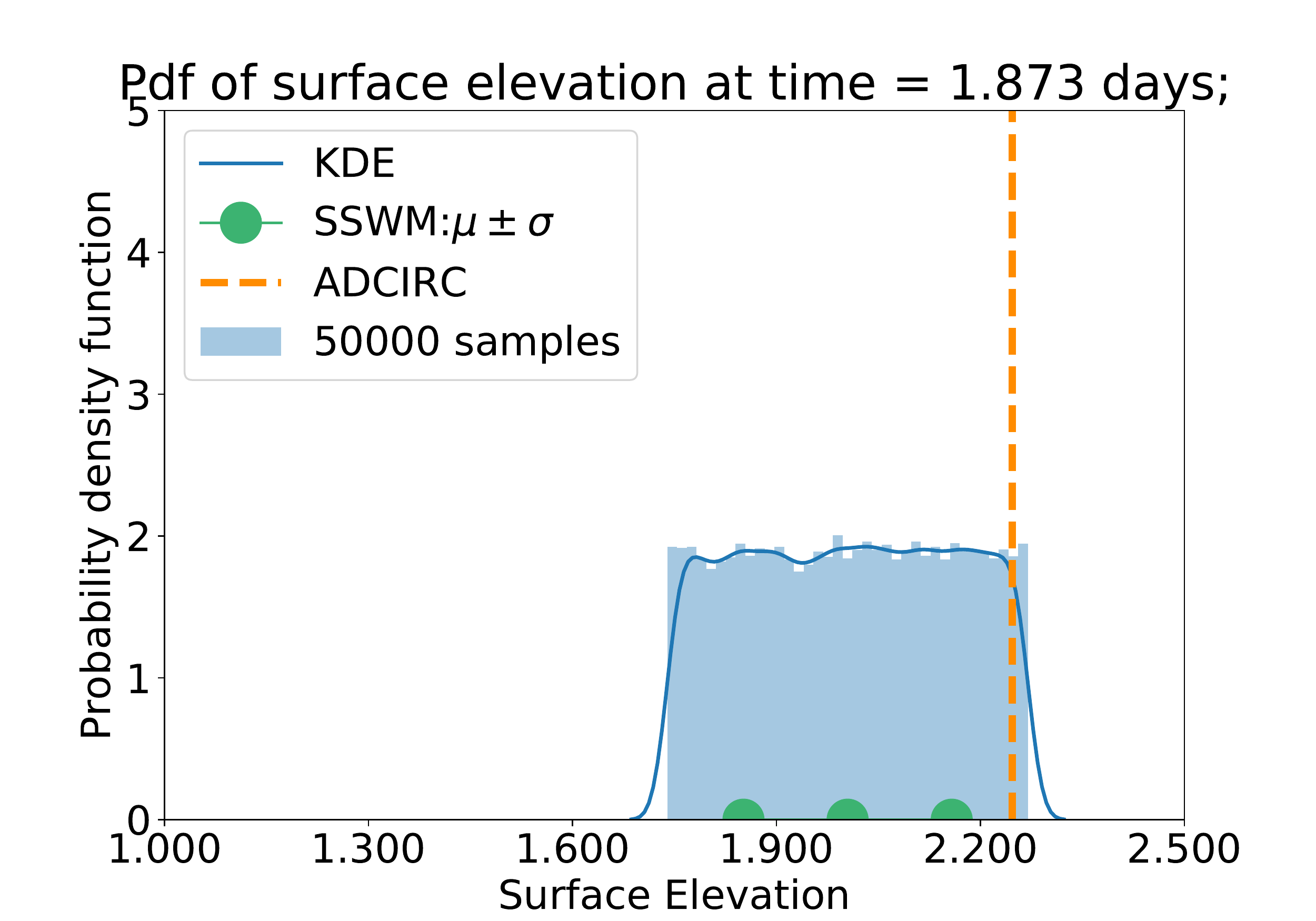}}\hfill
	\subfigure{\includegraphics[width=0.4\columnwidth]{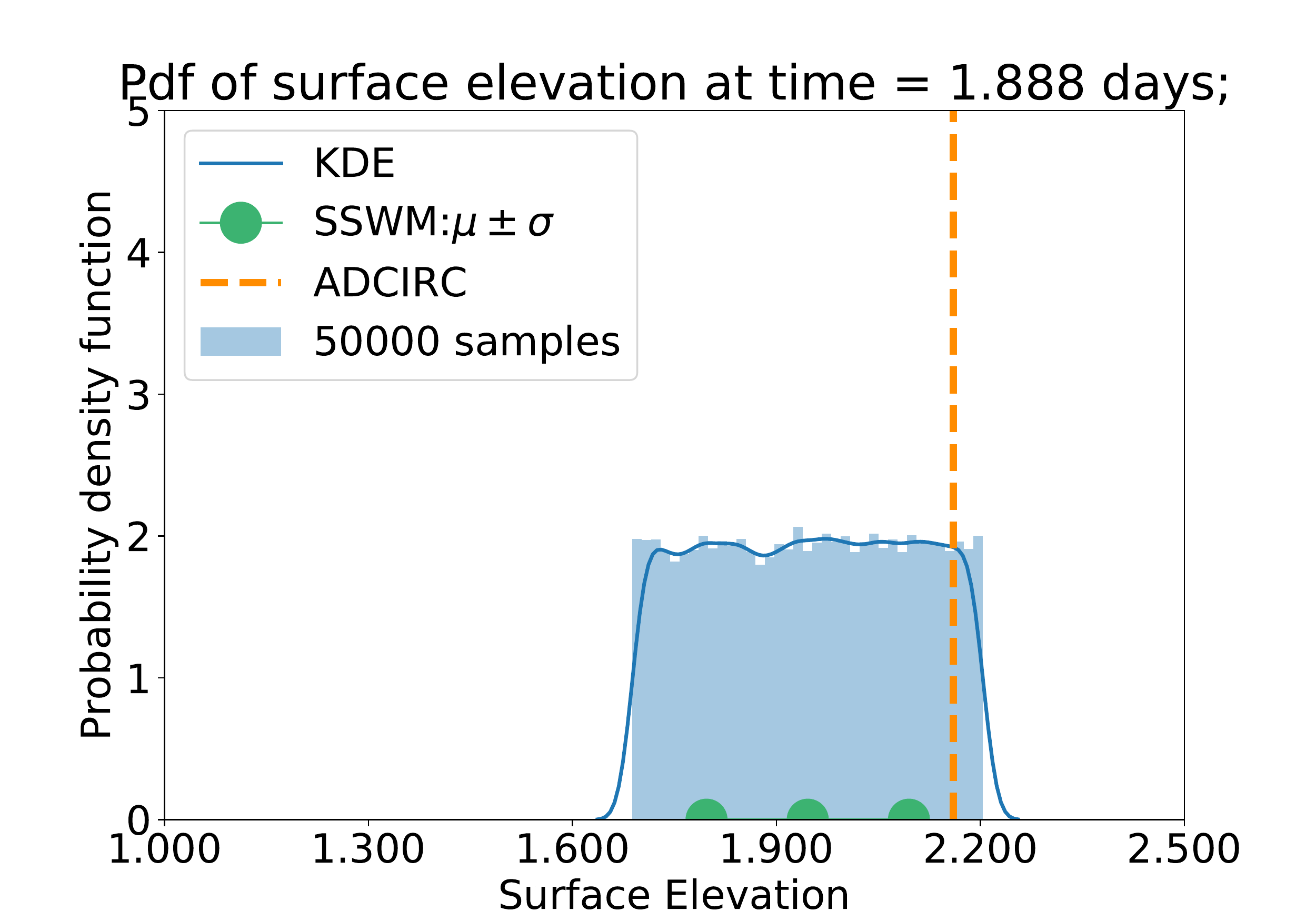}}\hfill
	\subfigure{\includegraphics[width=0.4\columnwidth]{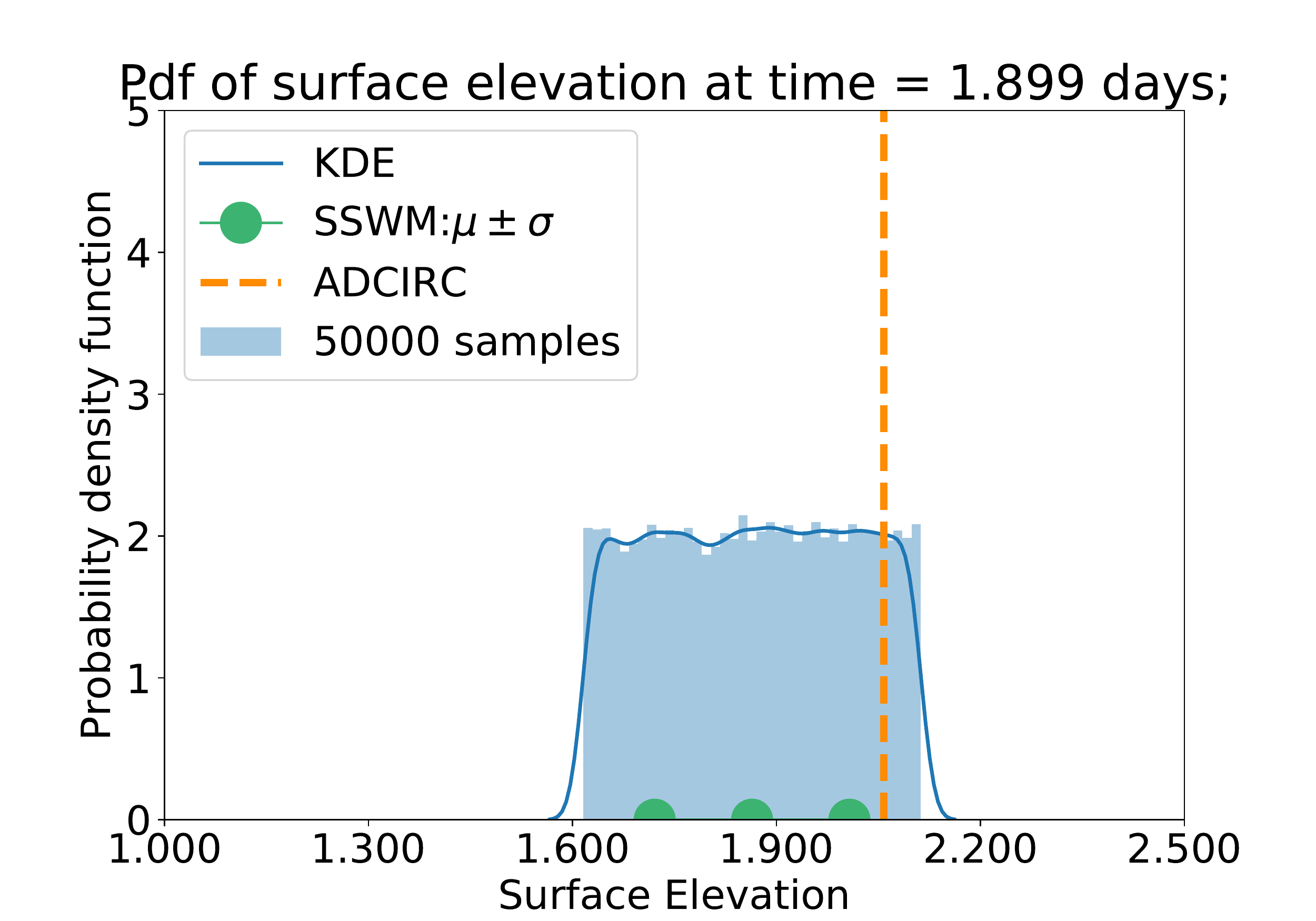}}\hfill
	\caption{Predicted PDF of surface elevation compared to ADCIRC at point 5171 during Hurricane Ike.}
	\label{fig:ike5171}
\end{figure}

\begin{figure}[h!]
	\centering
	\subfigure{\includegraphics[width=0.4\columnwidth]{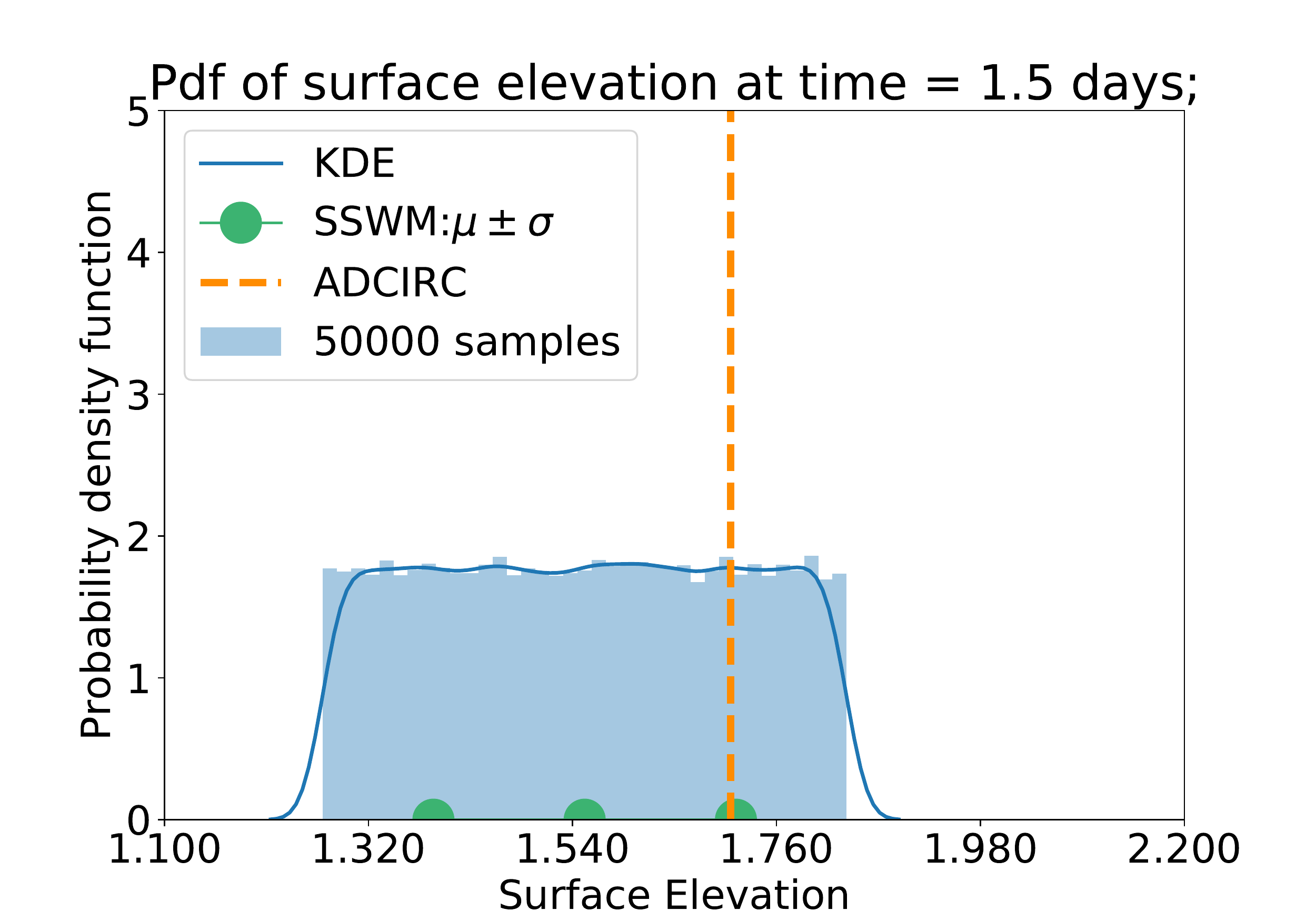}}\hfill
	\subfigure{\label{fig:ike3516widerange}\includegraphics[width=0.4\columnwidth]{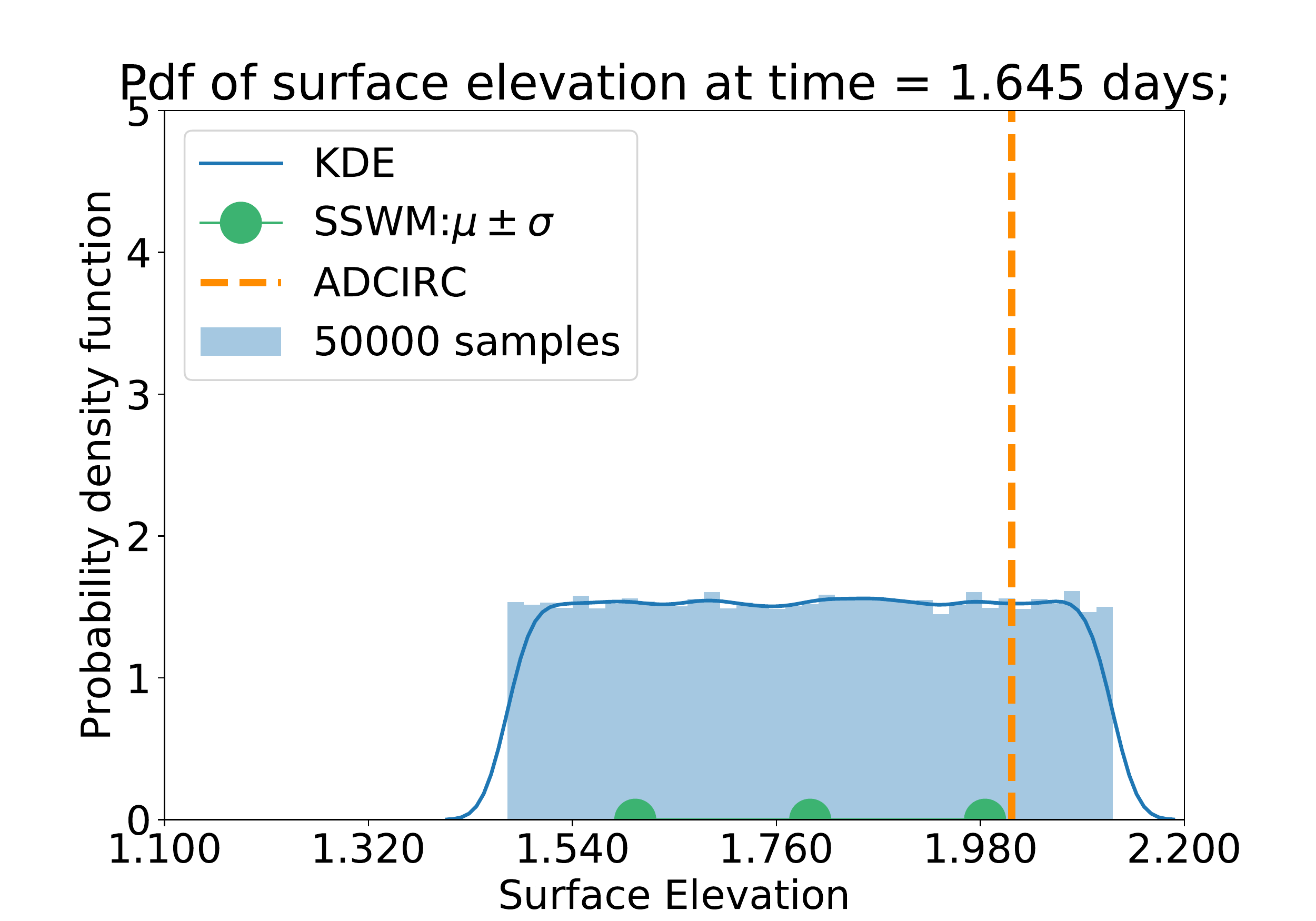}}\hfill
	\subfigure{\includegraphics[width=0.4\columnwidth]{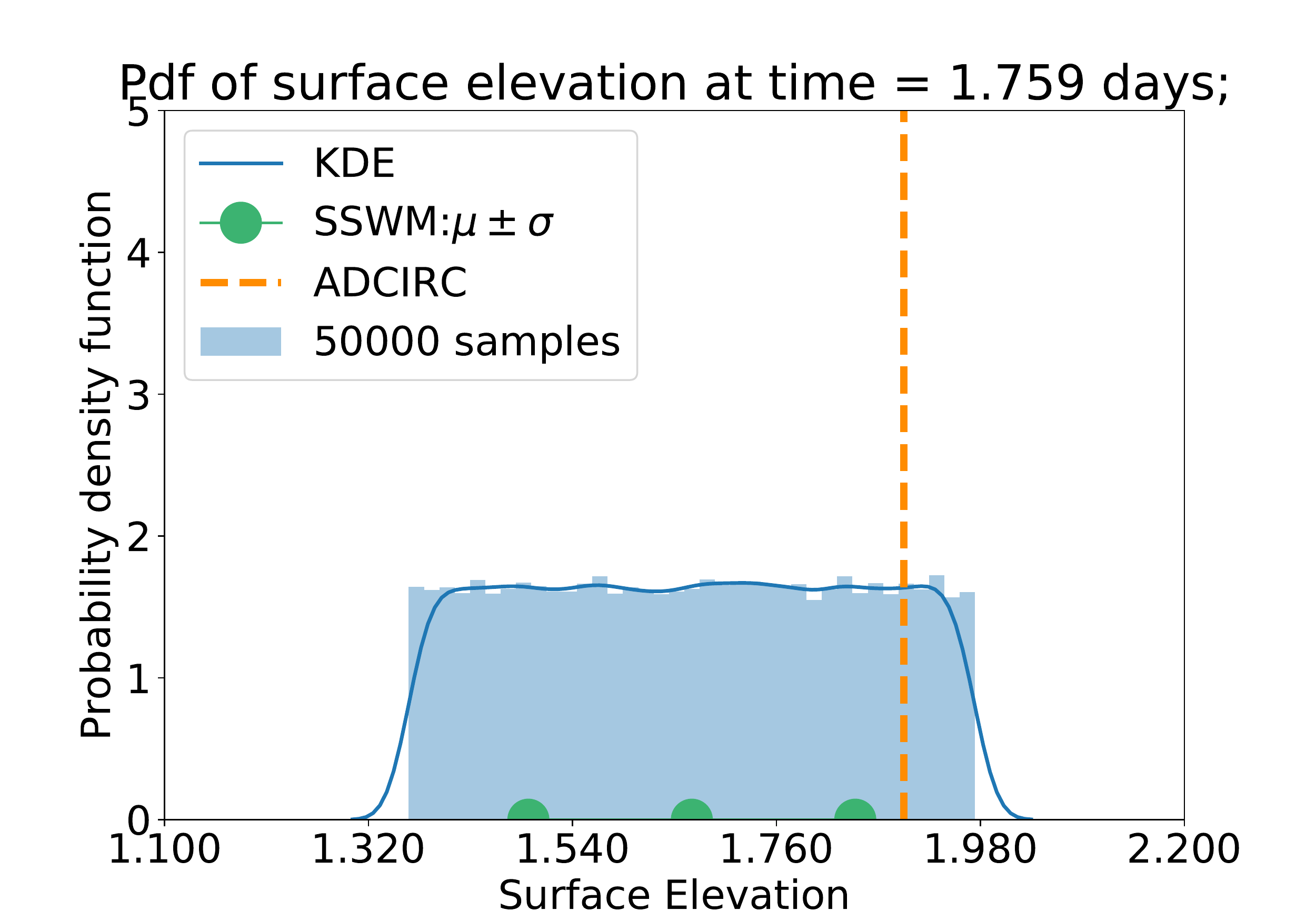}}\hfill
	\subfigure{\includegraphics[width=0.4\columnwidth]{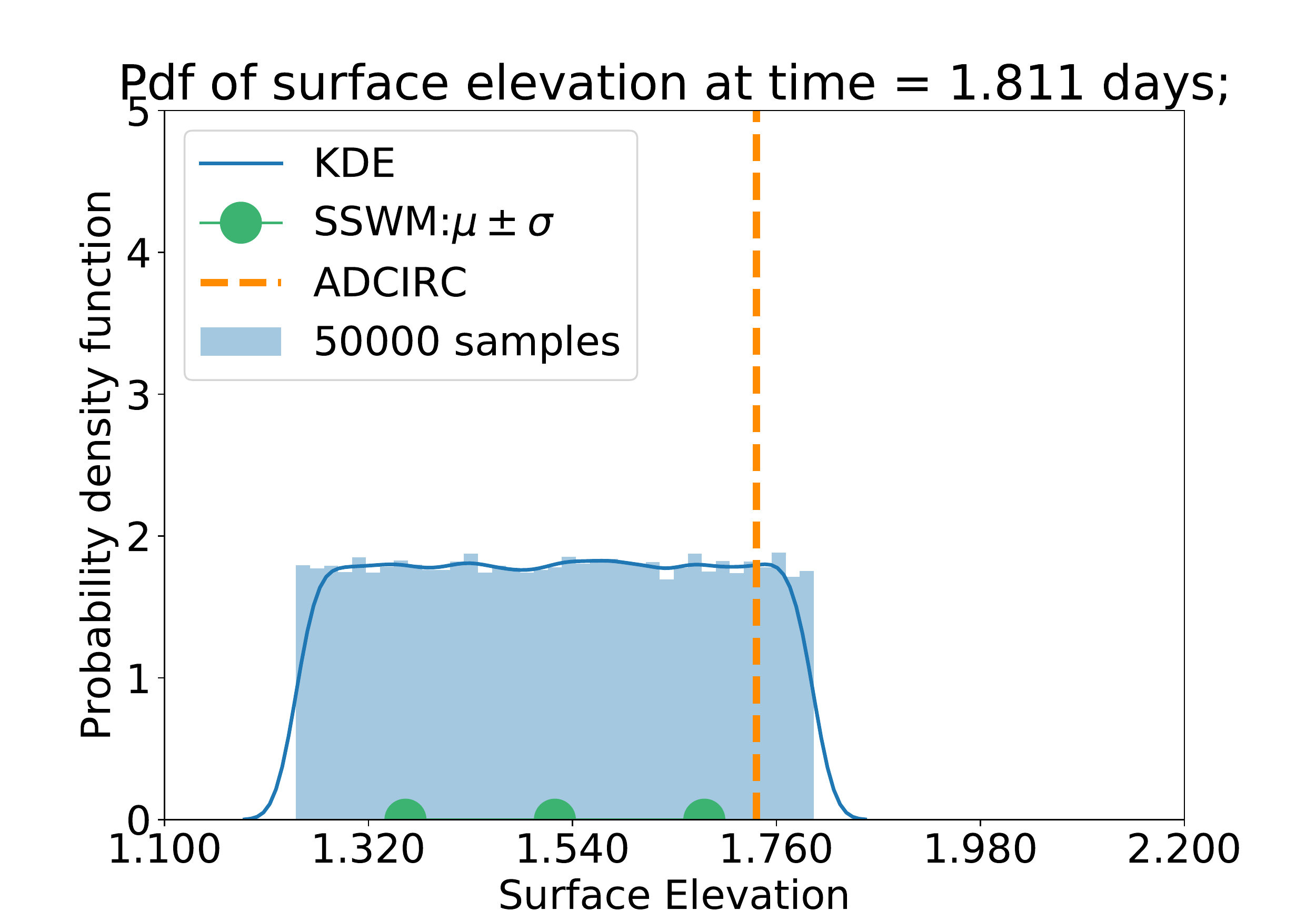}}\hfill
	\caption{Predicted PDF of surface elevation compared to ADCIRC at point 3516 during Hurricane Ike.}
	\label{fig:ike3516}
\end{figure}

\begin{figure}[h!]
	\centering
	\subfigure{\includegraphics[width=0.4\columnwidth]{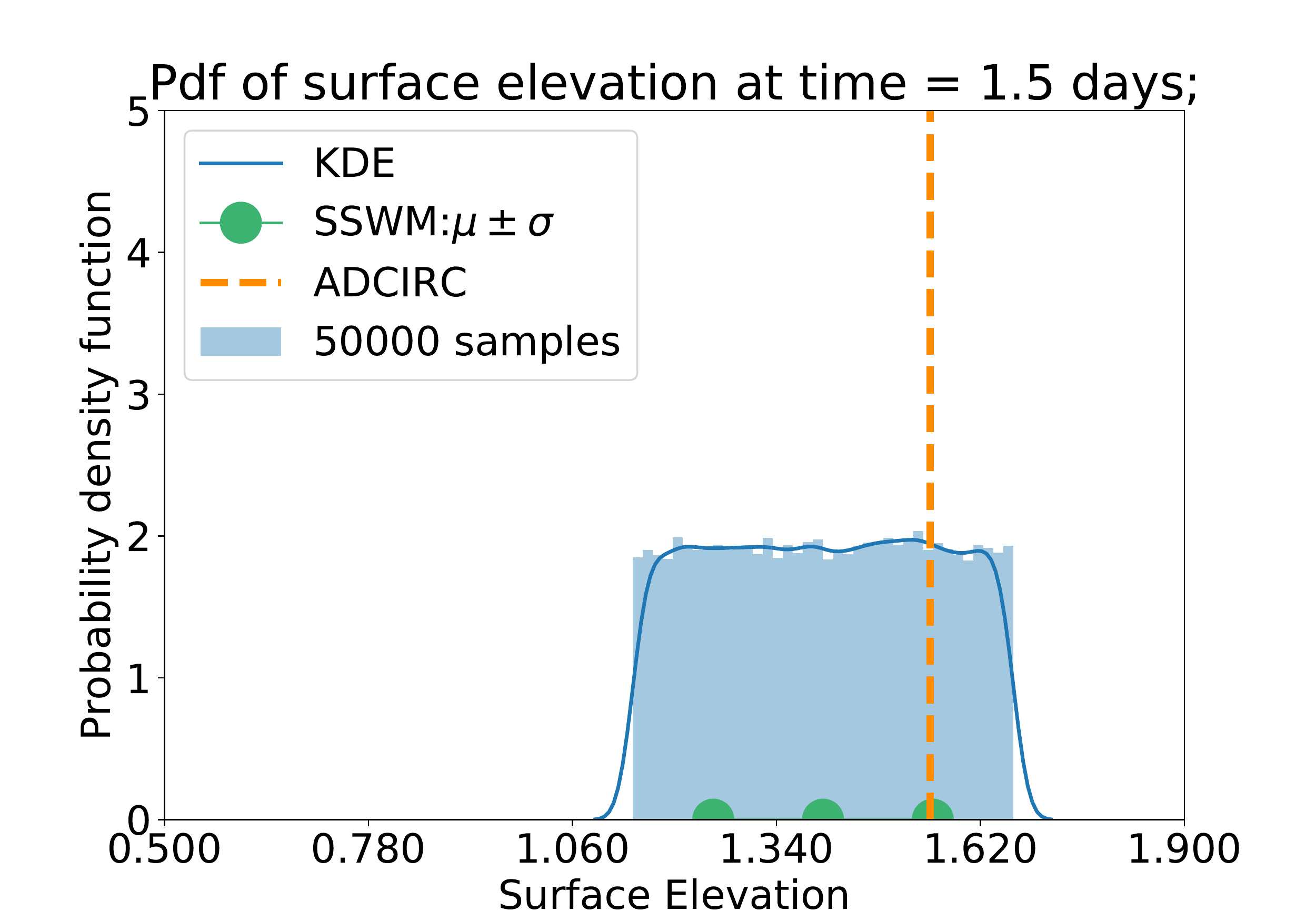}}\hfill
	\subfigure{\includegraphics[width=0.4\columnwidth]{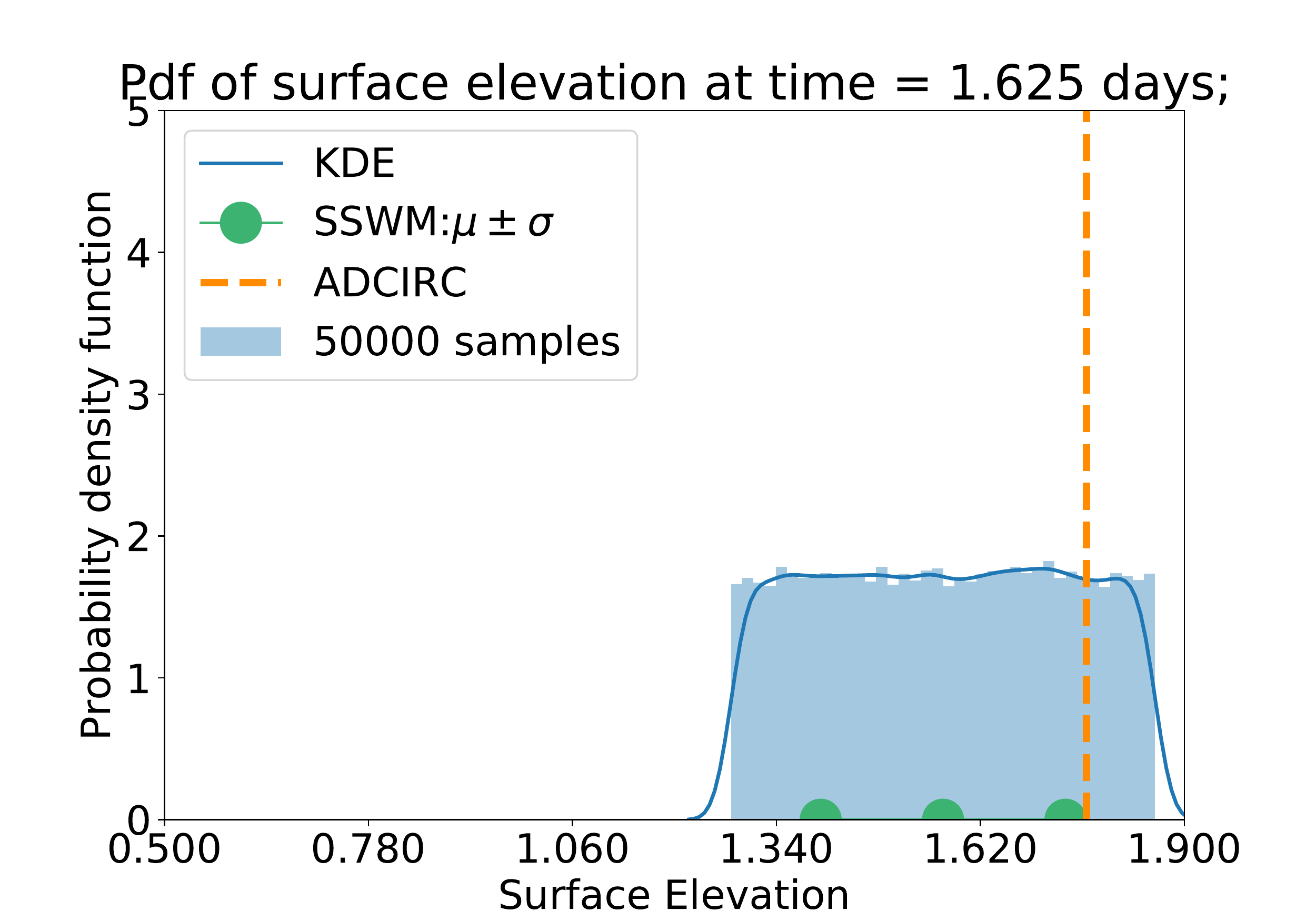}}\hfill
	\subfigure{\includegraphics[width=0.4\columnwidth]{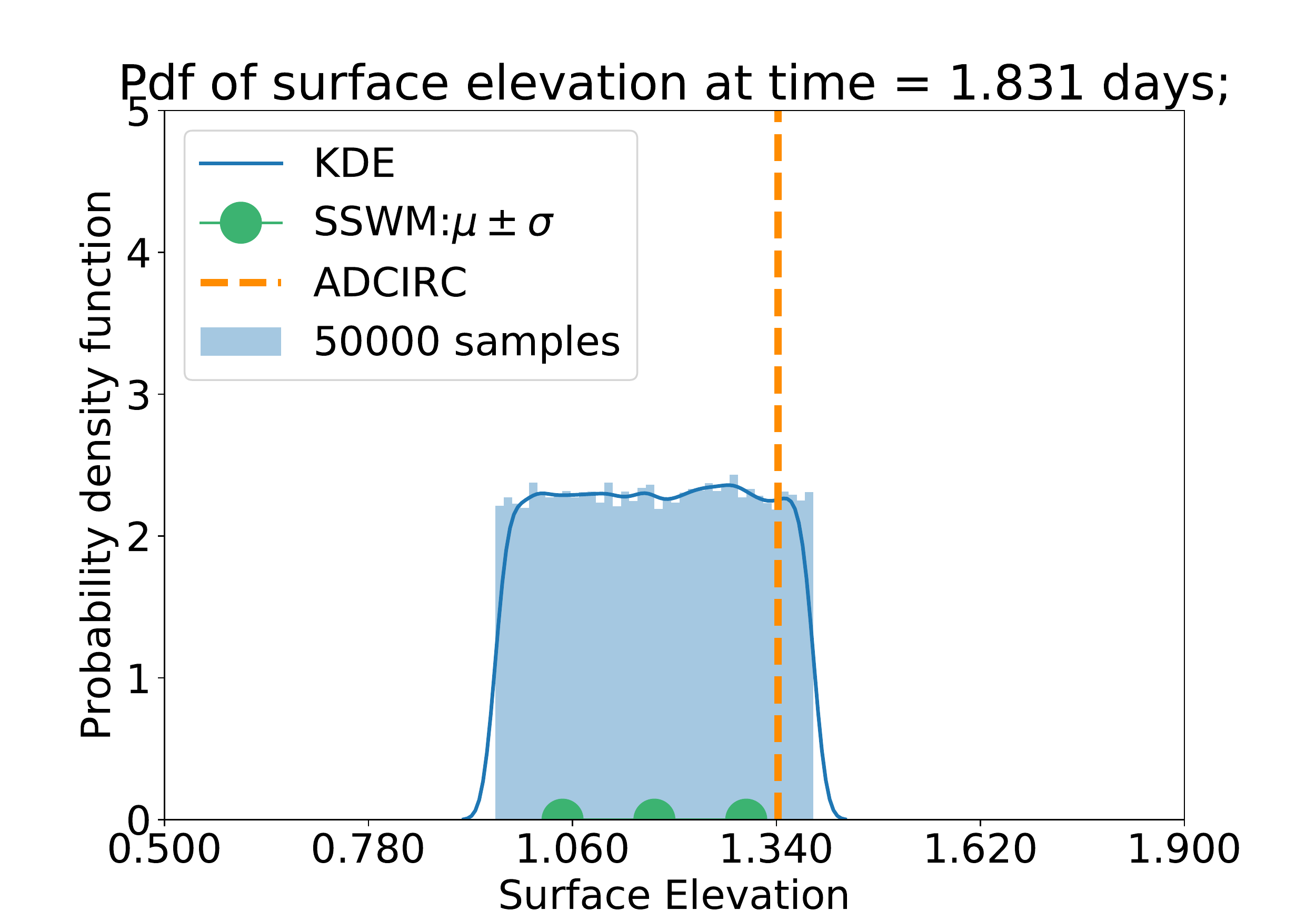}}\hfill
	\subfigure{\includegraphics[width=0.4\columnwidth]{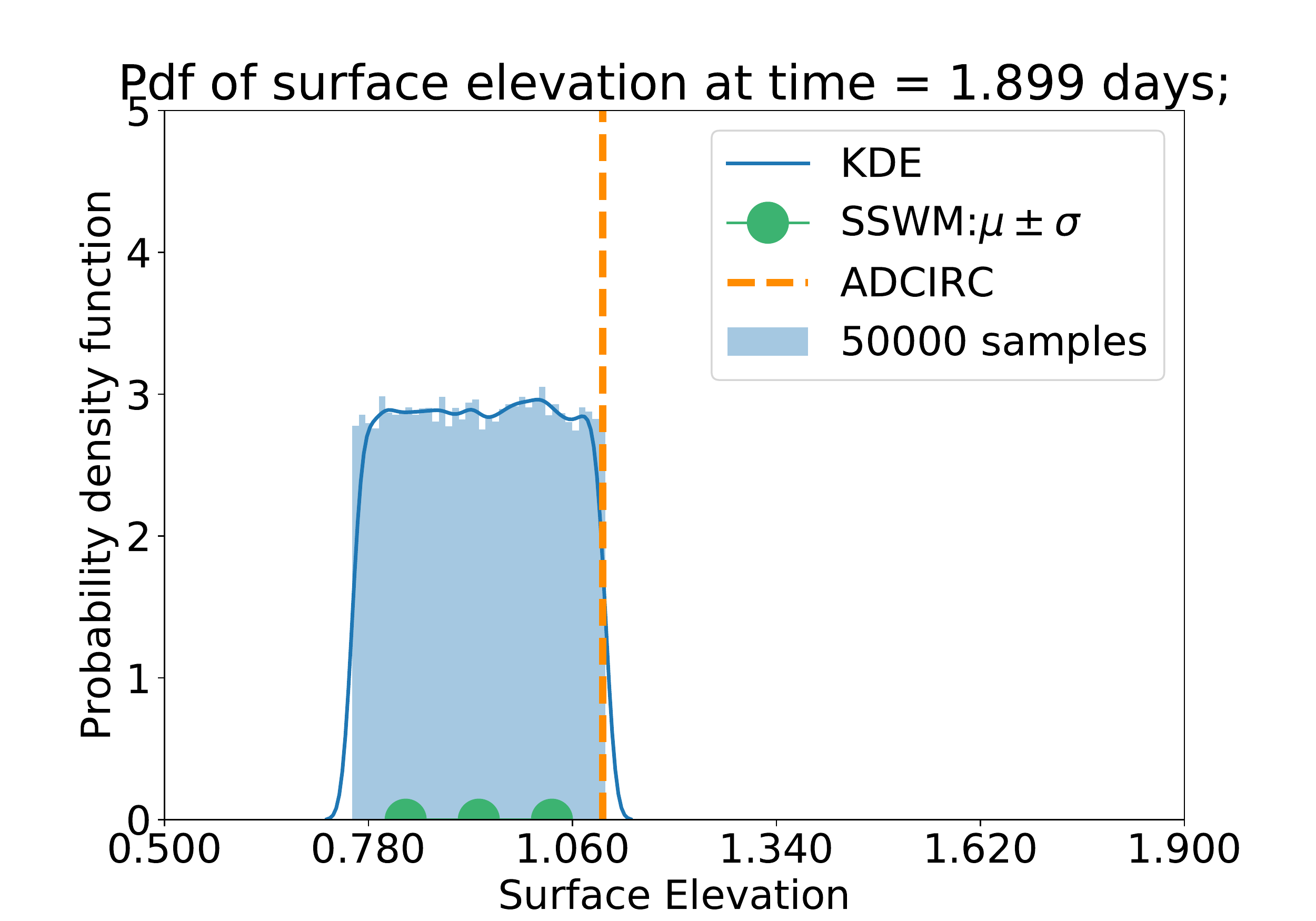}}\hfill
	\caption{Predicted PDF of surface elevation compared to ADCIRC at point 3702 during Hurricane Ike.}
	\label{fig:ike3702}
\end{figure}

We consider the Hurricane Ike case and select three points in locations that underwent significant surge, i.e., 3702, 3516 and 5171 (see Figure~\ref{fig:ikechoose}). In Figures~\ref{fig:ike5171},~\ref{fig:ike3516}, and~\ref{fig:ike3702}, we present the predicted PDFs, the ADCIRC solution, the 50,000 samples, and the kernel estimated PDF, i.e., the Kernel Density Estimate (KDE). 
As shown in these figures, the PDFs for surface elevation during the hurricanes resemble  uniform distributions. Hence, for each point, the SSWM provides a predicted range (i.e., the interval covered by the blue histogram at the horizontal axis) within which the values have an almost equal chance of occurrence. We also observe that in most cases the benchmark ADCIRC solution is within the predicted range. In other words, given a range for the uncertain input $C_d$, the SSWM is able to provide a predicted range that includes the benchmark solution. 
Inspection of the results in Figures~\ref{fig:ike5171} through~\ref{fig:ike3702}, leads to the observation that the widest predicted range ($1.45m$ to $2.11m$) across the two scenarios is found in Figure~\ref{fig:ike3516widerange}. This means that, with an offset from the true wind drag coefficient by scaling the truth 0.8 to 1.2 times (i.e., $C_d = \xi_1 C_{d}^{truth, ADCIRC}$, with $\xi_1 \sim U[0.8, 1.2]$) we expect that the peak surge will be offset from the truth by 0.66$m$. 
This further implies that the peak surge value is very sensitive to the wind drag coefficient, which is in fact a well known phenomenon that is critical in the calibration and validation of ADCIRC models.


\clearpage

\subsection{Maximum Surface Elevation Comparison} \label{sec:max_surge_comp}


In an operational storm surge prediction system, it is computationally intractable  to draw a predicted PDF at each point and time. Thus, to predict surge in real-time when a hurricane is forecast under uncertain wind drag coefficients, a reliable predictor from the SSWM is needed. Therefore, we propose a predictor from the SSWM and demonstrate its effectiveness at predicting the maximum surface elevation under the consideration of an uncertain wind drag coefficient. Here, we propose a safe predictor $\mu+\sigma$, for the purpose of real-time prediction under the situation of an uncertain wind drag coefficient. 

\begin{figure}[h!]
	\centering
	\includegraphics[width=\textwidth]{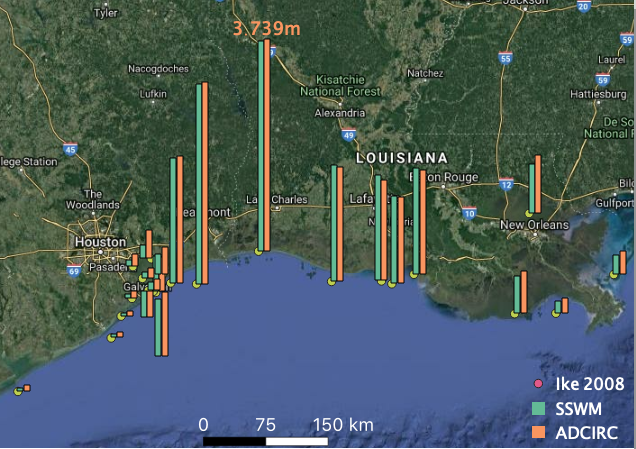}
	\caption{Maximum surface elevation comparison between ADCIRC and SSWM for Hurricane Ike.}
	\label{fig:ikeover}
\end{figure}

To show the effectiveness of the proposed predictor $\mu+\sigma$, we select 23 spatial points on the Texas and Louisiana coast and \more{consider  Hurricane Ike (Hurricane Harvey results are relegated to \ref{sec:hurricane_prediction2})}. In Figure~\ref{fig:ikeover}, we present the comparison of the maximum surface elevation between ADCIRC and the SSWM proposed predictor. 
These results show that the proposed indicator $\mu+\sigma$ underestimates the maximum surface elevation inside the Galveston Bay area for Hurricane Ike.  Otherwise, close agreement is observed  \more{throughout the coast}. This suggests that the proposed predictor $\mu+\sigma$ given by the SSWM is reliable for real-time prediction of maximum surface elevation, under the present condition of a uniformly distributed uncertain wind drag coefficient.


\subsection{Notes on the Visualization and Analysis of the SSWM Outputs } \label{sec:sswm_notes}

Based on the statistical analysis and visual representations in this section, we  draw \one{the following} conclusions: $i)$ the variance in the model outputs increases as the variances in the model inputs increase and $ii)$ we have explored the pattern of the distributions given by the surrogate at different locations and different time. We noticed strong similarity between the predicted probability density functions between the output quantities, over space and time in each test case. We also note that the \one{shapes of the} predicted PDFs vary under different uncertain sources. \moreR{Finally,} $iii)$ we have observed that maximum
variance occurs at the extreme mean for both surface elevation and water velocity over space and time. 

For the two considered hurricanes, we further showed the reliability of the SSWM and proposed a reliable predictor for the SSWM to use in real-time by considering mean outputs as well as predicted PDFs. \one{We also observe for this application that the maximum variance occurs at the extreme mean for both surface elevation and water velocity over space and over time. However, generally this may not be true. } 

\two{An important goal of the presented work is the development of a real-time prediction system under uncertain resources. While it is not tractable to perform draw PDFs for large number of samples as in Section~\ref{sec:com_ADC_PDF_storm}, the use of the predictor shown in Section~\ref{sec:max_surge_comp} represents a viable alternative. \twoR{ The computational framework utilized herein and run using a laptop computer with 8GB RAM and 1.8GHz CPU is able to present a prediction for $\mu + \sigma$ in a time frame of a couple hours.} We consider this to be acceptable in a forecasting scenario in which well established and validated finite element meshes are used for predictions  } \twoR{since forecasts for hurricanes typically start a few days before expected landfall.} 


\section{Concluding Remarks} \label{sec:conclusion}

In this paper, we have developed and extensively verified a stabilized stochastic shallow water model, i.e., the SSWM and conducted a comprehensive statistical analysis on the resulting SSWM surrogate outputs. We have also conducted a validation exercise for hindcasting of two past hurricanes, Ike and Harvey. We also propose a safe and reliable predictor $\mu + \sigma$ to show that the SSWM can be used for real-time predictions during hurricanes under uncertain resources. The effectivity of the predictor is demonstrated through hindcasting of maximum surge in Hurricane Ike and Harvey, respectively.

It is our  hope that this SSWM can be used to enhance the reliability of  current state of-the-art hurricane storm surge prediction systems. For future works, we  note there is gap in theory regarding the stability of a stochastic system. And, although we have provided one approach to stabilize such a system, different stabilization techniques are still needed for solving multiscale problems under more extreme conditions, such as wetting and drying of elements. Additionally, while the uniformly distributed uncertain input lead to uncertain outputs of other distributions, a natural question of how to determine the output distribution before running the stochastic model arises. This opens a new field of studying the internal mechanism in the relationship between input distribution and output distribution.
\two{The current developed framework was implemented using serial computations and we expect the use of parallel computing will further increase its computational tractability.}

\section*{Acknowledgements}
The authors would like to gratefully acknowledge the use of the "ADCIRC" allocation at the Texas Advanced Computing Center at the University of Texas at Austin.
This work has been supported by the United States National Science Foundation - NSF PREEVENTS Track 2 Program, under NSF Grant Number 1855047. Author Eirik Valseth would also like to gratefully acknowledge the support from the Marie Skłodowska-Curie Actions grant "HYDROCOUPLE" grant number 101061623.

\clearpage

\appendix
\section{Verification of the Deterministic Part of the Stochastic Model} \label{sec:verify_dswm}

To verify the deterministic part of the stochastic model, we first check the rate of convergence of the deterministic model for the slosh test case of Section~\ref{sec:sloshtest} \two{and the well balanced property of our model by considering the hump test case from Section~\ref{sec:humptest}.} Furthermore, for comparison against two existing hydraulic models, ADCIRC~\cite{luettich1992adcirc,Pringle:2020} and ADH~\cite{trahan2018formulation}, we consider the hump  test case of Section~\ref{sec:humptest} and Inlet test case of Section~\ref{sec:inletttest}, respectively. Finally, we consider Hurricane Ike in the Gulf of Mexico and compare the deterministic SSWM solution against the ADCIRC model. \two{In this verification process, we compare the outputs from our model to ADCIRC and ADH as these two models have been extensively validated. In particular, ADH was validated for the Galveston Bay area which contains complex inlet features in~\cite{trahan2018formulation}. The ADCIRC model is currently used in forecasting of hurricane storm surge~\cite{dietrich2013real} and was validated for Hurricane Ike in~\cite{hope2013hindcast} as well as for Hurricane Harvey in~\cite{goff2019outflow}. For completeness, we also consider a verification against experimentally measured data from literature.   }

%

\subsubsection{Convergence of the Deterministic SSWM} \label{app:conv_stud}

To assess the convergence properties of our deterministic model, we consider an analytical solution to the slosh test case from~\cite{wang2009verification}:
\begin{equation} \label{Eqn: theory eqn}
\begin{aligned}
& \eta(x, y, t) = a \cos \left(  \frac{\pi}{L} x  \right)  \cos  \left(   \frac{\pi \sqrt{gD} }{L}  t  \right), \\
& u(x, y, t) =  \frac{a \sqrt{gD} }{D}   \sin \left(  \frac{\pi}{L} x  \right)  \sin  \left(   \frac{\pi \sqrt{gD} }{L}  t  \right), \\
& v(x, y, t) = 0,
\end{aligned}
\end{equation}
where $D$ denotes bathymetry. We consider convergence of the FE error in the standard $L_2$ norm:
\begin{equation}
\Vert  e   \Vert_{\Omega} =  \sqrt{    \int_{\Omega}  (\phi - \phi_h)^2  \mathnormal{d} \Omega    },
\end{equation}
where $\phi$ is the analytic solution, and $\phi_h$ is the FE solution. We compute the FE solution with increasing mesh resolution of size $h = \{10m, 5m, 2.5m, 1.25m, 0.625m\}$, we obtain the rate of convergence for both surface elevation and water velocity and present the results in Table~\ref{tab:sloshrate}. 
\begin{table}[h!]
	\centering
	\begin{tabular}{  C{1.1cm}  C{0.8cm}  C{3cm}  C{2cm} | C{3cm}  C{2cm}  }
		\hline\xrowht{12pt}
		\multirow{2}{*}{Mesh} & \multirow{2}{*}{$h$} & \multicolumn{2}{c|}{\textit{Surface elevation}} & \multicolumn{2}{c}{\textit{Water velocity}} \\ \cline{3-6}  \xrowht{12pt}
		           & & $\Vert  e   \Vert_{L_2} $ & $p=1$  & $\Vert  e   \Vert_{L_2} $  &  $p=2$ \\ \hline
		h1       & 10		 & $4.0929 \times 10^{-2}$             & -                     &$3.8166 \times 10^{-4}$            & -              \\ 
		h2      & 5  	 	& $1.0268 \times 10^{-2}$ 		       &1.9949         & $1.2576 \times 10^{-4}$              & 1.6016         \\ 
		h3      & 2.5  	   & $2.5694 \times 10^{-3}$  		       &1.9987         & $4.2418 \times 10^{-5}$              & 1.5680         \\ 
		h4      & 1.25    & $6.4249 \times 10^{-4}$  		       &1.9997         & $1.4591 \times 10^{-5}$              & 1.5396         \\ 
		h5      & 0.625  & $1.6063 \times 10^{-4}$  		       &1.9999         & $5.0834 \times 10^{-6}$              & 1.5212          \\ \hline
	\end{tabular}
	\caption{Slosh test case: $L_2$ convergence rate.}
	\label{tab:sloshrate}
\end{table}
%
As shown in this  table, the error surface elevation reaches its optimal $L_2$ convergence rate of order $2$. However, the rate of convergence for the velocity is sub optimal. The reason for this is found in the IPCS scheme which neglects certain high order terms. 

\subsubsection{\two{Lake-At-Rest}}
\begin{figure}[h!]
	\centering
	\subfigure[][$\eta$ surface elevation at $(500.0m, 100.0m)$.]{\includegraphics[width=0.45\textwidth]{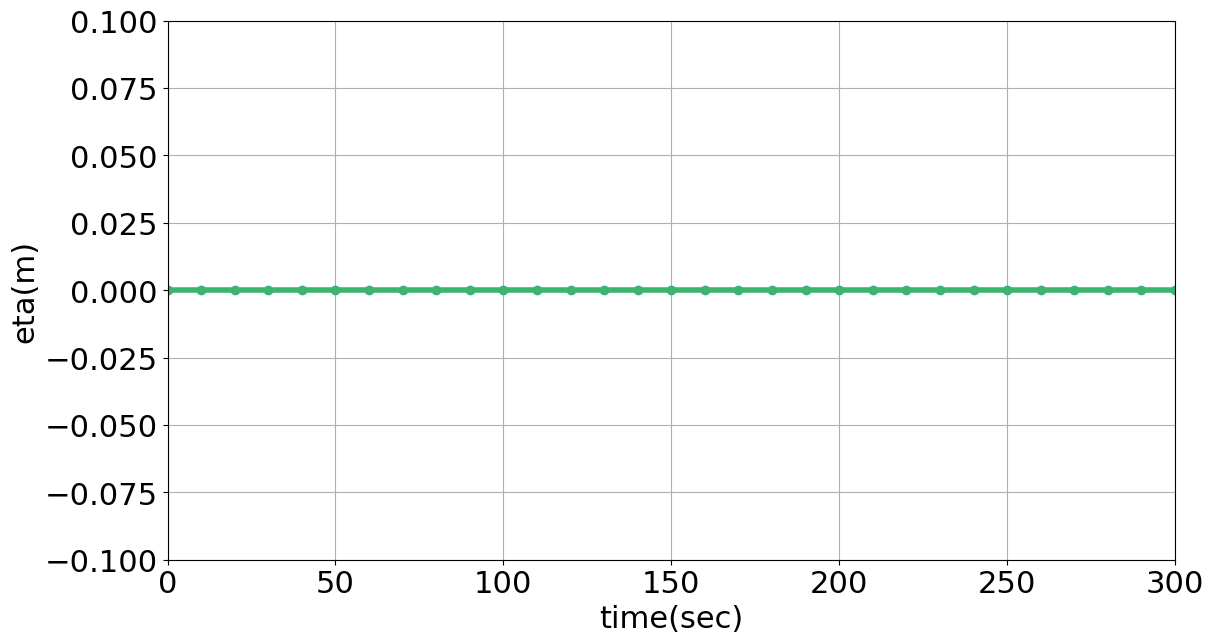}}\hfill
	\subfigure[][$x$-direction velocity at $(500.0m, 100.0m)$.]{\includegraphics[width=0.45\textwidth]{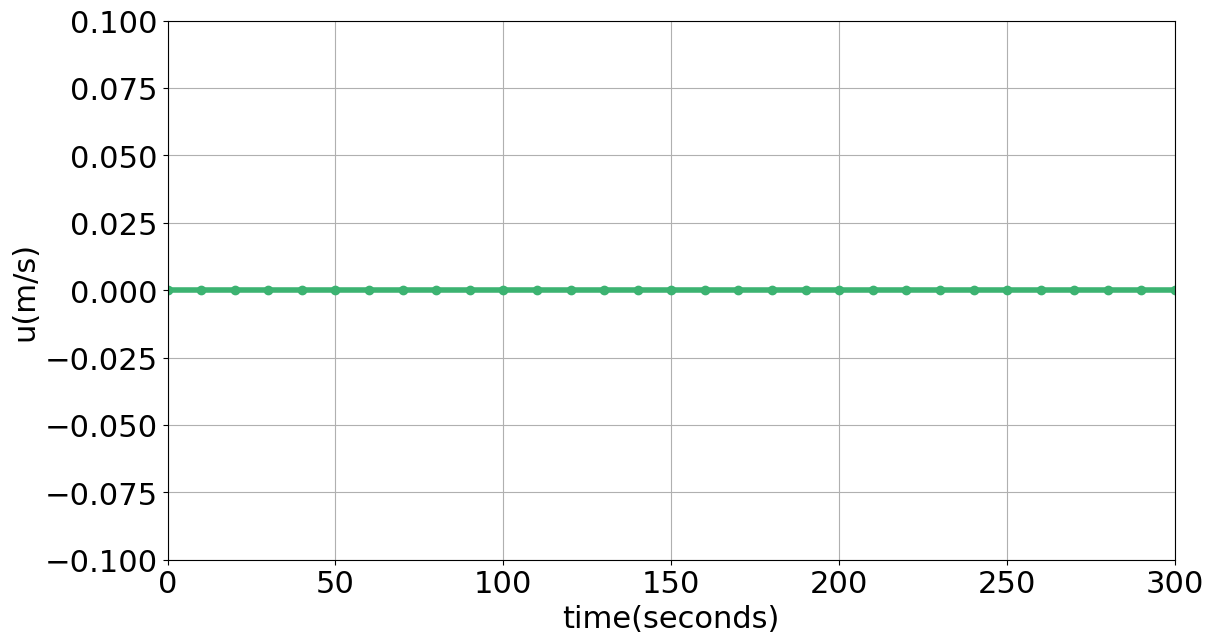}}\hfill
	\caption{$\eta$ and $x$-direction velocity for lake-at-rest test case.}
	\label{fig:well_balanced}
\end{figure}

\two{ To ensure the well balanced property of our model, we consider a lake-at-rest version of the hump test case of Section~\ref{sec:humptest}. In this modified test case, all boundaries have no-normal flow conditions and we set the initial conditions to be homogeneous. In Figure~\ref{fig:well_balanced}, we present the evolution of the surface elevation and $x$-direction velocity throughout the simulation at a selected point. As no flow is induced in this case, we concluded that the model is well balanced. }


\subsubsection{Comparison With the ADCIRC Model}

\begin{figure}[h!]
	\centering
	\subfigure[][Surface elevation.]{\includegraphics[width=0.425\textwidth]{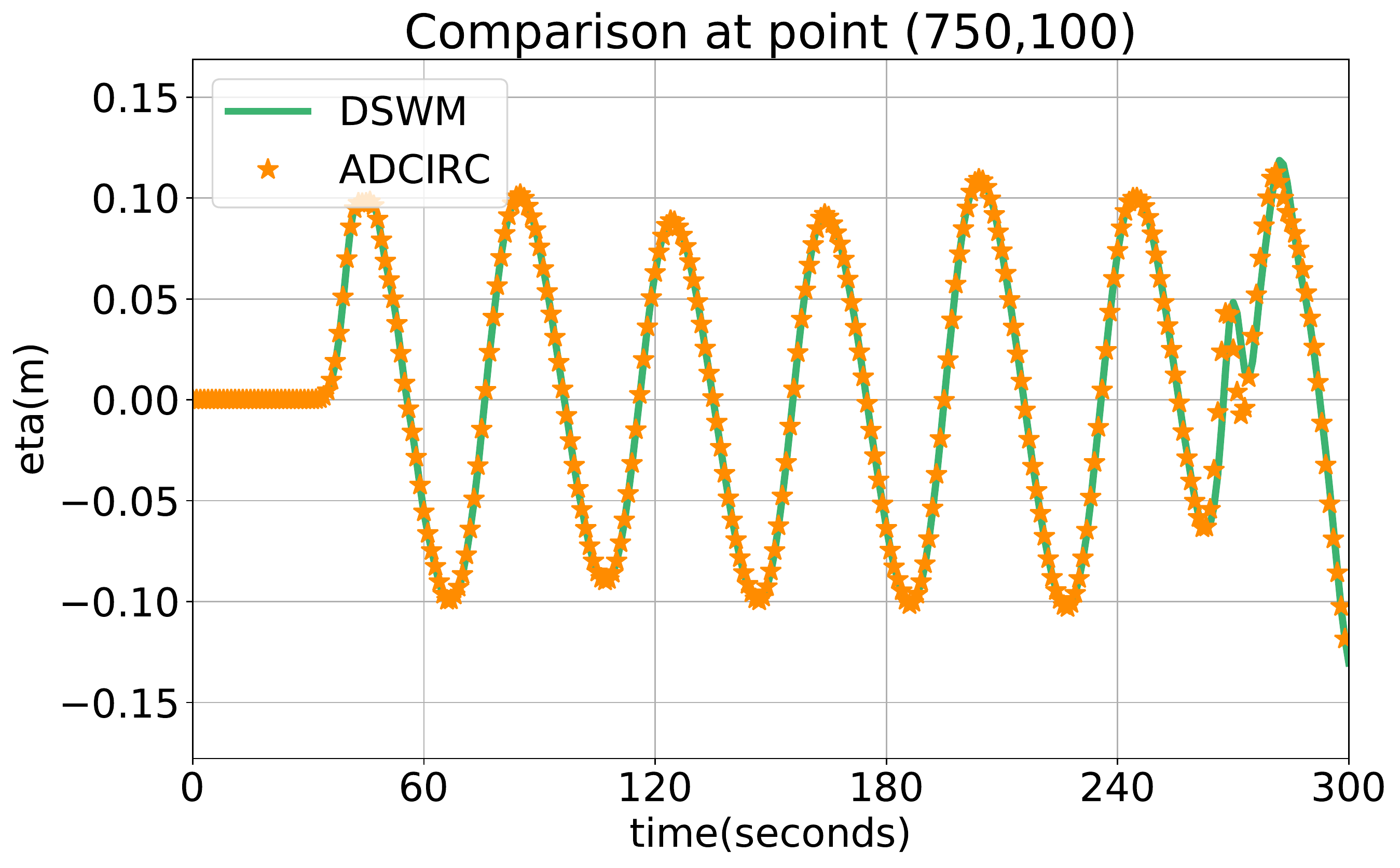}}\hfill
	\subfigure[][$x$-direction velocity.]{\includegraphics[width=0.425\textwidth]{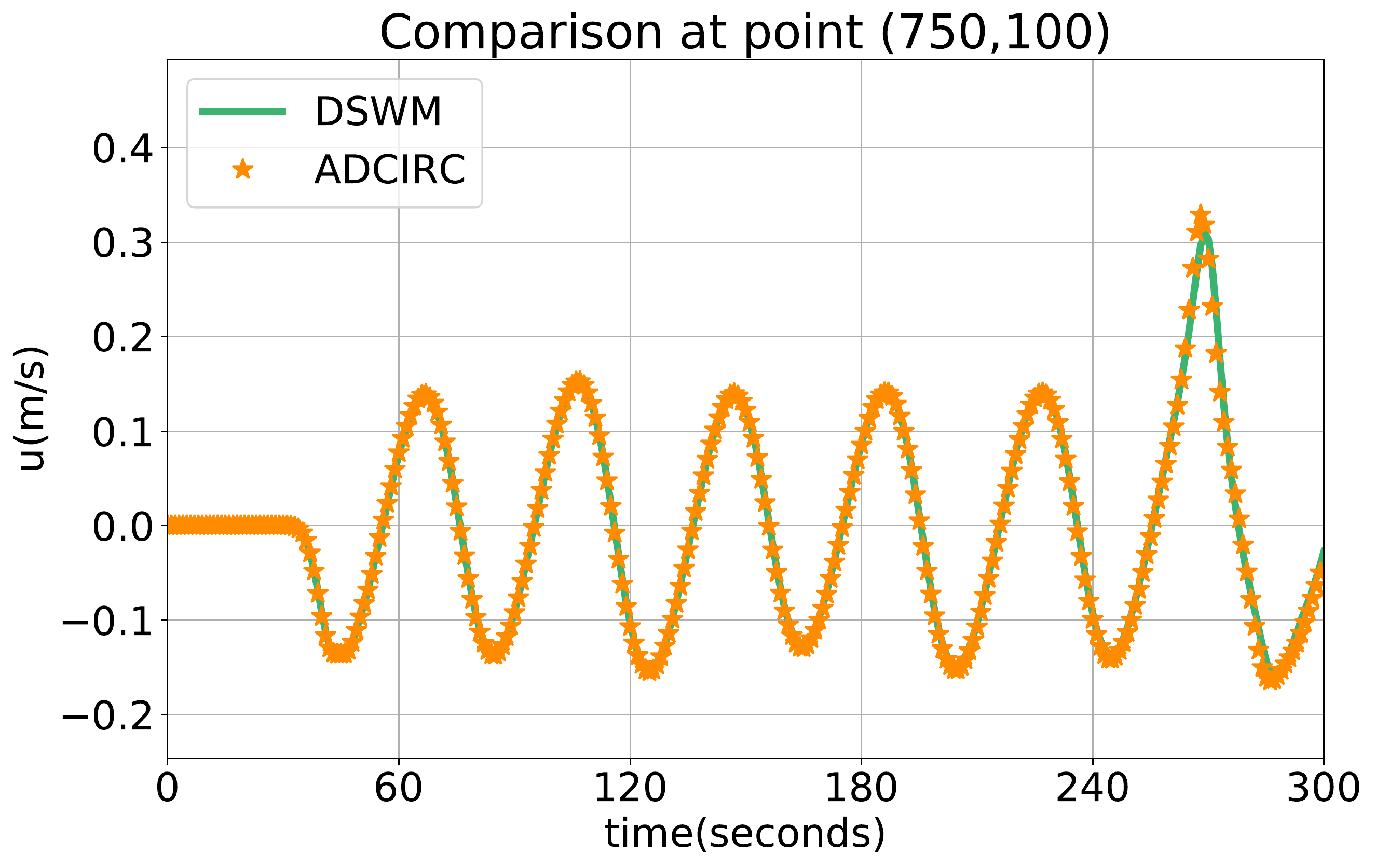}}\hfill
	\subfigure[][Surface elevation.]{\includegraphics[width=0.425\textwidth]{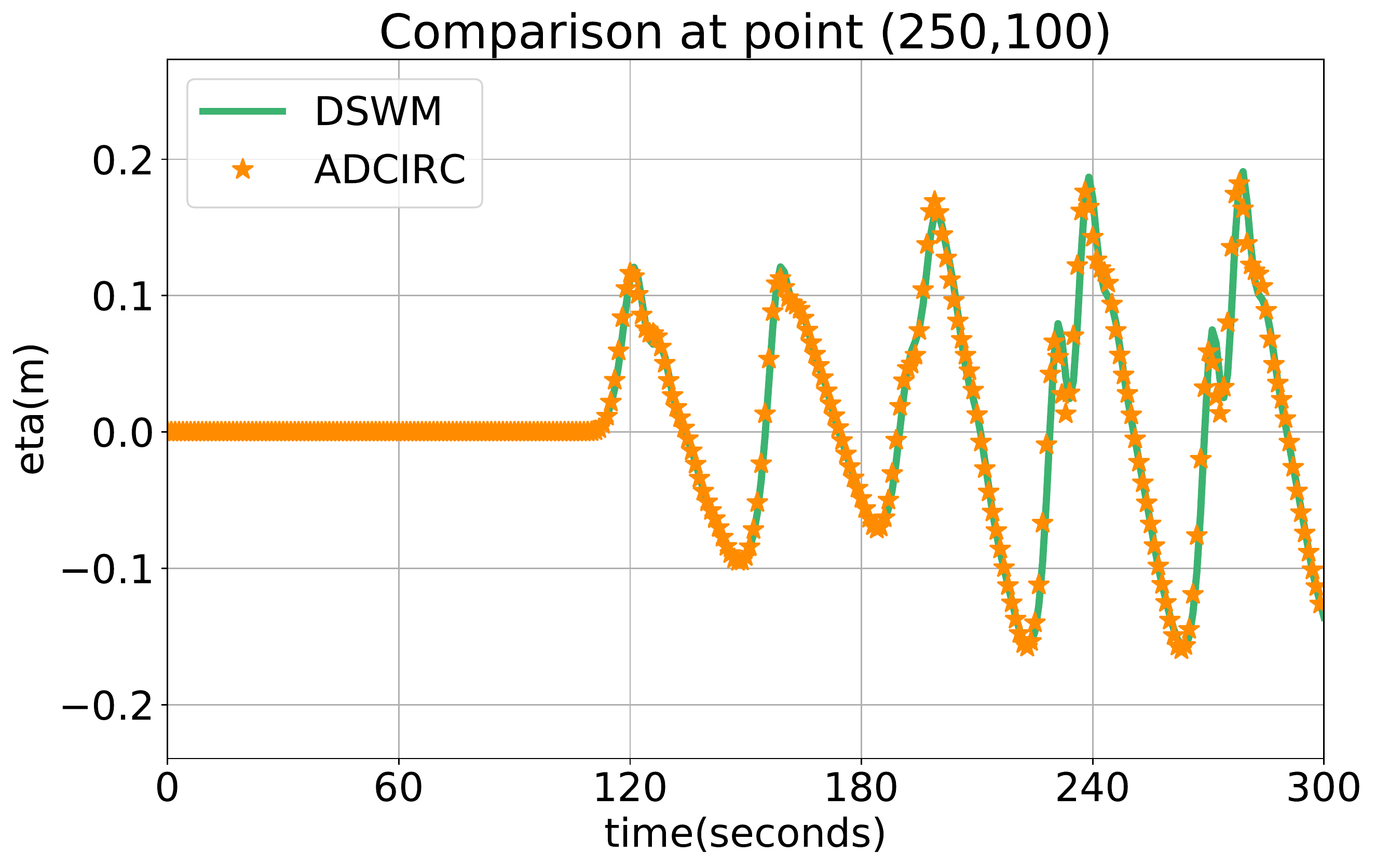}}\hfill
	\subfigure[][$x$-direction velocity.]{\includegraphics[width=0.425\textwidth]{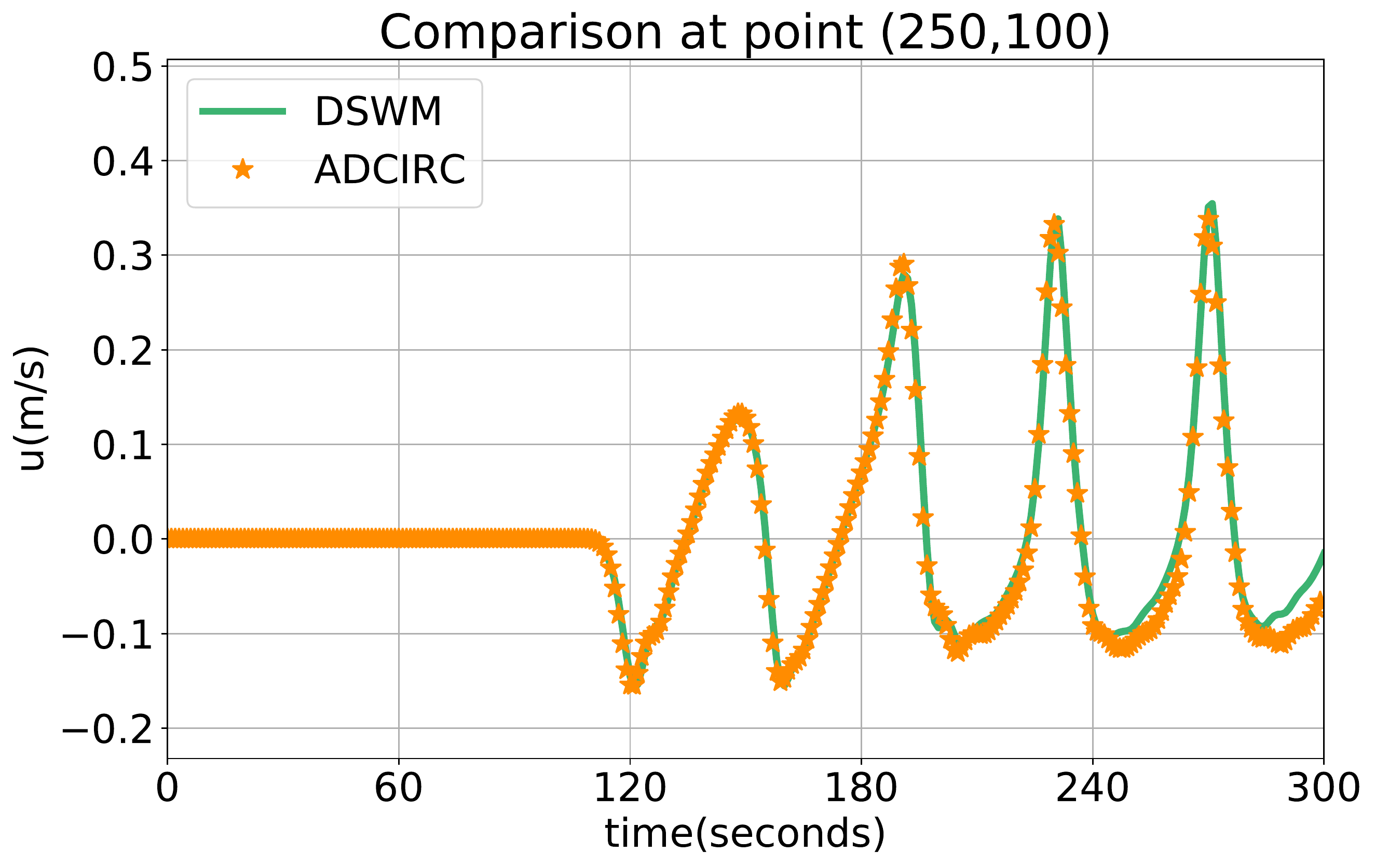}}\hfill
	\caption{Comparison between the new model and ADCIRC.}	\label{fig:humpline2}
\end{figure}

The convergence properties considered in \ref{app:conv_stud} as well as further studies presented in~\cite{chenthesis} give us confidence in the approximation properties of our deterministic SSWM. We now further verify our model by comparison against results from the ADCIRC model for the hump test case described in Section~\ref{sec:humptest}. To this end, we consider a comparison of the time series of the elevation and the $x$-direction component of the velocity field at the  points $(250.0m, 100.0m)$ and  $(750.0m, 100.0m)$. The comparison is presented in Figure~\ref{fig:humpline2}, where the agreement between the two models is very good for both solution variables throughout the  simulation.

\subsubsection{Comparison With the ADH Model}

\begin{figure}[h!]
	\centering
	\includegraphics[width=0.8\textwidth]{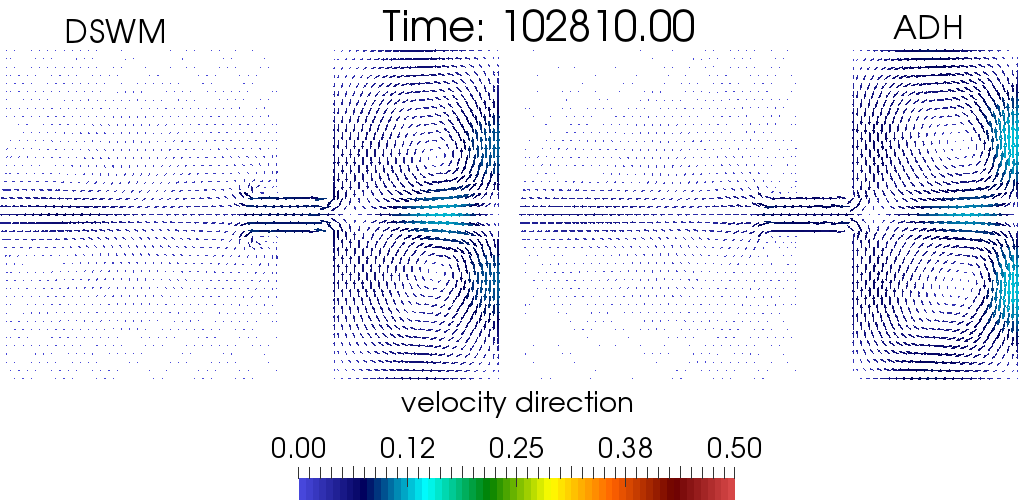}\hfill
	\subfigure{\includegraphics[width=0.8\textwidth]{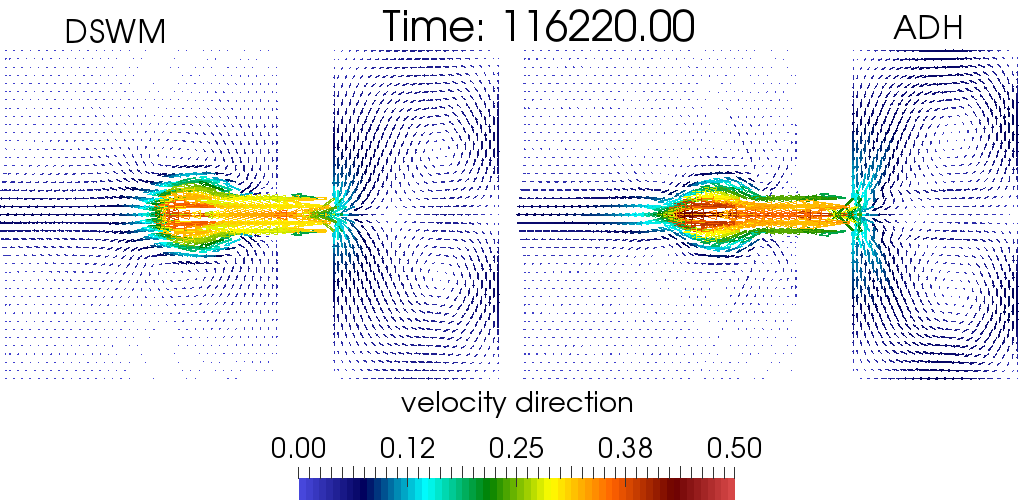}}\hfill
	\caption{Water velocity field in the idealized inlet test case.}
	\label{fig:inletvelocity1}
\end{figure}

\begin{figure}[h!]
	\centering
	\subfigure[Time $102810s$.]{\includegraphics[width=0.4125\textwidth]{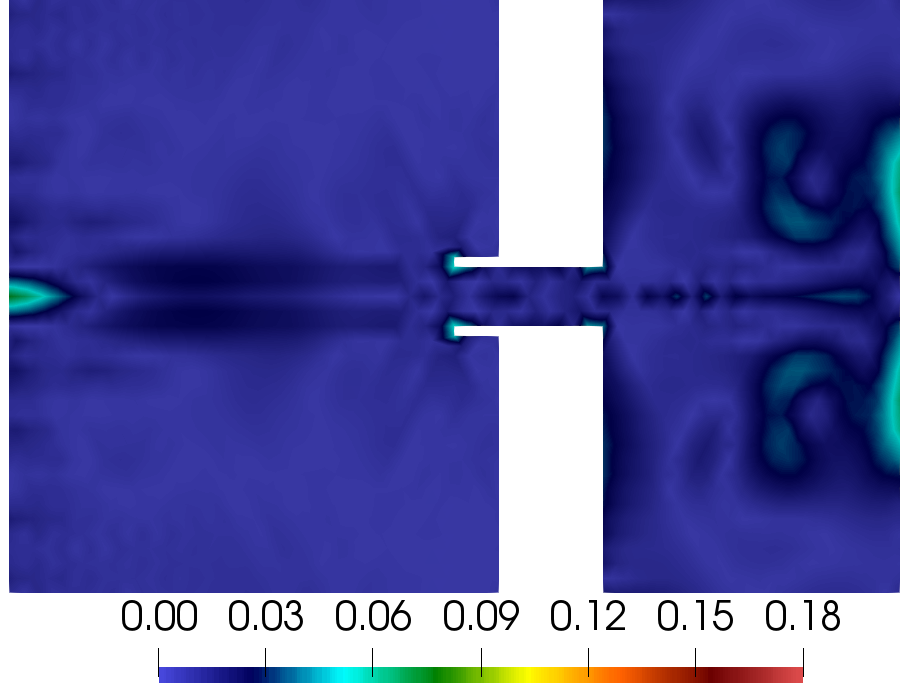}}\hfill
	\subfigure[Time $116220s$.]{\includegraphics[width=0.4125\textwidth]{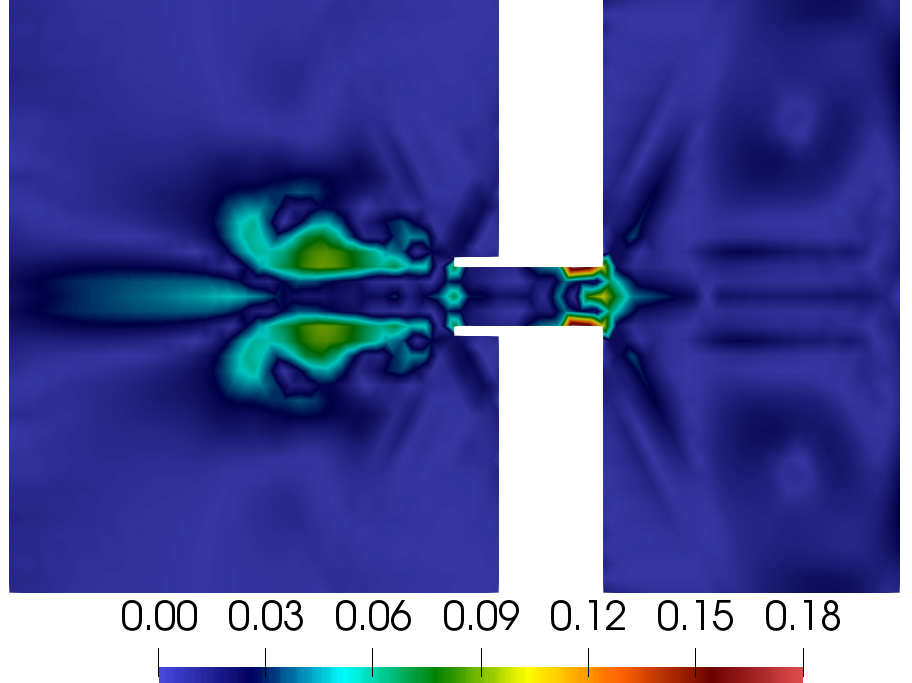}}\hfill
	\caption{Absolute error of the magnitude of water velocity.}
	\label{fig:inletvelocityabsError}
\end{figure}

As another verification of the deterministic model, we consider the inlet test case and compare  our results to the ADH model. This test is challenging due to the shocks that form at the exit of the narrow channel. In Figure~\ref{fig:inletvelocity1}, we present a visual comparison between the two models for the velocity field at two selected times. In Figure~\ref{fig:inletvelocityabsError}, the corresponding absolute errors in velocity magnitude are shown. Overall, the two models agree on the flow characteristics with certain localized discrepancies. 

\begin{figure}[h!]
	\centering
	\subfigure[][$x$-direction water velocity at $(-250m,0m)$.]{\includegraphics[width=0.5\textwidth]{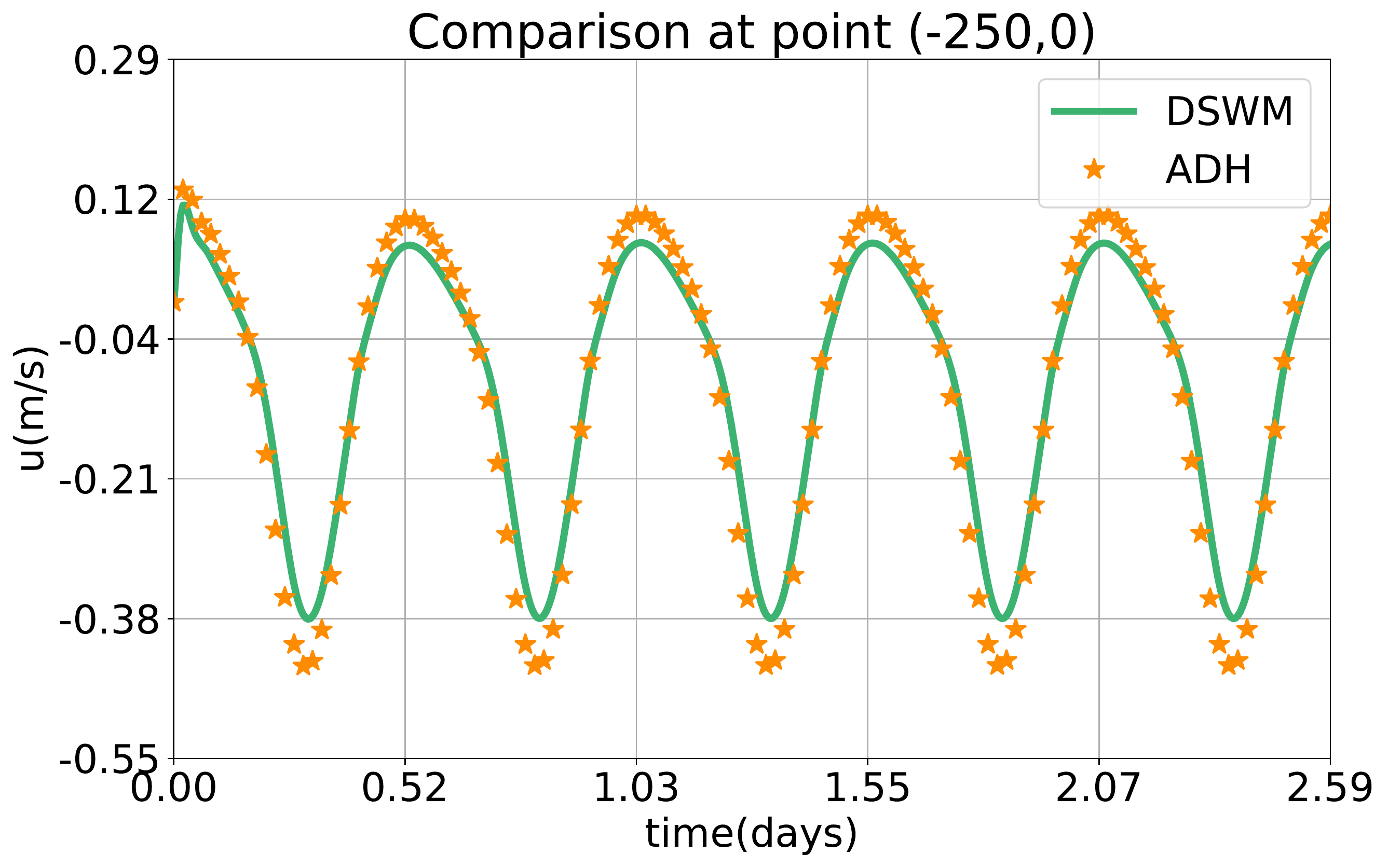}}\hfill
	\subfigure[][$x$-direction water velocity at $(750m,0m)$.]{\includegraphics[width=0.5\textwidth]{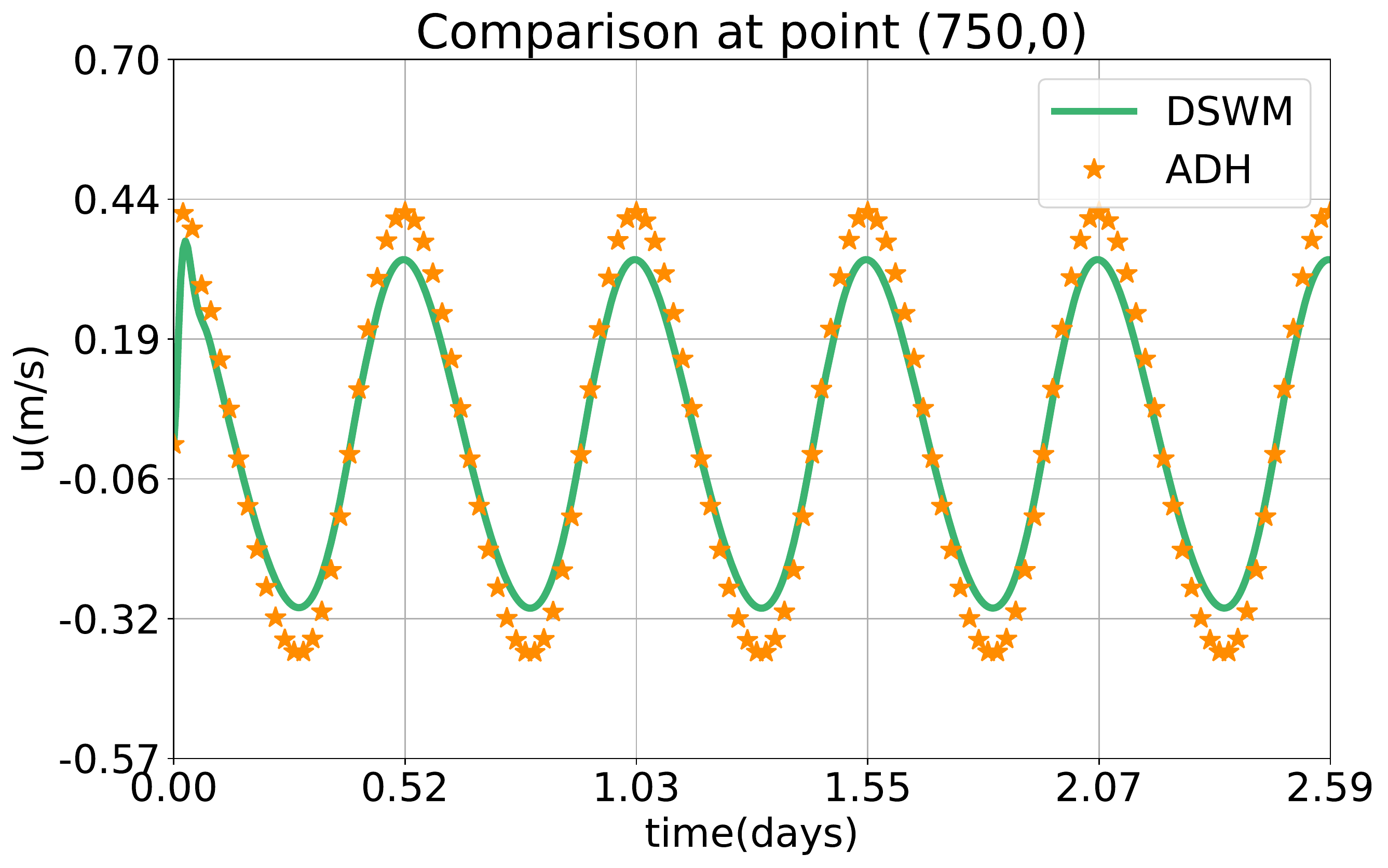}}\hfill
	\caption{Idealized inlet test case: $x$-direction velocity.}
	\label{fig:inletuveltime}
\end{figure}

\begin{figure}[h!]
	\centering
	\subfigure[][surface elevation at $(-250m,0m)$.]{\includegraphics[width=0.5\textwidth]{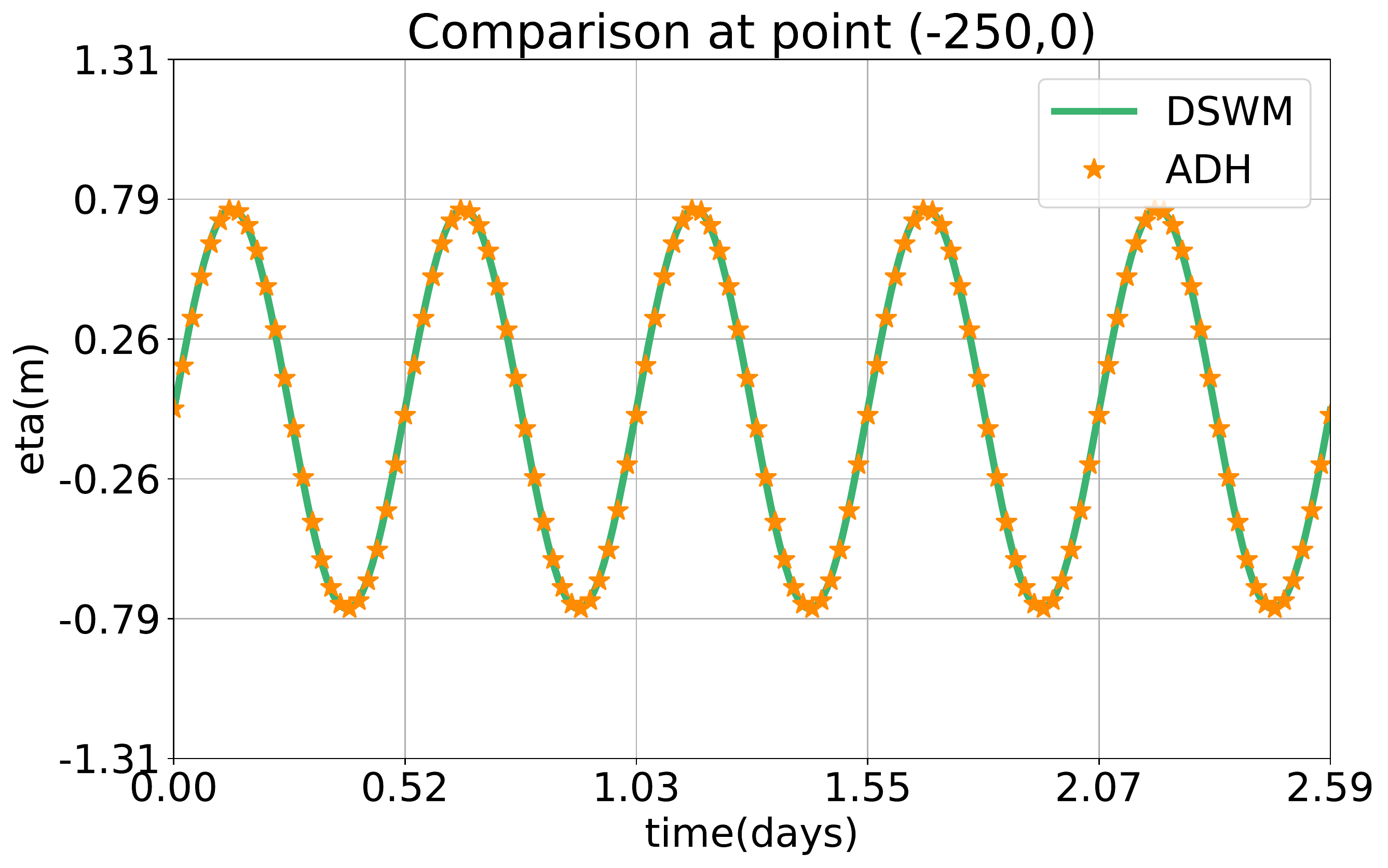}}\hfill
	\subfigure[][surface elevation at $(750m,0m)$.]{\includegraphics[width=0.5\textwidth]{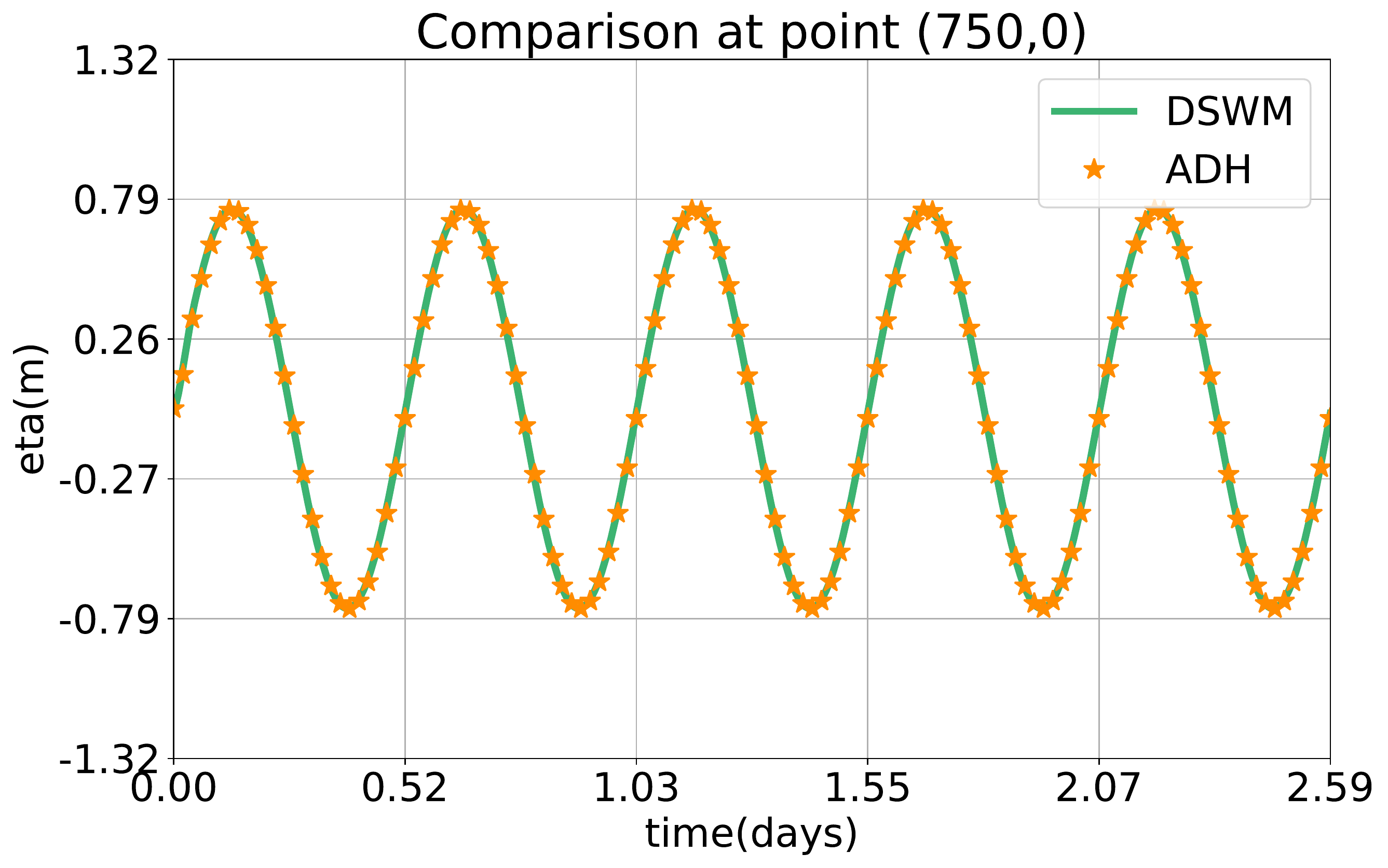}}\hfill
	\caption{Idealized inlet test case: Surface elevation.}
	\label{fig:inlettime}
\end{figure}

To quantify these localized discrepancies, we provide a comparison of the ADH and our models at two points over time, $(-250m, 0m)$ and $(750m, 0m)$; the first at the boundary of the ebb shoal and the second is at the channel exit. The corresponding plots are shown in Figure~\ref{fig:inletuveltime}.
In this figure, we observe that the ADH model produces a velocity magnitude that is approximately $0.085m/s$ greater than our model. This difference is likely due to the different approximation schemes in the two models. Since the models share the same periods and trends, we consider this discrepancy within acceptable tolerance. 
%
\begin{figure}[H]
	\centering
	\subfigure{\includegraphics[width=0.8\textwidth]{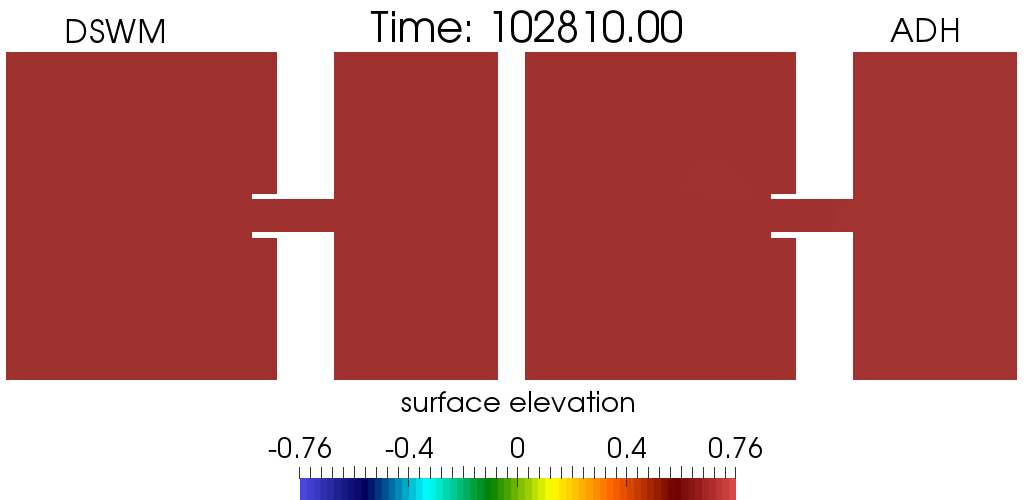}}
	\subfigure{\includegraphics[width=0.8\textwidth]{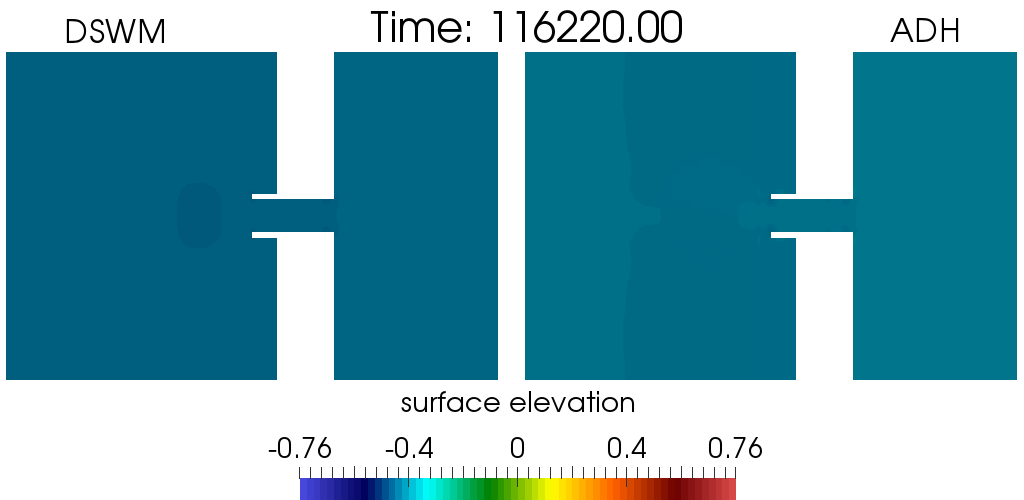}} 
	\caption{Idealized inlet test case: Surface elevation.}
	\label{fig:inletele}
\end{figure}
On the other hand, the surface elevations shown in Figure~\ref{fig:inletele} are nearly indistinguishable between the two models and differ only about $4cm$.
A good match can be also observed in a comparison between two models over time, in Figure~\ref{fig:inlettime} where we select the points $(-250m, 0m)$ and $(750m, 0m)$  as  examples.

\subsubsection{Comparison for a Hurricane Event}

\begin{figure}[H]
	\centering
	\subfigure{\includegraphics[width=0.75\textwidth]{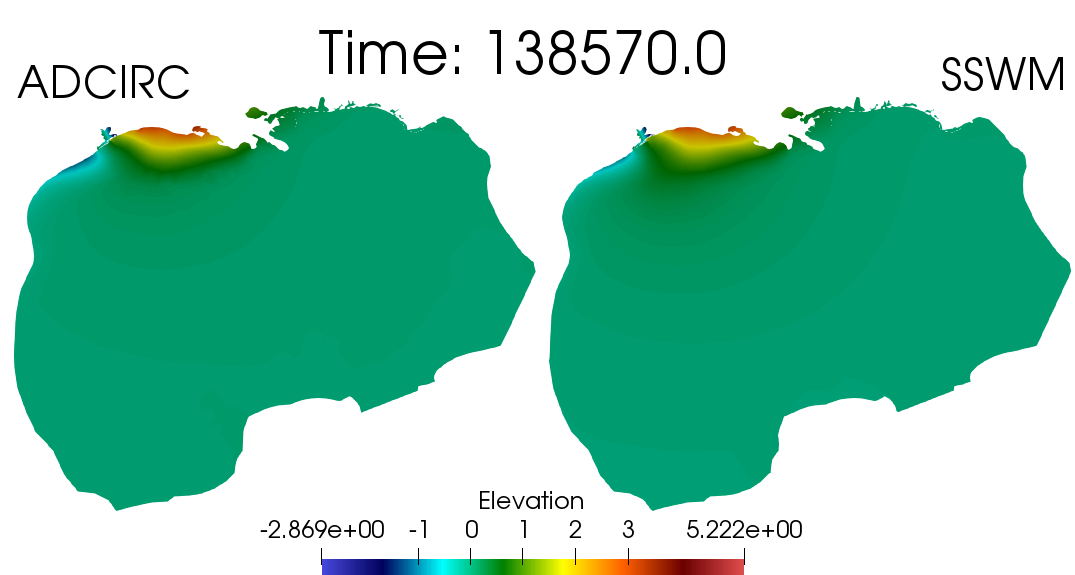}}\\
	\subfigure{\includegraphics[width=0.75\textwidth]{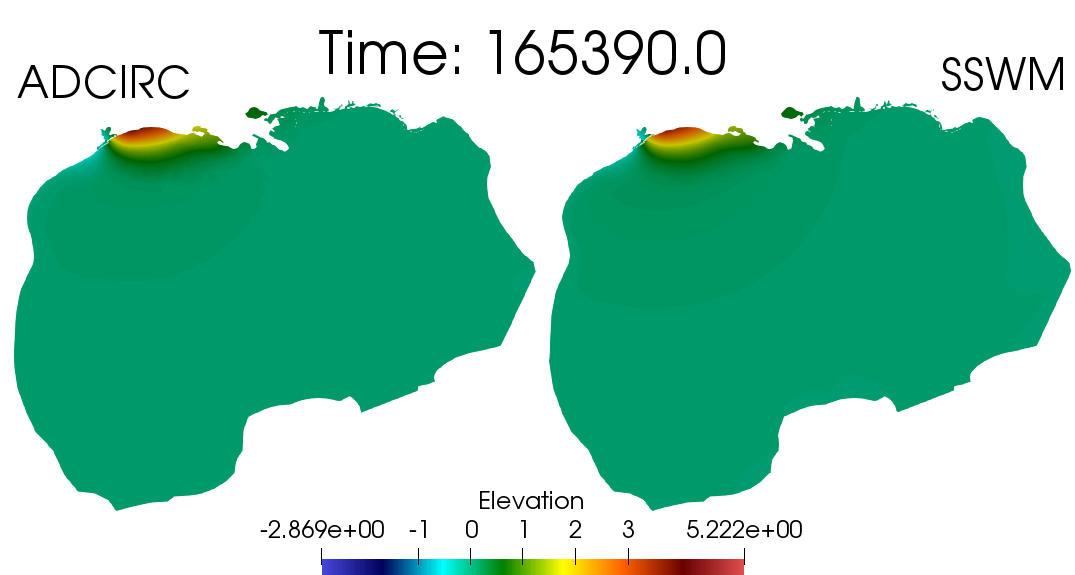}}\\
	\caption{Surface elevation comparisons during Hurricane Ike 2008.}
	\label{fig:ikeeta}
\end{figure}

\begin{figure}[H]
	\centering
	\subfigure[Time $138570s$]{\includegraphics[width=0.35\textwidth]{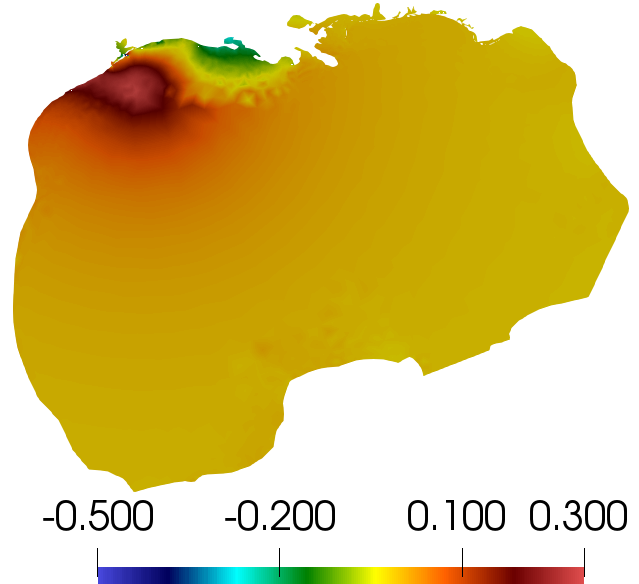}}
	\subfigure[Time $165390s$]{\includegraphics[width=0.35\textwidth]{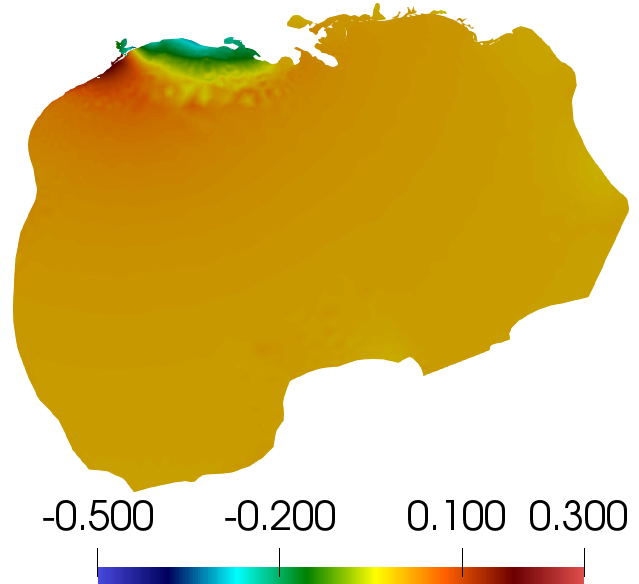}}
	\vspace{-0.3cm}
	\caption{Difference in surface elevation during Hurricane Ike 2008.}
	\label{fig:ikeetaerror}
\end{figure}

\two{Next, we consider Hurricane Ike to compare our deterministic model to ADCIRC results. To perform this comparison,} we select two time steps near the hurricane landfall on the Texas coast. In Figures~\ref{fig:ikeeta} and~\ref{fig:ikeetaerror}, the surface elevation and difference in surface elevation between the two models are shown, respectively. The results show good agreement in the maximum surge near Houston and the maximum absolute difference is $0.5m$ throughout the simulation. In the thesis~\cite{chenthesis}, further comparisons for the deterministic model are included and we refer interested readers to it.

\subsubsection{Comparison With Spur Dike Experimental Data} \label{sec:experiment_comare}
\begin{figure}[h!]
	\centering
	\includegraphics[width=0.9\textwidth]{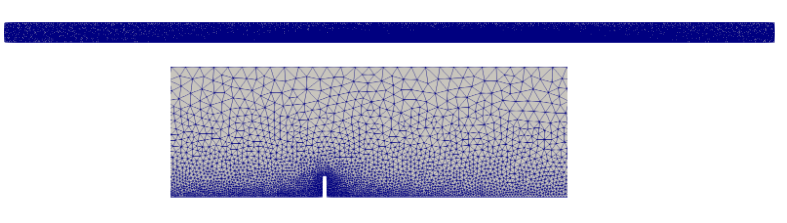}
	\caption{Dike Mesh.}
	\label{fig:dikemesh}
\end{figure}

\two{As a final verification of the deterministic model, we consider a case with experimentally measured data called a Spur Dike experiment.
A Spur dike is a man made obstacle placed on the side of a river with one end attached to the bank of river and another end intruded into river. Spur dikes are used to alter the flow fields in rivers to protect river banks from erosion, see,~\cite{Rajaratnam1983Flow}.
Using experimentally measured data from~\cite{Rajaratnam1983Flow}, we will further verify the deterministic model.  In this experiment, a rectangular domain of length $L = 37m$, width $W = 0.92m$, and constant initial water depth $H = 0.189m$, is considered. Within this  domain, the dike has width $B=0.152m$ and thickness $0.03m$ and is inserted at $x=14.0m$, perpendicular to the southern boundary of domain, as shown in Figure~\ref{fig:dikemesh} along with the computational mesh near the spur. A constant inflow of $Q=0.0453m^3/s$ is supplied at the western boundary at $x=0m$ and at the eastern boundary the surface elevation is kept fixed at $0m$. Along the north and south boundaries we apply no-normal flow boundary conditions. Initially, the water is assumed to be at rest, we fix the bottom friction coefficient $C_b=0.0015$ and the kinematic viscosity to $\nu = 10^{-6}$. No other external forcing is applied, the total simulation time is $800s$ and use the ICPS with a $5s$ time step.  }
%

\begin{figure}[h!]
	\centering
	\includegraphics[width=0.9\textwidth]{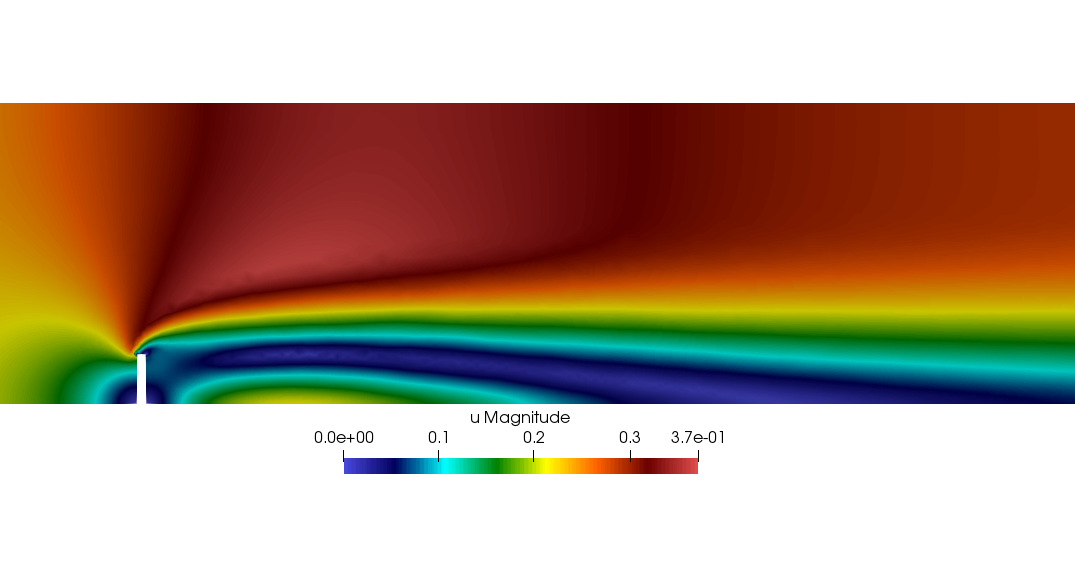}
	\caption{Dike steady state velocity magnitude.}
	\label{fig:dikesteadystate}
\end{figure}

\begin{figure}[h!]
	\centering
	\includegraphics[width=0.9\textwidth]{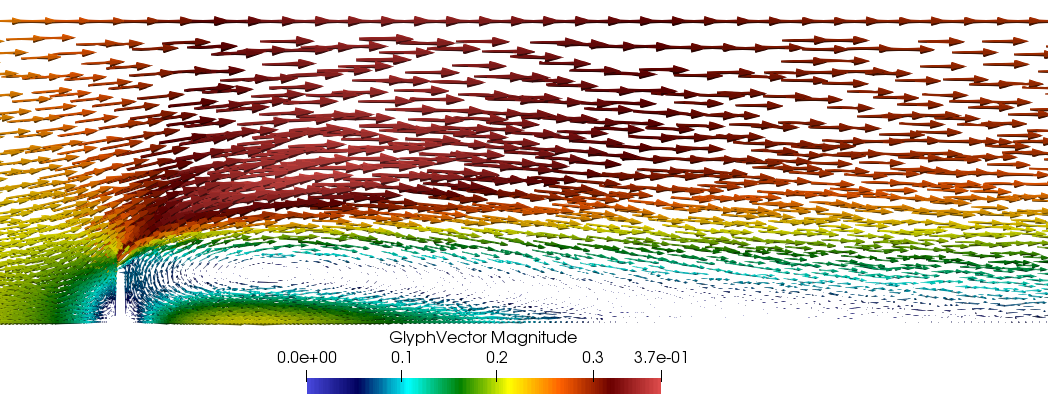}
	\caption{Dike steady state velocity direction.}
	\label{fig:dikedirection}
\end{figure}

\two{At steady state, around $600s$, a vortex appears downstream of the dike as shown in Figures~\ref{fig:dikesteadystate} and~\ref{fig:dikedirection}. There is a location along this vortex where the $x-$direction velocity changes sign from positive to negative. This location is called reattachment point, and the distance from the dike to the reattachment point is the reattachment length. The measured reattachment length reported in~\cite{Rajaratnam1983Flow} is  $12$ times the width of the dike. From our model simulation, we obtain a reattachment length of $12.85$ times the width of dike at steady state.}

\begin{figure}[h!]
	\centering
	\subfigure[][$d/B=2$.]{\includegraphics[width=0.425\textwidth]{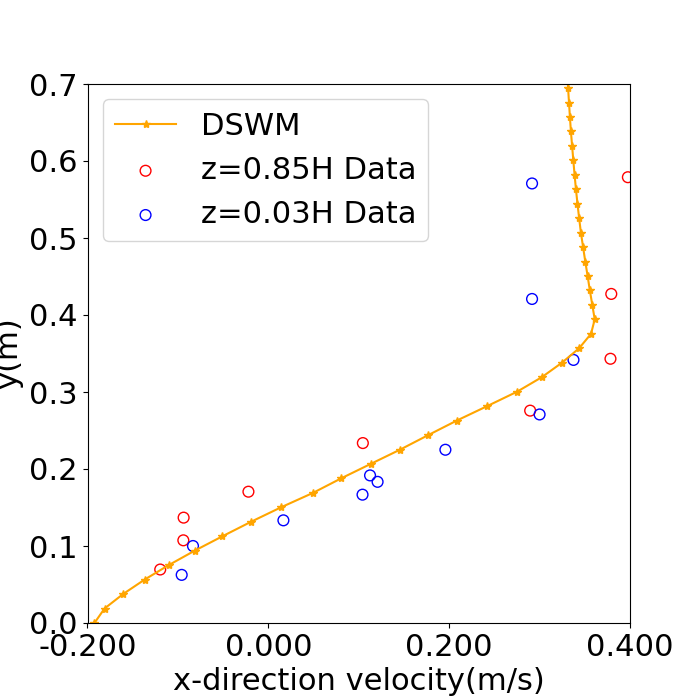}}\hfill
	\subfigure[][$d/B=4$.]{\includegraphics[width=0.425\textwidth]{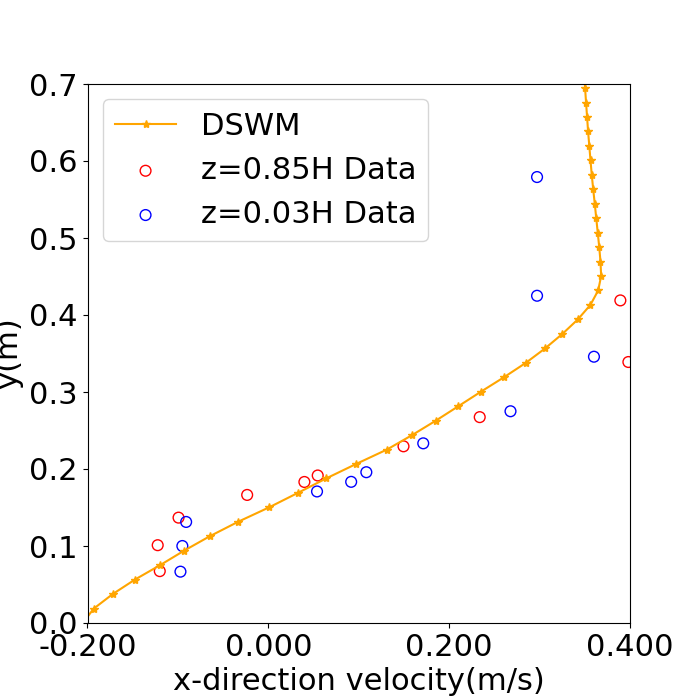}}\hfill
	\subfigure[][$d/B=6$.]{\includegraphics[width=0.425\textwidth]{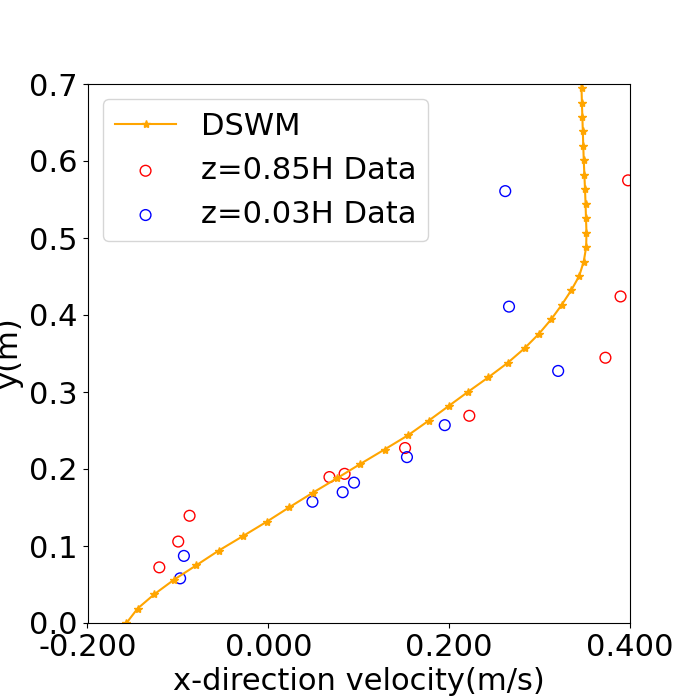}}\hfill
	\subfigure[][$d/B=8$.]{\includegraphics[width=0.425\textwidth]{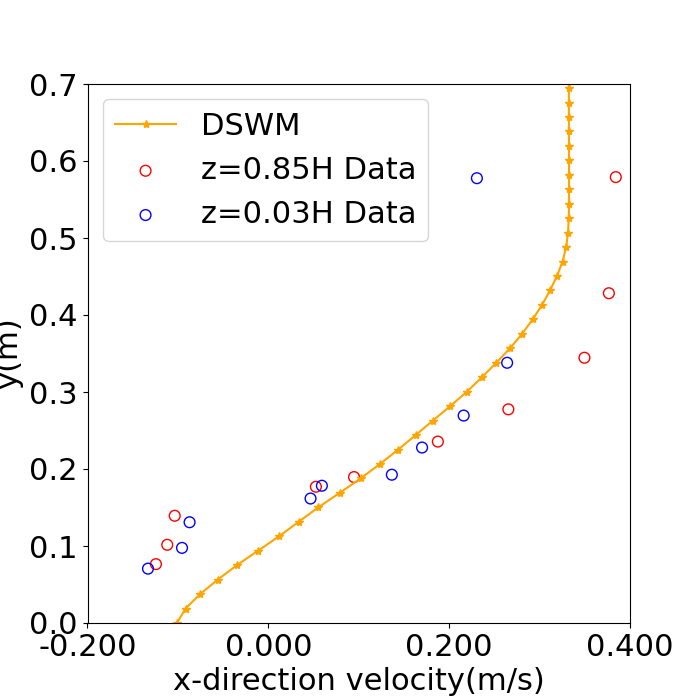}}\hfill
	\caption{Spur dike experimental data comparison.}
	\label{fig:experimentData}
\end{figure}

\two{To further compare the model results with reported experiment, we consider the  $x-$component velocity profile at different cross-sections downstream of the vortex. The eight profile locations measured in \cite{Rajaratnam1983Flow} are given by $d={2,4,6,8}B$, and $z={0.03, 0.85}H$, where $d$ is the distance between the downstream cross-section and the dike, $z$ is the vertical height where the measurement are taken. Since our model is two-dimensional and the computed velocities are an averages in the vertical direction, only four velocity profiles can be provided by DSWM for comparison, see Figure~\ref{fig:experimentData}. In this figure, we note that the simulated velocity profile mainly falls into the middle of the $z=0.03H$ and $z=0.85H$ measurements. Because we do not expect exact matches between numerical result and experimental data pointwise, we conclude that the results in Figure~\ref{fig:experimentData} shows reasonable agreement overall. }

\section{Verification of the Stochastic Part of the Stochastic Model} \label{sec:verify_sswm}
In this section, we provide the supplemental pointwise comparison for each test cases as an additional support for the verification of SSWM. 

\subsection{Pointwise comparison with uncertain initial condition - Slosh test case}
\begin{figure}[h!]
	\centering
	\subfigure[Surface elevation at $(0.8, 1.0)$.]{\includegraphics[width=0.45\textwidth]{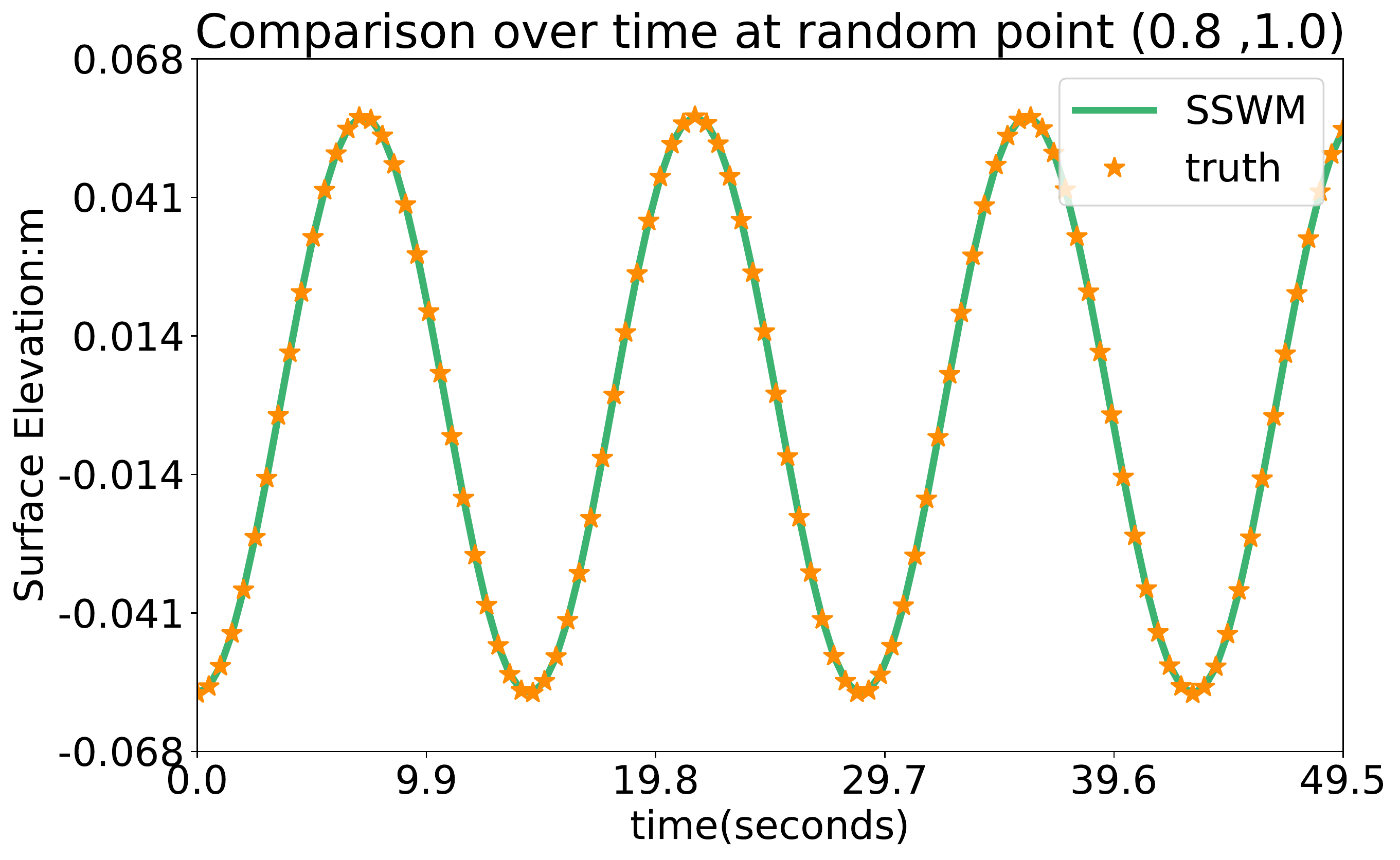}}\hfill
	\subfigure[Surface elevation at $(1.2, 2.0)$.]{\includegraphics[width=0.45\textwidth]{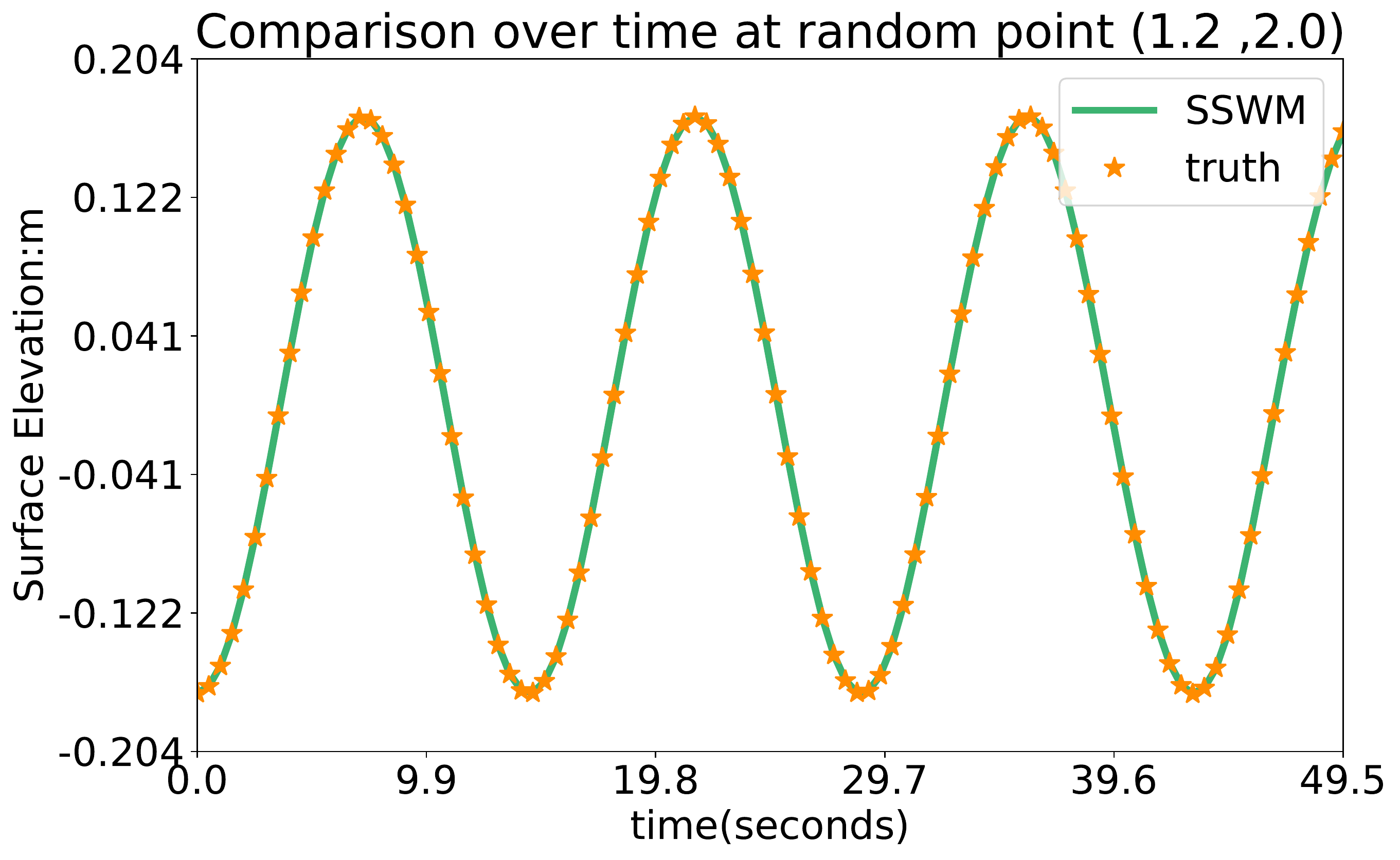}}\hfill
	\subfigure[$x$-direction water velocity at $(0.8, 1.0)$.]{\includegraphics[width=0.5\textwidth]{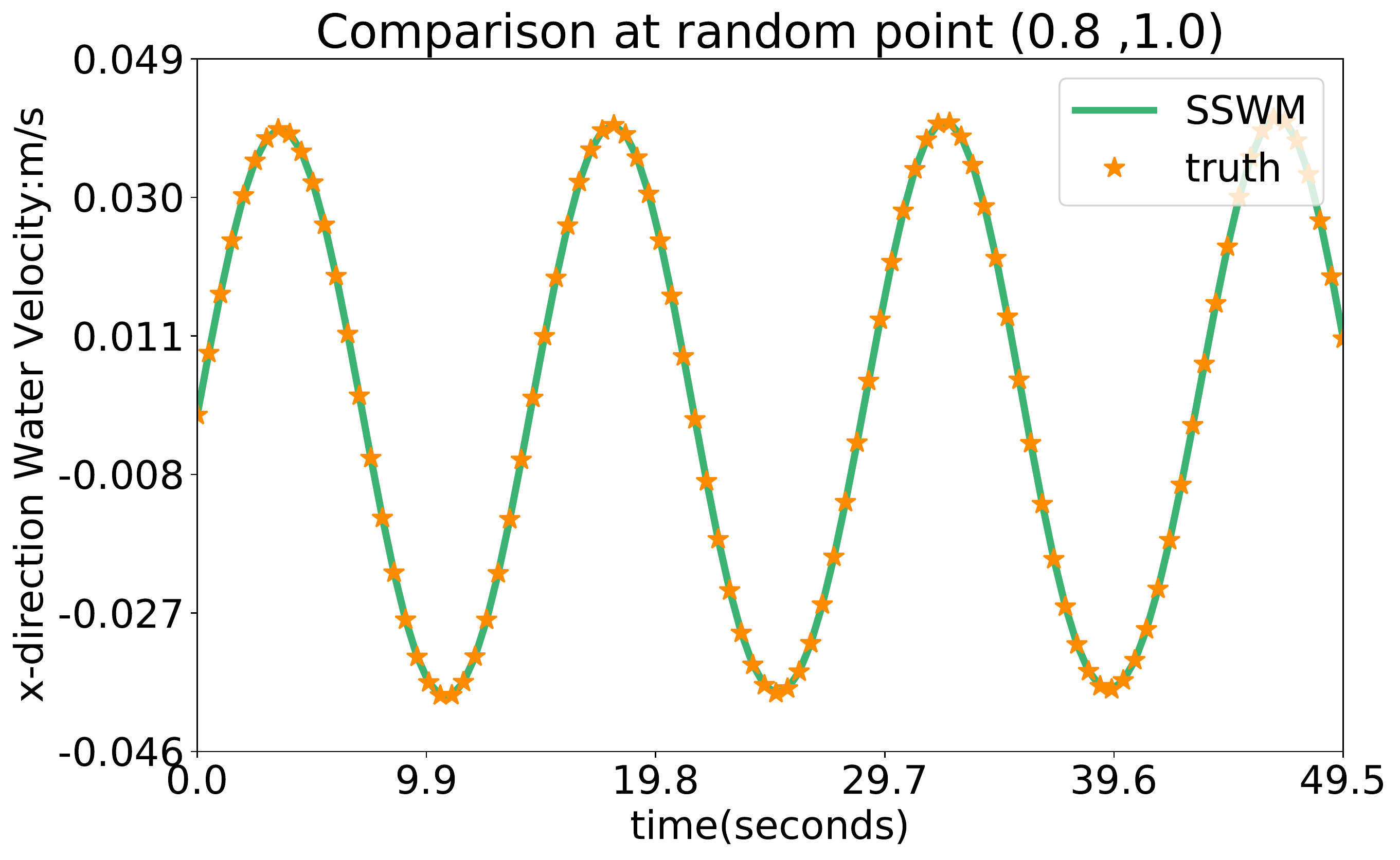}}\hfill
	\subfigure[$x$-direction water velocity at $(1.2, 2.0)$.]{\includegraphics[width=0.5\textwidth]{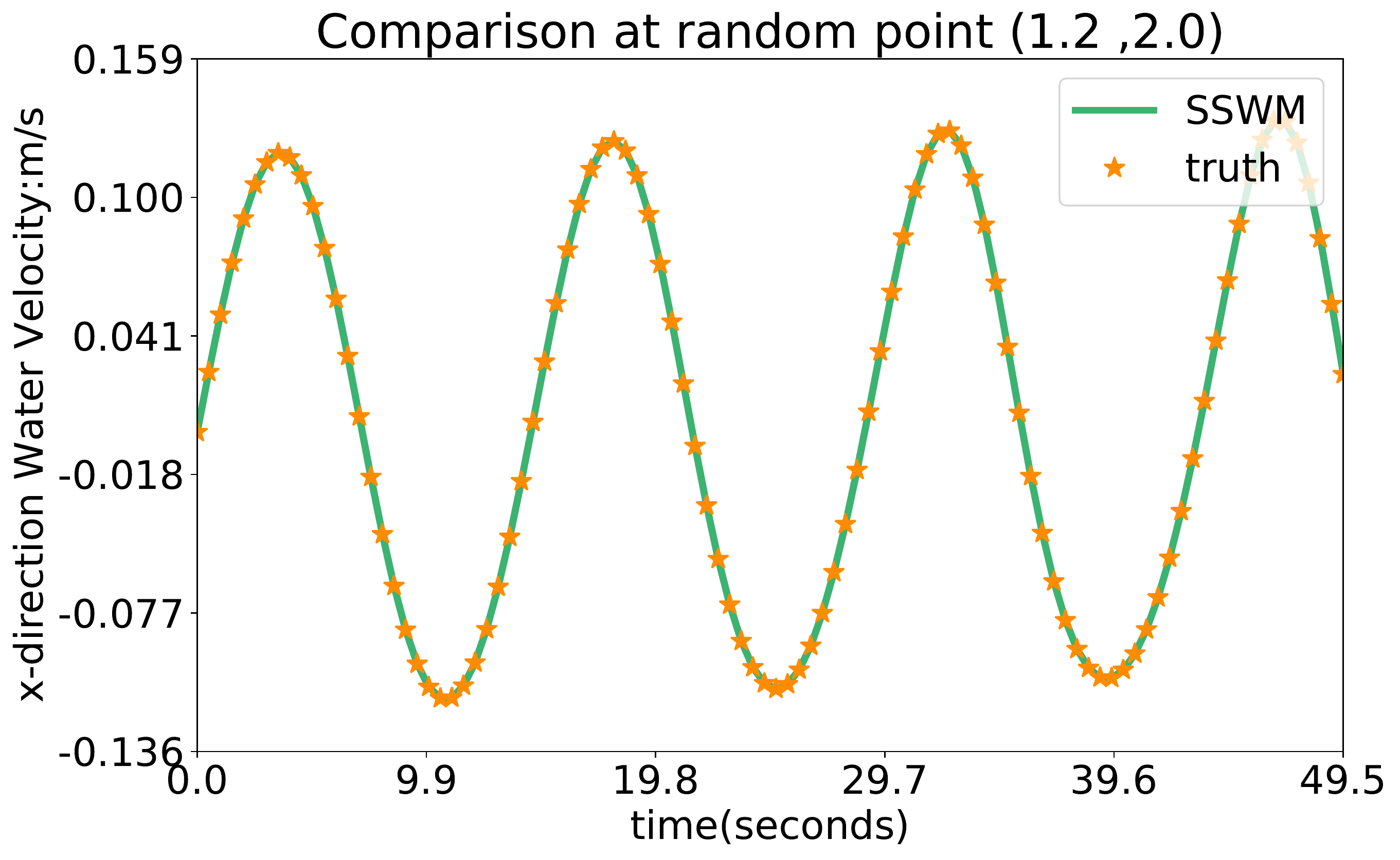}}\hfill
	\caption{Elevation surrogate and $x$-direction velocity surrogate compared at the spatial point $(75.0m, 25.0m)$.}
	\label{fig:sloshetatime}
\end{figure}

We plot the surrogate and benchmark at the spatial point $(75.0m, 25.0m)$ on two sample points $(0.8, 1.0), (1.2, 2.0)$ over all time steps for surface elevation and $x$-direction water velocity in Figure~\ref{fig:sloshetatime}. \moreR{Here} we observe good agreement for both surface elevation and $x$-direction velocity component.

%
\subsection{Pointwise comparison with uncertain bathymetry - Hump test case}
\oneR{The final comparison} we present for this test case is a time series comparison of the surface elevation and $x$-direction component of the velocity field at the spatial points $(250.0m, 100.0m)$ and $(750.0m, 100.0m)$ and two sample points $(0.8, 0.9)$ and $(1.2, 1.1)$.  
\begin{figure}[h!]
	\centering
	\subfigure[$(250.0m, 100.0m)$.]{\includegraphics[width=0.5\textwidth]{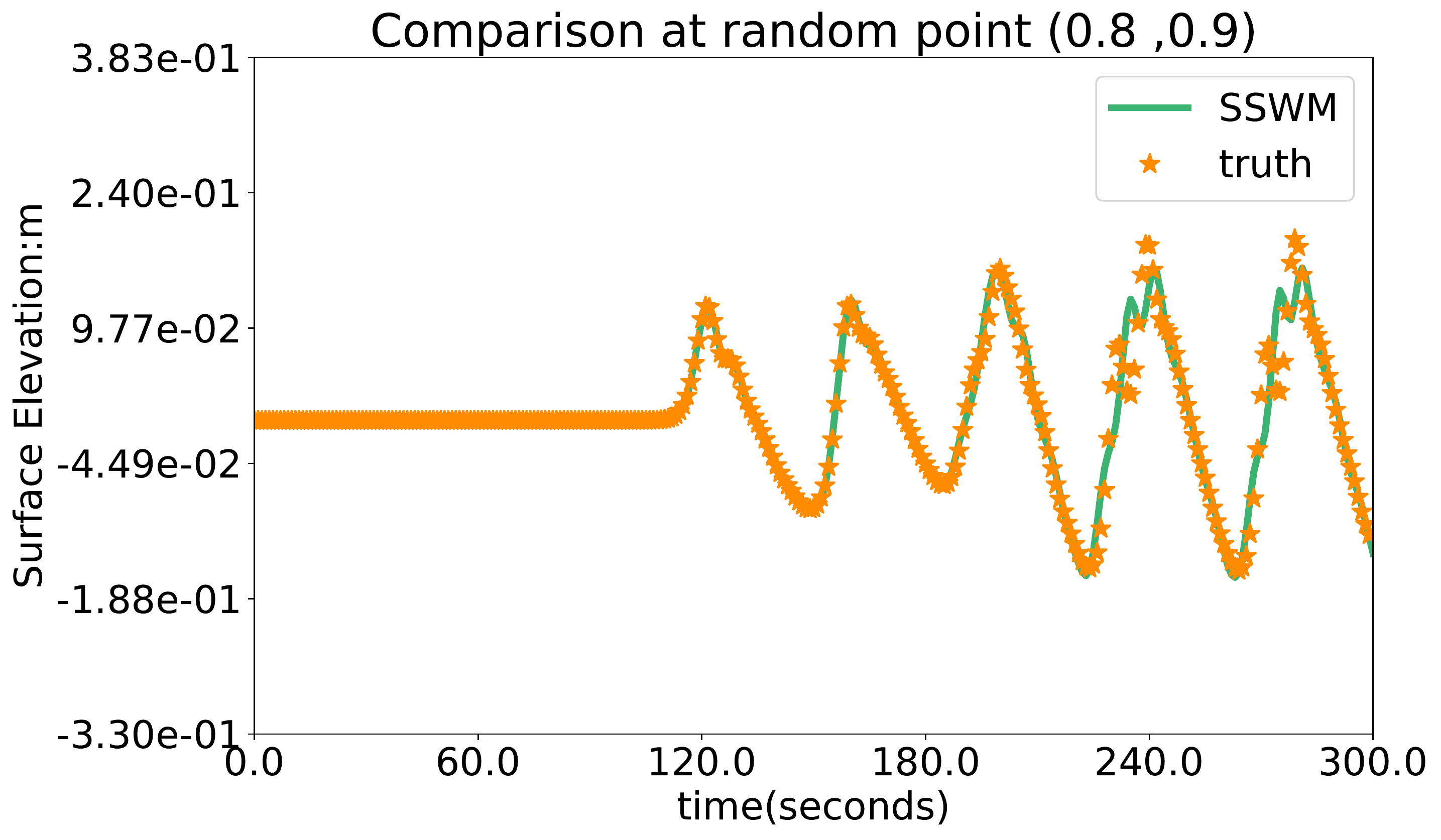}}\hfill
	\subfigure[$(750.0m, 100.0m)$.]{\includegraphics[width=0.5\textwidth]{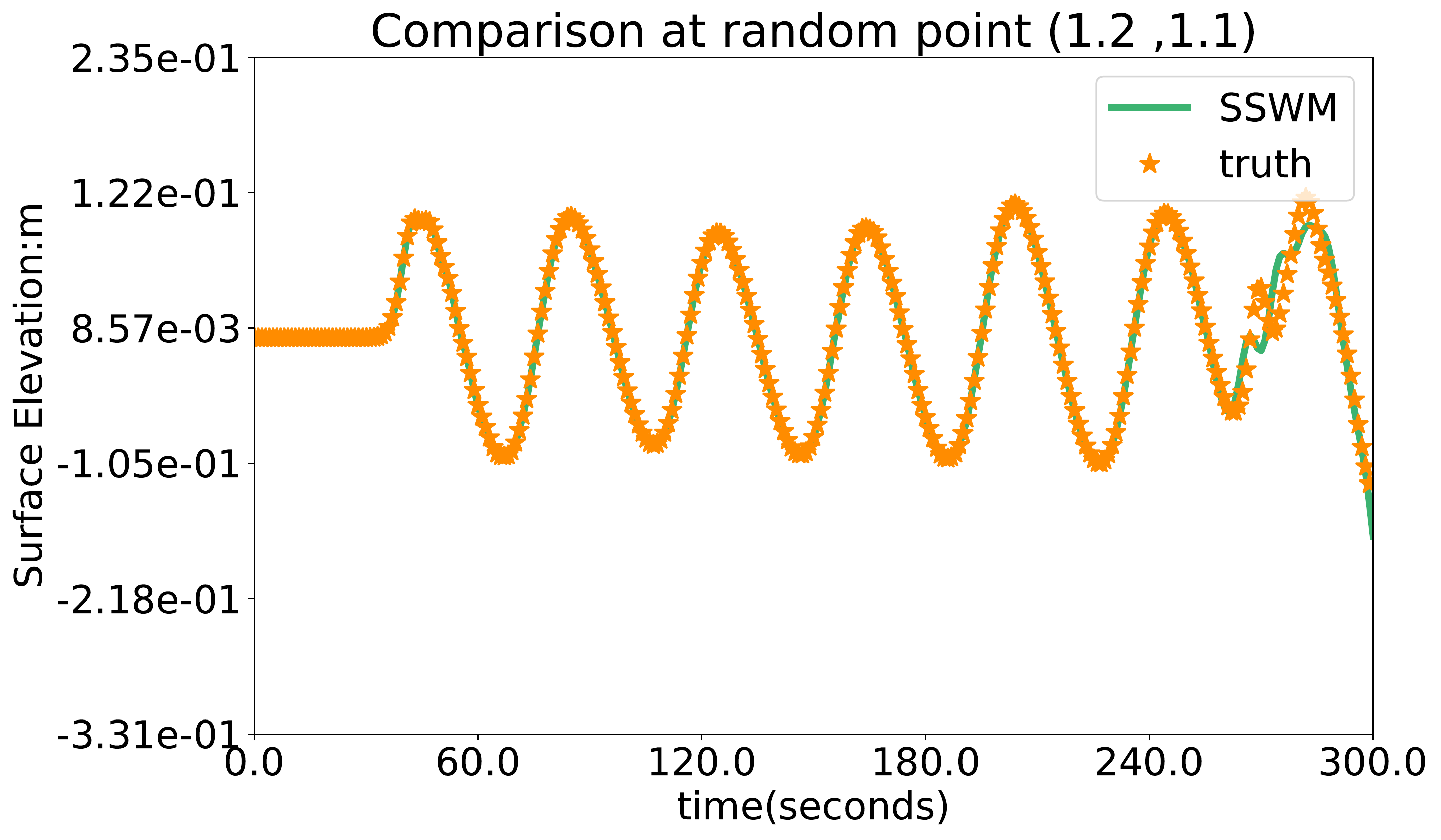}}\hfill
	\subfigure[$(250.0m, 100.0m)$.]{\includegraphics[width=0.5\textwidth]{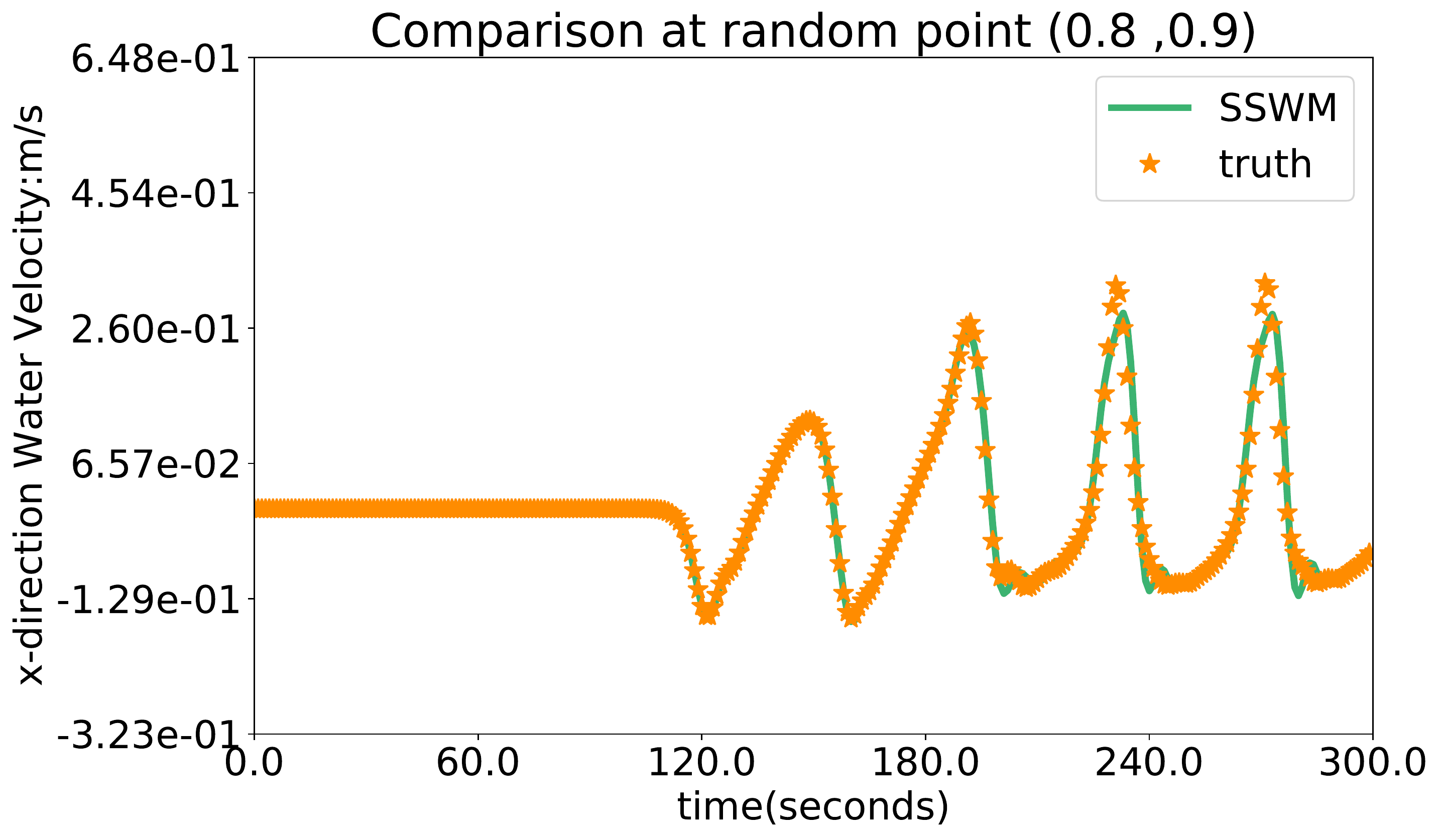}}\hfill
	\subfigure[$(750.0m, 100.0m)$.]{\includegraphics[width=0.5\textwidth]{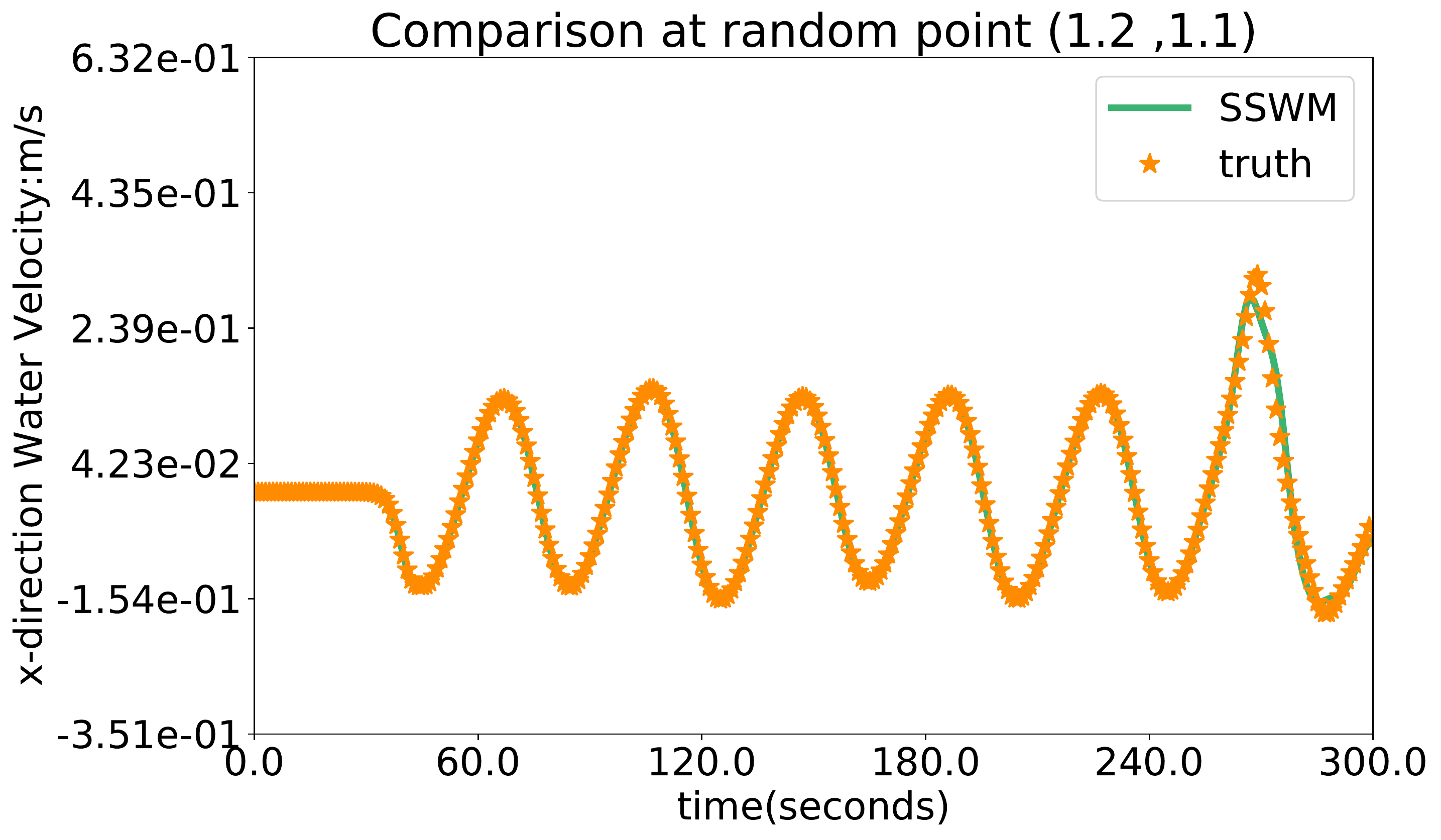}}\hfill
	\caption{Elevation \more{and velocity} surrogates at different spatial and sample points.}
	\label{fig:humpetatime}
\end{figure}
%
%
%
%
This comparison is in \one{Figure~\ref{fig:humpetatime}}, where we observe good agreement with the \one{ensembled deterministic} benchmark solution. The agreement is near perfect in most of the simulation with an exception near the final time. We attribute this discrepancy to accumulation of time discretization error. Fortunately, since the phase, amplitude  and frequency in both models are nearly identical at the previous time steps, we conclude that the surrogate function can well represent the random space.

\subsection{Pointwise comparison with uncertain boundary condition - Inlet test case}
\begin{figure}[h!]
	\centering
	\subfigure[Elevation at $\xi_1=1.0$.]{\includegraphics[width=0.5\textwidth]{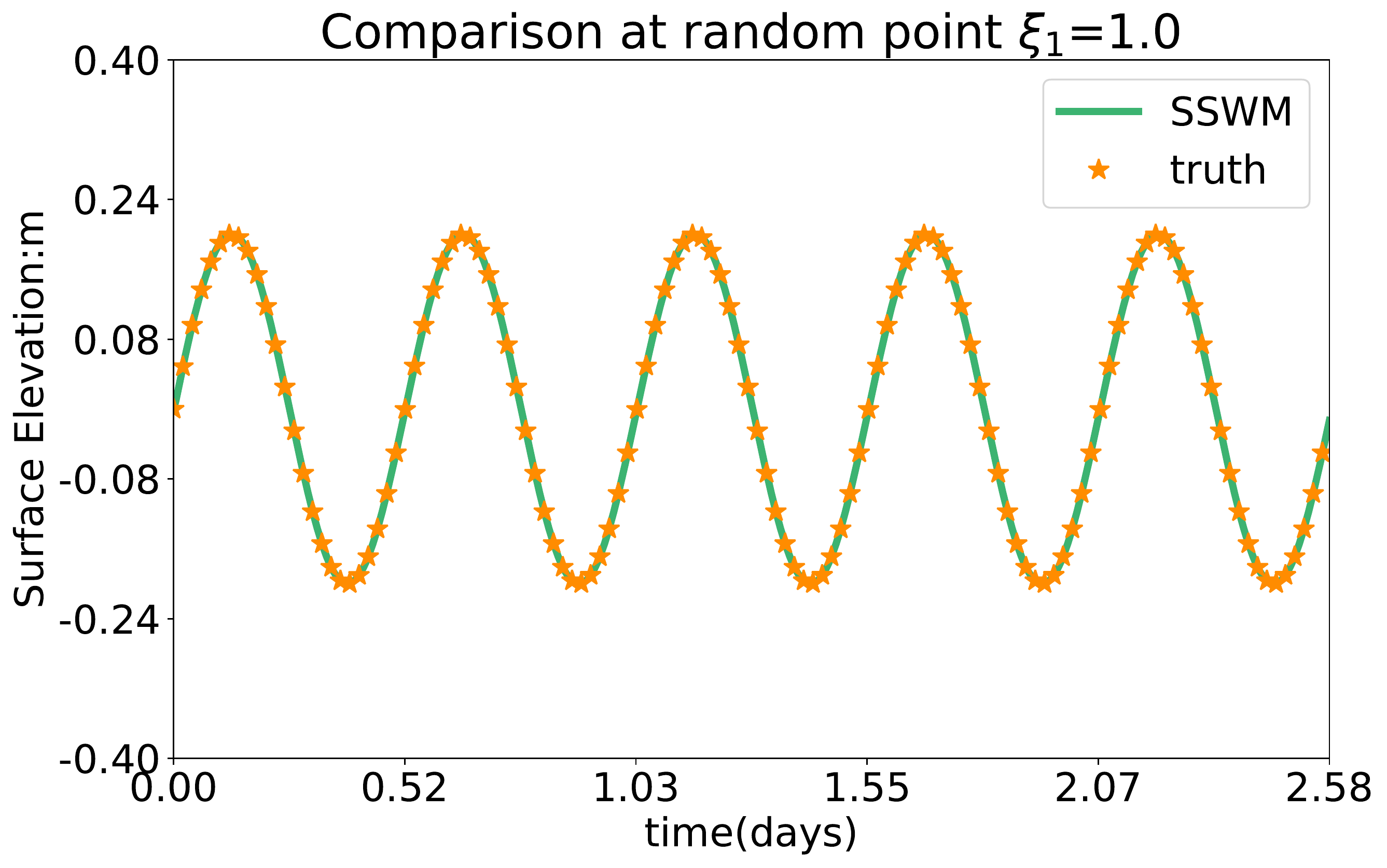}}\hfill
	\subfigure[Elevation at $\xi_1=2.0$.]{\includegraphics[width=0.5\textwidth]{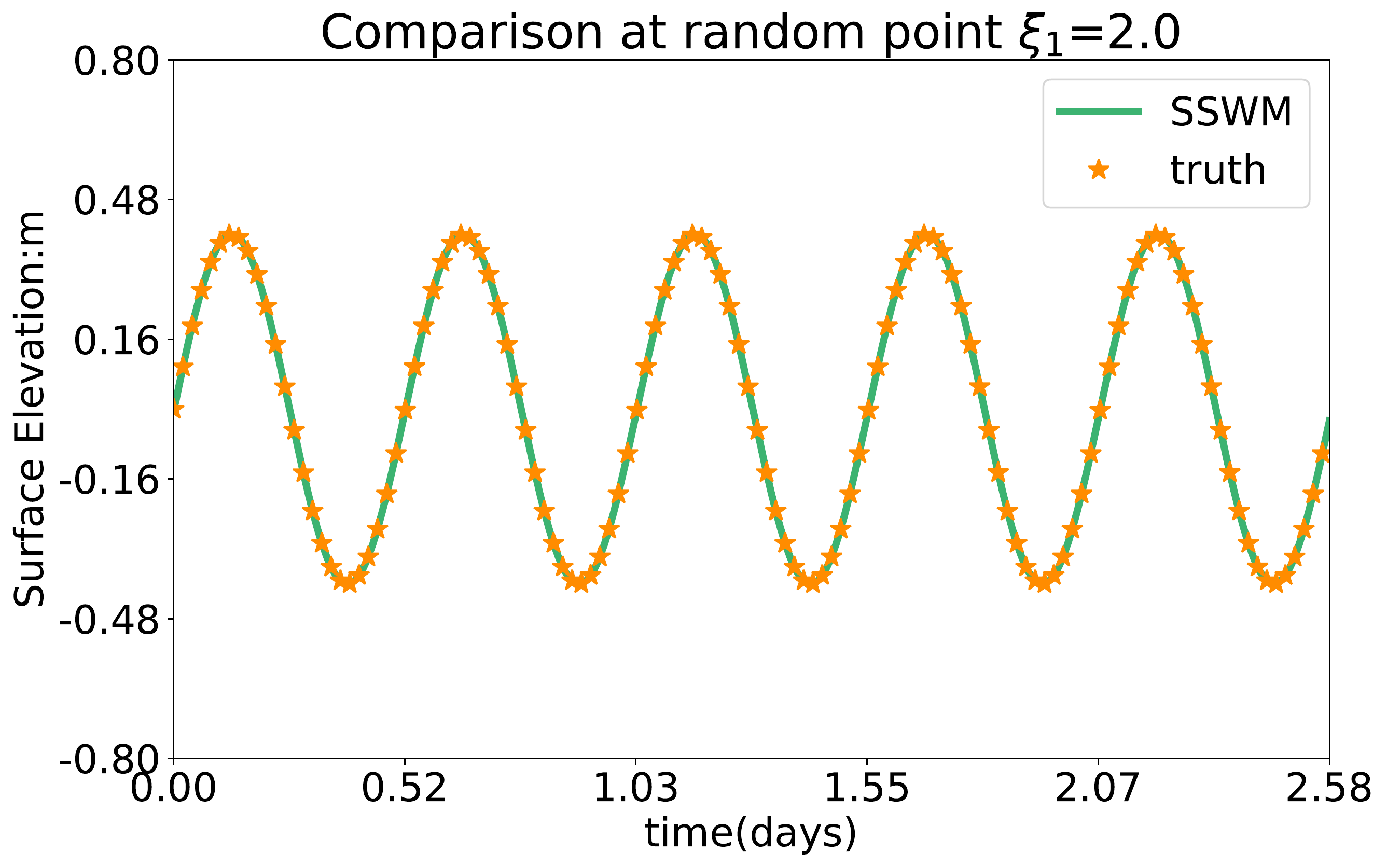}}\hfill
	\subfigure[$x$-velocity at $\xi_1=1.0$.]{\includegraphics[width=0.5\textwidth]{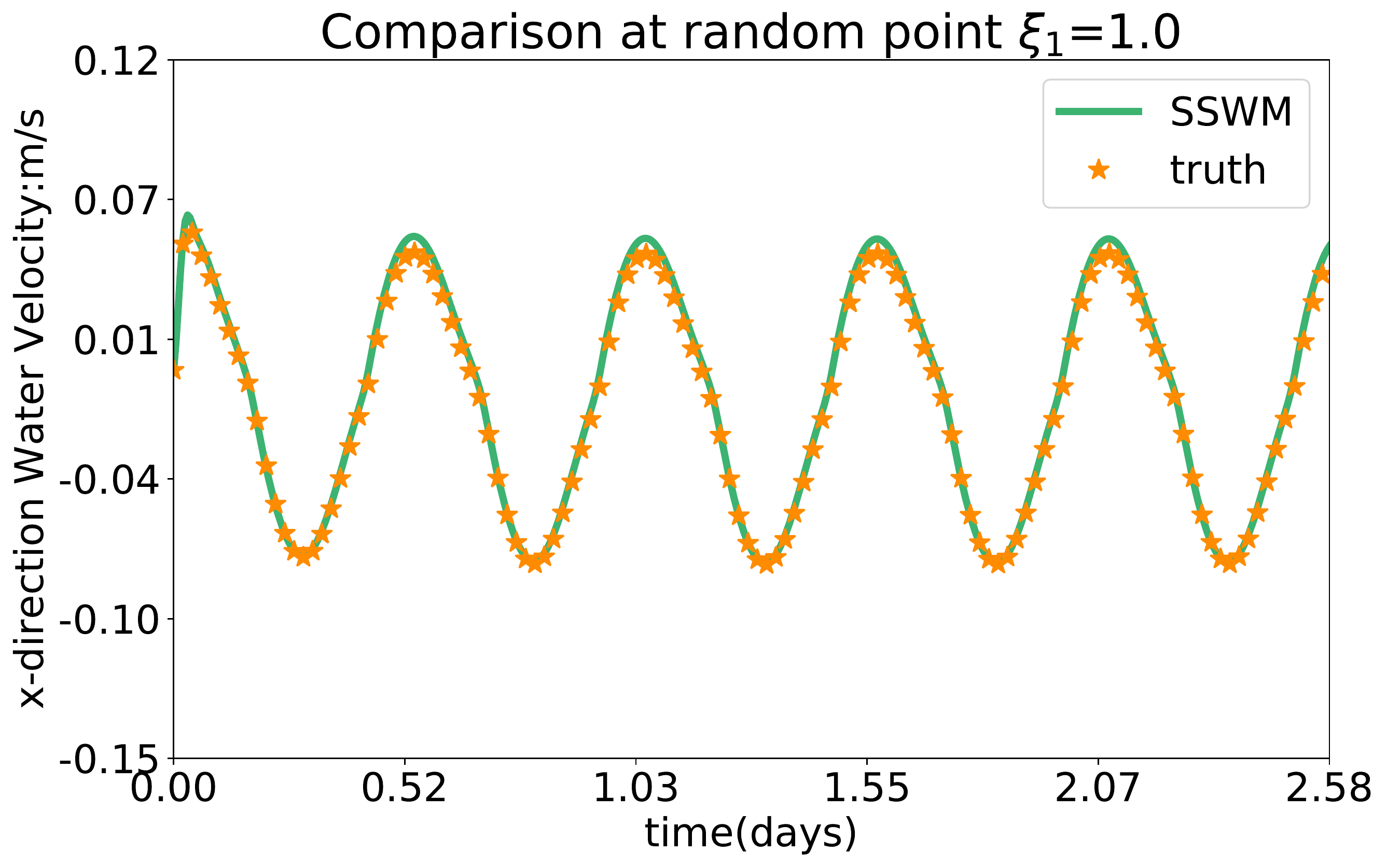}}\hfill
	\subfigure[$x$-velocity at $\xi_1=2.0$.]{\includegraphics[width=0.5\textwidth]{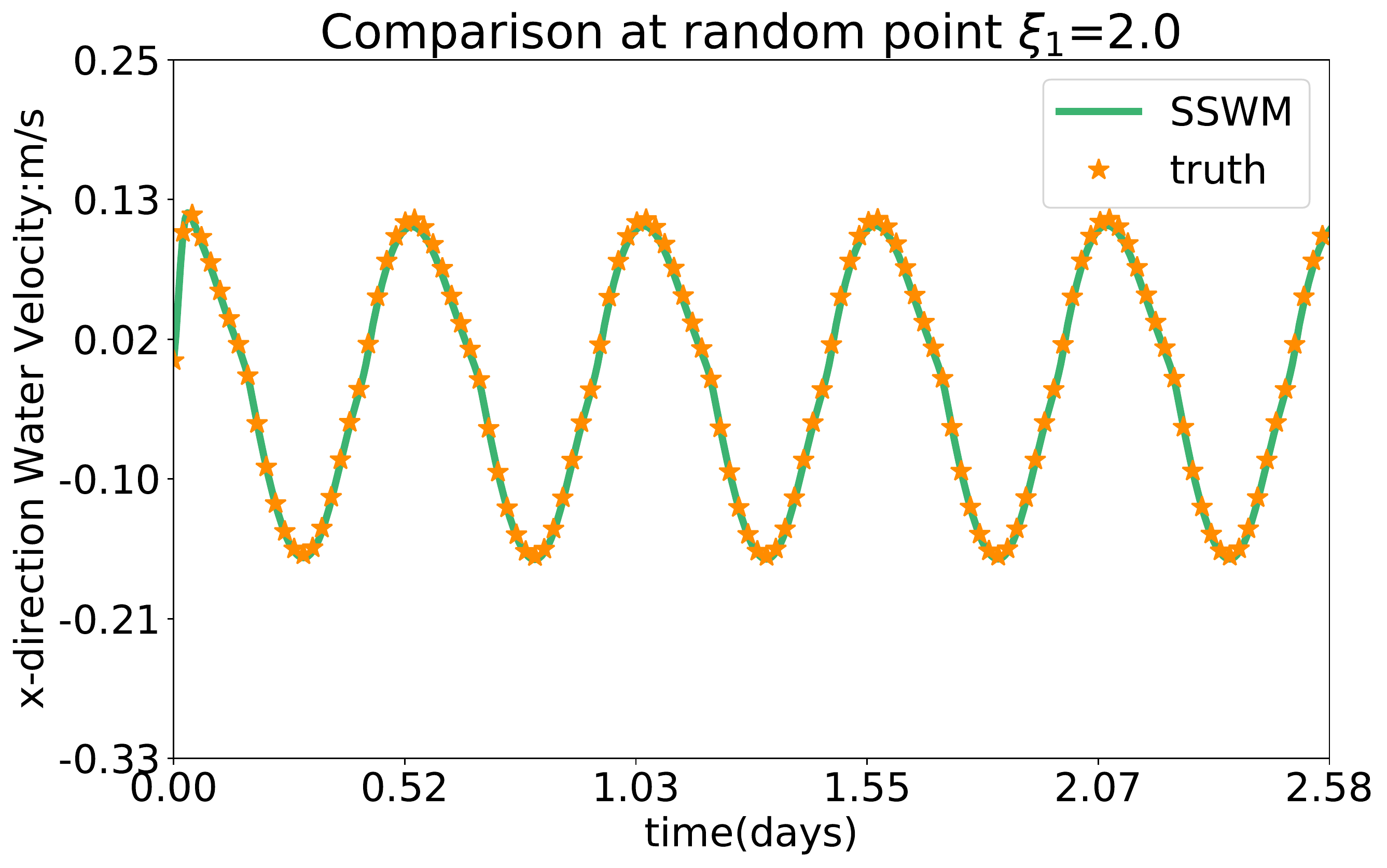}}\hfill
	\caption{Uncertain boundary condition in the inlet test: Elevation surrogate at the spatial point $(-250.0m, 0.0m)$.}
	\label{fig:inletetatime}
\end{figure}

Next, we consider the temporal distribution of  surface elevation and velocity at $(-250.0m, 0.0m)$. Two sample grids are selected in order to draw a general conclusion: we select the minimum and maximum samples, i.e., $\xi_1=1.0,2.0$ and present the results in Figure~\ref{fig:inletetatime}. Here, the close agreement between the two models is again apparent. 
\begin{figure}[h!]
	\centering
	\subfigure[Elevation surrogate at $1.061$ days.]{\includegraphics[width=0.45\textwidth]{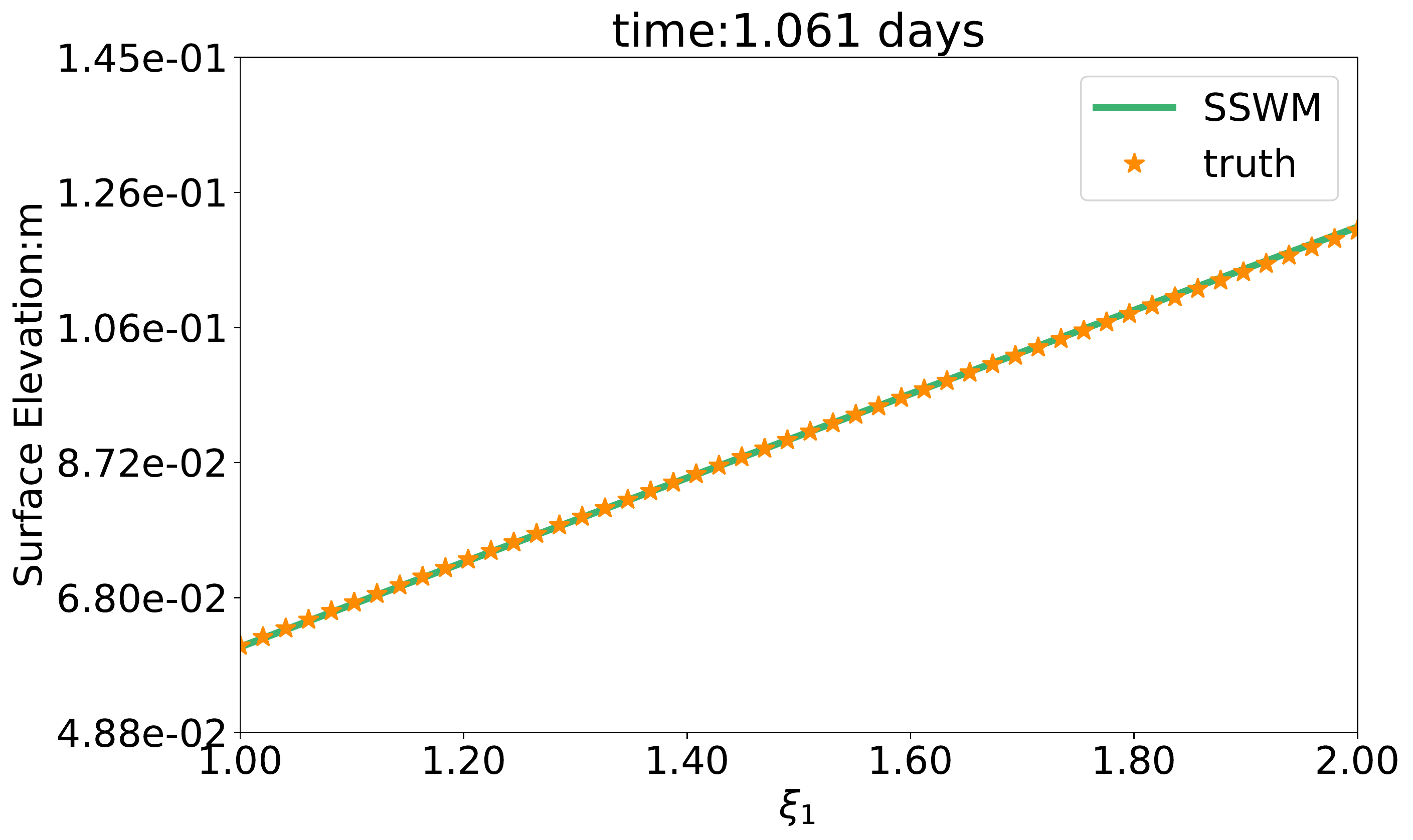}}\hfill
	\subfigure[Elevation surrogate at $2.095$ days.]{\includegraphics[width=0.45\textwidth]{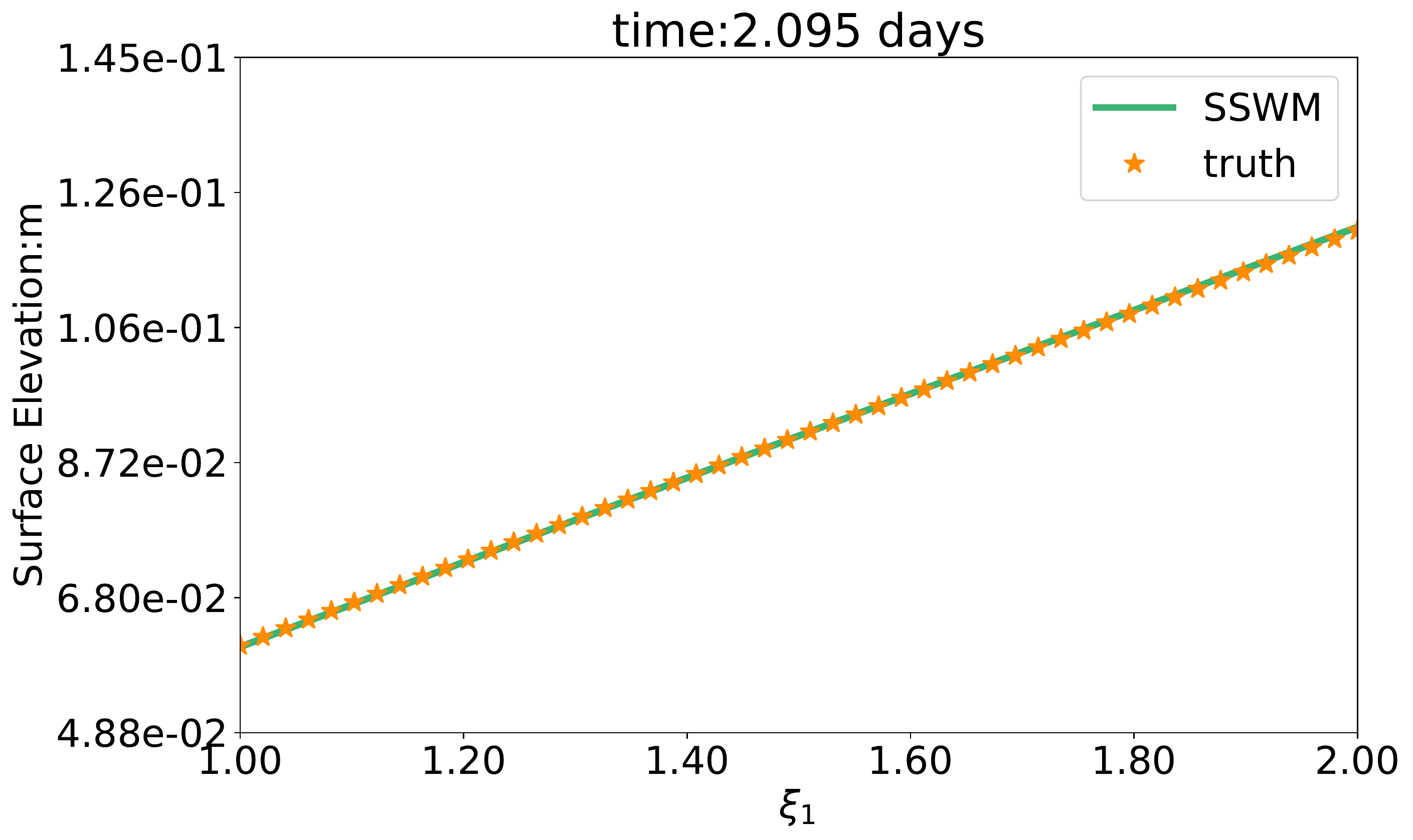}}\hfill
	\subfigure[$x$-velocity surrogate at $1.061$ days.]{\includegraphics[width=0.45\textwidth]{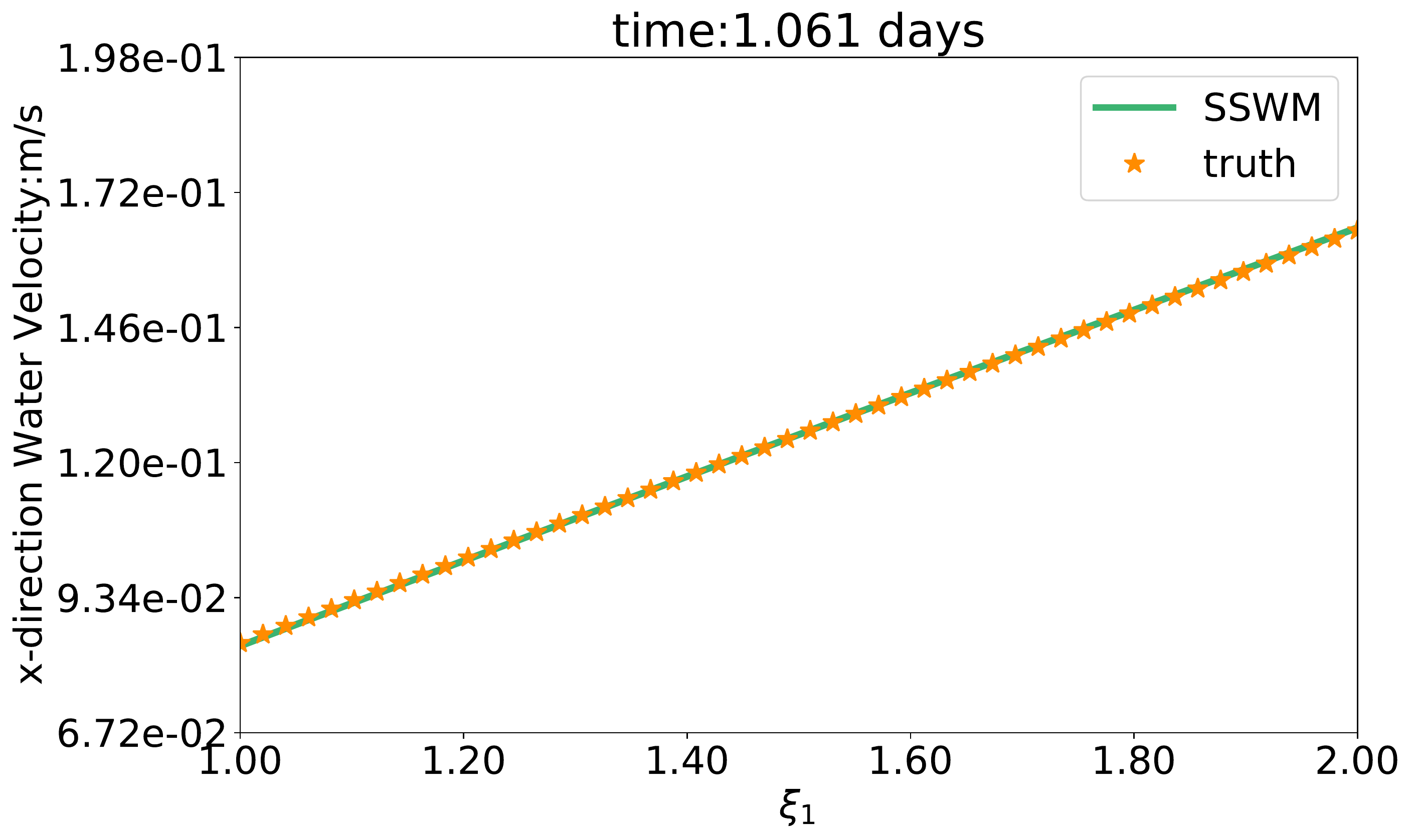}}\hfill
	\subfigure[$x$-velocity surrogate at $2.095$ days.]{\includegraphics[width=0.45\textwidth]{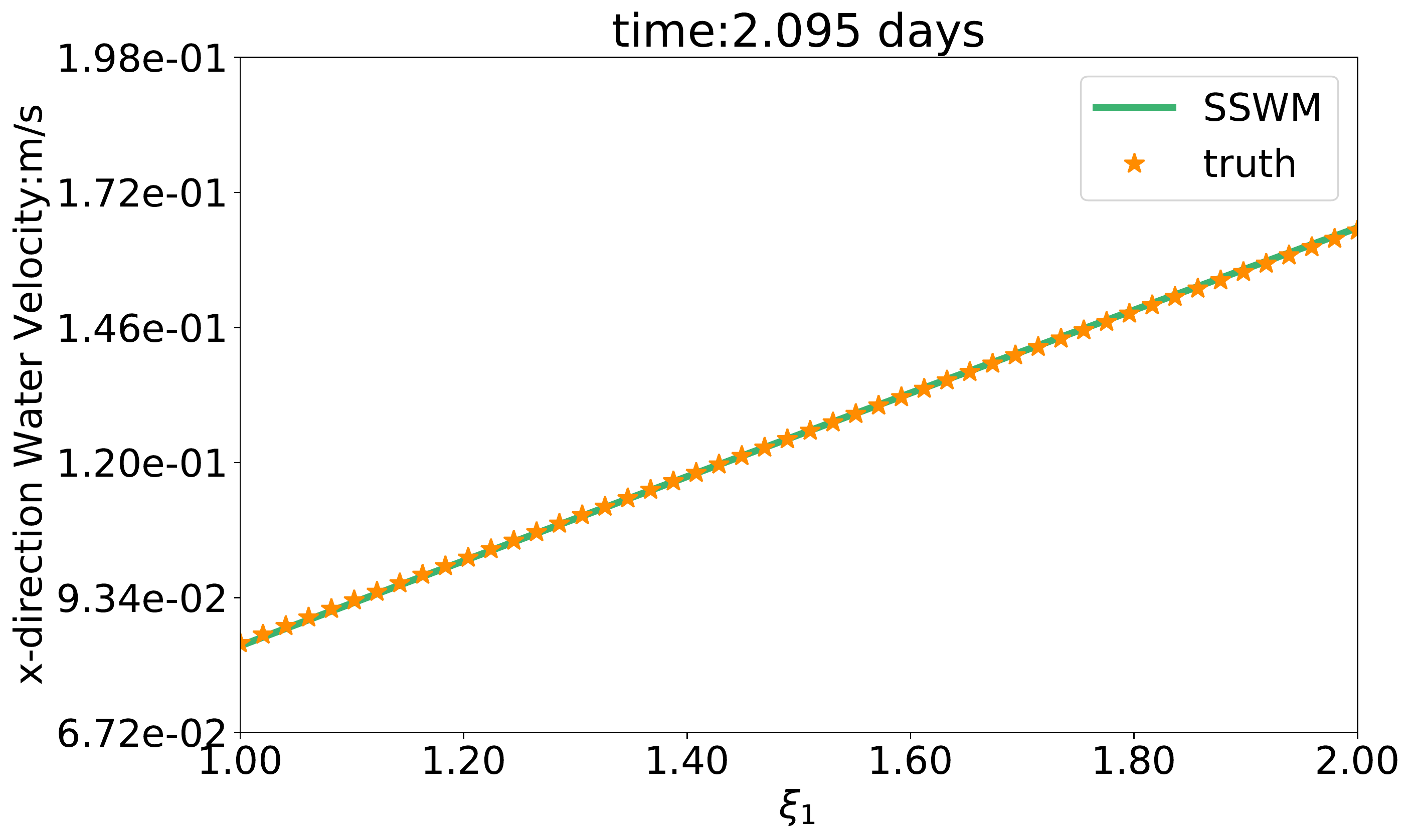}}\hfill
	\caption{Uncertain boundary condition in the inlet test: elevation surrogate at the spatial point $(0.0m, 0.0m)$ over the random space.}
	\label{fig:inleteta}
\end{figure}
%
%
%
%

\subsection{Pointwise comparison with uncertain wind drag parameter - Hurricane Harvey test case }
\begin{figure}[h!]
	\centering
	\subfigure[Elevation surrogate at $t=1.319$ days.]{\includegraphics[width=0.5\textwidth]{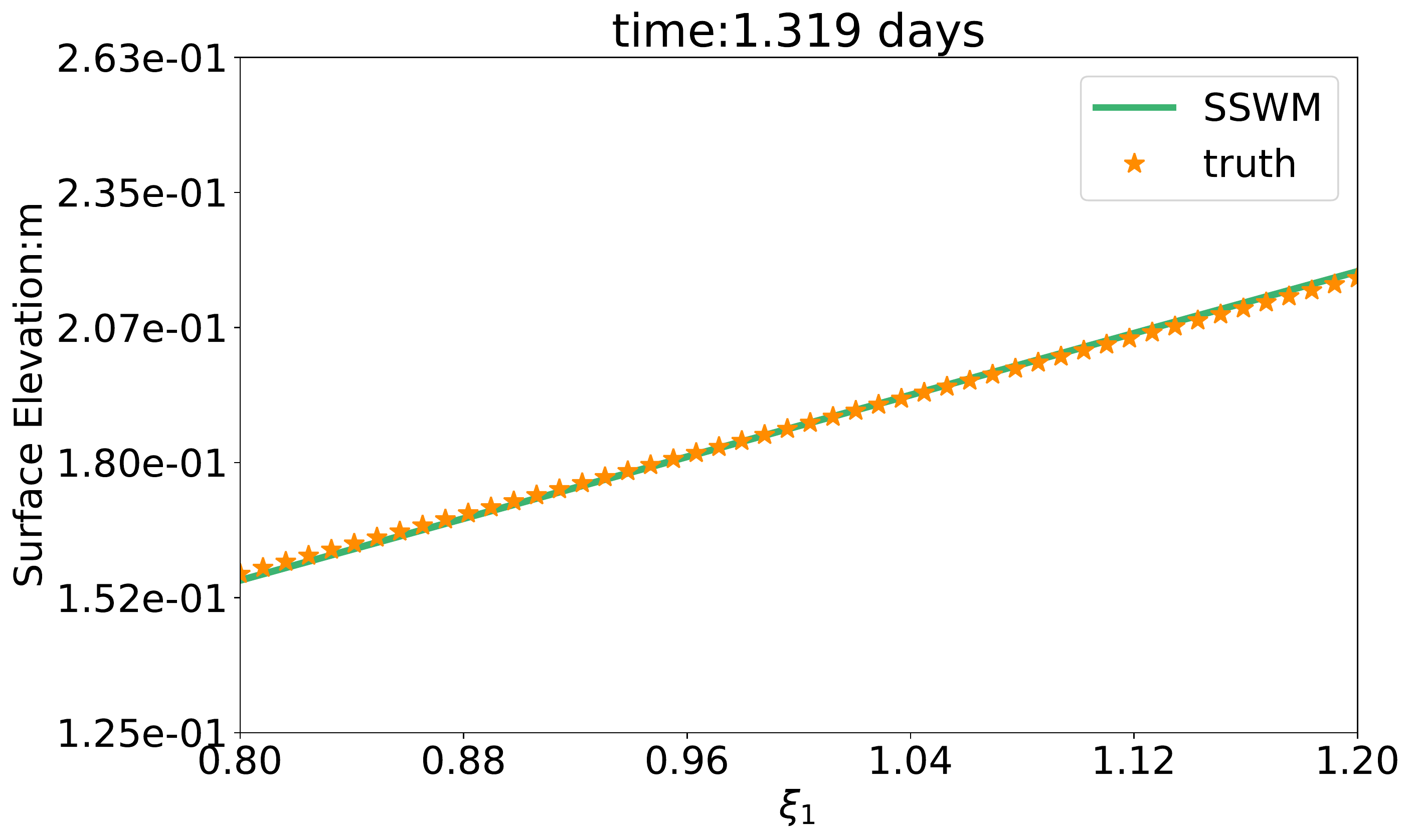}}\hfill
	\subfigure[Elevation surrogate at $t=4.139$ days.]{\includegraphics[width=0.5\textwidth]{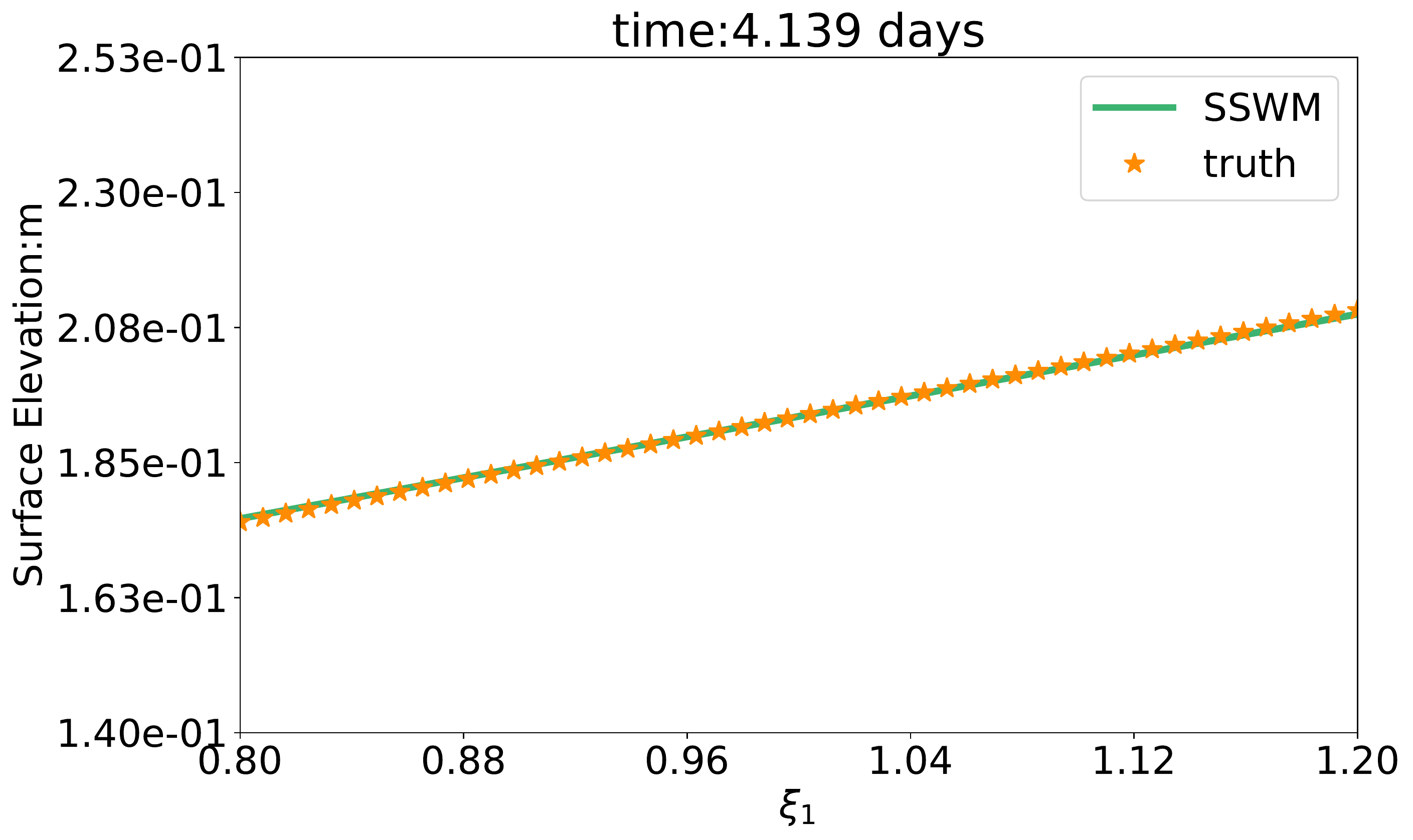}}\hfill
	\subfigure[$x$-velocity surrogate at $t=1.319$ days.]{\includegraphics[width=0.5\textwidth]{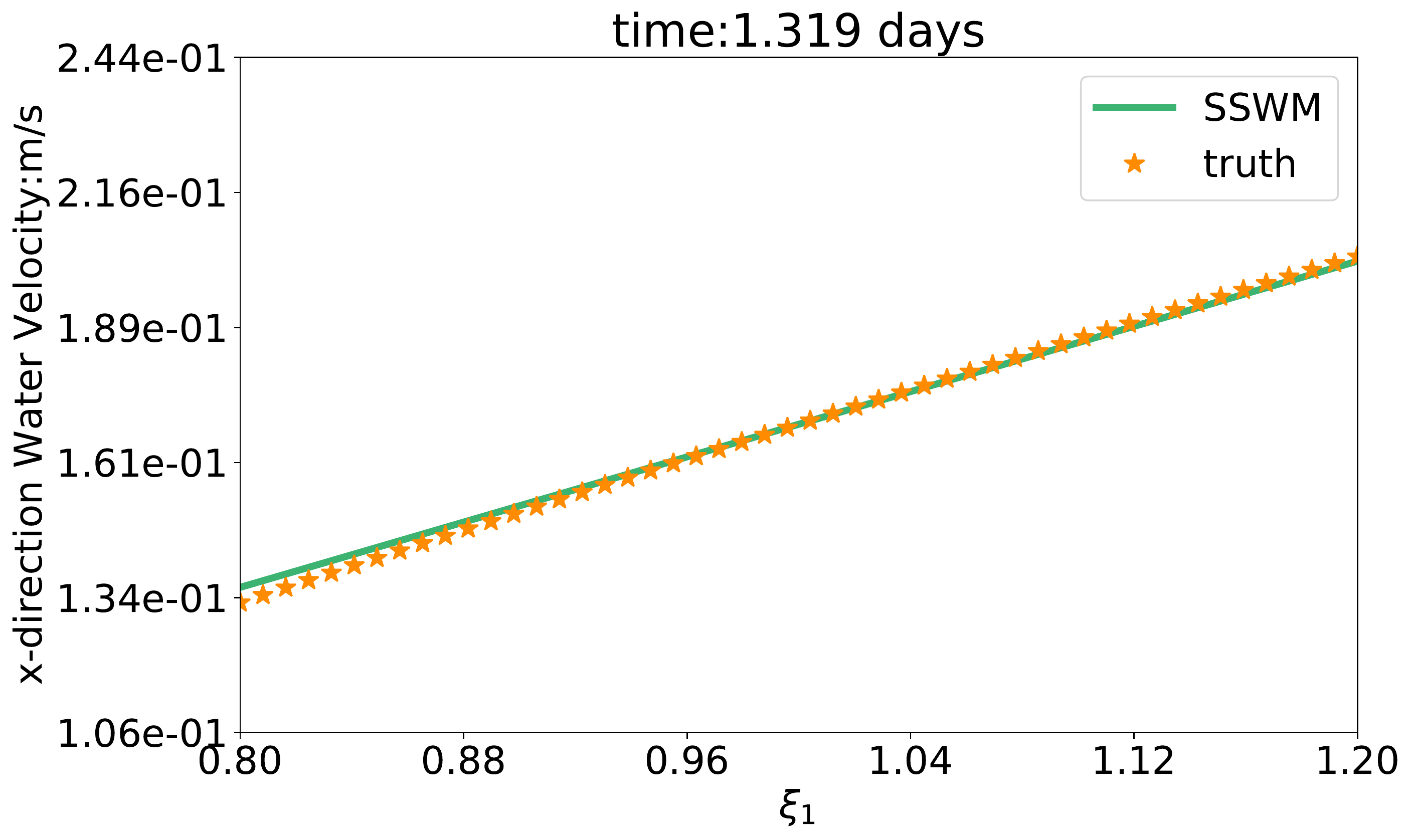}}\hfill
	\subfigure[$x$-velocity surrogate at $t=4.139$ days.]{\includegraphics[width=0.5\textwidth]{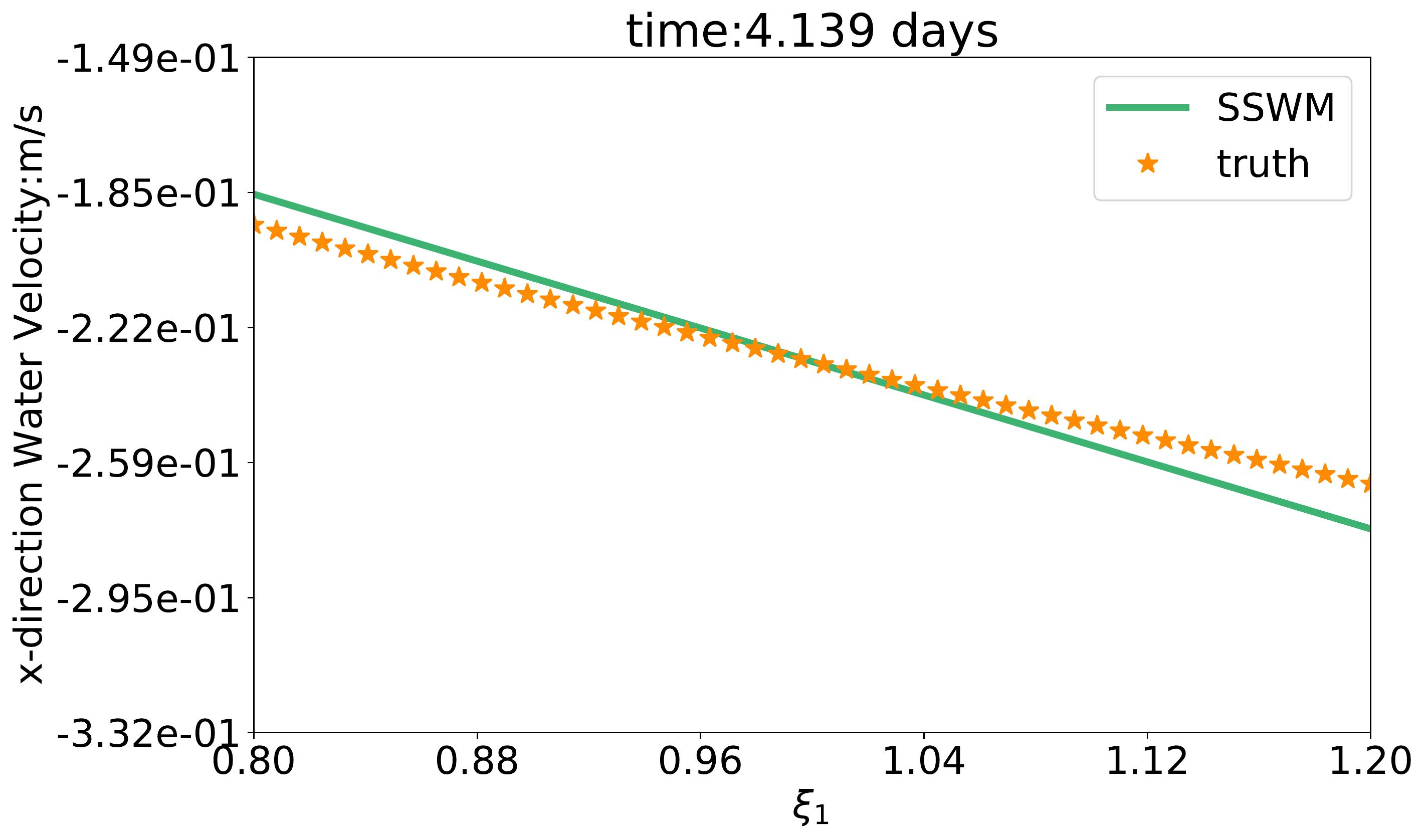}}\hfill
	\subfigure[$y$-velocity surrogate at $t=1.319$ days.]{\includegraphics[width=0.5\textwidth]{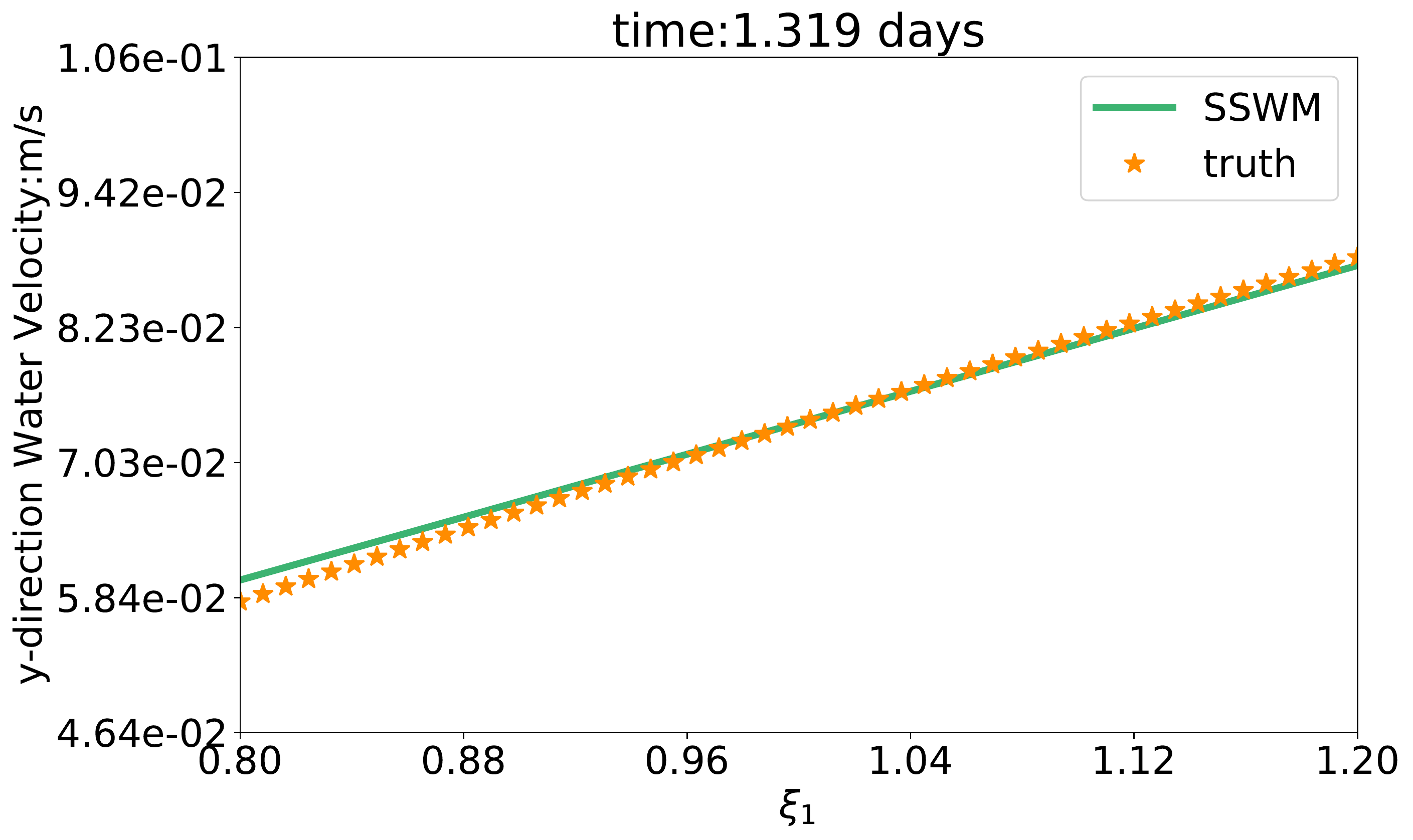}}\hfill
	\subfigure[$y$-velocity surrogate at $t=4.139$ days.]{\includegraphics[width=0.5\textwidth]{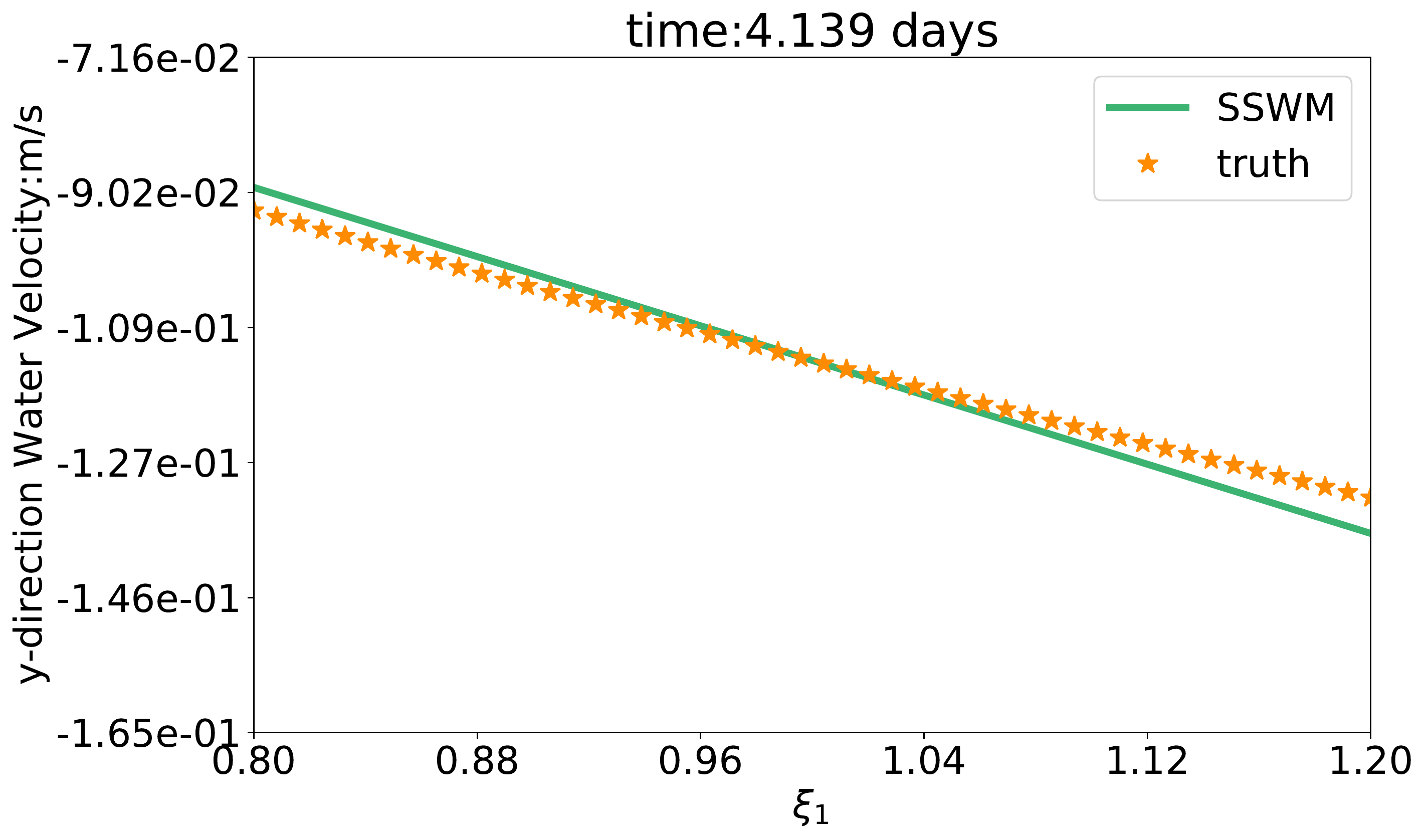}}\hfill
	\caption{Surrogate comparison for Hurricane Harvey at spatial point $(-95.24^{\circ},  28.85^{\circ})$}
	\label{fig:ikeeta2}
\end{figure}

\begin{figure}[h!]
	\centering
	\subfigure[Elevation at $(-95.24^{\circ}, 28.85^{\circ})$.]{\includegraphics[width=0.5\textwidth]{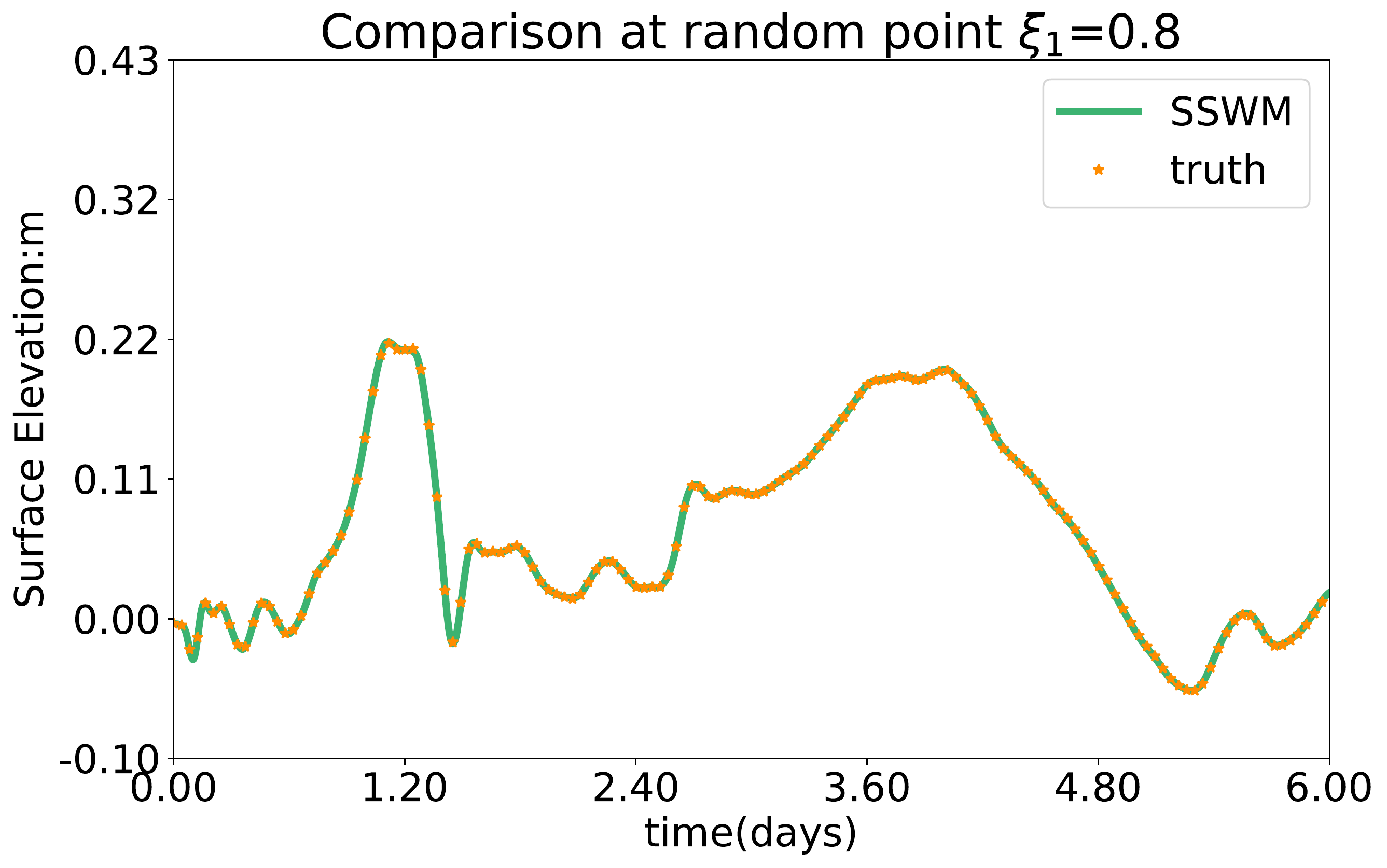}}\hfill
	\subfigure[Elevation at $(-95.24^{\circ}, 28.85^{\circ})$.]{\includegraphics[width=0.5\textwidth]{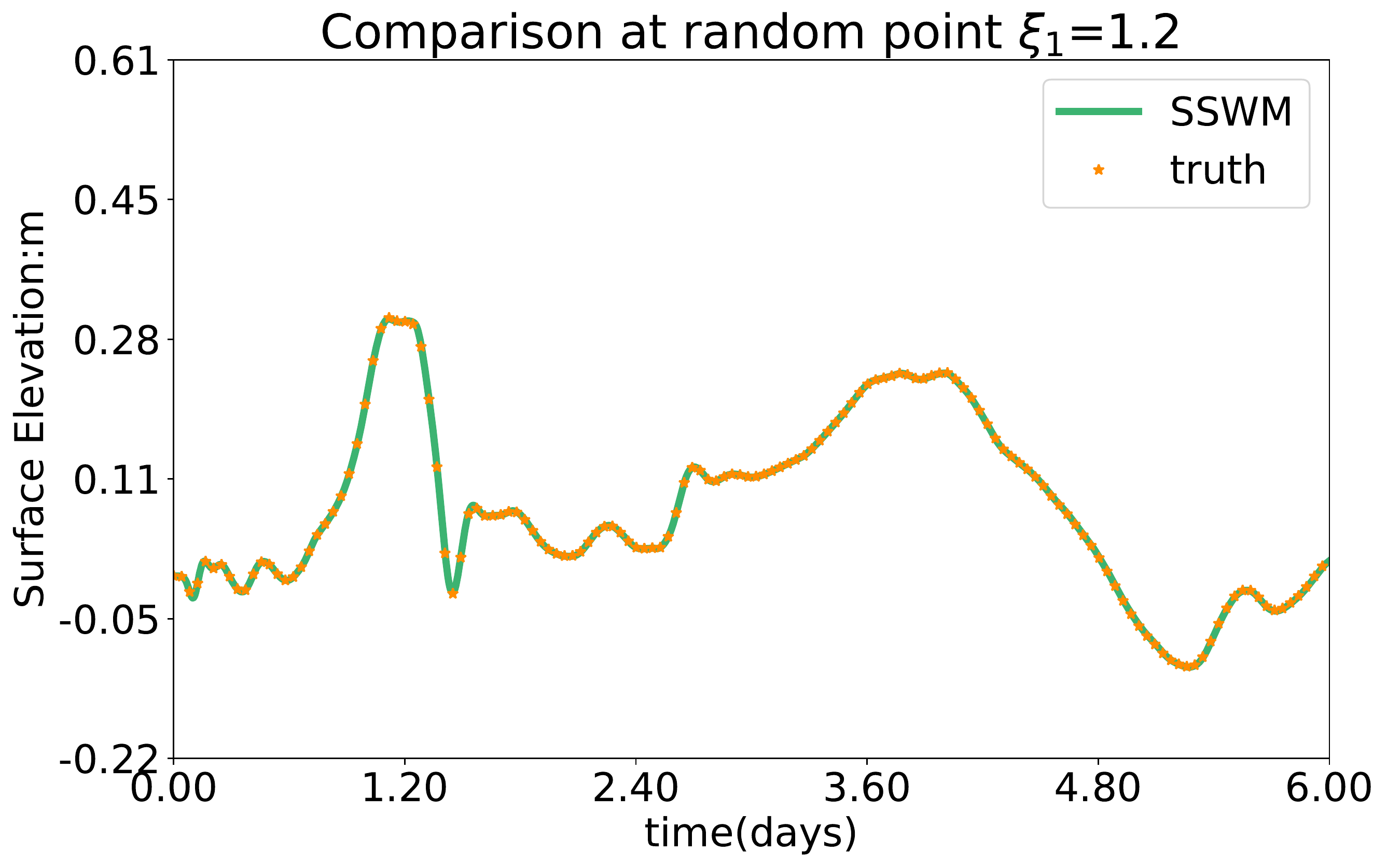}}\hfill
	\subfigure[$x$-velocity at $(-94.51^{\circ}, 29.43^{\circ})$.]{\includegraphics[width=0.5\textwidth]{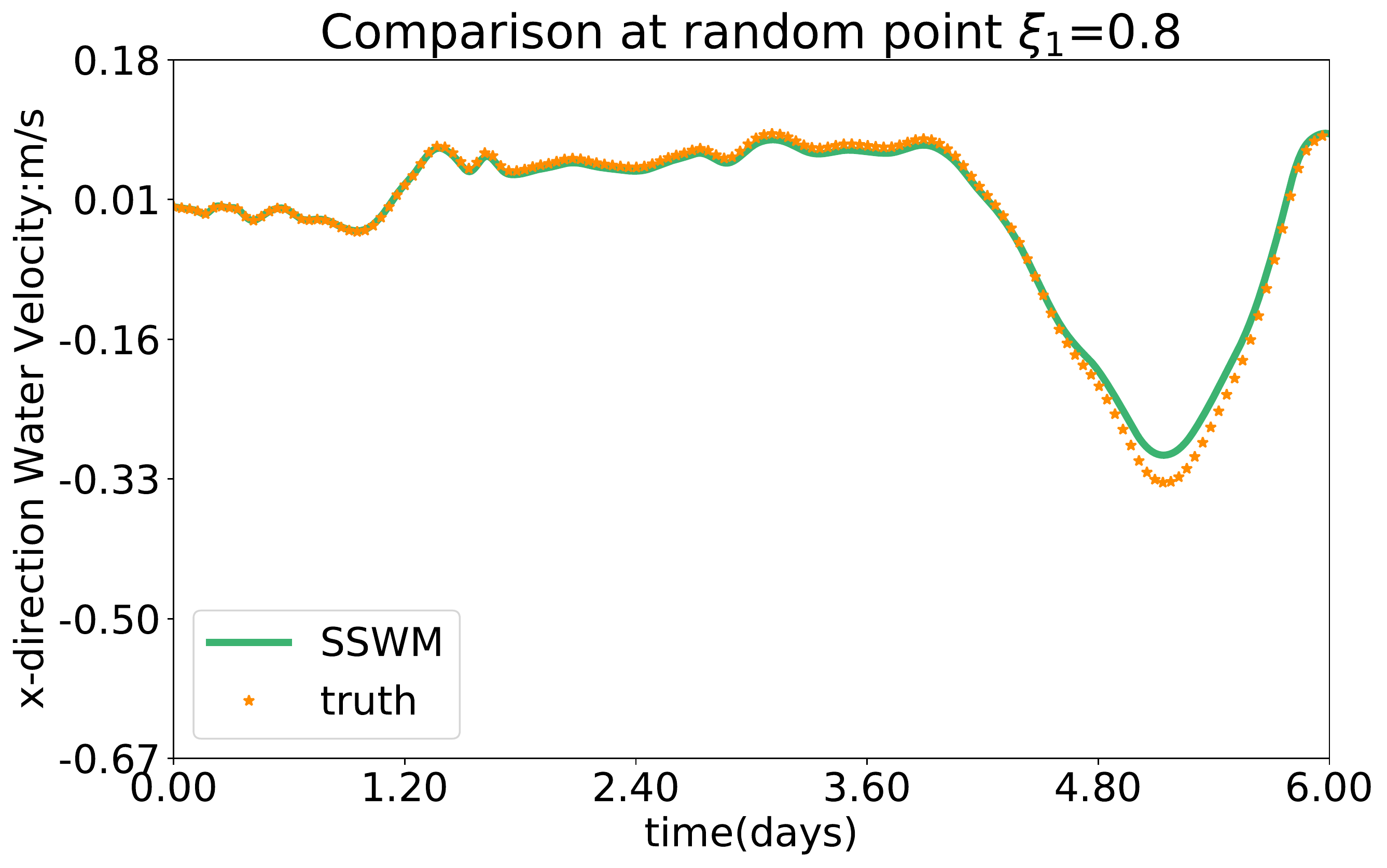}}\hfill
	\subfigure[$x$-velocity at $(-94.51^{\circ}, 29.43^{\circ})$.]{\includegraphics[width=0.5\textwidth]{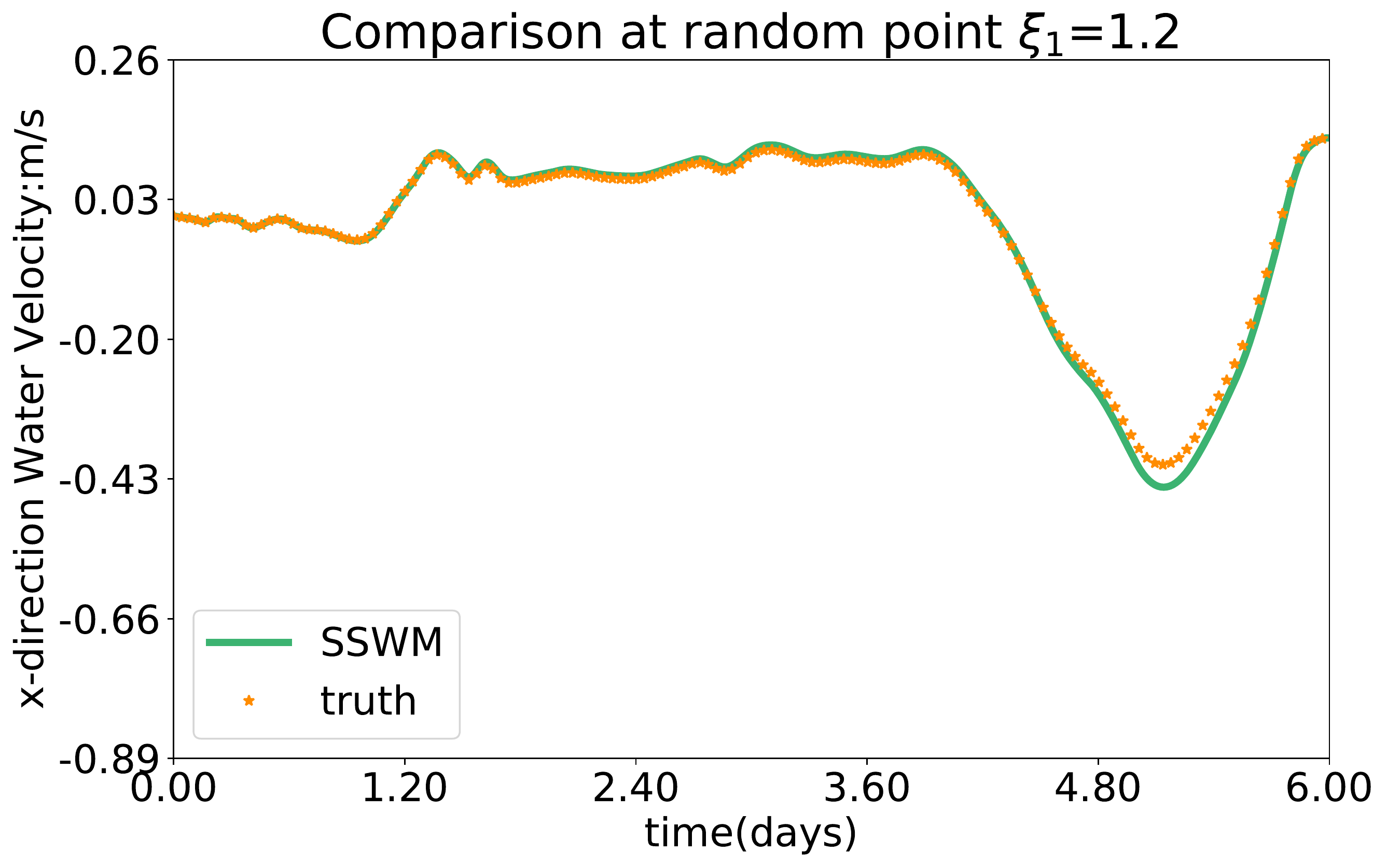}}\hfill
	\subfigure[$y$-velocity at $(-94.72^{\circ}, 29.34^{\circ})$.]{\includegraphics[width=0.5\textwidth]{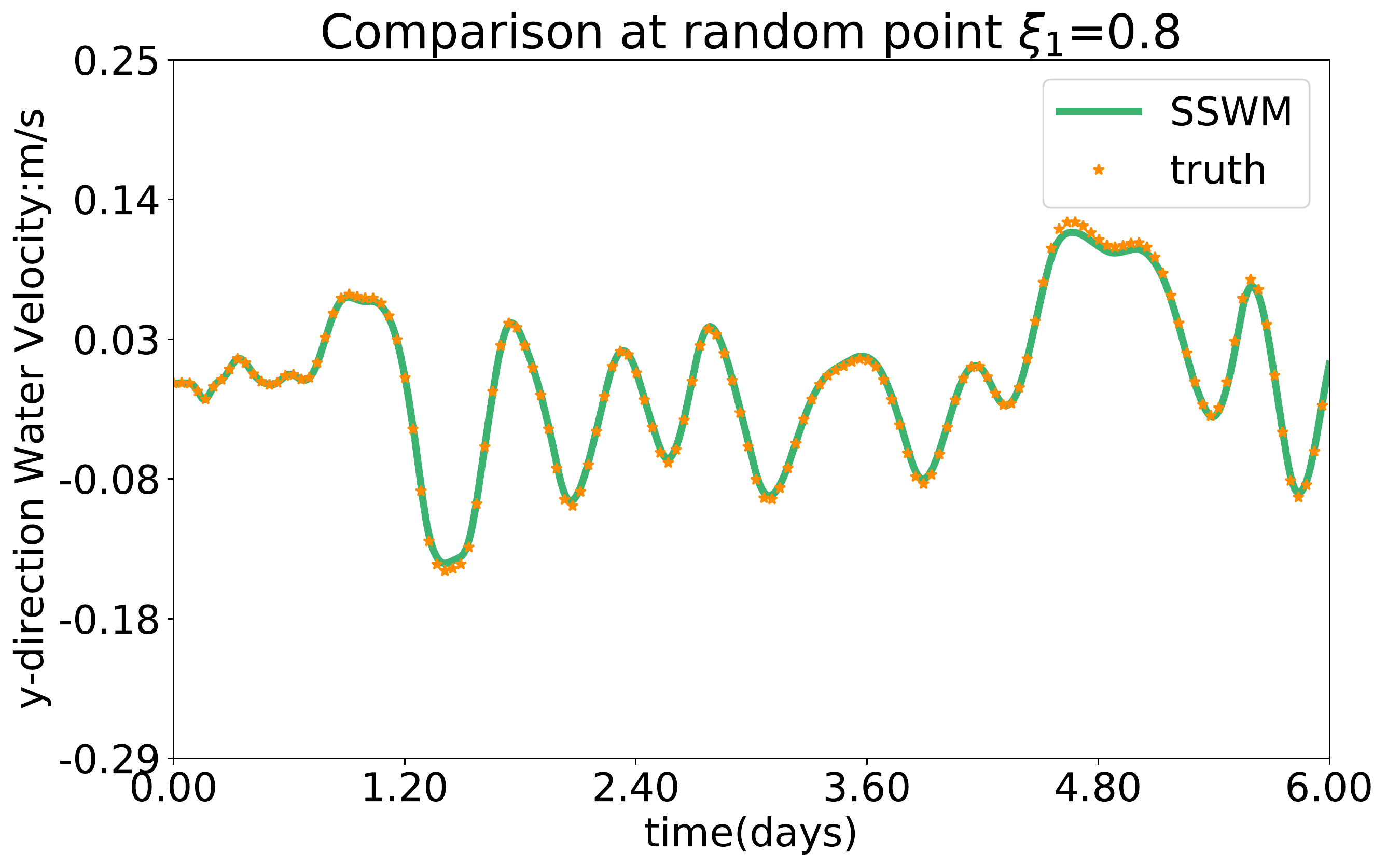}}\hfill
	\subfigure[$y$-velocity at $(-94.72^{\circ}, 29.34^{\circ})$.]{\includegraphics[width=0.5\textwidth]{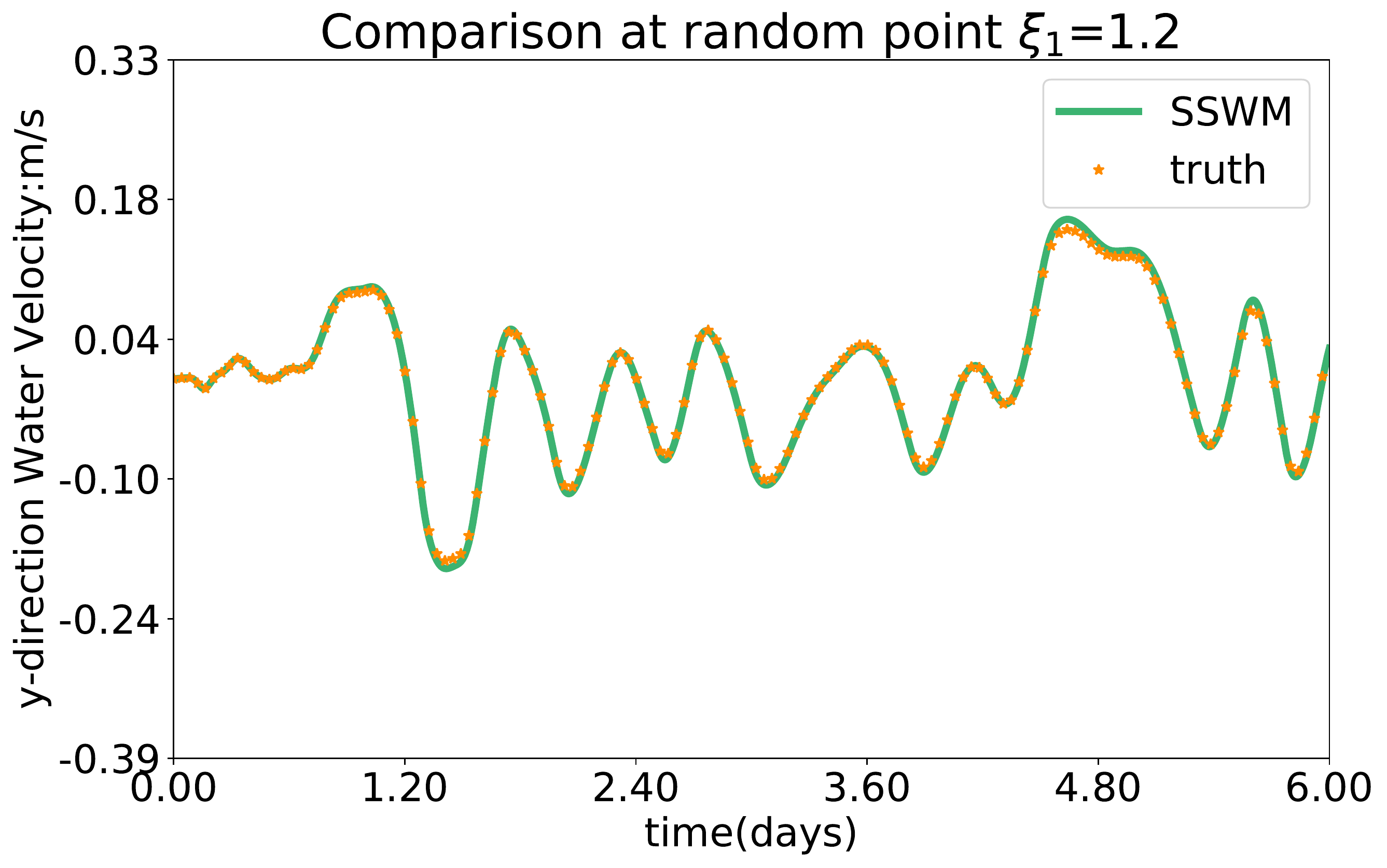}}\hfill
	\caption{Time series surrogate comparison at three spatial points.}
	\label{fig:ikeetatime}
\end{figure}

Lastly, We present the comparison of elevation, $x$-direction velocity, and $y$-direction velocity over the random space with respect to $\xi_1$ in Figure~\ref{fig:ikeeta2}. The surface elevation agrees very well in the random space and we only observe minor discrepancies for both velocity components in this figures.
%
%
Lastly, we consider surface elevation and water velocity against time $t$ at three spatial points $(-95.24^{\circ}, 28.85^{\circ}), ~  (-94.51^{\circ}, 29.43^{\circ})$, and $(-94.72^{\circ}, 29.34^{\circ})$. In Figure~\ref{fig:ikeetatime}, we present the comparisons of our surrogate and the benchmark.
In this figure, we see very close agreement with a small discrepancy in the $x$-direction velocity component which occurs at approximately five days.  

\section{The Variation of Variance} \label{sec:Variation_pdf}
\begin{figure}[h!]
	\centering
	\subfigure[][Deviation of $\eta$ with $\xi_2 \sim U(1.0, 2.0)$.]{\includegraphics[width=0.45\columnwidth]{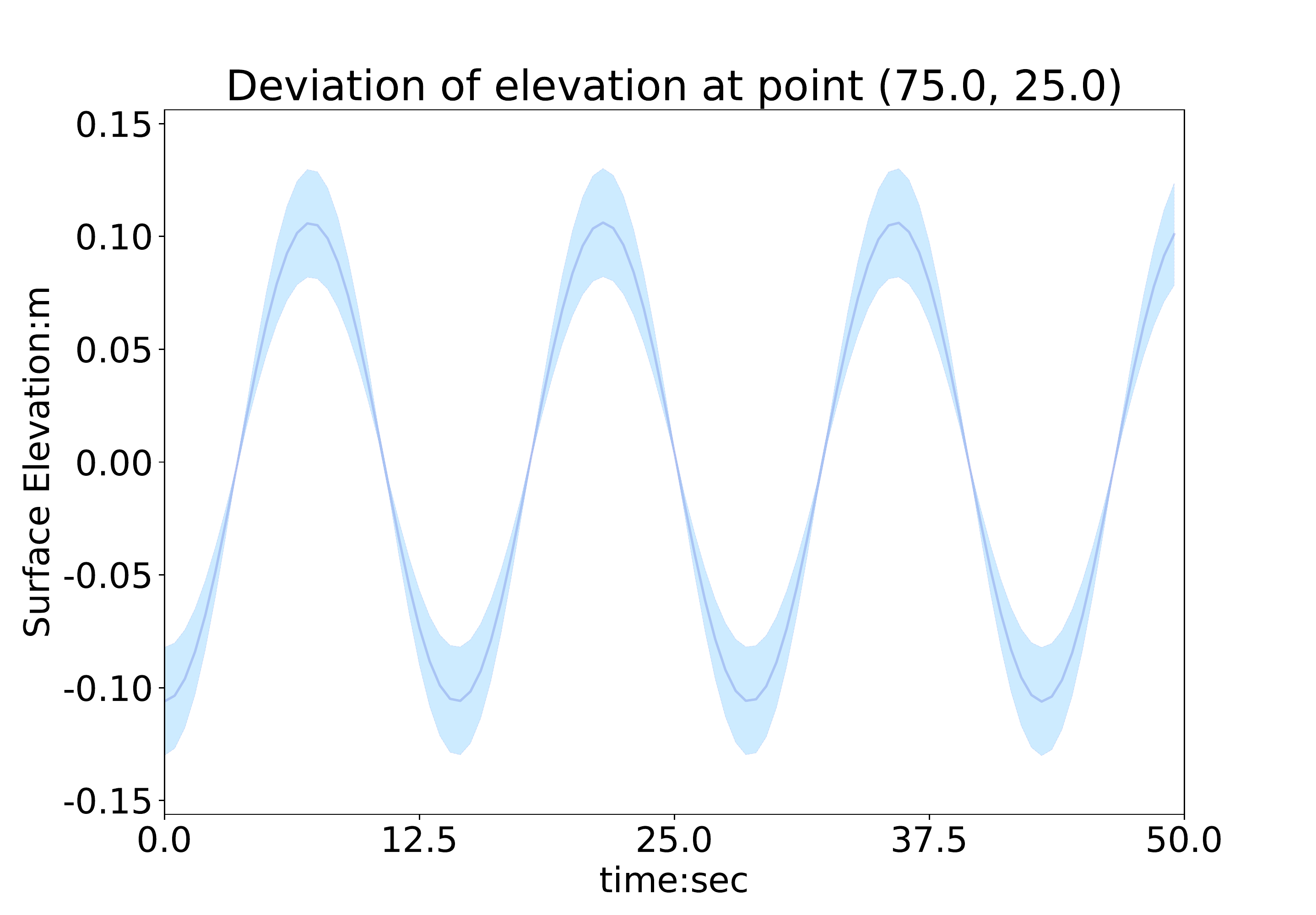}}\hfill
	\subfigure[][Deviation of $u$ with $\xi_2 \sim U(1.0, 2.0)$.]{\includegraphics[width=0.45\columnwidth]{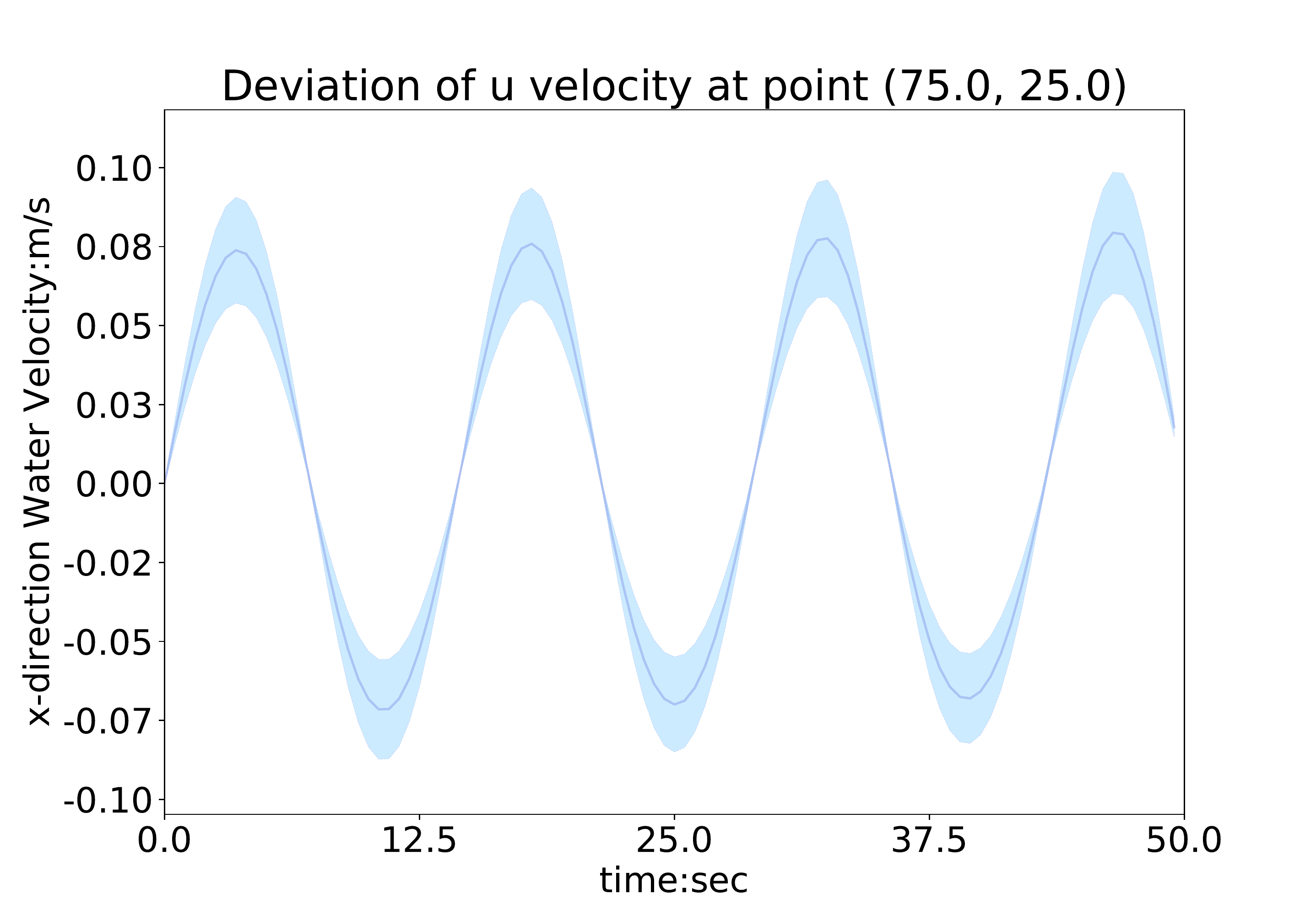}}\hfill
	\subfigure[][Deviation of $\eta$ with $\xi_2 \sim U(0.5, 2.5)$.]{\includegraphics[width=0.45\columnwidth]{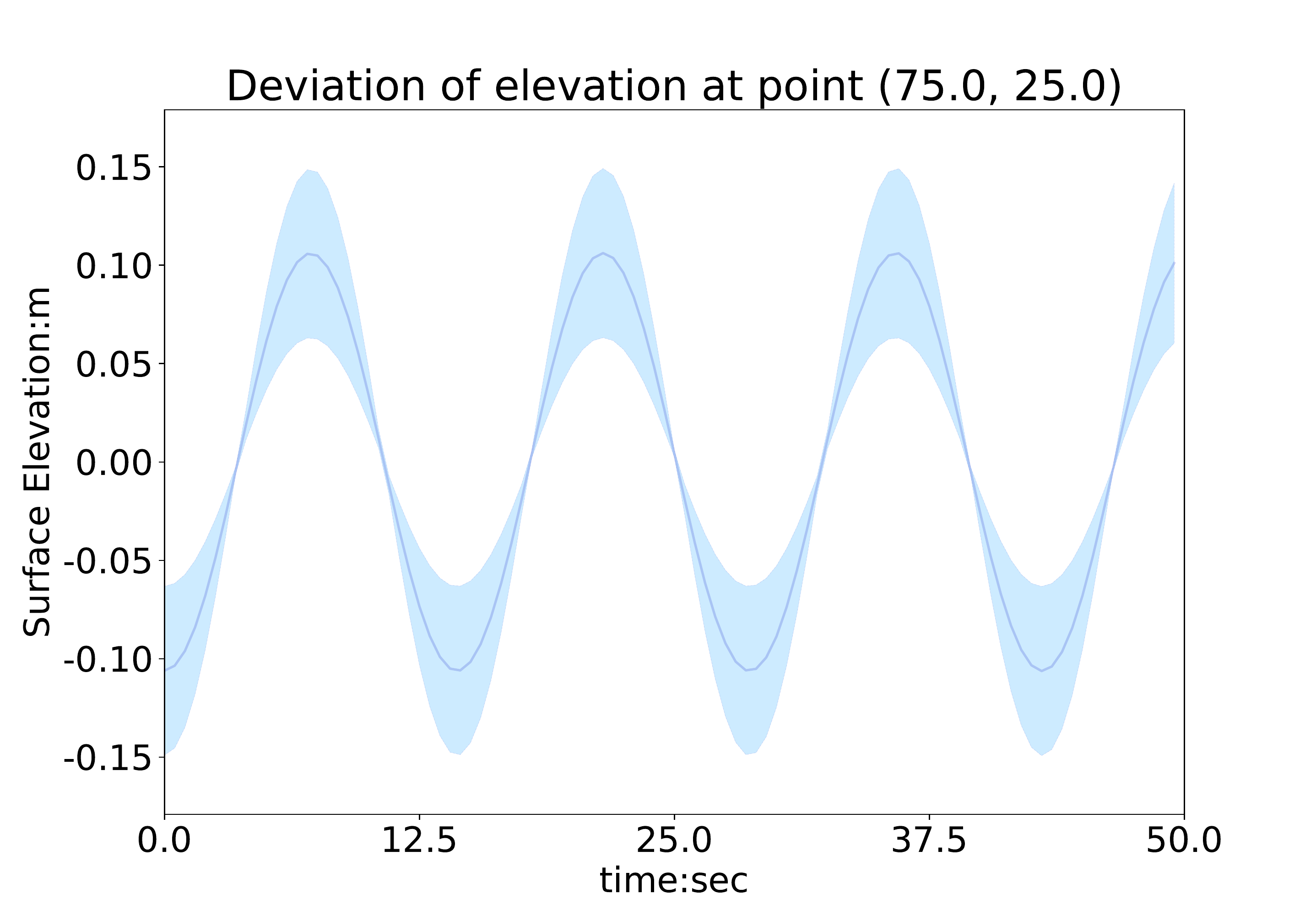}}\hfill
	\subfigure[][Deviation of $u$ with $\xi_2 \sim U(0.5, 2.5)$.]{\includegraphics[width=0.45\columnwidth]{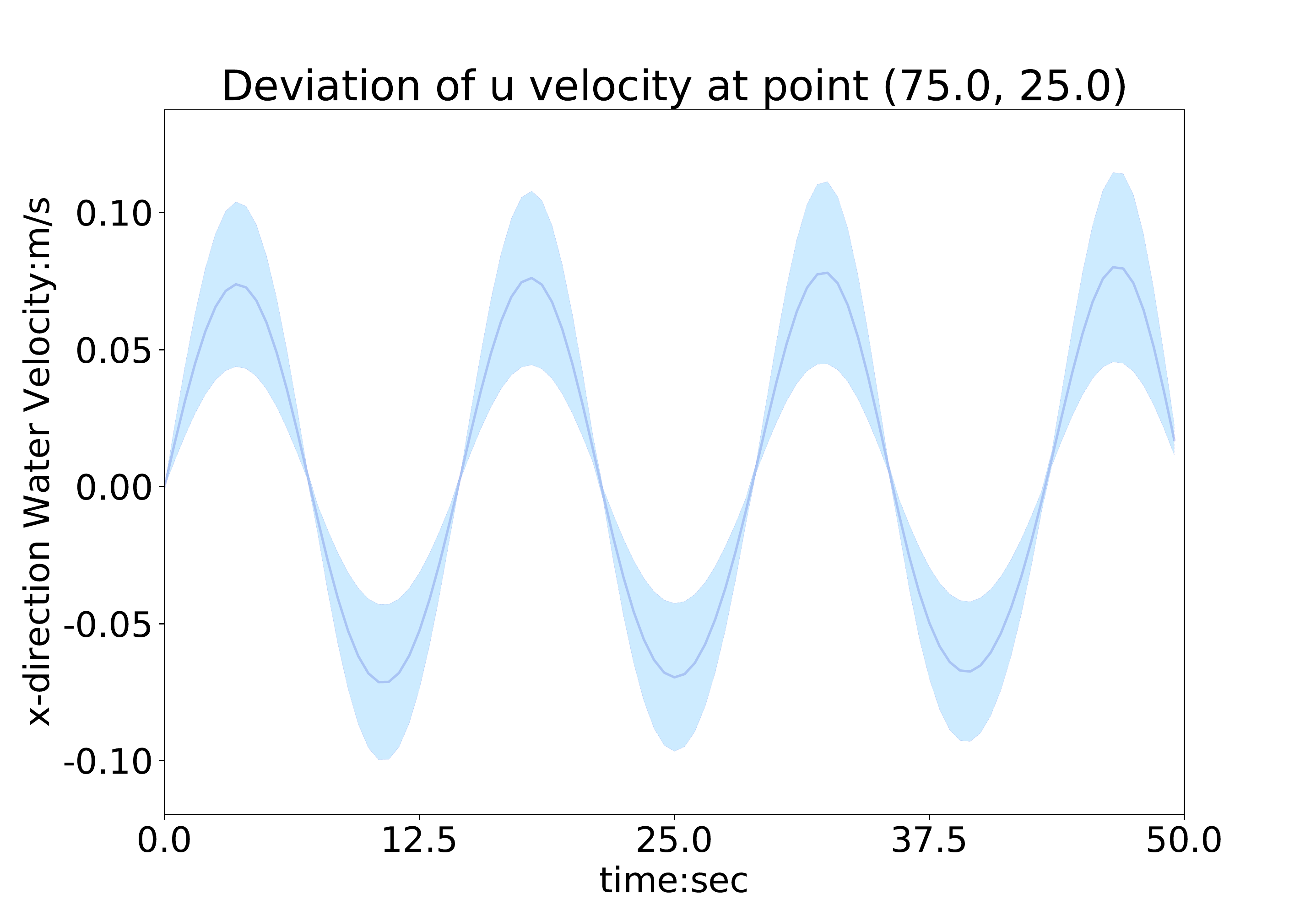}}\hfill
	\subfigure[][Deviation of $\eta$ with $\xi_2 \sim U(0.25, 2.75)$.]{\includegraphics[width=0.45\columnwidth]{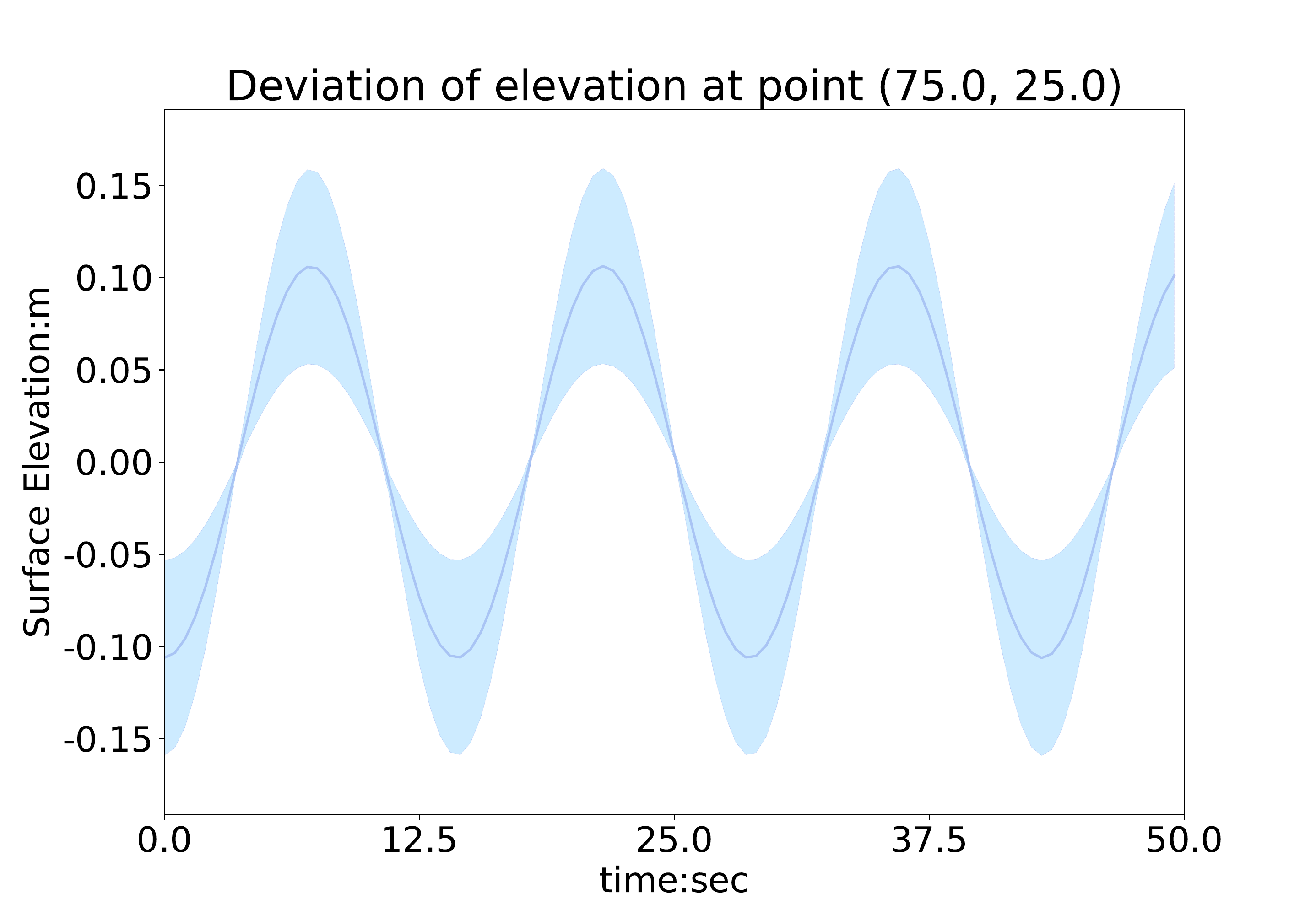}}\hfill
	\subfigure[][Deviation of $u$ with $\xi_2 \sim U(0.25, 2.75)$.]{\includegraphics[width=0.45\columnwidth]{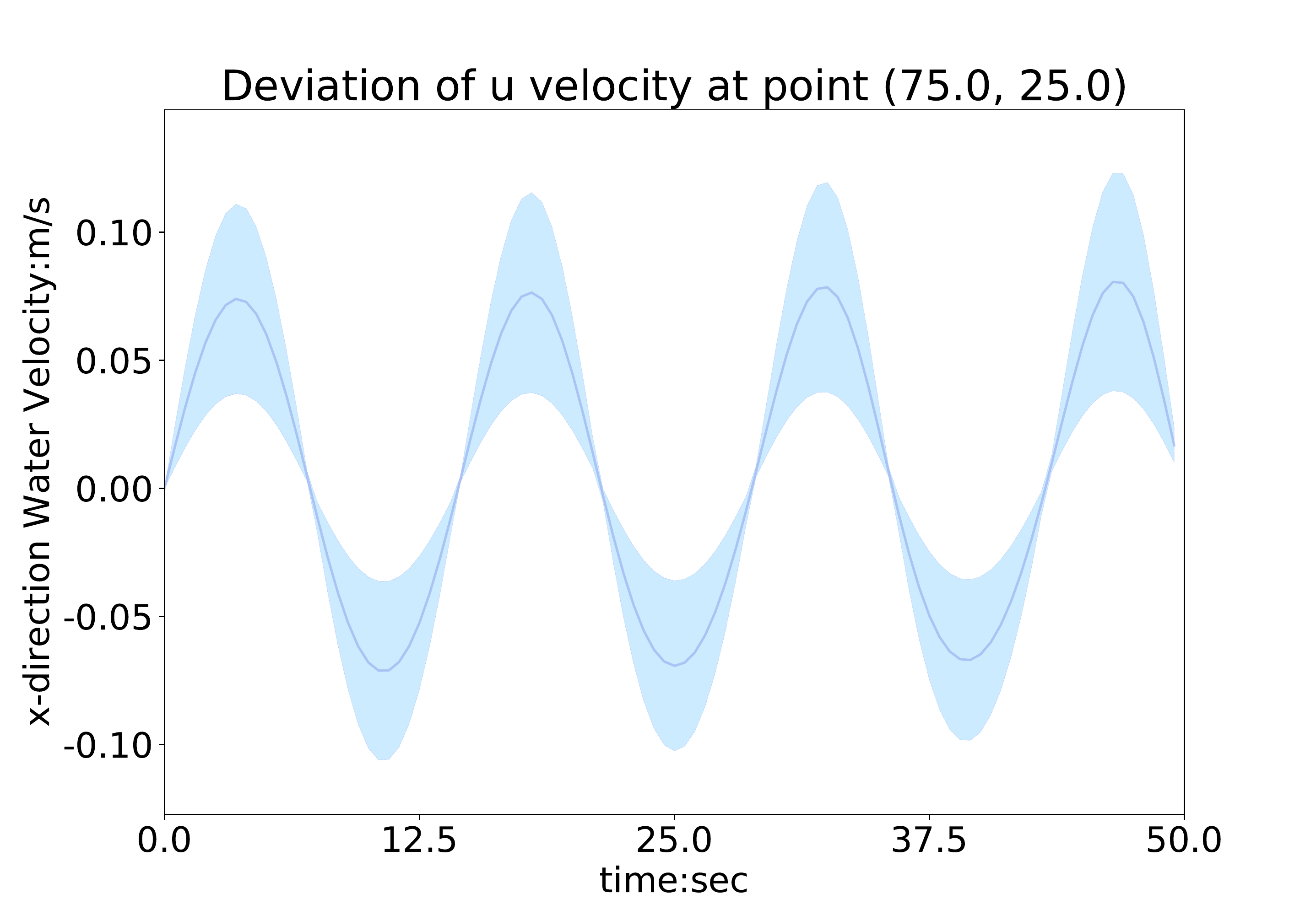}}\hfill
	\caption{Deviation of surface elevation and $x$-direction water velocity at spatial points $(75.0m, 25.0m)$.}
	\label{fig:sloshvariationcompareu}
\end{figure}

To supplement the results presented in Section \ref{sec:var_of_variance}, we \oneR{provide another} spatial point $(75.0m, 25.0m)$ to show the variation of variance in both surface elevation and  $x$-direction component of water velocity in Figure~\ref{fig:sloshvariationcompareu}. In this figure, the blue shaded area corresponds to one standard deviation at that spatial point and the central blue line represents the mean of the model solution.
We observe in Figures~\ref{fig:sloshvariationcompareu}  that the variance in both surface elevation and water velocity increases as the uncertain range of $\xi_2$ extends. Hence, we see further evidence of the conclusion that the variance of output increases as the variance of inputs increase.

\section{The time-varying probability density function} \label{sec:timevar_pdf}

\subsection{Uncertain Initial Condition} \label{sec:sloshtest_PDF}

\begin{figure}[h!]
	\centering
	\subfigure[]{\includegraphics[width=0.475\columnwidth]{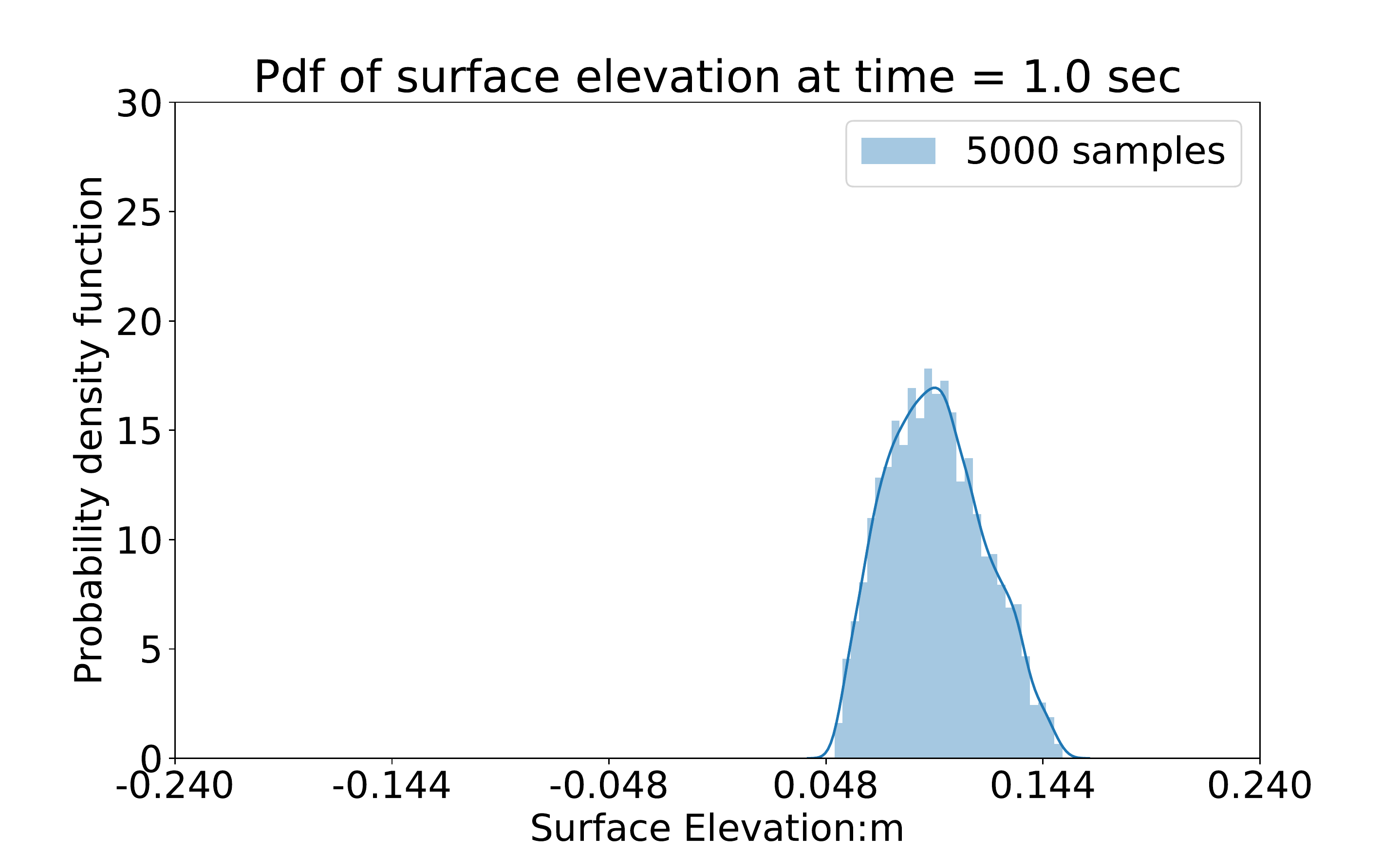}}\hfill
	\subfigure[]{\includegraphics[width=0.475\columnwidth]{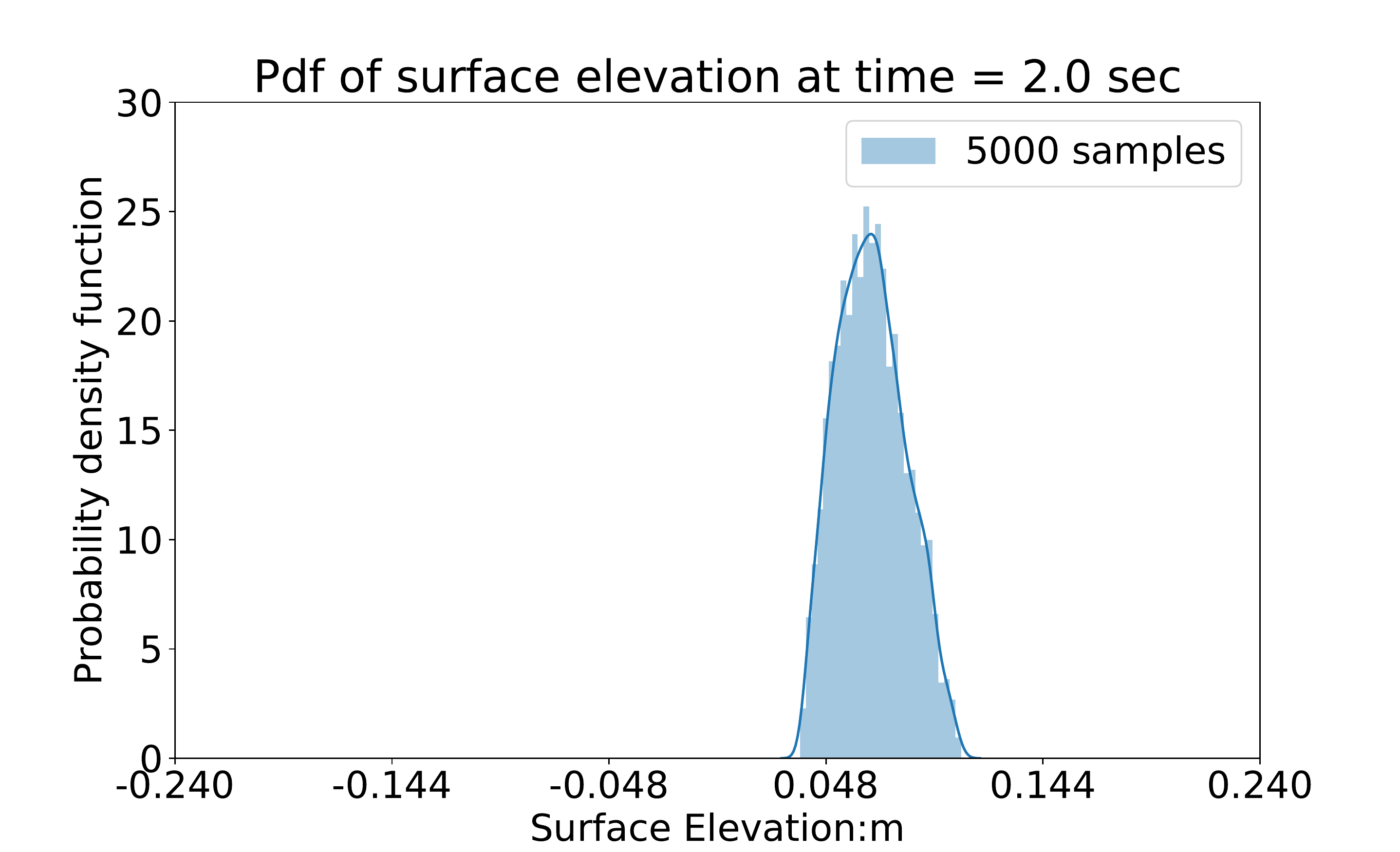}}\hfill
	\subfigure[]{\includegraphics[width=0.475\columnwidth]{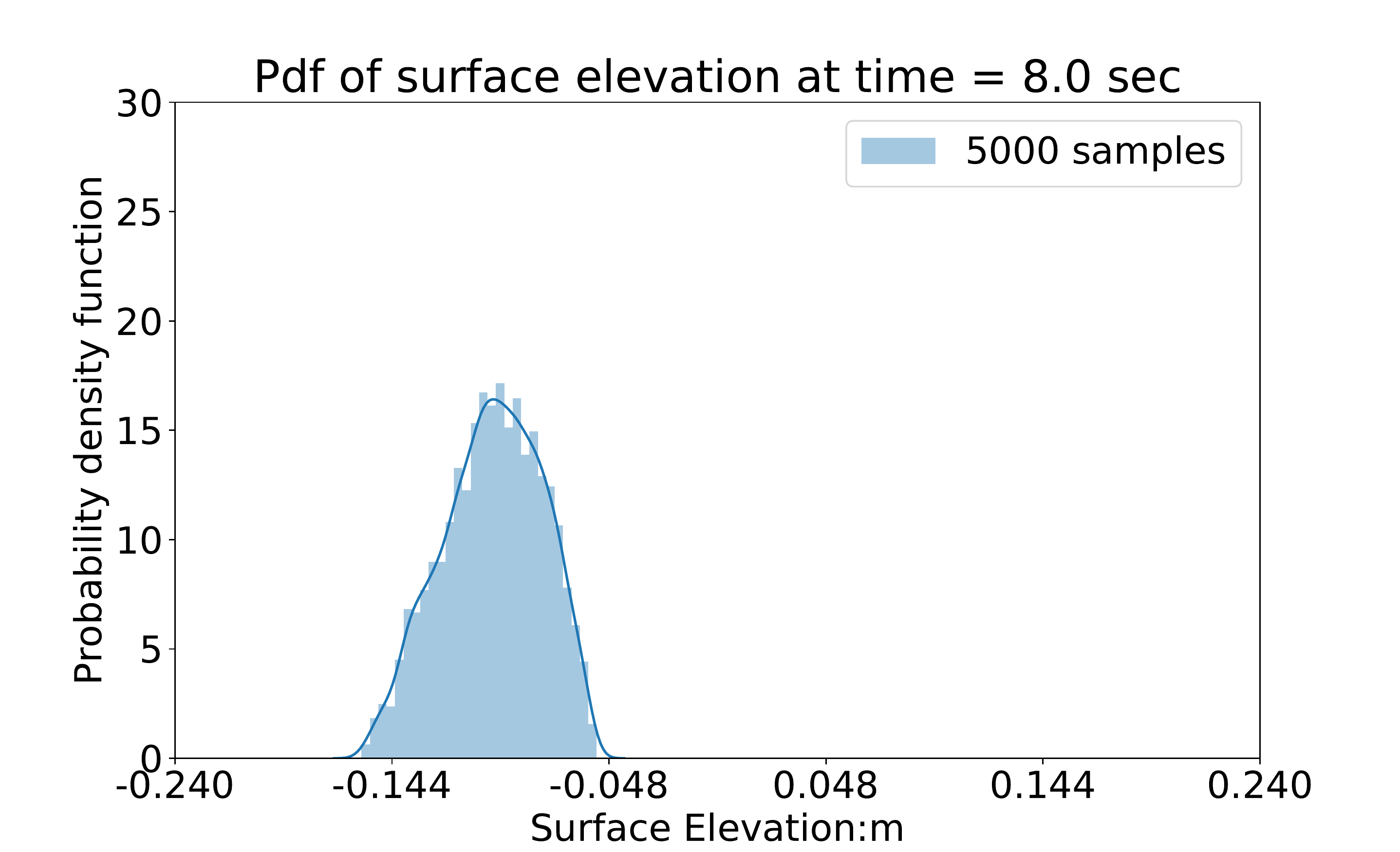}}\hfill
	\subfigure[]{\includegraphics[width=0.475\columnwidth]{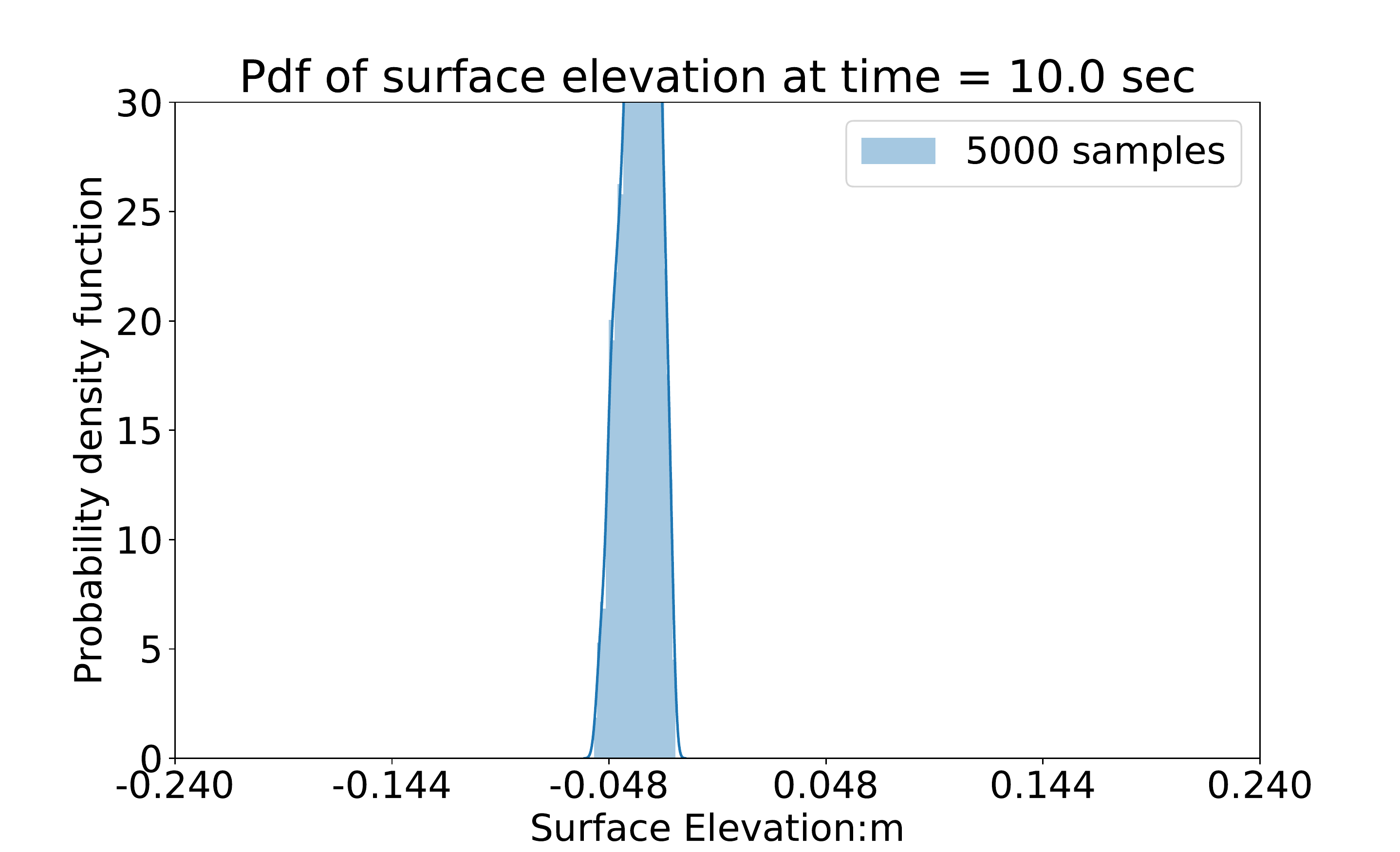}}\hfill
	\caption{PDF of the surface elevation at $(25.0m, 25.0m)$.}
	\label{fig:sloshpdfeta1}
\end{figure}

\begin{figure}[h!]
	\centering
	\subfigure[]{\includegraphics[width=0.475\columnwidth]{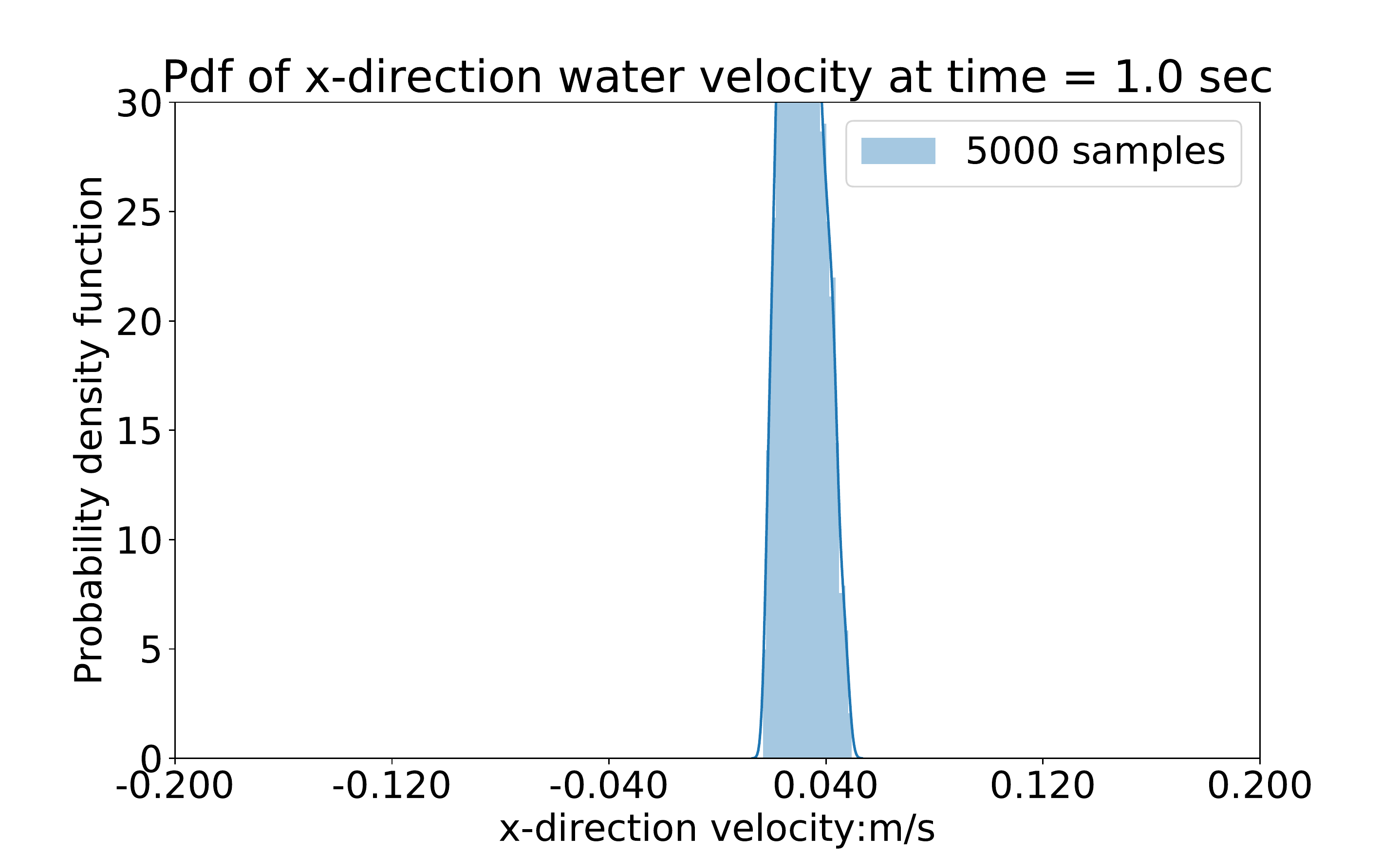}}\hfill
	\subfigure[]{\includegraphics[width=0.475\columnwidth]{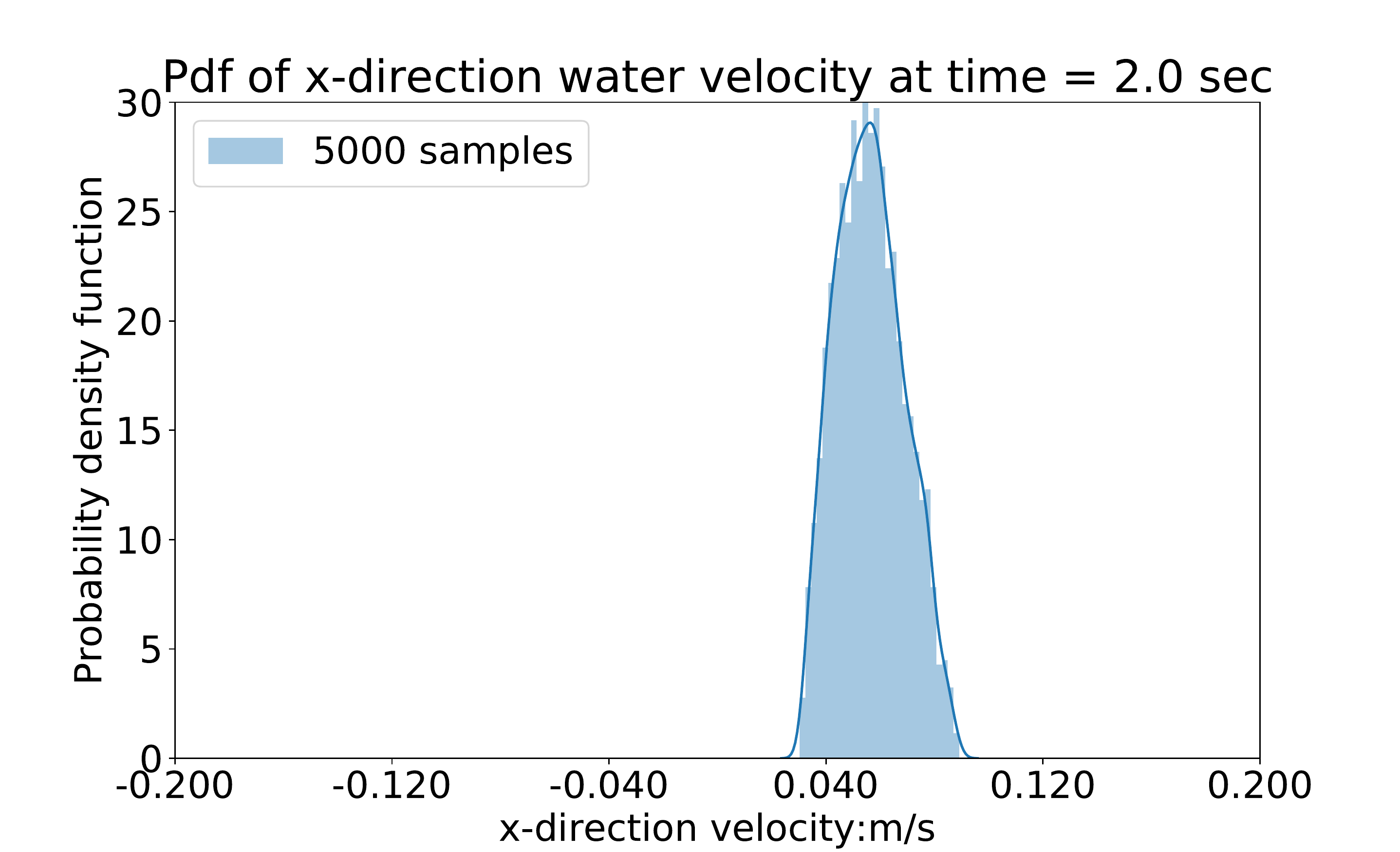}}\hfill
	\subfigure[]{\includegraphics[width=0.475\columnwidth]{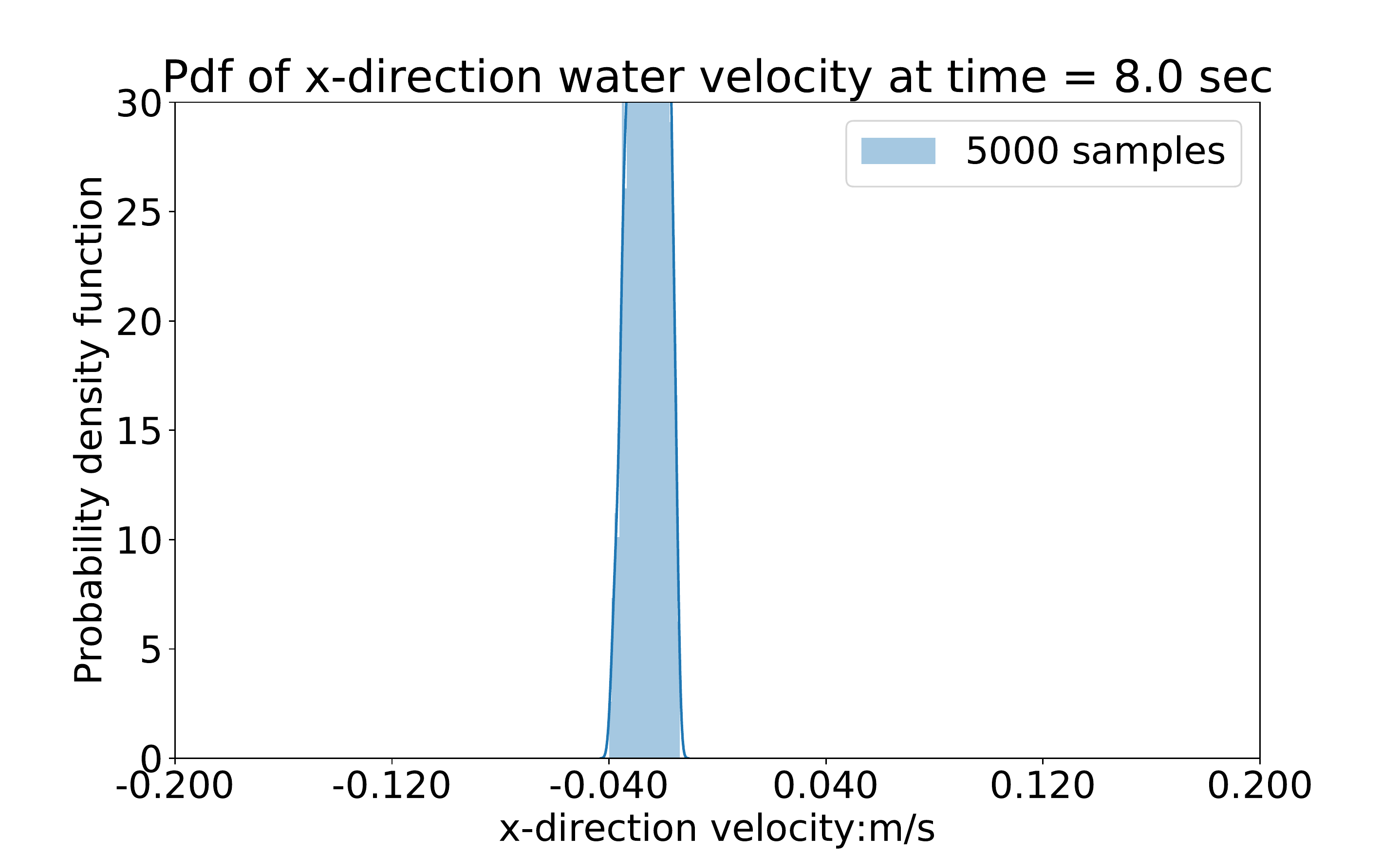}}\hfill
	\subfigure[]{\includegraphics[width=0.475\columnwidth]{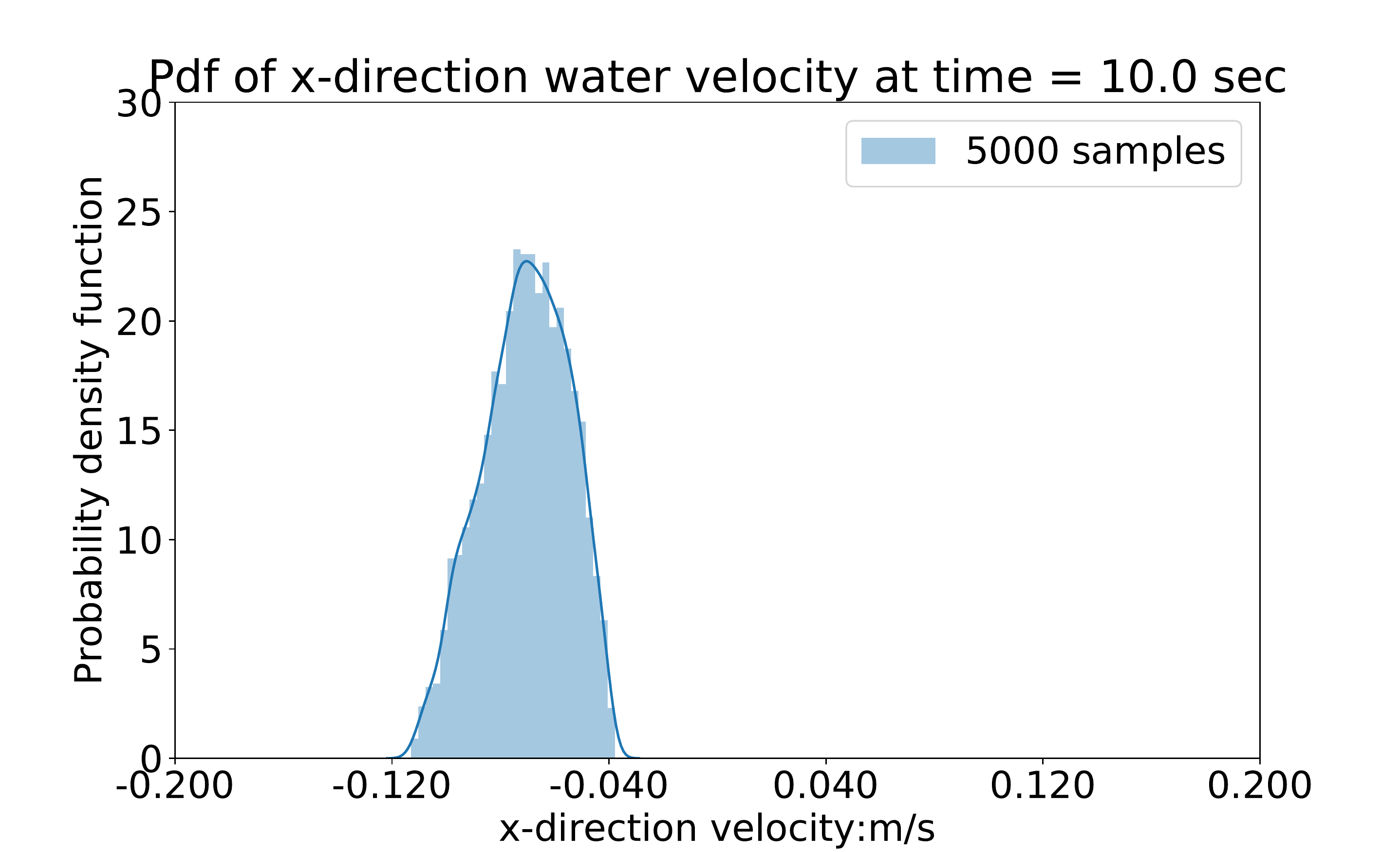}}\hfill
	\caption{PDF of the $x$-direction velocity at $(25.0m, 25.0m)$.}
	\label{fig:sloshpdfu1}
\end{figure}

We again consider the slosh test case with an uncertain initial condition, and select a spatial point $(25.0m, 25.0m)$ at which we visualize its predicted PDFs at four specific times. The PDFs of the surface elevation and the water velocity at that spatial point at the selected time steps are shown in Figures~\ref{fig:sloshpdfeta1} and~\ref{fig:sloshpdfu1}. 
In this case, these figures reveal that the PDF of both the surface elevation and the water velocity \one{are of similar shape and appear to mimic the behaviour of a} log-normal distribution with space-time varying means and variances.

\subsection{Uncertain Boundary Condition} \label{sec:inlettest_PDF}

\begin{figure}[h!]
	\centering
	\subfigure[]{\includegraphics[width=0.5\columnwidth]{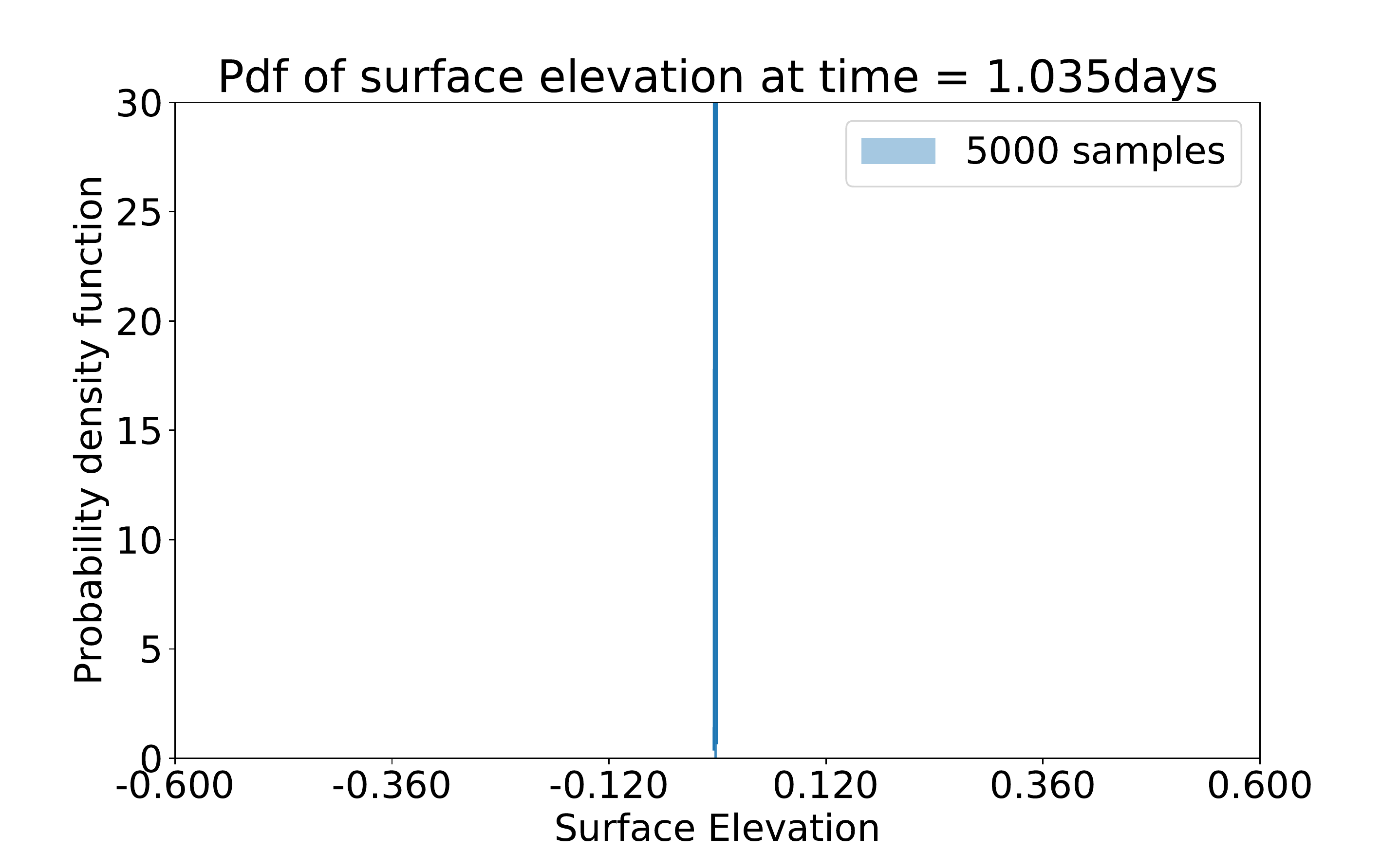}}\hfill
	\subfigure[]{\includegraphics[width=0.5\columnwidth]{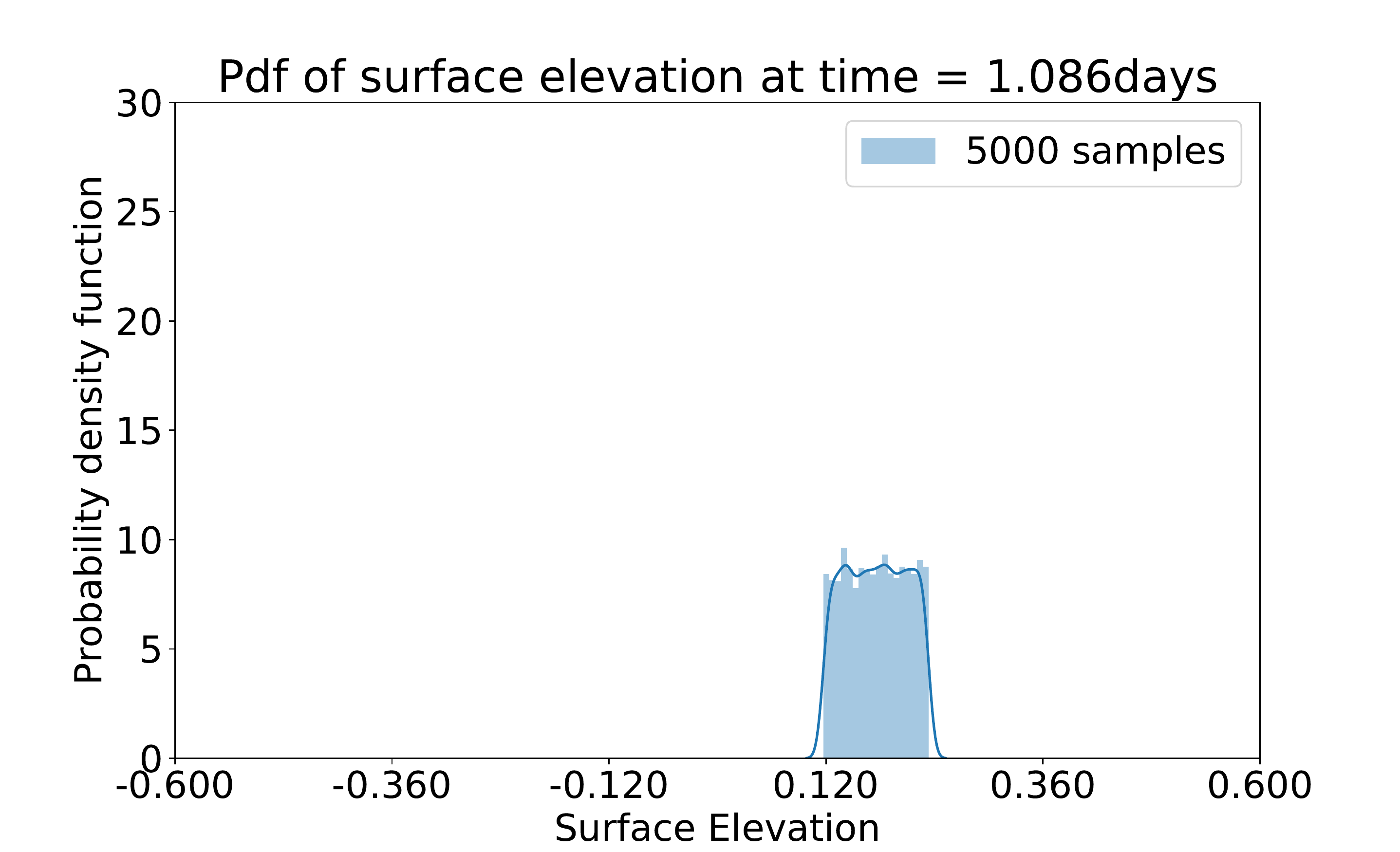}}\hfill
	\subfigure[]{\includegraphics[width=0.5\columnwidth]{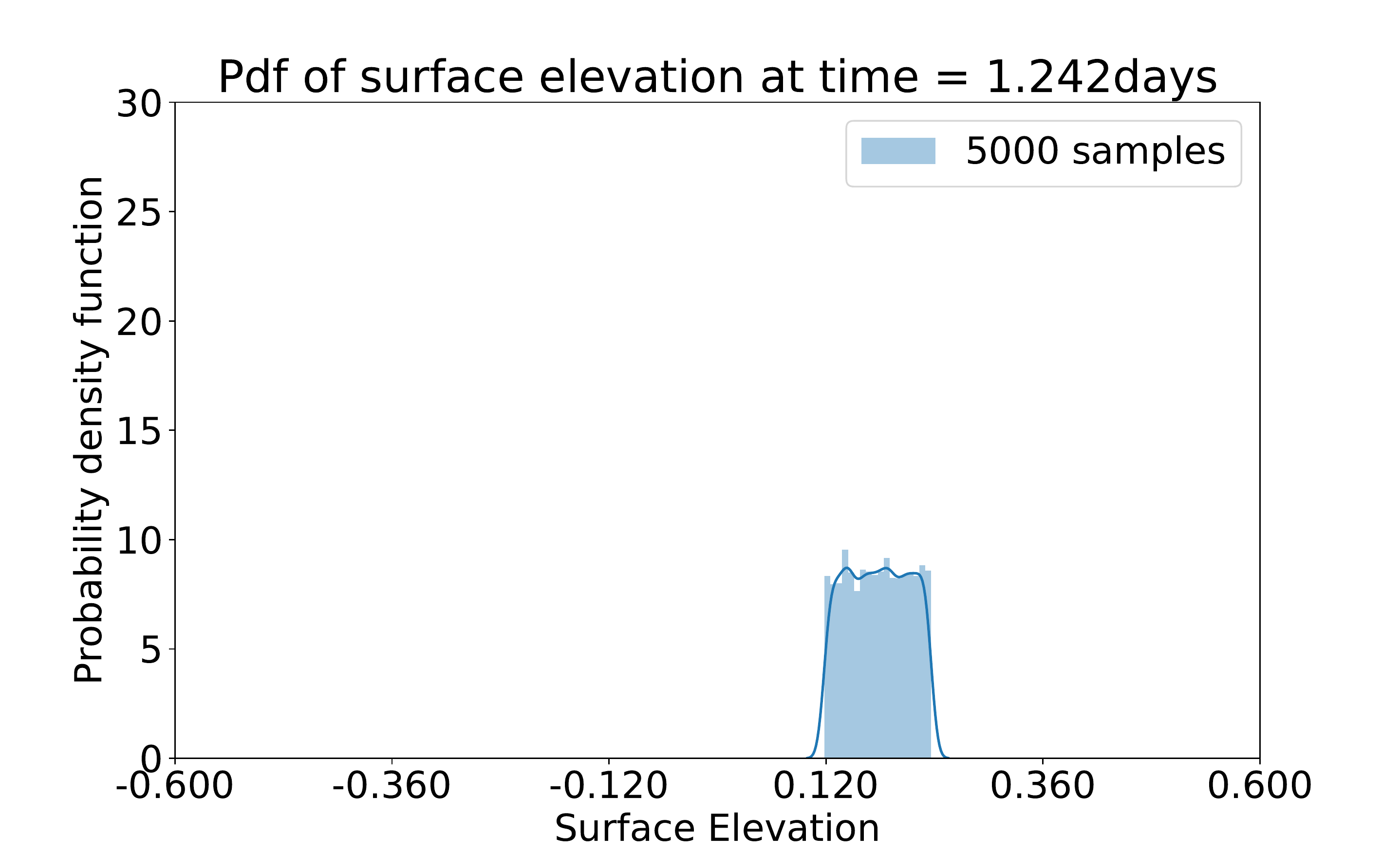}}\hfill
	\subfigure[]{\includegraphics[width=0.5\columnwidth]{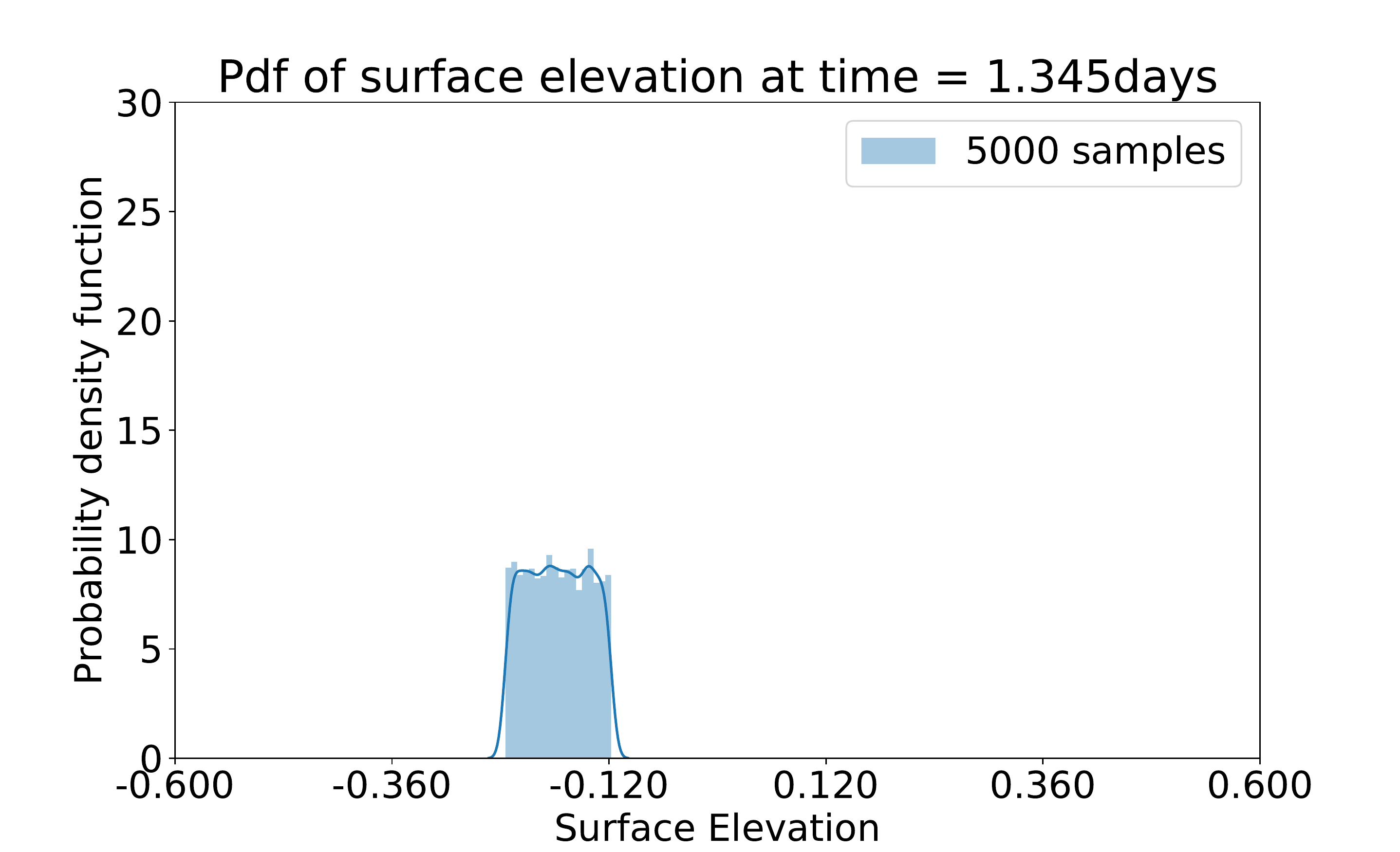}}\hfill
	\subfigure[]{\includegraphics[width=0.5\columnwidth]{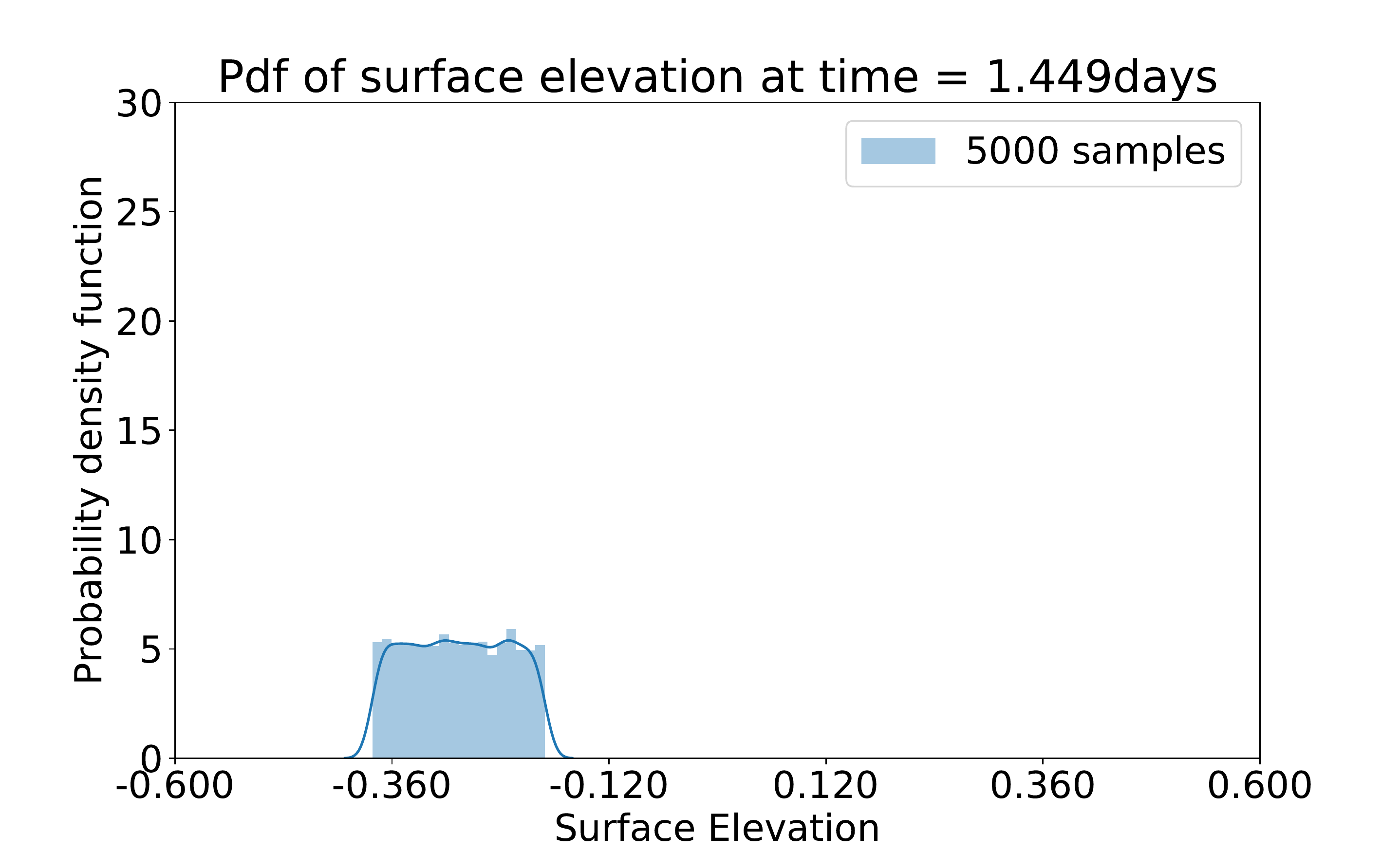}}\hfill
	\subfigure[]{\includegraphics[width=0.5\columnwidth]{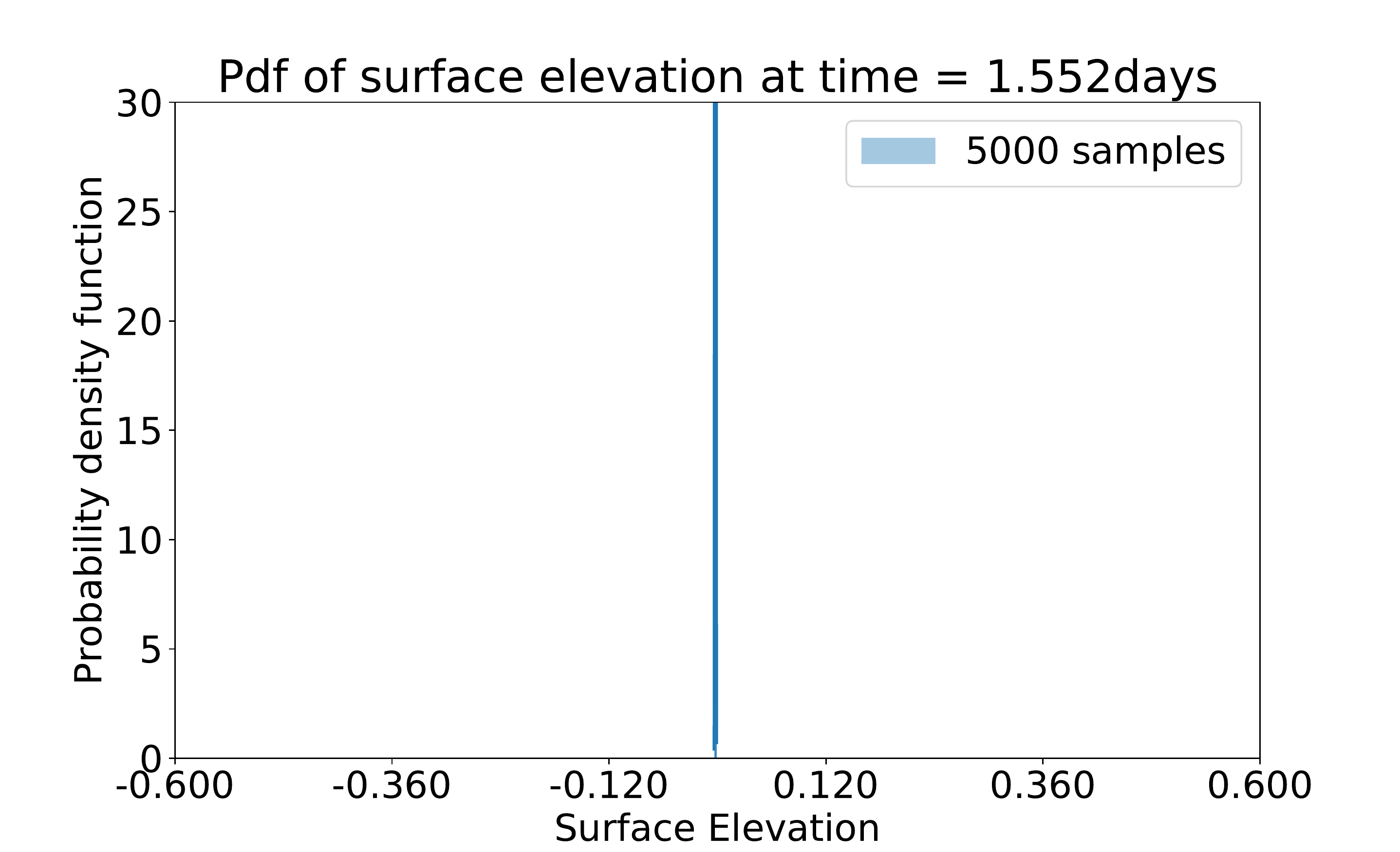}}\hfill
	\caption{Uncertain boundary condition in the inlet test: The probability density of elevation at (0.0m, 0.0m).}
	\label{fig:inletpdfeta}
\end{figure}

\begin{figure}[h!]
	\centering
	\subfigure[]{\includegraphics[width=0.5\columnwidth]{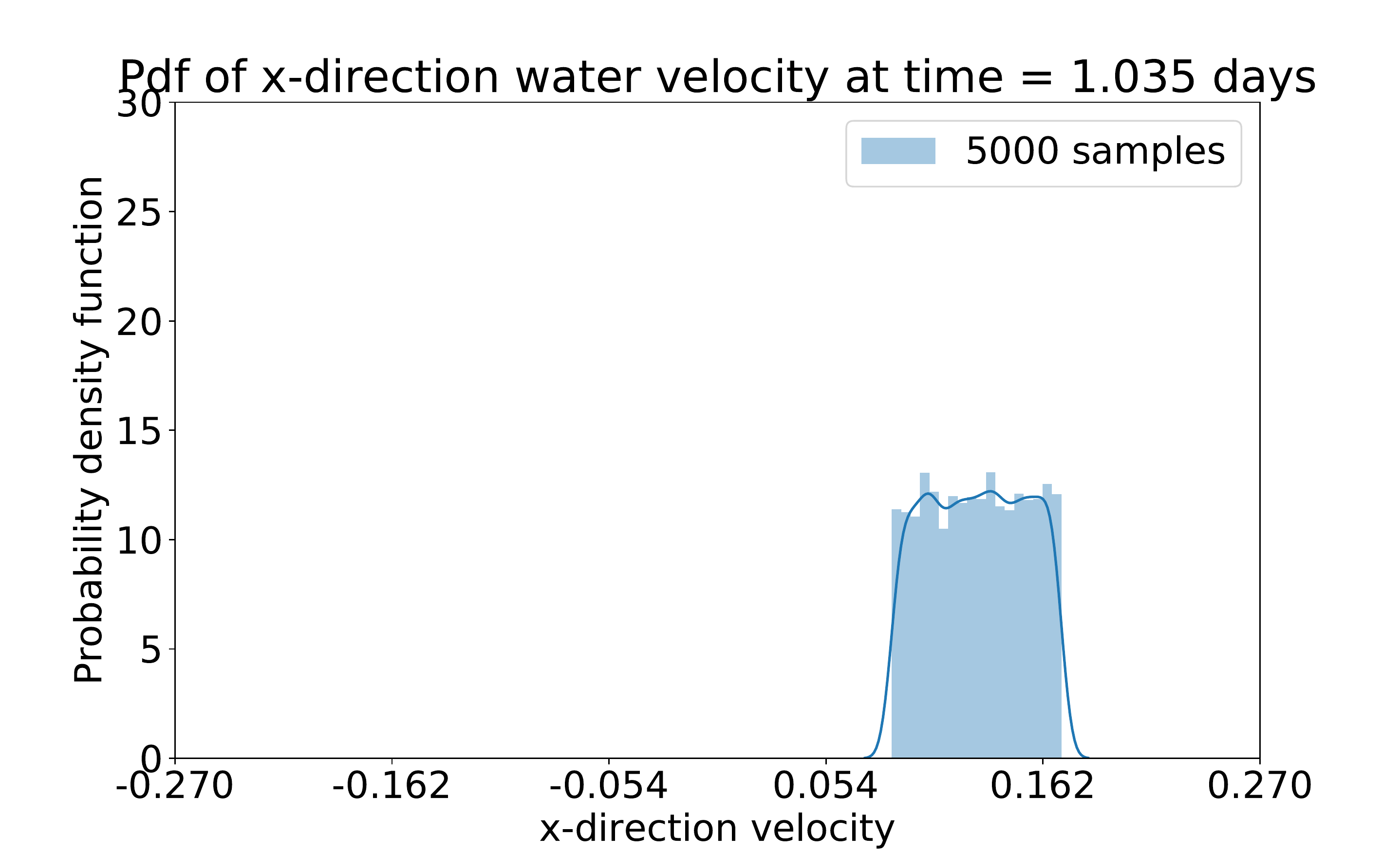}}\hfill
	\subfigure[]{\includegraphics[width=0.5\columnwidth]{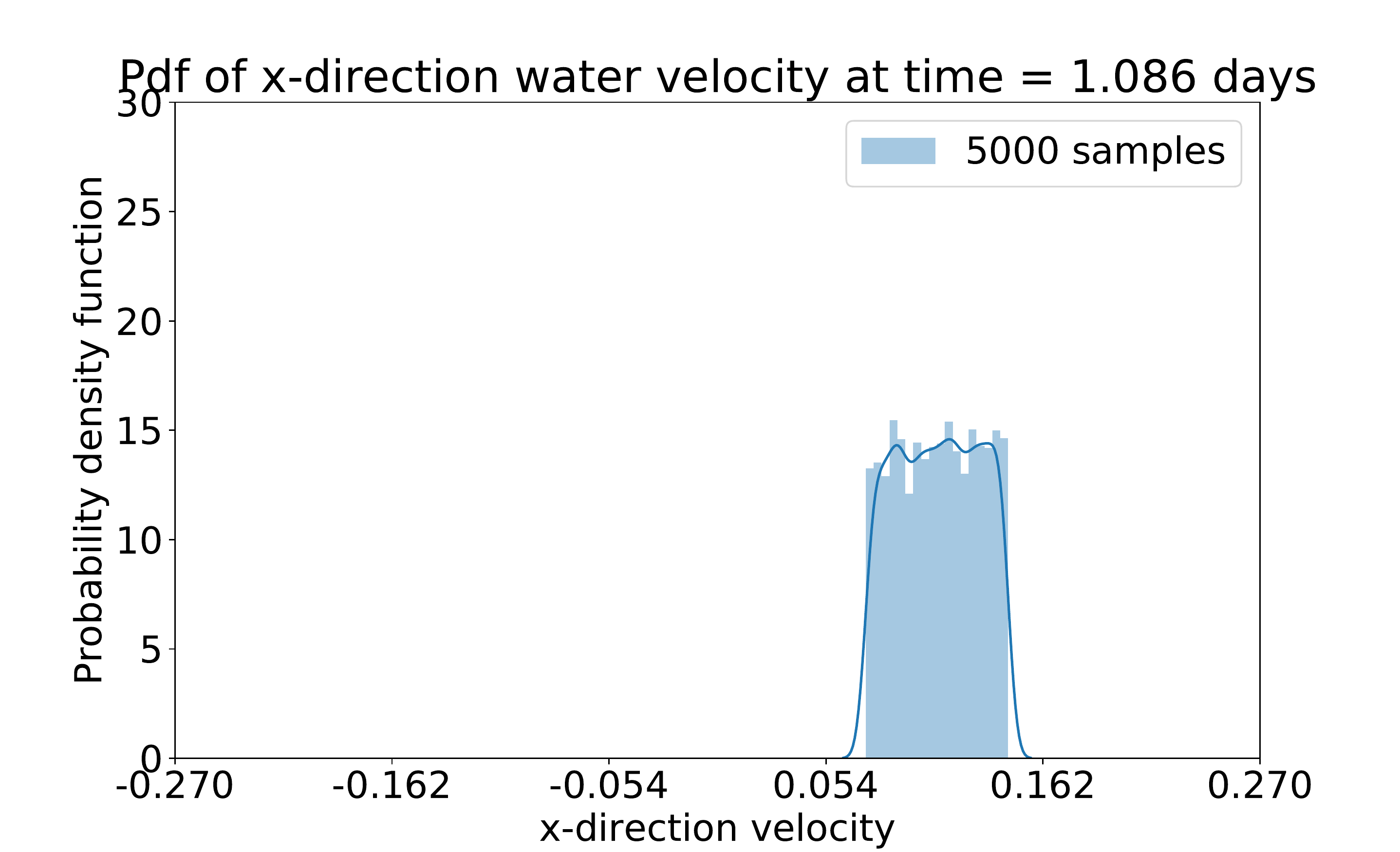}}\hfill
	\subfigure[]{\includegraphics[width=0.5\columnwidth]{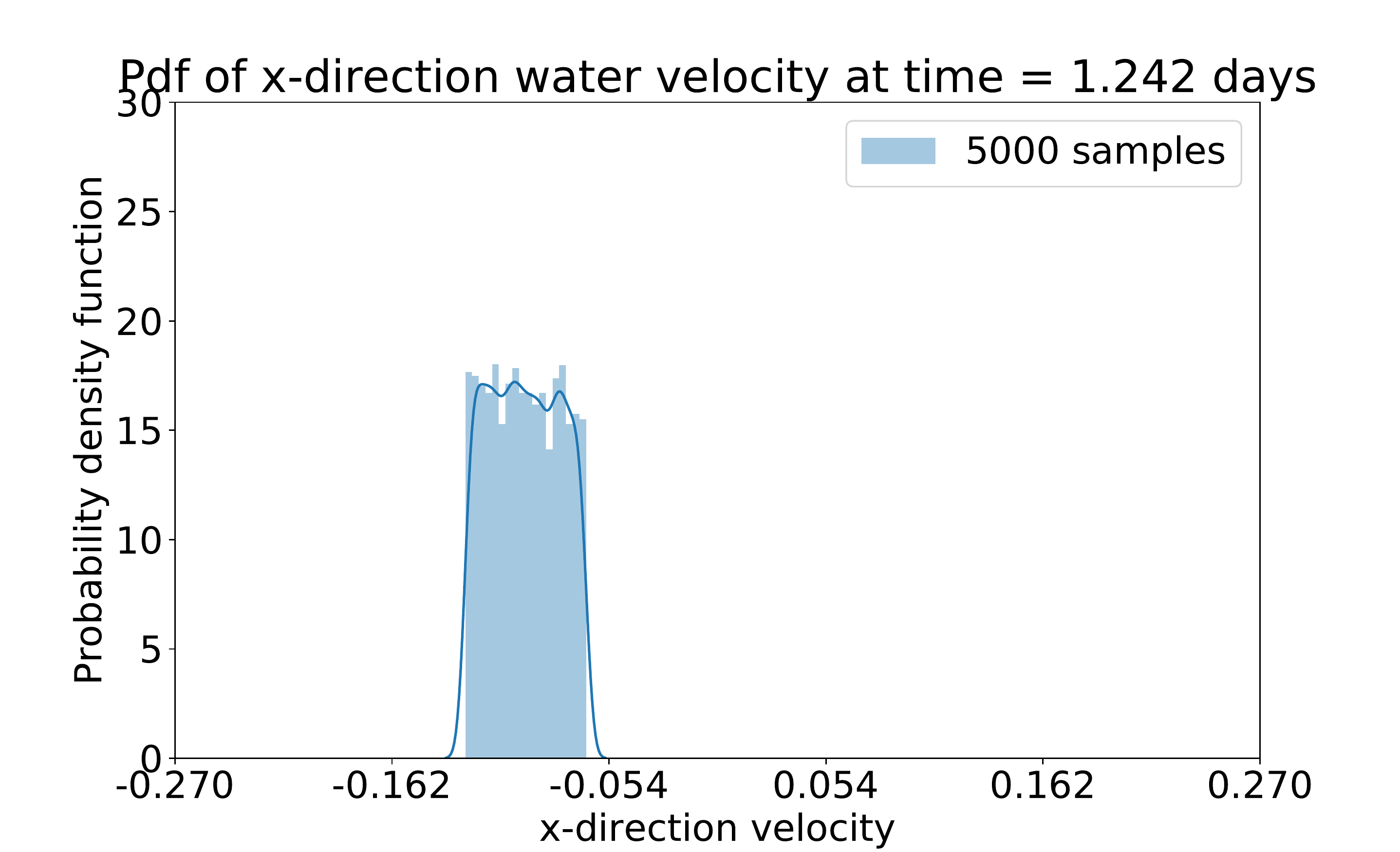}}\hfill
	\subfigure[]{\includegraphics[width=0.5\columnwidth]{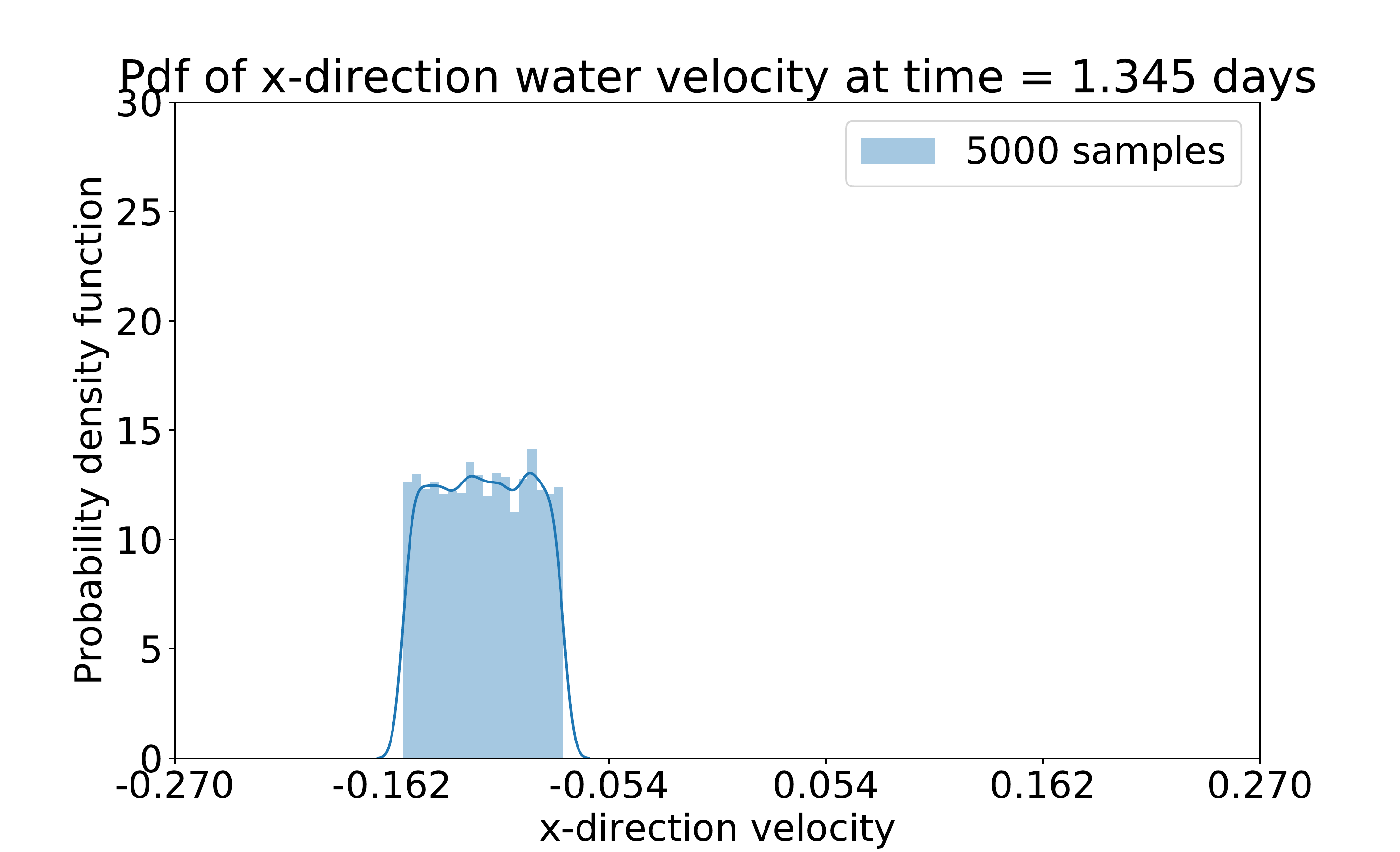}}\hfill
	\subfigure[]{\includegraphics[width=0.5\columnwidth]{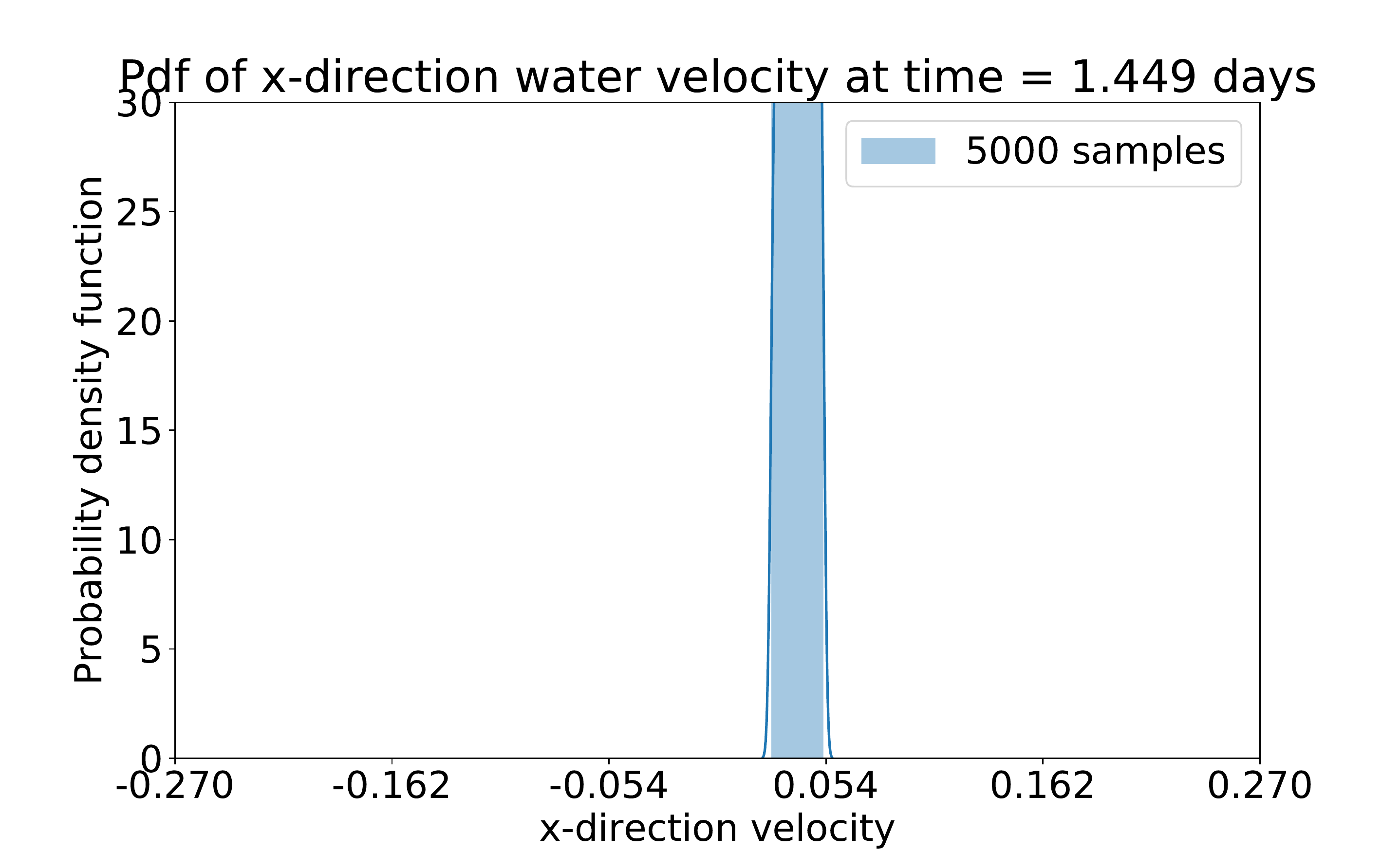}}\hfill
	\subfigure[]{\includegraphics[width=0.5\columnwidth]{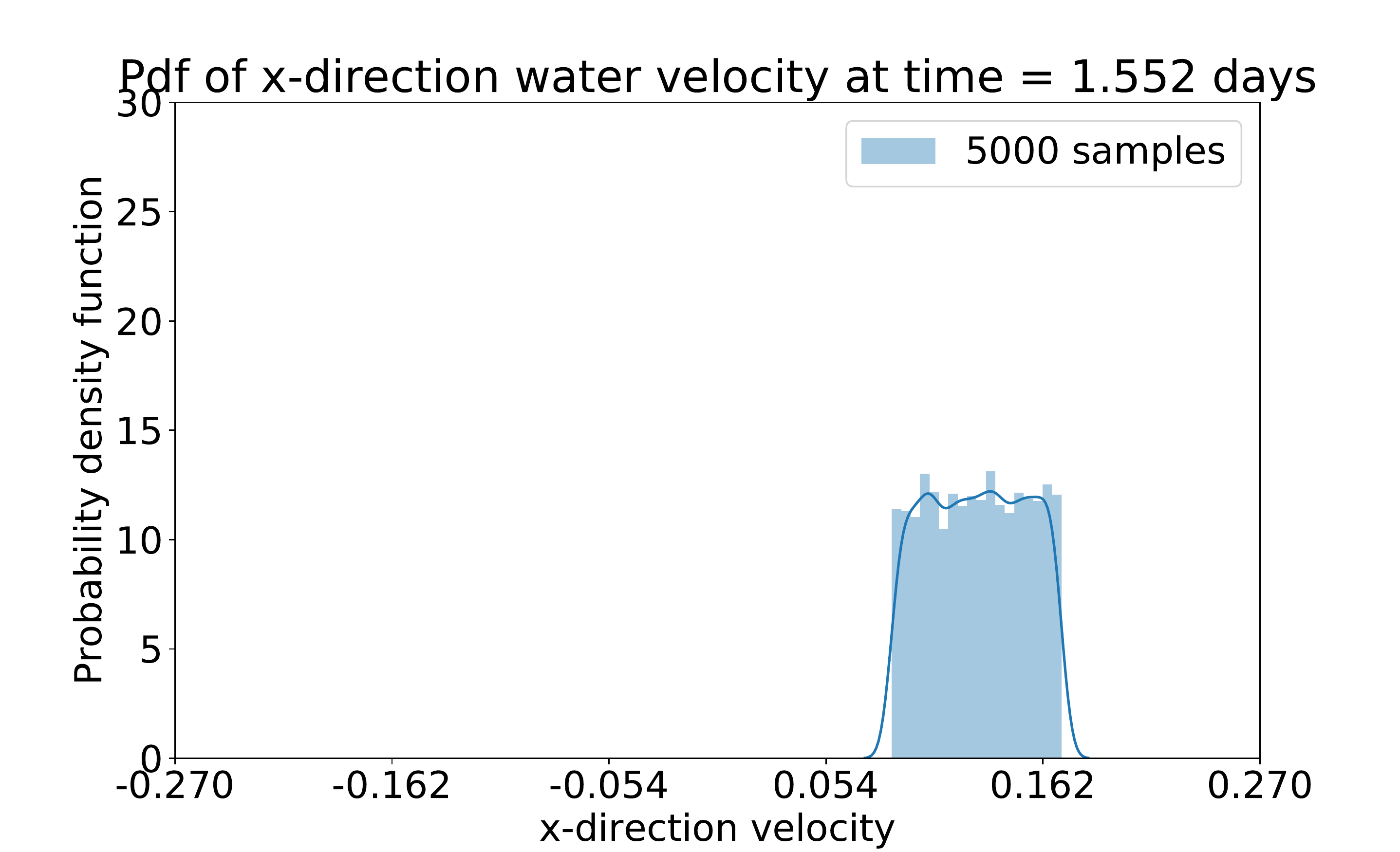}}\hfill
	\caption{Uncertain boundary condition in the inlet test: The probability density of x-direction velocity at (0.0m, 0.0m).}
	\label{fig:inletpdfu}
\end{figure}

In the inlet test case,  the elevation  boundary condition represents the uncertain source. In this test case, we choose a spatial point $(0.0m, 0.0m)$ at the entrance of the inlet in the domain and present the PDFs of the surface elevation and the water velocity at six times in Figures~\ref{fig:inletpdfeta} and~\ref{fig:inletpdfu}. 
In this case, the PDF \one{responses resemble}  uniform distributions for both the surface elevation and the water velocity.
\clearpage

\section{Prediction of Hurricane Storm Surge under Uncertain Wind Drag Coefficient \more{During Hurricane Harvey}} \label{sec:hurricane_prediction2}
\begin{figure}[h!]
	\centering
	\includegraphics[width=0.9\textwidth]{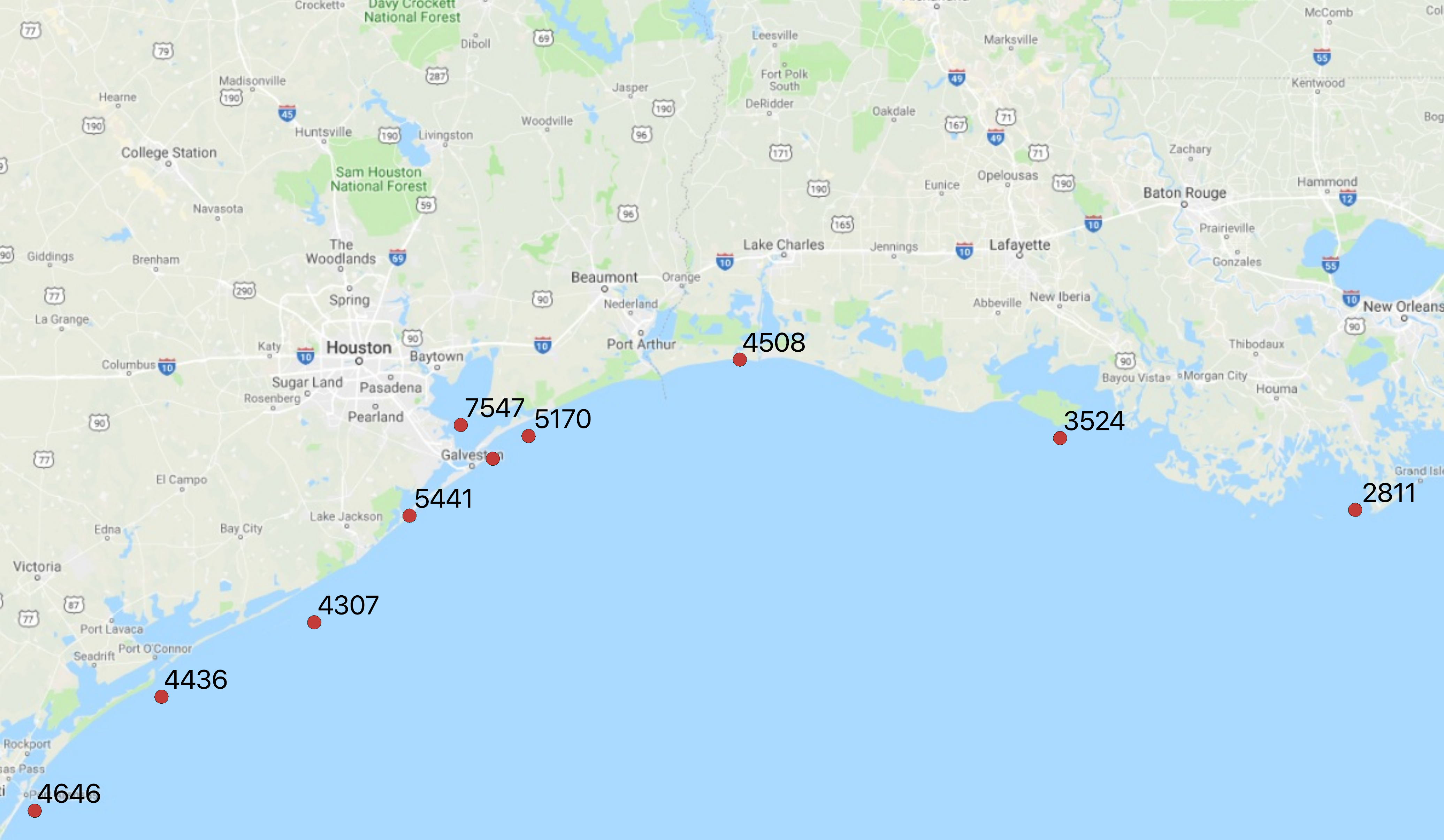}
	\caption{Spatial points for model comparison during Hurricane Harvey 2017.}
	\label{fig:harchoose}
\end{figure}

\begin{figure}[h!]
\includegraphics[width=\textwidth]{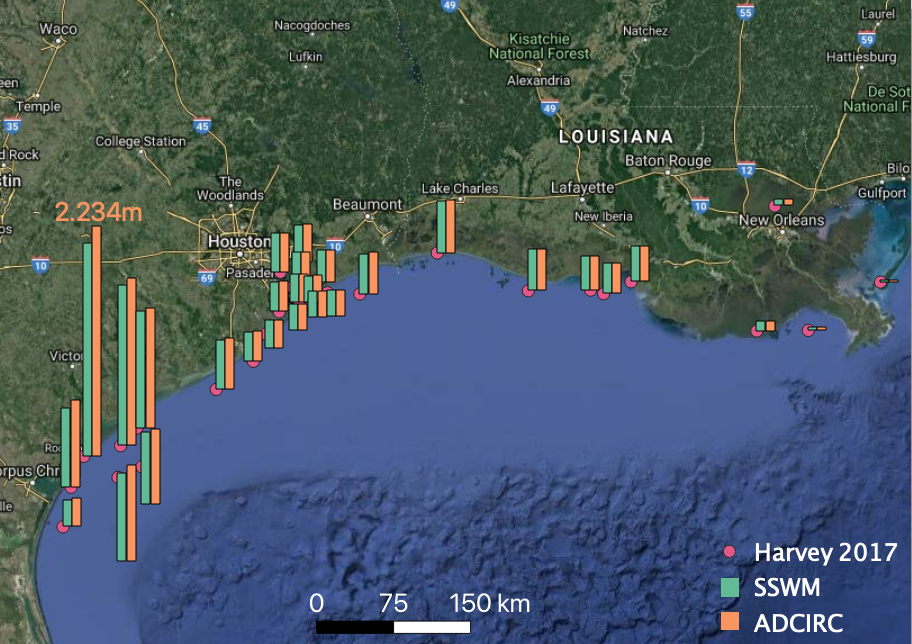}
	\caption{Maximum surface elevation comparison between ADCIRC and SSWM for Hurricane Harvey.}
	\label{fig:harover}
\end{figure}

To again show the effectiveness of the proposed predictor $\mu+\sigma$, 30 spatial points on the Texas and Louisiana coasts for Hurricane  Harvey, see Figure~\ref{fig:harchoose}. In Figure~\ref{fig:harover}, we present the comparison of the maximum surface elevation between ADCIRC and the SSWM proposed predictor. 
These results show that the proposed indicator $\mu+\sigma$ underestimates the maximum surface elevation along the western coast of Texas. Otherwise, close agreement is observed. This again suggests that the proposed predictor $\mu+\sigma$ given by the SSWM is reliable for real-time prediction of maximum surface elevation, under the present condition of a uniformly distributed uncertain wind drag coefficient.


\clearpage
\nocite{*}
\bibliographystyle{elsarticle-num} 
\bibliography{chenbib_CrossMode}


\end{document}